\documentclass{article}
 \usepackage{JOEconfig}

\usepackage[T1]{fontenc}

\usepackage{graphicx}

\usepackage{hyperref}
\hypersetup{
    linktocpage,  
    colorlinks,
    citecolor=black,
    filecolor=blue,
    linkcolor=blue,
    urlcolor=blue
}

\usepackage{accents} 
\usepackage{bm} 

\usepackage{amsmath, amsthm} 
\usepackage{array}
\usepackage{longtable,tabularx} 
\usepackage{enumitem} 
\setlist[itemize,1]{left=3ex,labelsep=0.75ex,itemsep=0pt,topsep=1pt,label={\footnotesize{\textbullet}}, after={\vspace{1ex}}}
\setlist[itemize,2]{itemsep=0pt,topsep=0pt}
\setlist[enumerate,1]{left=3ex,labelsep=0.75ex,itemsep=0pt,topsep=1pt}
\setlist[enumerate,2]{itemsep=0pt,topsep=0pt}

\usepackage{stackengine}
\usepackage{xfrac}
\usepackage{tensor}
\usepackage{mathtools} 
\usepackage{bigints}
\usepackage[svgnames]{xcolor}
\usepackage{cancel} 

\usepackage{cleveref} 

 \usepackage[slantedGreek,libertine,vvarbb]{newtx}
    
 \DeclareRobustCommand{\textsb}[1]{\small{\textbf{#1}}} 

\newsavebox\fooboxx

\usepackage[frak=euler,scr=cm]{mathalpha} 

\usepackage{upgreek} 

\usepackage[artemisia]{textgreek}

\DeclareMathSizes{10}{9.5}{7}{5.5} 
\DeclareMathSizes{9}{9.5}{7}{5}
 \DeclareMathSizes{8}{8}{6}{5}
 \DeclareMathSizes{7}{7}{5}{4}
 
\title{\vspace{-1.0cm}Orbit Elements from Kepler Solutions in Projective Coordinates} 
\author{Joseph T.A. Peterson\footnote{Department of Aerospace Engineering, Texas A\&M University, College Station, Texas.},
Manoranjan Majji\footnote{Edward "Pete" Aldridge Endowed Professor of Aerospace Engineering, Director of LASR Laboratory, Texas A\&M University, College Station, Texas.}, and
John L.~Junkins\footnote{Distinguished Professor of Aerospace Engineering, Director of The Hagler Institute for Advanced Study, Texas A\&M University, College Station, Texas.}
}

\begin{document}
\date{}
\maketitle



\newcommand{\pgrf}[1]{\noindent\textbf{#1}}
\newcommand\blfootnote[1]{%
  \begingroup
  \renewcommand\thefootnote{}\footnote{#1}%
  \addtocounter{footnote}{-1}%
  \endgroup
}

\newcommand{\cf}[0]{f}%
\newcommand{\cff}[0]{{\sl{f}}}

\newcommand{\tx}[1]{{\text{#1}}}
\newcommand{\txi}[1]{{\textnormal{\textit{#1}}}}
\newcommand{\tbf}[1]{{\textbf{\textup{\textrm{#1}}}}}
\newcommand{\trm}[1]{{\textup{\textrm{#1}}}}
\newcommand{\txnonb}[1]{{\textmd{#1}}} 
\newcommand{\txup}[1]{{\textup{#1}}} 
\newcommand{\bfi}[1]{{\textbf{\textit{#1}}}}
\newcommand{\Bfi}[1]{{ \scalebox{1.18}{\textbf{\textit{#1}}}  }}

\newcommand{\rmsb}[1]{{\textup{\textsb{#1}}}}%
\newcommand{\itsb}[1]{{\textit{\textsb{#1}}}}

\renewcommand{\sl}[1]{{\textnormal{\textrm{\textsl{#1}}}}}
\newcommand{\slb}[1]{{\textbf{\textsl{#1}}}}

\newcommand{\txsfb}[1]{\textbf{\textit{\textsf{#1}}}}
\newcommand{\sfi}[1]{{\textit{\textsf{#1}}}}%
\newcommand{\sfup}[1]{{\textup{\textsf{#1}}}}
\newcommand{\sfbup}[1]{ {\textbf{\textup{\textsf{#1}}}} }%

\renewcommand{\sc}[1]{{\textsc{#1}}}
\newcommand*{\rmsc}[1]{{\textrm{\textup{\textsc{#1}}}}}
\newcommand*{\bsc}[1]{{\textbf{\textsc{#1}}}}
\newcommand*{\rmbsc}[1]{{\textbf{\textrm{\textsc{#1}}}}}
\newcommand*{\itsc}[1]{{\textit{\textrm{\textsc{#1}}}}}
\newcommand*{\bisc}[1]{{\textit{\textbf{\textsc{#1}}}}}
\newcommand*{\sfsc}[1]{\textsf{\textit{\textsc{#1}}}}
\newcommand*{\sfbsc}[1]{\textbf{\textsf{\textit{\textsc{#1}}}}}
\newcommand{\scg}[0]{{\trm{\textsc{g}}}}
\newcommand{\sck}[0]{{\trm{\textsc{k}}}}
\newcommand{\scm}[0]{{\trm{\textsc{m}}}}
\newcommand{\scb}[0]{{\trm{\textsc{b}}}}
\newcommand{\kconst}[0]{\sl{k}}

\newcommand{\isc}[1]{{\scriptstyle{\textit{\textrm{\textsc{#1}}}}}} 
\newcommand{\ssc}[1]{{\scalebox{0.6}{$#1$}}} 

\newcommand{\bemph}[1]{\itsb{#1}} 
 \newcommand{\sbemph}[1]{\rmsb{#1}} 


\newcommand{\smsize}[1]{ {\text{\small{#1}}} }
\newcommand*{\fnsize}[1]{ {\text{\footnotesize{#1}}} }%
\newcommand{\scrsize}[1]{ {\text{\scriptsize{#1}}} }
\newcommand{\nssize}[1]{ {\text{\normalsize{#1}}} }

\newcommand*{\nmsz}[1]{{\displaystyle{#1}}}
\newcommand*{\scrsz}[1]{{\scriptstyle{#1}}}
\newcommand*{\ssz}[1]{{\scriptstyle{#1}}}
\newcommand*{\ii}[1]{{\scriptscriptstyle{#1}}} 
\renewcommand*{\ss}[1]{ {\scalebox{0.7}{$#1$}} }

\newcommand*{\smsz}[1]{\text{\small{$#1$}} }
\newcommand*{\fnsz}[1]{\text{\footnotesize{$#1$}} }
\newcommand*{\ns}[1]{{\scalebox{1.15}{$#1$}} }
\renewcommand*{\lg}[1]{{\scalebox{1.25}{$#1$}} }

\newcommand{\mscale}[2][.7]{ {\text{\scalebox{#1}{$#2$}}} } 

\newcommand{\ttfrac}[2]{{\scriptstyle{\frac{\bs{#1}}{\bs{#2}}}}}


\newcommand{\til}[1]{\tilde{#1}}
\newcommand{\wt}[1]{ \widetilde{#1}  }
\newcommand{\wh}[1]{\widehat{#1}}
\newcommand{\ol}[1]{\overline{#1}}

\newcommand{\uln}[1]{{\underline{\smash{#1}\mkern-1mu}{\mkern1mu}}}
\newcommand{\ubar}[1]{ \underbar{$#1\hspace{-0.15mm}$}\hspace{0.15mm} }
\newcommand{\ub}[1]{{\underaccent{\text{—}}{\smash{#1}}}} 

\newcommand{\dt}[1]{\accentset{\mbox{\small\textbf{.}}}{#1}}
\newcommand*{\ddt}[1]{ \accentset{\mbox{\small\bfseries..}}{#1} }
\newcommand*{\Dt}[1]{\accentset{\mbox{\normalsize\bfseries .}}{#1}}
\newcommand*{\DDt}[1]{ \accentset{\mbox{\bfseries .\hspace{-0.03ex}.}}{#1} }

\newcommand{\rng}[1]{ \mathring{#1} }
\newcommand{\rrng}[1]{{\rlap{$\,\mathring{\vphantom{#1}}\mkern4.5mu\mathring{\vphantom{#1}}$}}#1}
\newcommand{\rring}[1]{ \accentset{\circ\circ}{#1} }
\newcommand{\ringp}[1]{\accentset{\circ'}{#1} }
\newcommand{\rringp}[1]{\accentset{\circ\circ'}{#1}}

\newcommand{\pdt}[1]{ \acute{#1} }
\newcommand{\pddt}[1]{ \rlap{$\;\,\acute{\vphantom{#1}}\!\acute{\vphantom{#1}}$}#1 }
\newcommand{\pdot}[1]{{\'{#1}}}
\newcommand{\pddot}[1]{{\H{#1}}}

\newcommand{\tridot}[1]{ \accentset{\triangle}{#1} }
\newcommand{\triddot}[1]{ \accentset{\triangle\triangle}{#1} }
\newcommand{\tridt}[1]{ \accentset{\blacktriangle}{#1} }
\newcommand{\triddt}[1]{ \accentset{\blacktriangle\!\blacktriangle}{#1} }
\newcommand{\deldot}[1]{ \accentset{\triangledown}{#1} }
\newcommand{\delddot}[1]{ \accentset{\triangledown\triangledown}{#1} }

\newcommand{\bxdot}[1]{ \accentset{\square}{#1} }
\newcommand{\bxddot}[1]{ \accentset{\square\square}{#1} }
\newcommand{\sqdt}[1]{ \accentset{\blacksquare}{#1} }
\newcommand{\sqddt}[1]{ \accentset{\blacksquare\blacksquare}{#1} }

\newcommand{\hdot}[1]{ \accentset{\mbox{\hspace{0.1mm}.}}{#1} }
\newcommand{\hddot}[2][.2ex]{\ddot{\raisebox{0pt}[\dimexpr\height+#1][\depth]{$#2$}}}
\newcommand{\ggvec}[2][-2pt]{\vec{\raisebox{0pt}[\dimexpr\height+#1][\depth]{$#2$}}}

\newcommand{\harp}[1]{ {\accentset{\rightharpoonup}{#1}} }
\newcommand{\varvec}[1]{ {\accentset{\to}{#1}} }


\let\oldcdot\cdot
\renewcommand*{\cdot}{{\mkern1.5mu\oldcdot\mkern1.5mu}}
\newcommand{\slot}[0]{{\oldcdot}} 
\newcommand{\cdt}[0]{ \pmb{\cdot} }
\newcommand{\bdt}[0]{ \bs{\cdot} }

\newcommand{\txblt}[0]{{\text{\small{\textbullet\hspace{.7ex}}}}}
\newcommand*{\sblt}[0]{ \raisebox{0.25ex}{$\scriptscriptstyle{\bullet}$}  }
\newcommand*{\bltt}[0]{ \text{\large{${\,\,\bullet\,\,}$}} }
\newcommand*{\nmblt}[0]{ \raisebox{0.15ex}{$\scriptstyle{\bullet}$}  }

\newcommand{\cddt}[0]{{\mkern2mu\pmb{\cdot}\mkern2mu}}
\newcommand{\cddot}[0]{ {\mkern2mu\uln{\pmb{\cdot}}\mkern2mu} }

\newcommand*{\otms}[0]{ {\raisebox{0.15ex}{$\scriptstyle{\,\otimes\,}$}}  }
\newcommand*{\opls}[0]{ \raisebox{0.15ex}{$\scriptstyle{\,\oplus\,}$}  }
\newcommand*{\wdg}[0]{ \raisebox{0.15ex}{$\scriptstyle{\,\wedge\,}$}  }
\newcommand*{\tms}[0]{ \raisebox{0.1ex}{$\scriptstyle{\,\times\,}$}  }
\newcommand{\botimes}[0]{ \scalebox{1.3}{$\otimes$} }
\newcommand{\boplus}[0]{ \scalebox{1.3}{$\oplus$} }
\newcommand{\bwedge}[1]{ {\scalebox{1}[1.3]{$\wedge$}}^{\!#1} }

\newcommand*{\ssqr}[0]{ \raisebox{0.15ex}{$\scriptstyle{\blacksquare}$}  }
\newcommand*{\iisqr}[0]{ \raisebox{0.25ex}{$\,\scriptscriptstyle{\blacksquare}\,$}  }
\newcommand*{\sbx}[0]{ \raisebox{0.15ex}{$\scriptstyle{\square}$}  }
\newcommand*{\iibx}[0]{ \raisebox{0.25ex}{$\,\scriptscriptstyle{\square}\,$}  }

\newcommand{\dimd}[0]{\diamond}%
\newcommand*{\tri}[0]{\triangle}%
\newcommand*{\trid}[0]{\triangledown} %
\newcommand*{\rtri}[0]{\triangleright} %
\newcommand*{\ltri}[0]{\triangleleft} %
\newcommand*{\stri}[0]{ \raisebox{0.25ex}{$\scriptstyle{\,\triangle\,}$} }
\newcommand*{\strid}[0]{\raisebox{0.25ex}{$\scriptstyle{\,\triangledown\,}$}}%
\newcommand*{\srtri}[0]{ \raisebox{0.25ex}{$\scriptstyle{\,\triangleright\,}$} }
\newcommand*{\slri}[0]{\raisebox{0.25ex}{$\scriptstyle{\,\triangleleft\,}$}}%

\newcommand*{\rnlarrow}[0]{ \overset{\nLeftarrow\;}{\Rightarrow}}
\newcommand*{\Rnlarrow}[0]{ \overset{\Leftarrow \!/ \!=\,}{\Longrightarrow}}
\newcommand*{\lnrarrow}[0]{ \overset{\,\nRightarrow}{\Leftarrow}}
\newcommand*{\Lnrarrow}[0]{ \overset{=\! /\!\Rightarrow\,}{\Longleftarrow}}
\newcommand{\nDownarrow}[0]{\rotatebox[origin=c]{-90}{$\nRightarrow$}}
\newcommand{\nUparrow}[0]{\rotatebox[origin=c]{90}{$\nRightarrow$}}


\newcommand*{\shrp}[0]{{\scalebox{0.55}{$\sharp$}}}
\newcommand*{\flt}[0]{{\scalebox{0.6}{$\flat$}}}

\newcommand{\iio}[0]{{\scalebox{0.45}{$\bs{\circ}$}} }
\newcommand{\iix}[0]{{\scriptscriptstyle{\times}} }
\newcommand{\iistr}[0]{{\scriptscriptstyle{\star}} }
\newcommand{\str}[0]{{\scriptscriptstyle{\star}} }
\newcommand{\drg}[1]{ #1^\ss{\dagger} } 

\newcommand{\hdge}[1]{ #1^{\scriptscriptstyle{\star}}  }
\newcommand{\hhdge}[1]{ #1^{\scriptscriptstyle{\star\star}}  }
\newcommand{\hodge}[0]{\star}
\newcommand{\varhodge}[0]{ {\,^{\scriptscriptstyle{\bar{\star}}}} }
\newcommand{\ax}[1]{ {#1^{\scriptscriptstyle{\times}}} }
\newcommand{\axx}[1]{ {#1^{\scriptstyle{\times}}} }

\newcommand{\inv}[1]{ #1^{\scriptscriptstyle{-\!1}} }
\newcommand{\negg}[0]{{\textnormal{-}}}

 \newcommand{\trn}[1]{ #1^{\ii{\textsf{\textup{T}}}}  }
\newcommand{\invtrn}[1]{ #1^{\ii{\textsf{\textup{--T}}}} }
 
\newcommand*{\zer}[0]{ {\textrm{\tiny{\textit{0}}}} } %
\newcommand*{\zr}[0]{ {\ss{0}} }%
\newcommand{\nozer}[0]{ {\scriptscriptstyle{\emptyset}} }

\newcommand*{\sodot}[0]{ \raisebox{0.12ex}{$\scriptstyle{\odot}$}  }

\newcommand*{\nc}[0]{{\ii{\mrm{nc}}}}


\newcommand*{\mrm}[1]{ {\mathrm{#1}} } 
\newcommand*{\iimrm}[1]{ {\ii{\mathrm{#1}}} }
\newcommand*{\iirm}[1]{ {\ii{\textrm{#1}}} }

\newcommand*{\mbf}[1]{{\mathbf{#1}}} 

\newcommand{\rmb}[1]{\tbf{#1}}


\newcommand{\bs}[1]{{\boldsymbol{#1}}}
\newcommand{\bsi}[1]{\boldsymbol{\mit{#1}}} 
\newcommand{\accalign}[2]{ \mkern2mu #1{\mkern-2mu #2} }
\newcommand{\tbs}[1]{ {\accalign{\tilde}{\bs{#1}}} }%
\newcommand{\wtbs}[1]{ {\accalign{\widetilde}{\bs{#1}}} }%
\newcommand{\hbs}[1]{ {\accalign{\hat}{\bs{#1}}} }%
\newcommand{\whbs}[1]{ {\accalign{\widehat}{\bs{#1}}} }%
\newcommand{\barbs}[1]{ {\accalign{\bar}{\bs{#1}}} }%
\newcommand{\vecbs}[1]{ {\accalign{\vec}{\bs{#1}}} }%
\newcommand*{\dtbs}[1]{ {\accalign{\dt}{\bs{#1}}} }%
\newcommand*{\ddtbs}[1]{ {\accalign{\ddt}{\bs{#1}}} }%

\newcommand{\tbm}[1]{\tilde{\bm{#1}}}
\newcommand{\wtbm}[1]{\widetilde{\bm{#1}}}
\newcommand{\hbm}[1]{\hat{\bm{#1}}}
\newcommand{\whbm}[1]{\widehat{\bm{#1}}}
\newcommand{\barbm}[1]{\bar{\bm{#1}}}

\newcommand*{\nbs}[1]{ \scalebox{1.15}{$\boldsymbol{#1}$} }
\newcommand*{\lbs}[1]{ \scalebox{1.4}{$\boldsymbol{#1}$} }
\newcommand*{\nbm}[1]{ \scalebox{1.2}{$\bm{#1}$} }
\newcommand*{\lbm}[1]{ \scalebox{1.4}{$\bm{#1}$} }


\newcommand{\msfb}[1]{{\mathsfb{#1}}} 

\newcommand{\sfb}[1]{{\txsfb{#1}}} 

\renewcommand{\accalign}[2]{ \mkern1.5mu #1{\mkern-1.5mu #2} }
\newcommand{\tsfb}[1]{ {\accalign{\tilde}{\sfb{#1}}} }%
\newcommand{\wtsfb}[1]{ {\accalign{\widetilde}{\sfb{#1}}} }%
\newcommand{\hsfb}[1]{ {\accalign{\hat}{\sfb{#1}}} }%
\newcommand{\whsfb}[1]{ {\accalign{\widehat}{\sfb{#1}}} }%
\newcommand{\barsfb}[1]{ {\accalign{\bar}{\sfb{#1}}} }%
\newcommand{\bsfb}[1]{ {\accalign{\bar}{\sfb{#1}}} }%
\newcommand{\vecsfb}[1]{ {\accalign{\vec}{\sfb{#1}}} }%
\newcommand*{\dtsfb}[1]{ {\accalign{\dt}{\sfb{#1}}} }%
\newcommand*{\ddtsfb}[1]{ {\accalign{\ddt}{\sfb{#1}}} }%

\newcommand*{\sfg}[0]{{\bs{g}}}


\newcommand{\tn}[1]{{\sfb{#1}}} 

\newcommand*{\mbb}[1]{ {\mathbb{#1}} } 
\newcommand*{\tmbb}[1]{ \tilde{\mbb{#1}} }
\newcommand*{\wtmbb}[1]{ \widetilde{\mbb{#1}} }
\newcommand*{\hmbb}[1]{ \hat{\mbb{#1}} }
\newcommand*{\whmbb}[1]{ \widehat{\mbb{#1}} }
\newcommand*{\bbk}[0]{\Bbbk} 

\newcommand*{\mfrak}[1]{ {\mathfrak{#1}} }
\newcommand*{\bfrak}[1]{ {\bs{\mathfrak{#1}}} }
\newcommand*{\tmfrak}[1]{ \tilde{\mfrak{#1}} }
\newcommand*{\wtmfrak}[1]{ \widetilde{\mfrak{#1}} }
\newcommand*{\hmfrak}[1]{ \hat{\mfrak{#1}} }
\newcommand*{\whmfrak}[1]{ \widehat{\mfrak{#1}} }


\newcommand*{\mscr}[1]{{\mathscr{#1}}}%
\newcommand*{\bscr}[1]{ \bs{\mathscr{#1}} }
\newcommand*{\tscr}[1]{\tilde{\mathscr{#1}}}
\newcommand*{\wtscr}[1]{\widetilde{\mathscr{#1}}}
\newcommand*{\hscr}[1]{ \hat{\mathscr{#1}} }
\newcommand*{\whscr}[1]{ \widehat{\mathscr{#1}} }
\newcommand*{\barscr}[1]{ \bar{\mathscr{#1}} }
\newcommand*{\olscr}[1]{ \overline{\mathscr{#1}} }

\newcommand*{\mcal}[1]{ {\mathcal{#1}} }
\newcommand*{\bcal}[1]{ \bs{\mathcal{#1}} }
\newcommand*{\tcal}[1]{\tilde{\mathcal{#1}}}
\newcommand*{\wtcal}[1]{\widetilde{\mathcal{#1}}}
\newcommand*{\hcal}[1]{ \hat{\mathcal{#1}} }
\newcommand*{\whcal}[1]{ \widehat{\mathcal{#1}} }
\newcommand*{\barcal}[1]{ \bar{\mathcal{#1}} }
\newcommand*{\olcal}[1]{ \overline{\mathcal{#1}} }

\newcommand{\iical}[1]{{\scriptscriptstyle{\mcal{#1}}}}
\newcommand{\sscal}[1]{{{}^{{}_{\mathcal{#1}}}}} 
\newcommand{\iiscr}[1]{ {\scriptscriptstyle{\mscr{#1}}} }
\newcommand*{\sscr}[1]{ {\scalebox{0.7}{$\mscr{#1}$}} } 



\newcommand{\en}[0]{{\itsc{n}}}
\newcommand{\emm}[0]{{\itsc{m}}} 
\newcommand{\envec}[0]{ \hbe_\en }
\newcommand{\enform}[0]{\,\hat{\!\bs{\epsilon}}^\en}
\newcommand{\tilp}[0]{{\mkern2mu \til{\mkern-2mu\smash{p}}}}

\newcommand{\pu}[0]{\mu}  
\newcommand{\varpu}[0]{\til{\pu}} %

\newcommand*{\tp}[1]{{\vphantom{#1}\uln{#1}}} 
\newcommand*{\btp}[1]{ {\bar{\tp{#1}}} }
\newcommand*{\bartp}[1]{ {\bar{\tp{#1}}} }
\newcommand*{\tiltp}[1]{ {\til{\tp{#1}}} }
\newcommand*{\ttp}[1]{ {\til{\tp{#1}}} }
\newcommand*{\wttp}[1]{ {\wt{\tp{#1}}} }
\newcommand*{\htp}[1]{ {\hat{\tp{#1}}} }
\newcommand*{\whtp}[1]{ {\wh{\tp{#1}}} }
\newcommand{\dottp}[1]{{\dot{\tp{#1}}} }

\newcommand*{\tup}[1]{{\bs{#1}}}
 \newcommand*{\bartup}[1]{ {\bar{\tup{#1}}} } 
\newcommand*{\tiltup}[1]{ {\til{\tup{#1}}} }
\newcommand*{\ttup}[1]{ {\til{\tup{#1}}} }
\newcommand*{\wttup}[1]{ {\wt{\tup{#1}}} }
\newcommand*{\htup}[1]{ {\hat{\tup{#1}}} }
\newcommand*{\whtup}[1]{ {\wh{\tup{#1}}} }
\newcommand{\dottup}[1]{{\dot{\tup{#1}}} }
\newcommand{\nrmtup}[1]{ {#1} } %


\newcommand*{\pt}[1]{{#1}}
\newcommand{\tilpt}[1]{ {\til{\pt{#1}}} }
\newcommand*{\barpt}[1]{ {\bar{\pt{#1}}} }
\newcommand{\hatpt}[1]{ {\hat{\pt{#1}}} }
\newcommand{\hpt}[1]{ {\hat{\pt{#1}}} }
\newcommand*{\ptvec}[1]{ {\vec{\pt{#1}}} }
\newcommand*{\dtpt}[1]{\dt{\pt{#1}}}


\renewcommand{\d}[0]{\mrm{d}}
\newcommand*{\rmd}[0]{{\text{\rmsb{d}}}}
\newcommand*{\dif}[0]{ {\textbf{\textsf{\textsl{d}}}} }
\newcommand*{\Dif}[0]{{ \scriptstyle{\textbf{\textsf{\textsl{D}}}} }}

\newcommand{\difbar}[0]{{ \txsfb{\kern0.6ex\ooalign{\hidewidth\raisebox{0.63ex}{--}\hidewidth\cr{\kern-0.6ex\textsf{d}}\cr}} \vphantom{d} }}
\newcommand{\itdifbar}{%
   {\mkern4.5mu \txsfb{\ooalign{\hidewidth\raisebox{0.65ex}{--}\hidewidth\cr$\mkern-4.5mu \bs{d}$\cr}} \vphantom{d} }%
}

\newcommand*{\hdif}[0]{  \hspace{1mm}\hat{\hspace{-1mm}\dif} }
\newcommand*{\tdif}[0]{  \hspace{1mm}\tilde{\hspace{-1mm}\dif} }
\newcommand*{\deldif}[0]{\bs{\delta}}

\newcommand{\tinywedge}[0]{ \scalebox{0.45}{$\pmb{\wedge}$}}
\newcommand{\exd}[0]{\dif_{\hspace{-0.25mm}{\tinywedge}}}
\newcommand{\exdbar}[0]{\difbar_{\hspace{-0.25mm}{\tinywedge}}}

\newcommand{\codif}[0]{ {\exd^{\scriptscriptstyle{\star}}} }

\newcommand*{\fibd}[0]{{\sfb{F}}}%

\newcommand{\pderiv}[2]{ \tfrac{\uppartial #1}{\uppartial #2} }
\newcommand{\ppderiv}[3]{\tfrac{\uppartial^2 #1}{\uppartial #2 \uppartial #3}}

\newcommand{\pder}[2]{ \tfrac{ \displaystyle{\uppartial #1}}{\displaystyle{\uppartial #2}} }  
\newcommand{\pderr}[2]{ \tfrac{ \textstyle{\uppartial #1}}{\textstyle{\uppartial #2}} }  
\newcommand{\ppder}[3]{ \tfrac{\displaystyle{\uppartial^2 #1}}{\displaystyle{\uppartial #2 \uppartial #3}} }

\newcommand{\Pderiv}[2]{\frac{\uppartial #1}{\uppartial #2}}
\newcommand{\PPderiv}[3]{\frac{\uppartial^2 #1}{\uppartial #2 \uppartial #3}}

 \newcommand*{\diff}[2]{ \tfrac{\mathrm{d} #1}{\mathrm{d} #2} }
\newcommand{\ddiff}[2]{ \tfrac{\mathrm{d}^2{#1}}{\mathrm{d} {#2}^2} }
\newcommand{\Diff}[2]{ \frac{\mathrm{d}#1}{\mathrm{d}\,#2} }
\newcommand{\DDiff}[2]{ \frac{\mathrm{d}^2{#1}}{\mathrm{d}\,{#2}^2} }

\newcommand{\fdiff}[1]{ \,\tfrac{ {\!\!\!\!}^{ \text{\footnotesize{$#1$}} }\d }{\;\d\, t} }
\newcommand{\fDiff}[3]{ \,\frac{ \!\!\!\!^{ \text{\footnotesize{$#3$}} }\d #1 }{\;\d\, #2} }
\newcommand{\fddiff}[1]{ \,\tfrac{ \!\!\!\!^{ \text{\scriptsize{$#1$}} }\d^2 }{\;\d\, t^2} }
\newcommand{\fDDiff}[3]{ \,\frac{ \!\!\!\!^{ \text{\footnotesize{$#3$}} }\d^2 #1 }{\;\d\, #2^2} }

\newcommand{\mdiff}[2]{ \tfrac{ \mathrm{D} #1}{\mathrm{d} #2}   }
\newcommand{\nabdiff}[2]{ \tfrac{\nabla #1}{\mathrm{d} #2}   }
\newcommand{\nabpderiv}[2]{ \tfrac{\nabla #1}{\uppartial #2}   }

\newcommand{\difscale}[1]{\scalebox{0.72}{$#1$}}
\newcommand{\lderiv}[1]{ {\textit{\textrm{\pounds}}}_{\difscale{#1}} } 
\newcommand{\nab}[1][\mkern1mu]{ \nabla_{\mkern-4mu \difscale{#1}\mkern1mu} }
\newcommand{\del}[1]{ \nabla_{\hspace{-0.6mm} {#1}}  }


\newcommand{\pmat}[1]{ \begin{pmatrix} #1 \end{pmatrix} }
\newcommand{\bmat}[1]{ \begin{bmatrix} #1 \end{bmatrix} }
\newcommand{\fnpmat}[1]{ \fnsz{\begin{pmatrix} #1 \end{pmatrix}} }
\newcommand{\smpmat}[1]{ \smsz{\begin{pmatrix} #1 \end{pmatrix}} }
\newcommand{\fnbmat}[1]{ \fnsz{\begin{bmatrix} #1 \end{bmatrix}} }

\newcommand*{\kd}[0]{1}
\newcommand*{\lc}[0]{{\itsc{e}}}
\newcommand{\ibase}[0]{\hat{\pmb{\oldstylenums{1}}}}
\newcommand*{\imat}[0]{{\text{1}}\hspace{-.2mm}}
\newcommand*{\jmat}[0]{J}

\newcommand{\Id}[0]{{\mathrm{Id}}}%
\newcommand{\iden}[0]{\sfb{1}}
\newcommand*{\gvol}[0]{\epsilon}
\newcommand*{\spvol}[0]{\sigma} 
\newcommand*{\lagvol}[0]{\varsigma}
\newcommand*{\spform}[0]{\theta}  
\newcommand*{\spformup}[0]{\thetaup}
\newcommand*{\lagform}[0]{\vartheta} 
\newcommand*{\lagformup}[0]{\varthetaup}

\newcommand{\vf}[0]{v} 
\newcommand{\pf}[0]{p} 
\newcommand{\vlin}[0]{{\scalebox{0.7}{$\mathcal{V}$}}}
\newcommand{\plin}[0]{{\pi}} 
\newcommand{\uplin}[0]{{\piup}} 
\newcommand{\tilplin}[0]{{\mkern2mu \til{\mkern-2mu\smash{\pi}}}} 


\newcommand{\rmat}[1]{ \mathrm{M}^{#1}_\ii{\mbb{R}} }
\newcommand{\mats}[1]{ \rmat{#1} }
\newcommand{\mat}[2]{ \mathrm{M}^{#1}_\ii{#2} }
\newcommand{\lgrp}[3]{\mathrm{#1}^{#2}_\ii{\mbb{#3}}}
\newcommand{\lalg}[3]{\mathfrak{#1}^{#2}_\ii{\mbb{#3}}}
\newcommand{\Afmat}[1]{ \mathrm{Af}^{#1}_\ii{\mbb{R}} }
\newcommand{\afmat}[1]{ \mathfrak{af}^{#1}_\ii{\mbb{R}} }
\newcommand{\Glmat}[1]{ \mathrm{Gl}^{#1}_\ii{\mbb{R}} }
\newcommand{\Slmat}[1]{ \mathrm{Sl}^{#1}_\ii{\mbb{R}} }
\newcommand{\glmat}[1]{ \mathfrak{gl}^{#1}_\ii{\mbb{R}} }
\newcommand{\slmat}[1]{ \mathfrak{sl}^{#1}_\ii{\mbb{R}} }
\newcommand{\Spmat}[1]{ \mathrm{Sp}^{#1}_\ii{\mbb{R}} }
\newcommand{\spmat}[1]{ \mathfrak{sp}^{#1}_\ii{\mbb{R}} }
\newcommand{\Omat}[1]{ \mathrm{O}^{#1}_\ii{\mbb{R}} }
\newcommand{\Somat}[1]{ \mathrm{SO}^{#1}_\ii{\mbb{R}} }
\newcommand{\omat}[1]{ \mathfrak{o}^{#1}_\ii{\mbb{R}} }
\newcommand{\somat}[1]{ \mathfrak{so}^{#1}_\ii{\mbb{R}} }
\newcommand{\Umat}[1]{ \mathrm{U}^{#1}_\ii{\mbb{C}} }
\newcommand{\Sumat}[1]{ \mathrm{SU}^{#1}_\ii{\mbb{C}} }
\newcommand{\umat}[1]{ \mathfrak{u}^{#1}_\ii{\mbb{C}} }
\newcommand{\sumat}[1]{ \mathfrak{su}^{#1}_\ii{\mbb{C}} }
\newcommand{\Emat}[1]{ \mathrm{E}^{#1}_\ii{\mbb{R}} }
\newcommand{\Semat}[1]{ \mathrm{SE}^{#1}_\ii{\mbb{R}} }
\newcommand{\emat}[1]{ \mathfrak{e}^{#1}_\ii{\mbb{R}} }
\newcommand{\semat}[1]{ \mathfrak{se}^{#1}_\ii{\mbb{R}} }

\newcommand{\Aften}[0]{ \mathrm{Af} }
\newcommand{\aften}[0]{ \mathfrak{af} }
\newcommand{\Glten}[0]{ \mathrm{Gl} }
\newcommand{\Slten}[0]{ \mathrm{Sl} }
\newcommand{\glten}[0]{ \mathfrak{gl} }
\newcommand{\slten}[0]{ \mathfrak{sl} }
\newcommand{\Spten}[0]{ \mathrm{Sp}}
\newcommand{\spten}[0]{ \mathfrak{sp} }
\newcommand{\Oten}[0]{ \mathrm{O} }
\newcommand{\Soten}[0]{ \mathrm{SO} }
\newcommand{\oten}[0]{ \mathfrak{o} }
\newcommand{\soten}[0]{ \mathfrak{so} }
\newcommand{\Eten}[0]{ \mathrm{E} }
\newcommand{\Seten}[0]{ \mathrm{SE} }
\newcommand{\eten}[0]{ \mathfrak{e} }
\newcommand{\seten}[0]{ \mathfrak{se} }

\newcommand{\Spism}[0]{\mathscr{S}\mkern-1mu p}
\newcommand{\Dfism}[0]{\mathscr{D}\mkern-2mu f}
\newcommand{\Hmism}[0]{\mathscr{H}m }
\newcommand{\Ctism}[0]{\mathscr{C}t}
\newcommand{\Isom}[0]{\mathcal{I}\mkern-3mu s}


\newcommand{\man}[1]{ \mathcal{#1} }  
\newcommand{\tman}[1]{ \til{\man{#1}} } 
\newcommand{\barman}[1]{ \bar{\man{#1}} }
\newcommand{\bman}[1]{ \bar{\man{#1}} }
\newcommand{\hman}[1]{ \hat{\man{#1}} }
\newcommand{\whman}[1]{ \widehat{\man{#1}} }
\newcommand{\sman}[1]{\sl{#1}} 
\newcommand{\tint}[0]{\mathcal{I}} 

\newcommand{\chart}[2]{ \man{#1}_{\hspace{-0.2mm}\textup{\tiny(} {\scriptscriptstyle{#2}} \textup{\tiny)}} }
\newcommand{\rchart}[2]{ \mathbb{#1}_{\hspace{-0.2mm}\textup{\tiny(} {\scriptscriptstyle{#2}} \textup{\tiny)}} }
\newcommand{\chrt}[2]{ {#1}_{\hspace{-0.2mm}\textup{\tiny(} {\scriptscriptstyle{#2}} \textup{\tiny)}} }

\newcommand{\vsp}[1]{ \mathbb{#1} } 
\newcommand{\aff}[1]{\man{#1}} 
\newcommand{\tvsp}[1]{ \tilde{\vsp{#1}} } 
\newcommand{\bvsp}[1]{ \bar{\vsp{#1}} } 
\newcommand{\taff}[1]{ \tilde{\aff{#1}} } 
\newcommand{\baff}[1]{ \bar{\aff{#1}} } 

\newcommand{\affE}[0]{\man{E}}
\newcommand{\vecE}[0]{\vsp{E}}
\newcommand{\Eaf}[0]{\man{E}}
\newcommand{\Evec}[0]{\vsp{E}}
\newcommand{\emet}[0]{I} 
\newcommand{\bemet}[0]{\sfb{I}}
\newcommand{\rfun}[0]{{r}} 
\newcommand{\lang}[0]{{\ell}} 
\newcommand{\Lang}[0]{{L}}
\newcommand{\langb}[0]{\bm{\ell}} 
\newcommand{\Langb}[0]{\sfb{L}} 
\newcommand{\slang}[0]{{\ell}} 
\newcommand{\Slang}[0]{{L}} 
\newcommand{\slangb}[0]{\bm{\ell}}
\newcommand{\Slangb}[0]{\sfb{L}}

\newcommand{\blang}[0]{\langb} 
\newcommand{\bLang}[0]{\Langb} 
\newcommand{\bslang}[0]{\slangb} 
\newcommand{\bSlang}[0]{\Slangb}

\newcommand*{\fun}[0]{{\scalebox{0.85}[1]{$\mathcal{F}\mkern-2mu$}}}  
\newcommand*{\tens}[0]{ \mathscr{T} } 
\newcommand*{\forms}[0]{ \Lambdaup }
\newcommand{\formsex}[0]{ \Lambdaup_\ii{\mrm{ex}\!}}
\newcommand{\formscl}[0]{ \Lambdaup_\ii{\mrm{cl}\!}}

\newcommand*{\veckl}[0]{ \mathfrak{X}_{\scriptscriptstyle{\!\mathfrak{i\!s}}\!} }
\newcommand*{\vect}[0]{ \mathfrak{X} } 
\newcommand*{\vecsp}[0]{ \mathfrak{X}_{\scriptscriptstyle{\!\mathfrak{s\!p}}\!\!} } 
\newcommand*{\vechm}[0]{ \mathfrak{X}_{\scriptscriptstyle{\!\mathfrak{h\!\!m}}\!\!} } 
\newcommand{\ham}[1]{{\mathscr{#1}}}
\newcommand{\tham}[1]{\widetilde{\mathscr{#1}}}

 \newcommand{\Tspacefont}[1]{ {\mathrm{#1}} } 
\newcommand{\Tan}[0]{ \Tspacefont{T}   }
\newcommand{\Tanh}[0]{  \Tspacefont{H}  }
\newcommand{\Tanv}[0]{  \Tspacefont{V}  }
\newcommand{\tsp}[1][\mkern2mu]{ \Tspacefont{T}_{\mkern-2mu #1} }
\newcommand{\tspv}[1][\mkern2mu]{\Tspacefont{V}_{\!#1}}
\newcommand{\tsph}[1][]{\Tspacefont{H}_{#1}}
\newcommand{\tspn}[1][\hspace{.5mm}]{\Tspacefont{N}_{\hspace{-.2mm}#1}}
\newcommand{\cotsp}[1][\hspace{1mm}]{ {\Tspacefont{T}^* {\hspace{-1ex}}_{\hspace{-.4mm}#1\hspace{0.5mm}}} } 
\newcommand{\cotspv}[1][\hspace{1mm}]{ \Tspacefont{V}^* {\hspace{-1ex}}_{\!#1\hspace{0.5mm}} }
\newcommand{\cotsph}[1][\hspace{1mm}]{ \Tspacefont{H}^* {\hspace{-1ex}}_{#1\hspace{0.1mm}} }
\newcommand{\cotspn}[1][\hspace{1mm}]{ \Tspacefont{N}^* {\hspace{-1ex}}_{#1\hspace{0.1mm}} }

\newcommand{\prj}[0]{ \scalebox{0.7}{$\Pi$} }  
 \newcommand{\tpr}[0]{ \scalebox{0.5}[0.68]{$\bs{\mcal{T}}$} }
 \newcommand{\copr}[0]{{ \mkern1mu\hat{\mkern-1mu\smash{\tpr}} }}

\newcommand{\vfun}[1][]{{V^{#1}}}
\newcommand{\pfun}[1][]{{P^{#1}}}
\newcommand{\varpfun}[1][]{{l^{#1}}}

\newcommand*{\tlift}[0]{ T\hspace{-.3mm} }
\newcommand*{\colift}[0]{ \hat{T}\hspace{-.3mm} }
\newcommand*{\uptlift}[0]{ \mrm{T}\hspace{-.3mm} }
\newcommand*{\upcolift}[0]{ \hat{\mrm{T}}\hspace{-.3mm} }

\newcommand{\formsh}[0]{\Lambdaup_\ii{\mrm{h}\!}} 
\newcommand{\formsv}[0]{\Lambdaup_\ii{\mrm{v}\!}} 
\newcommand{\vectv}[0]{\mathfrak{X}_\ii{\mrm{v}\!}} 
\newcommand{\vecth}[0]{\mathfrak{X}_\ii{\mrm{h}\!}} 
\newcommand{\formsbh}[0]{\Lambdaup_\ii{\mrm{bh}\!}} 
\newcommand{\vectbv}[0]{\mathfrak{X}_\ii{\mrm{bv}\!}}

\newcommand{\lft}[1]{ #1^{\hspace{-0.2mm}\scriptscriptstyle\upharpoonright} } 
\newcommand{\lift}[1]{ #1^{\hspace{-0.2mm}\scriptscriptstyle\uparrow} } 
\newcommand{\cotlft}[1]{  #1^{\hspace{-0.2mm}\hat{\scriptscriptstyle{\uparrow}}} }
\newcommand{\invlift}[1]{ #1^{\hspace{-0.2mm}\scriptscriptstyle\downarrow} }

\newcommand{\vlift}[1]{ #1^{\hspace{-0.2mm}\scriptscriptstyle\uparrow} }
\newcommand{\Vlift}[1]{  #1^{\scriptscriptstyle\Uparrow} }
\newcommand{\coVlift}[1]{  #1^{\hat{\scriptscriptstyle{\Uparrow}}} }%

\newcommand{\covlift}[1]{ #1^{\hat{\textup{\tiny{\textsf{V}}}}}}
\newcommand{\verlift}[1]{ #1^{\textup{\tiny{\textsf{V}}}} }%
\newcommand{\coverlift}[1]{ #1^{\hat{\textup{\tiny{\textsf{V}}}}}}

\newcommand{\hlift}[1]{ #1^{\textup{\tiny{\textsf{H}}}} }%
\newcommand{\horlift}[1]{ #1^{\textup{\tiny{\textsf{H}}}} }%
\newcommand{\cohlift}[1]{ #1^{\hat{\textup{\tiny{\textsf{H}}}}} }%

 \newcommand{\dubuparrow}[0]{\rotatebox[origin=c]{90}{$\ii{\twoheadrightarrow}$}}
\newcommand{\tailuparrow}[0]{\rotatebox[origin=c]{90}{$\ii{\rightarrowtail}$}}
\newcommand{\lftt}[1]{#1^{\dubuparrow}}
\newcommand{\llft}[1]{#1^{\tailuparrow}}

\newcommand{\rhk}[0]{\hookrightarrow}
\newcommand{\lhk}[0]{\hooklefttarrow}
\newcommand{\subman}[0]{ \,\accentset{\scriptstyle{\subset}}{\scriptstyle{\hookrightarrow}}\, }

\newcommand{\pbrak}[2]{\{#1  ,  #2 \}}
\newcommand{\Pbrak}[2]{\big\{#1\;,\;#2\big\}}
\newcommand{\ppbrak}[2]{\{ \hspace{-0.8mm} \{#1 ,  #2 \} \hspace{-0.8mm} \}}

\newcommand{\lbrak}[2]{ [ #1 , #2 ] }
\newcommand{\Lbrak}[2]{ \big[ #1\;,\;#2 \big] }
\newcommand{\llbrak}[2]{ [\hspace{-0.8mm}[#1 ,  #2 ]\hspace{-0.8mm}] }

\newcommand{\altbrak}[2]{ {\lfloor #1 ,  #2 \rfloor} }
\newcommand{\aaltbrak}[2]{ {\lfloor \hspace{-0.8mm} \lfloor #1 ,  #2 \rfloor \hspace{-0.8mm} \rfloor} }
\newcommand{\lagbrak}[2]{ {\lfloor #1 ,  #2 \rfloor}  }
\newcommand{\llagbrak}[2]{ {\lfloor \hspace{-0.8mm} \lfloor #1 ,  #2 \rfloor \hspace{-0.8mm} \rfloor} }

\newcommand{\inner}[2]{\langle#1  , #2\rangle}
\newcommand{\iinner}[2]{\langle\!\langle#1\,,\,#2\rangle\!\rangle}

\newcommand{\ang}[1]{\langle #1 \rangle}
\newcommand{\Ang}[1]{\big\langle #1 \big\rangle }


\newcommand{\abs}[1]{ \big| #1 \big| }
\newcommand{\Abs}[1]{ \left| #1 \right| }
\renewcommand{\mag}[1]{ |\!| #1 |\!| }
\newcommand{\nrm}[1]{{\lvert #1 \rvert} }
\renewcommand{\det}[0]{ {\textrm{\textup{\footnotesize{det}}}\hspace{0.5mm}} }
\renewcommand{\dim}[0]{ {\textrm{\textup{\footnotesize{dim}}}\hspace{0.5mm}} }
\newcommand{\codim}[0]{ {\textrm{\textup{\footnotesize{codim}}}\hspace{0.5mm}} } %
\newcommand{\rnk}[0]{ {\textrm{\textup{\footnotesize{rnk}}}\hspace{0.5mm}} }
\newcommand{\sgn}[0]{ {\textrm{\textup{\footnotesize{sgn}}}\hspace{0.5mm}} }
\newcommand{\tr}[0]{ {\textrm{\textup{\footnotesize{tr}}}\hspace{0.5mm}} }
\renewcommand{\ker}[0]{ {\textrm{\textup{\footnotesize{ker}}}\hspace{0.5mm}} }
\newcommand{\img}[0]{ {\textrm{\textup{\footnotesize{im}}}\hspace{0.5mm}} }
\newcommand{\spn}[0]{ {\textrm{\textup{\footnotesize{span}}}\hspace{0.5mm}} }
\renewcommand{\div}[1][\,]{ {{\textrm{\textup{\footnotesize{div}}}}_{\!#1}}}

\newcommand{\snn}[1][]{\sin{{\mkern-2mu}#1}}
\newcommand{\csn}[1][]{\cos{{\mkern-2mu}#1}}

\newcommand{\crd}[2]{ \tensor*[^{\ss{#2}}]{[#1]}{} }
\newcommand{\crdl}[2]{ \tensor*[^{\ss{#2}}]{#1}{} }
\newcommand{\cordl}[2]{  #1_{\scriptscriptstyle{\!#2}}  } 
\newcommand{\cord}[2]{  #1_{\scriptscriptstyle{#2}}  }

\newcommand{\eval}[2]{\left.{#1}\right|_{#2}}

\newcommand{\pdup}[0]{\uppartial} %
\newcommand{\upd}[0]{{\rotatebox[origin=t]{10}{$\partial$} \mkern-2mu}}
\newcommand{\pd}[0]{{\rotatebox[origin=t]{15}{$\partial$} \mkern-2mu}}


\newcommand{\mypd}[0]{\rotatebox[origin=t]{10}{$\partial$}} 

\newcommand*{\ffsize}[1]{{\scalebox{0.7}{$#1$}}}

\newcommand*{\bpart}[1]{\bs{\mypd}_{\hspace{-.3mm}#1}}
\newcommand*{\bpartup}[1]{\bs{\mypd}^{#1}}
\newcommand*{\tbpart}[1]{\tilde{\bs{\mypd}}_{\hspace{-.3mm}#1}}
\newcommand*{\tbpartup}[1]{{\tilde{\bs{\mypd}}\vphantom{l}^{#1}}}
\newcommand*{\hbpart}[1]{\hbs{\pd}_{\hspace{-.3mm}#1}}
\newcommand*{\hbpartup}[1]{{\hbs{\pd}\vphantom{l}^{#1}}}

\newcommand*{\pdii}[1]{ \bs{\mypd}_{\ffsize{\!#1}}}
\newcommand*{\pdupii}[1]{ \bs{\mypd}^{\ffsize{#1}} }
\newcommand*{\pdiiup}[1]{ \bs{\mypd}^{\ffsize{#1}} }
\newcommand*{\tpdii}[1]{ \tilde{\bs{\mypd}}_{\hspace{-.5mm}\ffsize{#1}} }
\newcommand*{\tpdupii}[1]{ \tilde{\bs{\mypd}}{}^{\ffsize{#1}} }
\newcommand*{\hpdii}[1]{ \hbs{\pd}_{\ffsize{\!#1}} }
\newcommand*{\hpdiiup}[1]{ {\hbs{\pd}}^{\ffsize{#1}} }
\newcommand*{\barpdii}[1]{ \bar{\bs{\mypd}}_{\hspace{-.6mm}\ffsize{#1}} }
\newcommand*{\barpdiiup}[1]{ {\bs{\bar{\pd}}}^{\ffsize{#1}} }

\newcommand{\bpd}[1][]{ {\bs{\mypd}_{\ffsize{\!#1}}} }%
\newcommand{\bpdup}[1][]{ {\bs{\mypd}^{\ffsize{#1}}} }
\newcommand{\hbpd}[1][]{ {\hbs{\pd}_{\ffsize{\!#1}}} }%
\newcommand{\hbpdup}[1][]{ {\hbs{\pd}^{\ffsize{#1}}} }
\newcommand{\tbpd}[1][]{ {\tbs{\pd}_{\ffsize{\!#1}}} }%
\newcommand{\tbpdup}[1][]{ {\tbs{\pd}^{\ffsize{#1}}} }
\newcommand{\barbpd}[1][]{ {\barbs{\pd}_{\ffsize{\!#1}}} }%
\newcommand{\barbpdup}[1][]{ {\barbs{\pd}^{\ffsize{#1}}} }

\newcommand*{\bpdh}[1]{ {\bs{\mypd}_{\mkern-1mu\hat{#1}}} }
\newcommand*{\bpdhup}[1]{ {\bs{\mypd}^{\hat{#1}}} }
\newcommand*{\bpdt}[1]{ {\bs{\mypd}_{\mkern-1mu\til{#1}}} }
\newcommand*{\bpdtup}[1]{ {\bs{\mypd}^{\til{#1}}} }
\newcommand*{\bpdbar}[1]{ {\bs{\mypd}_{\mkern-1mu\bar{#1}}} }
\newcommand*{\bpdbarup}[1]{ {\bs{\mypd}^{\bar{#1}}} }
\newcommand*{\bdelh}[1]{ \bs{\delta}^{\hat{#1}}\vphantom{l} }
\newcommand{\bdelhdn}[1][]{ {\bs{\delta}_{\mkern-1mu\hat{#1}}\vphantom{l}} }
\newcommand*{\bdelt}[1]{ \bs{\delta}^{\til{#1}}\vphantom{l} }
\newcommand{\bdeltdn}[1][]{ {\bs{\delta}_{\mkern-1mu\til{#1}}\vphantom{l}} }
\newcommand*{\bdelbar}[1]{ \bs{\delta}^{\bar{#1}}\vphantom{l} }
\newcommand{\bdelbardn}[1][]{ {\bs{\delta}_{\mkern-1mu\bar{#1}}\vphantom{l}} }



\newcommand{\bdel}[1][]{ {\bs{\delta}^{\ffsize{#1}}} }%
\newcommand{\bdeldn}[1][]{ {\bs{\delta}_{\ffsize{\!#1}}} }
\newcommand{\hbdel}[1][]{ \hbs{\delta}^{\ffsize{#1}}\vphantom{l} }%
\newcommand{\hbdeldn}[1][]{ {\hbs{\delta}_\ffsize{\!#1}\vphantom{l}} } 
\newcommand{\tbdel}[1][]{ \tbs{\delta}^{\ffsize{#1}}\vphantom{l} }%
\newcommand{\tbdeldn}[1][]{ {\tbs{\delta}_\ffsize{\!#1}\vphantom{l}} }
\newcommand{\barbdel}[1][]{ \barbs{\delta}^{\ffsize{#1}}\vphantom{l} }%
\newcommand{\barbdeldn}[1][]{ {\barbs{\delta}_\ffsize{\!#1}\vphantom{l}} }

\newcommand{\bD}[1][]{{ \fnsz{\sfb{D}}_{\ffsize{\!#1}}} }
\newcommand{\bDel}[1][]{{ \fnsz{\bs{\Delta}}^{\mkern-1mu\ffsize{#1}}}}%
\newcommand{\bDeldn}[1][]{{ \fnsz{\bs{\Delta}}_{\ffsize{#1}}}}%

\newcommand{\ff}[2]{\sfb{#1}_{\ffsize{\!#2}}} 
\newcommand{\ffup}[2]{\sfb{#1}^{\ffsize{#2}}} 
\newcommand{\coff}[2]{\bs{#1}^{\ffsize{#2}}}
\newcommand{\coffdn}[2]{\bs{#1}_{\ffsize{\!#2}}}

\newcommand{\bi}[1][]{{\sfb{i}_{\ffsize{\!#1}}}}
\newcommand{\bio}[1][]{{\bs{\iota}^{\ffsize{#1}}} \vphantom{\iota}}
\newcommand{\hbi}[1][]{{\hsfb{\i}_{\ffsize{\!#1}}}} 
\newcommand{\hbio}[1][]{{\hbs{\iota}^{\ffsize{#1}}} \vphantom{\iota}}

\newcommand{\be}[1][]{{\sfb{e}_{\ffsize{\!#1}}}}
\newcommand{\beup}[1][]{{\sfb{e}^{\ffsize{#1}}}} 
\newcommand{\tbe}[1][]{ {\tsfb{e}_{\ffsize{\!#1}}} }
\newcommand{\hbe}[1][]{ {\hsfb{e}_{\ffsize{\!#1}}} }
\newcommand{\bep}[1][\!]{{\bs{\epsilon}^{\ffsize{#1}}}}
\newcommand{\bepdn}[1][]{{\bs{\epsilon}_{\ffsize{\!#1}}}}
\newcommand{\tbep}[1][\!]{\tbs{\epsilon}^{\ffsize{#1}} \vphantom{\epsilon}}
\newcommand{\hbep}[1][\!]{{\hbs{\epsilon}^{\ffsize{#1}}} \vphantom{\epsilon}}

\newcommand{\bt}[1][]{ \sfb{t}_{\ffsize{\!#1}} \vphantom{t}}
\newcommand{\tbt}[1][]{\tsfb{t}_{\ffsize{\!#1}} \vphantom{t}}
\newcommand{\hbt}[1][]{\hsfb{t}_{\ffsize{\!#1}} \vphantom{t}}
\newcommand{\btup}[1][\,]{ \sfb{t}^{\ffsize{#1}\!} \vphantom{t}}
\newcommand{\tbtup}[1][\,]{\tsfb{t}^{\ffsize{#1}\!} \vphantom{t}}
\newcommand{\hbtup}[1][\,]{\hsfb{t}^{\ffsize{#1}\!} \vphantom{t}}

\newcommand{\btau}[1][\!]{{\bm{\tau}^{\ffsize{#1}}}}
\newcommand{\tbtau}[1][\!]{{\tbm{\tau}^{\ffsize{#1}}}}
\newcommand{\hbtau}[1][\!]{{\hbm{\tau}^{\ffsize{#1}}}}
\newcommand{\btaudn}[1][]{{\bm{\tau}_{\ffsize{\!#1}}}}
\newcommand{\tbtaudn}[1][]{{\tbm{\tau}_{\ffsize{\!#1}}}}
\newcommand{\hbtaudn}[1][]{{\hbm{\tau}_{\ffsize{\!#1}}}}


\newcommand{\Zvar}{{ \text{\ooalign{\hidewidth\raisebox{0.25ex}{--}\hidewidth\cr$Z$\cr}}\vphantom{Z}} }

\newcommand{\zvar}{%
  \text{\ooalign{\hidewidth -\kern-.3em-\hidewidth\cr$z$\cr}} \vphantom{z}%
}

\renewcommand{\Zbar}{{\textnormal{\ooalign{\hidewidth\raisebox{0.3ex}{\footnotesize{--}}\hidewidth\cr$Z$\cr}} \vphantom{Z}} }

\newcommand{\zbar}{ \mkern0.5mu
  {\text{\ooalign{\hidewidth\raisebox{0.1ex}{\scriptsize{--}}\hidewidth\cr$\mkern-0.5muz$\cr}} \vphantom{z}}%
} 

\newcommand{\Jbar}{ \mkern2mu {\textnormal{\ooalign{\hidewidth\raisebox{0.4ex}{\scriptsize{--}}\hidewidth\cr$\mkern-2mu J$\cr}} \vphantom{J}} }

\newcommand{\Ibar}{{\textnormal{\ooalign{\hidewidth\raisebox{0.4ex}{\scriptsize{--}}\hidewidth\cr$I$\cr}} \vphantom{I}} }

\newcommand{\Sbar}{{\mkern0mu \textnormal{\ooalign{\hidewidth\raisebox{0.4ex}{\scriptsize{\textbf{---}}}\hidewidth\cr$\mkern-0mu S$\cr}} \vphantom{S}} }

\newcommand{\sbar}{ \mkern-0.5mu
  {\text{\ooalign{\hidewidth\raisebox{0.1ex}{\scriptsize{--}}\hidewidth\cr$\mkern0.5mu s$\cr}} \vphantom{s}}%
}

\newcommand{\Cbar}{{\mkern-7mu \textnormal{\ooalign{\hidewidth\raisebox{0.35ex}{\footnotesize{--}}\hidewidth\cr$\mkern7muC$\cr}} \vphantom{C}} }

\newcommand{\cbar}{%
  {\mkern-3mu\text{\ooalign{\hidewidth\raisebox{0.1ex}{\scriptsize{--}}\hidewidth\cr$\mkern3muc$\cr}} \vphantom{c}}%
}

\newcommand{\ldash}[0]{{\mkern-2mu \textit{\l}\mkern1mu}} 
\newcommand{\Ldash}[0]{{\textit{\L}}} 
\newcommand{\ldashb}[0]{{\itsb{\l}}}
\newcommand{\Ldashb}[0]{{\txsfb{\L}}} 
\newcommand{\Lbar}{ {\mkern-2mu \textnormal{\ooalign{\hidewidth\raisebox{0.35ex}{\scriptsize{--}}\hidewidth\cr$\mkern2mu L$\cr}} \vphantom{L}}  }

\newcommand{\lbar}{%
    {\textnormal{\ooalign{\hidewidth\raisebox{0.3ex}{-}\hidewidth\cr$l$\cr}} \vphantom{l}}%
} 

\newcommand{\blbar}{%
    {\textbf{\ooalign{\hidewidth\raisebox{0.4ex}{\scriptsize{--}}\hidewidth\cr$\bs{l}$\cr}} \vphantom{l}}%
} 

\newcommand{\kbar}{%
   {\mkern-1mu \text{\ooalign{\hidewidth\raisebox{0.8ex}{\scriptsize{--}}\hidewidth\cr$\mkern1mu k$\cr}} \vphantom{k} }%
}
\newcommand{\kbarup}{%
   {\mkern-4mu \textnormal{\ooalign{\hidewidth\raisebox{0.8ex}{\scriptsize{--}}\hidewidth\cr$\mkern4mu \mrm{k}$\cr}} \vphantom{k} }%
}
\newcommand{\txkbar}[0]{\kbarup}

\newcommand{\hvar}{%
   {\mkern-1mu \text{\ooalign{\hidewidth\raisebox{0.8ex}{\scriptsize{--}}\hidewidth\cr$\mkern1mu h$\cr}} \vphantom{h} }%
}
\newcommand{\hbarup}{%
   {\mkern-4mu \textnormal{\ooalign{\hidewidth\raisebox{0.8ex}{\scriptsize{--}}\hidewidth\cr$\mkern4mu \mrm{h}$\cr}} \vphantom{k} }%
}
\newcommand{\txhbar}[0]{\hbarup}

\newcommand{\bbar}{%
   {\mkern-1mu \text{\ooalign{\hidewidth\raisebox{0.8ex}{\scriptsize{--}}\hidewidth\cr$\mkern1mu b$\cr}} \vphantom{b} }%
}
\newcommand{\bbarup}{%
   {\mkern-4mu \textnormal{\ooalign{\hidewidth\raisebox{0.8ex}{\scriptsize{--}}\hidewidth\cr$\mkern4mu \mrm{b}$\cr}} \vphantom{b} }%
} 
\newcommand{\txbbar}[0]{\bbarup}

\newcommand{\dbar}{%
   {\mkern4.5mu \text{\ooalign{\hidewidth\raisebox{0.8ex}{\scriptsize{--}}\hidewidth\cr$\mkern-4.5mu d$\cr}} \vphantom{d} }%
}
\newcommand{\dbarup}{%
   {\mkern3mu \textnormal{\ooalign{\hidewidth\raisebox{0.8ex}{\scriptsize{--}}\hidewidth\cr$\mkern-3mu \mrm{d}$\cr}} \vphantom{d} }%
} 
\newcommand{\txdbar}[0]{\dbarup}

\newcommand{\Gambar}{%
   {\mkern-4mu \text{\ooalign{\hidewidth\raisebox{0.4ex}{\scriptsize{--}}\hidewidth\cr$\mkern4mu \Gamma$\cr}} \vphantom{h} }%
}


\renewcommand{\v}[0]{{\nu}} 

\newcommand{\txv}[0]{{\txi{v}}}
\newcommand{\altv}[0]{{\scriptstyle{\mathcal{V}}}}
\newcommand{\newv}[0]{{\scriptstyle{\mathscr{V}}}}

\newcommand{\one}[0]{{\ss{1}}}
\newcommand{\two}[0]{{\ss{2}}}
\newcommand{\three}[0]{{\ss{3}}}
\newcommand{\four}[0]{{\ss{4}}}
\newcommand{\six}[0]{{\ss{6}}}
\newcommand{\eight}[0]{{\ss{8}}}


\renewcommand{\a}[0]{{\alpha}}
\renewcommand{\b}[0]{{\ii{\beta}}}
\newcommand{\g}[0]{{\ss{\gamma}}}
\newcommand{\gam}[0]{{\gamma}}
\newcommand{\Gam}[0]{{\Gamma}}
\newcommand{\y}[0]{{\ss{\lambda}}}
\newcommand{\lam}[0]{{\lambda}}
\newcommand{\Lam}[0]{{\Lambda}}
\newcommand{\sig}[0]{{\sigma}}
\newcommand{\Sig}[0]{{\Sigma}}
\newcommand{\varsig}[0]{{\varsigma}}
\newcommand{\ep}[0]{{\epsilon}}
\newcommand*{\varep}[0]{{\varepsilon}}
\newcommand{\omg}[0]{{\omega}}
\newcommand{\Omg}[0]{{\Omega}}
\newcommand{\kap}[0]{{\kappa}}
\newcommand{\varkap}[0]{{\varkappa}}
\newcommand{\ro}[0]{{\varrho}}
\renewcommand{\th}[0]{{\theta}}
\newcommand{\Th}[0]{{\Theta}}
\newcommand{\varth}[0]{{\vartheta}}

\DeclareRobustCommand{\rchi}{{\mathpalette\irchi\relax}}
\newcommand{\irchi}[2]{\raisebox{\depth}{$#1\chi$}} 
\DeclareRobustCommand{\pup}{{\mathpalette\newp\relax}}
\newcommand{\newp}[2]{\raisebox{0.8\depth}{$#1p$}}
\DeclareRobustCommand{\jup}{{\mathpalette\newj\relax}}
\newcommand{\newj}[2]{\raisebox{0.8\depth}{$#1\jmath$}}

\newcommand{\upa}[0]{{\upalpha}}
\newcommand{\upb}[0]{{\upbeta}}
\newcommand{\upgam}[0]{{\upgamma}}
\newcommand{\upGam}[0]{{\upGamma}}
\newcommand{\Upgam}[0]{{\Upgamma}}
\newcommand{\uplam}[0]{{\uplambda}}
\newcommand{\upLam}[0]{{\upLambda}}
\newcommand{\Uplam}[0]{{\Uplambda}}
\newcommand{\updel}[0]{{\updelta}}
\newcommand{\upsig}[0]{{\upsigma}}
\newcommand{\upSig}[0]{{\upSigma}} 
\newcommand{\Upsig}[0]{{\Upsigma}}
\newcommand{\upomg}[0]{{\upomega}}
\newcommand{\upOmg}[0]{{\upOmega}}
\newcommand{\Upomg}[0]{{\Upomega}}
\newcommand{\upep}[0]{\upepsilon}
\newcommand{\upth}[0]{\uptheta}
\newcommand{\upvarth}[0]{\upvartheta}
\newcommand{\upTh}[0]{\upTheta}
\newcommand{\Upth}[0]{\Uptheta}
\newcommand{\upkap}[0]{\upkappa}
\newcommand{\upvark}[0]{\upvarkappa}
\newcommand{\upn}[0]{\upeta}
\newcommand{\upi}[0]{\upiota}

\newcommand{\aup}[0]{{\alphaup}}
\newcommand{\bup}[0]{{\betaup}}
\newcommand{\gamup}[0]{{\gammaup}}
\newcommand{\Gamup}[0]{{\Gammaup}}
\newcommand{\lamup}[0]{{\lambdaup}}
\newcommand{\Lamup}[0]{{\Lambdaup}}
\newcommand{\delup}[0]{{\deltaup}}
\newcommand{\sigup}[0]{{\sigmaup}}
\newcommand{\varsigup}[0]{{\varsigmaup}}
\newcommand{\Sigup}[0]{{\Sigmaup}}
\newcommand{\wup}[0]{{\omegaup}} 
\newcommand{\omgup}[0]{{\omegaup}}
\newcommand{\Omgup}[0]{{\Omegaup}}
\newcommand{\epup}[0]{\epsilonup}
\newcommand{\thup}[0]{\thetaup}
\newcommand{\varthup}[0]{\varthetaup}
\newcommand{\kup}[0]{\kappaup}
\newcommand{\kapup}[0]{\kappaup}
\newcommand{\varkup}[0]{\varkappaup}
\newcommand{\nup}[0]{\etaup}
\newcommand{\iup}[0]{\iotaup}

\newcommand{\txa}[0]{{\text{\textalpha}}}
\newcommand{\txb}[0]{{\text{\textbeta}}}
\newcommand{\txbet}[0]{{\text{\textbeta}}}
\newcommand{\txg}[0]{{\text{\textgamma}}}
\newcommand{\txgam}[0]{{\text{\textgamma}}}
\newcommand{\txGam}[0]{{\text{\textGamma}}}
\newcommand{\txy}[0]{{\text{\textlambda}}}
\newcommand{\txlam}[0]{{\text{\textlambda}}}
\newcommand{\txr}[0]{{\text{\textrho}}}
\newcommand{\txrho}[0]{{\text{\textrho}}}
\newcommand{\txsig}[0]{{\text{\textsigma}}}
\newcommand{\txSig}[0]{{\text{\textSigma}}}
\newcommand{\txw}[0]{{\text{\textomega}}}
\newcommand{\txomg}[0]{{\text{\textomega}}}
\newcommand{\txep}[0]{{\text{\textepsilon}}}
\newcommand{\txth}[0]{{\text{\texttheta}}}
\newcommand{\txvarth}[0]{{\text{\textvartheta}}}
\newcommand{\txphi}[0]{{\text{\textphi}}}
\newcommand{\txvarphi}[0]{{\text{\textvarphi}}}
\newcommand{\txpsi}[0]{{\text{\textpsi}}}
\newcommand{\txd}[0]{{\text{\textdelta}}}
\newcommand{\txdel}[0]{{\text{\textdelta}}}
\newcommand{\txiota}[0]{{\text{\textiota}}}
\newcommand{\txn}[0]{{\text{\texteta}}}
\newcommand{\txeta}[0]{{\text{\texteta}}}
\newcommand{\txxi}[0]{{\text{\textxi}}}
\newcommand{\txchi}[0]{{\text{\textchi}}}
\newcommand{\txpi}[0]{{\text{\textpi}}}
\newcommand{\txtau}[0]{{\text{\texttau}}}
\newcommand{\txz}[0]{{\text{\textzeta}}}
\newcommand{\txzeta}[0]{{\text{\textzeta}}}
\newcommand{\txnu}[0]{{\text{\textnu}}}
\newcommand{\txmu}[0]{{\text{\textmugreek}}} 
\newcommand{\txups}[0]{{\text{\textupsilon}}} 
\newcommand{\txk}[0]{{\text{\textkappa}}}
\newcommand{\txkap}[0]{{\text{\textkappa}}}

\newcommand{\tximu}[0]{ {\textit{\textmugreek}} } %
\newcommand{\txiu}[0]{ {\textit{\textupsilon}} } 
\newcommand{\txitau}[0]{ {\textit{\texttau}} } %
\newcommand{\txipi}[0]{ {\textit{\textpi}} }%
\newcommand{\txirho}[0]{ {\textit{\textrho}} } 
\newcommand{\txilam}[0]{ {\textit{\textlambda}} }

\newcommand{\wtxt}[0]{{\textomega}}
\newcommand{\ytxt}[0]{{\textlambda}}
\newcommand{\thtxt}[0]{{\texttheta}}
\newcommand{\btxt}[0]{{\textbeta}}
\newcommand{\atxt}[0]{{\textalpha}}
\newcommand{\gtxt}[0]{{\textgamma}}
\newcommand{\tautxt}[0]{{\texttau}}

\newcommand{\gvgr}[1]{\vec{\tbf{#1}}}
\newcommand{\uvgr}[1]{\hat{\tbf{#1}}}
\newcommand{\tvgr}[1]{\widetilde{\tbf{#1}}}
\newcommand{\bvgr}[1]{\bar{\tbf{#1}}}
\newcommand{\gvbb}[1]{\vec{\pmb{#1}}}
\newcommand{\uvbb}[1]{\hat{\pmb{#1}}}
\newcommand{\tvbb}[1]{\widetilde{\pmb{#1}}}
\newcommand{\bvbb}[1]{\bar{\pmb{#1}}}

\newcommand{\sigvec}[0]{\vec{\pmb{\upsigma}} }
 \newcommand{\sigtvec}[0]{\widetilde{\pmb{\upsigma}} }

\newcommand{\gvec}[1]{\vec{\bs{\mathrm{#1}}}}
\newcommand{\gv}[1]{\vec{\bm{#1}}}
\newcommand{\gvb}[1]{\vec{\bs{#1}}}
\newcommand*{\bsvec}[1]{\vec{\bs{#1}}}
\newcommand*{\sfvec}[1]{\,\vec{\!\sfb{#1}}}

\newcommand{\uvec}[1]{\hat{\bs{#1}}}
\newcommand{\uv}[1]{\hat{\bs{#1}}}
\newcommand{\ihat}[0]{\hat{\bs{\iota}}}
\newcommand{\ehat}[0]{\hat{\bs{e}}}
\newcommand{\bhat}[0]{\hat{\bs{b}}}
\newcommand{\shat}[0]{\hat{\bs{s}}}
\newcommand{\nhat}[0]{\uvec{n}}
\newcommand{\evec}[0]{\vec{\bs{e}}}
\newcommand{\epvec}[0]{\vec{\bs{\epsilon}}}

\newcommand{\uvecb}[1]{ {\hbs{#1}} }
\newcommand{\uvb}[1]{ {\hbs{#1}} }
\newcommand{\ihatb}[0]{\hat{\pmb{\upiota}}} 
\newcommand{\ehatb}[0]{\uvecb{e}} 
\newcommand{\bhatb}[0]{\!\!\uvecb{\;b}}
\newcommand{\hhatb}[0]{\!\!\uvecb{\;h}}
\newcommand{\shatb}[0]{\uvecb{s}}
\newcommand{\nhatb}[0]{\uvecb{n}}
\newcommand{\ohatb}[0]{\uvecb{o}}
\newcommand{\ahatb}[0]{\uvecb{a}}
\newcommand{\epvecb}[0]{\vec{\bs{\epsilon}}}
\newcommand{\evecb}[0]{ \hspace{0.3mm}\vec{\mbf{e}}\hspace{0.1mm} }

\newcommand{\tvb}[1]{\widetilde{\mbf{#1}}}
\newcommand{\tvecb}[1]{\bar{\bs{#1}}}
\newcommand{\bv}[1]{\bar{\bs{#1}}}
\newcommand{\bvb}[1]{\bar{\bs{\mathrm{#1}}}}


\newcommand{\rnote}[1]{\noindent{\footnotesize\textit{{\color{red}${\circ}$}  #1}}}
\newcommand{\red}[1]{{\color{red}#1}}

\newcommand{\bnote}[1]{\noindent{\footnotesize\textit{{\color{blue}${\circ}$} #1}}}

\newcommand{\gnote}[1]{\noindent{\footnotesize\textit{$\circ$  #1}}}

 \newcommand{\note}[1]{\noindent{\footnotesize\textit{{\color{darkgray} #1}}}}

\newcommand{\nsp}[0]{\!\!\!\!}    
\newcommand{\nquad}[0]{\hspace{-1em}} 
\newcommand{\nqquad}[0]{\hspace{-2em}} 


\theoremstyle{plain} 
\newtheorem{thrm}{Theorem}[section]
\newtheorem{defn}{Def.}[section]

\theoremstyle{plain} 
\newtheorem{remit}{Remark}[section]

\theoremstyle{definition} 
\newtheorem{remark}{Remark}[section]


\newtheoremstyle{remsans}
{8pt} 
{12pt} 
{\sffamily\slshape\small}
{}
{\rmfamily\small}
{.}
{.5em}
{}

\theoremstyle{remsans}
\newtheorem{remsf}{$\iisqr\,$\rmsb{remark}}[section]

\newtheoremstyle{remrmsmall}
{8pt} 
{12pt} 
{\rmfamily\small}
{}
{\rmfamily\small\slshape}
{.}
{.3em}
{}

\theoremstyle{remrmsmall}
\newtheorem{remrm}{$\iisqr\,${Remark}}[section]
\newtheorem{noether}[remark]{$\iisqr\,$\rmsb{Noether}}

\newtheoremstyle{remslant}
{8pt} 
{12pt} 
{\rmfamily\slshape\small}
{}
{\rmfamily\slshape\bfseries\small}
{.}
{.3em}
{}

\theoremstyle{remslant} 
\newtheorem{remsl}{Remark}[section]

\newtheoremstyle{remnopunc}
{8pt} 
{12pt} 
{\rmfamily\small}
{}
{\bfseries\small}
{}
{.2em}
{}
\theoremstyle{remnopunc} 
\newtheorem*{noteblt}{$\nmblt$}
\newtheorem*{notestr}{\raisebox{0.1ex}{$\star$}}
\newtheorem*{noteast}{$\bs{*}$}

\newenvironment{notesq}
    {\begin{small}
     \begin{itemize}[left=0pt,labelsep=0.25ex,topsep=10pt]
     \item[$\iisqr$] }
    {\end{itemize}
     \end{small} }

\newtheoremstyle{remslantnopunc}
{8pt} 
{12pt} 
{\slshape\rmfamily\small}
{}
{\slshape\bfseries\small}
{}
{.2em}
{}
\theoremstyle{remslantnopunc} 
\newtheorem*{notesl}{$\iisqr$}

\newtheoremstyle{remrmnonbold}
{} 
{} 
{\rmfamily\footnotesize}
{}
{\rmfamily\itshape\footnotesize}
{.}
{.2em}
{}
\theoremstyle{remrmnonbold}
\newtheorem*{notation}{Notation}



\newcommand{\eq}[1]{\text{$#1$}}

\newenvironment{eqn}
    {\begin{align} }
    {\end{align}}
    
\newenvironment{smeqn}
    {\begin{small}\begin{align} }
    {\end{align}\end{small} }

\newenvironment{flsmeqn}
    {\begin{small}\begin{flalign} }
    {\end{flalign}\end{small} }




\vspace{-4ex} 
\begin{abstract}
    Closed-form Kepler solutions in projective coordinates are used to define a corresponding set of eight orbit elements and obtain their governing equations for arbitrarily-perturbed two-body dynamics. The elements and their dynamics are singularity-free in all cases besides rectilinear motion (when angular momentum vanishes). The classic $J_2$-perturbed two-body problem is developed and used for numerical verification. 
\end{abstract}


\renewcommand{\contentsname}{}
\tableofcontents  


\section*{INTRODUCTION}
\addcontentsline{toc}{section}{INTRODUCTION}

 Projective coordinate transformations, 
 combined with an appropriate transformation of the evolution parameter\footnote{In particular, a parameter \eq{s} defined by \eq{\mrm{d} t = r^2 \mrm{d} s}, or a parameter \eq{\tau} defined by \eq{\mrm{d} t = (r^2/\slang) \mrm{d} \tau} where \eq{\slang} is the (specific) angular momentum magnitude, making \eq{\tau} the true anomaly up to an additive constant.},
 are known to transform the nonlinear  second-order dynamics of the $3$-dim Kepler problem to those of a $4$-dim linear oscillator \cite{burdet1969mouvement,Burdet+1969+71+84,vitins1978keplerian,silver1975short,schumacher1987results}. 
 Previous work by Ferrándiz et al.~\cite{ferrandiz1987general,ferrandiz1988extended,ferrandiz1992increased,ferrandiz1994extended} and Deprit et al.~\cite{deprit1994linearization} extended such dimension-raising point transformations of configuration coordinates to canonical transformations of phase space coordinates in the Hamiltonian framework.  We refer to  Ferrándiz's transformation as the Burdet-Ferrándiz (BF) transformation (following the convention of \cite{deprit1994linearization}) as it is a ``canonical extension'' of Burdet's initial projective point transformation \cite{burdet1969mouvement,Burdet+1969+71+84,silver1975short}.

The present authors developed their own canonically-extended projective transformation in \cite{peterson2025prjCoord,peterson2025phdThesis,peterson2022nonminimal,peterson2023regularized} (with a geometric formulation given in \cite{peterson2025prjGeomech}), which they feel is comparatively more simple and intuitive in both its final form and its derivation. In particular, it possesses an immediate connection to orbital reference frames and attitude kinematics, with the angular momentum matrix playing a prominent role.

This paper is an extension of the aforementioned work in projective-based linear regularization of orbital dynamics. Given that the Kepler problem has readily-available closed-form solutions in projective coordinates, it is only natural to use these solutions to define a corresponding set of orbit elements and obtain their governing equations of motion for the arbitrarily-perturbed Kepler problem. That is the focus of this work. It should be noted, however, that the orbit elements introduced here are not action-angle coordinates obtained from the Hamilton-Jacobi equation, and that our treatment of the elements (section \ref{sec:VOP_2}) deviates from the Hamiltonian framework in which the projective coordinate regularization was previously formulated in \cite{peterson2025prjCoord,peterson2025phdThesis,peterson2025prjGeomech} (as reviewed in section \ref{sec:prj_sum} of this work.). 
We further note that much of this work will use a set of ``quasi-canonical'' projective coordinates where one of the conjugate momenta coordinates is exchanged for a quasi-momenta coordinate. 


\paragraph{Outline.}
This paper is outlined as follows:
\begin{small}
\begin{enumerate}
    \item In section \ref{sec:prj_sum}, we review the projective coordinate canonical transformation developed in \cite{peterson2025prjCoord,peterson2025phdThesis,peterson2022nonminimal} (see \cite{peterson2025prjGeomech} or latter chapters of \cite{peterson2025phdThesis} for a more mathematical, geometric, formulation). The Hamiltonian dynamics for perturbed central-force motion are given, along with closed-form solutions for pure Kepler dynamics. We note that this linearizing transformation also involves a transformation of the evolution parameter. Two such parameter are given, both achieving linearization: (1) a parameter \eq{s} defined by \eq{\mrm{d} t = r^2 \mrm{d} s}, and (2) a parameter \eq{\tau} defined by \eq{\mrm{d} t = (r^2/\slang)\mrm{d} \tau} (making \eq{\tau} the true anomaly up to an additive constant).  
    \item In section \ref{sec:VOP_2}, we use the closed-form \eq{\tau}-parameterized Kepler solutions in projective coordinates — with one conjugate momentum coordinate exchanged for a quasi-momenta coordinate — to define a set of orbit elements which, in the unperturbed case, correspond to initial conditions. Variation of parameter is used to obtain the governing equations of motion, which are singularity-free in all cases besides rectilinear motion (when \eq{\slang=0}). The \eq{J_2}-perturbed two-body problem is presented and used for numerical verification. 
    \item[\textbullet] Appendix \ref{sec:ang_momentum} contains some useful relations involving the angular momentum, and its partial derivatives, in projective coordinates. 
    Appendix \ref{sec:prj_STM_new} contains state transition matrices and other matrix relations for Kepler dynamics in projective coordinates (these are used in section \ref{sec:VOP_2}). Appendix \ref{sec:VOP} contains a different approach to the orbit elements that was given previously by the authors. Appendix \ref{apx:dqp_ds_ext} outlines some properties of Hamiltonian systems with a transformation of the evolution parameter. 
\end{enumerate}
\end{small}

\paragraph{Notation \& Conventions.}

This work is written in the coordinate-dominant manner of classic analytical dynamics with all transformations framed as ``passive'' coordinate transformations. Our mathematical arena is only ever regarded as some \eq{\en}-dim real coordinate vector space, \eq{\mbb{R}^\en}, equipped with the usual structures, operations, and abuses of notation.
Some specific notation used in this work is collected below.

 \begin{footnotesize}
\begin{longtable}[htbp]{@{}p{0.13\textwidth} p{0.80\textwidth}@{}}
  $\imat_\en, \, \kd_{ij}, \, \lc_{i_1 \dots i_\en}$
  & $\en$-dim identity matrix, Kronecker delta \textit{symbol}, and Levi-Civita permutation \textit{symbol}, respectively. (usually, \eq{\en=3}). 
\\[1ex]
    $\ibase_i \in\mbb{R}^\en$
    & standard basis for $\mbb{R}^\en$. I.e., columns of \eq{\imat_\en}. 
\\[1ex]
       $ \Somat{\en}, \, \somat{\en} $ &  Lie group of real special orthogonal matrices and its Lie algebra of antisymmetric matrices.
\\[1ex]
       $ \Spmat{2\en}, \, \spmat{2\en} $ &  Lie group of real symplectic matrices and its Lie algebra of Hamiltonian matrices.
\\[2ex]
    $(\tup{r},\tup{\v})$ 
   &inertial cartesian position and velocity/momentum 
    coordinates, $(\tup{r},\tup{\v})\in\mbb{R}^{6}$.  
\\[1ex]
      $(\bartup{q},\bartup{p})$
   & projective coordinates, $(\bartup{q},\bartup{p})=(\tup{q},u,\tup{p},p_\ss{u})\in\mbb{R}^{6+2}$. Sometimes re-ordered as $(\tup{q},\tup{p},u,p_\ss{u})$.
\\[1ex]
    $w:= u^2 p_\ss{u}$ 
    &quasi-momenta coordinate in place of $p_\ss{u}$.
\\[1ex]
   $\tup{\slang},\; \hdge{\tup{\slang}}$ 
   & (specific) angular momentum coordinate vector $\tup{\slang}=\hdge{\tup{\v}}\cdot\tup{r} = \hdge{\tup{p}}\cdot\tup{q} \in\mbb{R}^3 $, and matrix $\hdge{\tup{\slang}}=\tup{r}\wdg \tup{\v} = \tup{q}\wdg \tup{p} \in \somat{3}$. 
\\[2ex]
   $\mscr{K},\, \mscr{H}$ 
   &``cartesian coordinate Hamiltonian'' $\mscr{K}$, and ``projective coordinate Hamiltonian''  $\mscr{H}$.
\\[1ex]
    $V^0,\; V^1$
    & central-force potential $V^0(r)$, and arbitrary perturbing potential $V^1$. 
\\[1ex]
    $\tup{a}^\nc \in\mbb{R}^3$
    & inertial cartesian components of any/all nonconservative perturbing forces.
\\[1ex]
    $\tup{F}:=-\pderiv{V^1}{\tup{r}}+\tup{a}^\nc$
    & inertial cartesian components of total perturbing forces.
\\[1ex]
    $\bartup{\alpha}\in\mbb{R}^4$
    & generalized nonconservative perturbing forces for projective coordinates. Split as $\bartup{\alpha}=(\tup{\alpha},\alpha_\ss{u})\in\mbb{R}^4$.
\\[1ex]
    $\bartup{f}:=-\pderiv{V^1}{\bartup{q}}+\bartup{\alpha}$
    & generalized total perturbing forces for projective coordinates. Split as $\bartup{f}=(\tup{f},f_\ss{u})\in\mbb{R}^4$.
\\[2ex]
    $\dot{\square}:=\diff{\square}{t}$ 
    &derivative ``with respect to'' \eq{t} (time).
\\[1ex]
    $\pdt{\square}:=\diff{\square}{s}$ 
    &derivative ``with respect to'' \eq{s}, where $\mrm{d} t = r^2 \mrm{d}s = u^{-2} \mrm{d}s$
\\[1ex]
    $\rng{\square}:=\diff{\square}{\tau}$
    &derivative ``with respect to'' \eq{\tau}, where $\mrm{d} t = (r^2/\slang)\mrm{d} \tau = 1/(\slang u^2) \mrm{d} \tau$.
\\[2ex]
    $a:=b,\, b=:a$ 
    & $a$ is defined as $b$.
\\[1ex]
    $a\simeq b$
    & $a=b$ under the condition/simplification that $\mag{\tup{q}}=1$ and \eq{\lambda=\htup{q}\cdot\tup{p}=0}. 
\\[2ex]
       ODE     & ordinary differential equation.
 \\[1ex]
      VOP    &  variation of parameters.
     \\[1ex]
      KS    &  Kustaanheimo-Stiefel (two people).
   \\[1ex]
      BF    &  Burdet-Ferrándiz (two people).
  \\[1ex]
     DEF   & Deprit-Elipe-Ferrer (three people).
    \\[1ex]
        LVLH     & local vertical, local horizontal (basis).
\end{longtable}
\end{footnotesize}

\noindent In addition, we note the following general notations and conventions: 
\begin{small}
\begin{itemize}
    \item Matrix notation is used frequently, along with occasional ``cartesian index notation''. For the latter, all indices appear as subscripts and summation is implied over any repeated indices. 
    Almost always, Latin indices such as \eq{i,j,k,l} range from \eq{1} to \eq{3}.
    \item Expressions such as \eq{\tup{u}\in \mbb{R}^\en} are often an abuse of notation indicating that \eq{\tup{u}} is an object that \textit{takes values in} \eq{\mbb{R}^\en}. For any such \eq{\tup{u}}, we make no distinction between \eq{\tup{u}} as an ordered \eq{\en}-tuple, \eq{(u_1,\dots,u_\en)}, and \eq{\tup{u}} as a column vector, \eq{\trn{[u_1 \dots u_\en]}}. 
    \item  The norm of some  \eq{\tup{u}\in \mbb{R}^\en} is denoted by 
    \eq{u:=\mag{\tup{u}}:= \sqrt{\tup{u}\cdot \tup{u}}}, and the normalized unit vector by \eq{\htup{u}:=\tfrac{1}{u}\tup{u}} such that \eq{\htup{u}\cdot\htup{u}=1}.
    \item \sloppy  The ``dot product'' denotes contraction in the usual sense on \eq{\mbb{R}^\en}. That is, for \eq{\tup{u},\tup{\v}\in \mbb{R}^\en}, then \eq{\tup{u}\cdot\tup{\v} = \kd_{ab}u_a \v_b = u_a \v_a} (for \eq{a,b= 1,\dots, \en}). Contraction with some  \eq{M \in\mbb{R}^{\en \times \en}} is denoted either with or without a dot. That is,   \eq{M\tup{u}=M\cdot\tup{u}=M_{ab}u_b \ibase_a} and \eq{\trn{M}\tup{u} = \trn{M}\cdot\tup{u}=\tup{u}\cdot M = M_{ab}u_a \ibase_b = M_{ba}u_b \ibase_a }.   
   \item  Tensor products, \eq{\otimes}, and exterior/wedge products, \eq{\wedge}, also have their usual interpretation on \eq{\mbb{R}^\en}. In particular, the tensor product of any \eq{\tup{u},\tup{\v}\in \mbb{R}^\en} is equivalent to the matrix \eq{\tup{u}\otms\tup{\v}=\tup{u}\trn{\tup{\v}}\in\mbb{R}^{\en\times\en}}, and the exterior product is equivalent to the antisymmetric matrix  \eq{\tup{u}\wdg\tup{\v} = \tup{u}\otms \tup{\v}-\tup{\v}\otms \tup{u} = \tup{u}\trn{\tup{\v}}-\tup{\v}\trn{\tup{u}} \in\somat{\en}\subset \mbb{R}^{\en\times\en}}.
    \item  \textit{\textbf{Hodge dual on \eq{\mbb{R}^3}}.} For any \eq{\tup{u}\in\mbb{R}^3}, the Hodge dual, \eq{\hdge{\tup{u}}\in\somat{3}}, is the  antisymmetric matrix with components given in terms of the 3-dim Levi-Civita permutation symbol by \eq{\hdge{u}_{ij} := \lc_{ijk}u_k}. For \eq{\mbb{R}^3}, this differs from the usual ``axial dual'' or ``cross product matrix'', \eq{\ax{\tup{u}}}, only by a sign: \eq{\hdge{\tup{u}} = -\ax{\tup{u}}}. We note the following relations, specific to \eq{\mbb{R}^3}:
     \begin{flalign} \label{hodge_cord}
     \;\;
      \begin{array}{lcll}
         \hdge{u}_{ij} := \lc_{ijk}u_k 
           &\leftrightarrow& u_i = \tfrac{1}{2}\lc_{ijk}\hdge{u}_{jk}
         \\[3pt]
           \hdge{(\tup{u}\tms \tup{\v})} =  \tup{u} \wdg \tup{\v} &\leftrightarrow &
           \hdge{(\tup{u} \wdg \tup{\v})} = \tup{u} \tms \tup{\v} 
     \end{array}
     &&,&&
     \begin{array}{lll}
          \hdge{\tup{u}}\cdot\tup{\v} =-\tup{u}\tms \tup{\v}  = \tup{\v}\tms \tup{u} = -\hdge{\tup{\v}}\cdot\tup{u} = \tup{u}\cdot\hdge{\tup{\v}}
     \\[3pt]
        \hdge{\tup{u}} \cdot \hdge{\tup{\v}} \cdot \tup{w} = \tup{u}\tms (\tup{\v}\tms  \tup{w}) 
     \end{array}
     &&,&& 
     \begin{array}{lll}
         \hhdge{\tup{u}} =\tup{u}
     \\[3pt]
        \hdge{\tup{u}}\cdot\hdge{\tup{\v}} = \tup{\v}\otms \tup{u} - (\tup{u}\cdot\tup{\v})\imat_3
     \end{array}
     \;\;
     \end{flalign}
     \item \textit{A note on ``\eq{y=y(x)}''.}  We (ab)use the notation \eq{y=y(x)}, or just \eq{y(x)}, to indicate there is something called \eq{y} that is a ``function of'' something called \eq{x}. Similarly, under some transformation \eq{x\leftrightarrow z}, the notation \eq{y(x)} and \eq{y(z)} should be interpreted, respectively, as ``\eq{y} expressed in terms of \eq{x}'' and ``\eq{y} expressed in terms of \eq{z}''. 
    \item \textit{A note on mass.} Throughout this work, the mass is scaled out of the Hamiltonian, kinetic/potential energy, forces, angular momentum, etc. All such quantities are given per unit mass (or, for two-body dynamics, per unit 
    \textit{reduced} mass, \eq{m:=\tfrac{m_a m_b}{m_a+m_b}}). Often, we do not state this explicitly. For instance, when we refer to the angular momentum we always mean the 
    \textit{specific}\footnote{The term ``specific angular momentum'' means the angular momentum per unit mass (or per unit reduced mass). The same applies to terms such as ``specific energy''. }
    angular momentum. 
    \item \textit{A note on inertial cartesian velocity vs.~momentum coordinates.} These are equivalent for the purposes of this work wherein the (reduced) mass is scaled out of everything. As such, there is no difference, in practice, between \textit{inertial cartesian} velocity and momentum coordinates.\footnote{Though, in mathematical/geometric terms, there \textit{is} still a difference:~velocity coordinates are functions on velocity phase space (a tangent bundle) whereas momentum coordinates are functions on phase space (a cotangent bundle) and they therefore cannot be mathematically equal, even in the inertial cartesian case. We ignore such details in this work.}   
\end{itemize}
\end{small}

\section{ORBITAL DYNAMICS IN PROJECTIVE COORDINATES} \label{sec:prj_sum}



 \noindent We present an overview of the ``canonically-extended''  projective coordinate transformation and its application to regularization and linearization of central-force particle dynamics — in particular, Kepler-type dynamics.
 Most of the following was detailed by the authors in \cite{peterson2025prjCoord,peterson2025phdThesis,peterson2025prjGeomech}, where the derivations and details can be found.
The main contribution of this work, the orbit elements, begins in section \ref{sec:VOP_2}.



\paragraph{Original Cartesian Coordinate Formulation.}
Let \eq{(\tup{r},\tup{\v})\in\mbb{R}^6} be cartesian position and velocity 
(or momentum\footnote{In this work, there is no difference, in practice, between inertial cartesian velocity and momentum coordinates since the mass is scaled out. We stress that this is true only for \textit{inertial cartesian} velocity and momentum coordinates. }) 
coordinates in an orthonormal inertial frame. 
We start with the Hamiltonian  and canonical equations of motion for a particle in Euclidean 3-space, subject to conservative forces corresponding to some central-force potential function, \eq{V^{0}(r)}, that depends only on the radial distance.
We also allow for any additional arbitrary perturbing forces with cartesian components \eq{\tup{F}\in\mbb{R}^3}.
The Hamiltonian dynamics in cartesian coordinates \eq{(\tup{r},\tup{\v})\in\mbb{R}^6} are then:
\begin{small}
\begin{align} \label{K0}
    &\mscr{K} \,=\, \tfrac{1}{2} \v^2 \,+\, V^{0}(r)
&& \begin{array}{ll}
      \dot{\tup{r}} \,=\, \pderiv{\mscr{K}}{\tup{\v}} \,=\, \tup{\v} 
&,\qquad 
    \dot{\tup{\v}} \,=\, -\pderiv{\mscr{K}}{\tup{r}} + \tup{F} 
    \,=\, -\pderiv{V^{0}}{r}\htup{r} + \tup{F} 
\end{array}
\end{align}
\end{small}
where \eq{r :=\mag{\tup{r}}} and \eq{\htup{r}:=\tup{r}/r}. 
The perturbing forces may be conservative and/or nonconservative, that is:
    \begin{small}
    \begin{align} \label{Fcart_0}
         \tup{F} \,=\, -\pderiv{V^1}{\tup{r}} + \tup{a}^\nc
    \end{align}
    \end{small}
    where \eq{V^1=V^1(\tup{r},t)} is a potential function modeling any arbitrary conservative forces (e.g., higher-order gravity terms or third body terms), and \eq{\tup{a}^\nc\in\mbb{R}^3} are the cartesian components of any \textit{non}conservative forces (e.g., thrust or drag).
\begin{small}
\begin{itemize}
    \item Conservative perturbations could (and usually would) be included simply by adding their potential, \eq{V^1}, to the Hamiltonian function.\footnote{With \eq{V^1} included in \eq{\mscr{K}}, Hamilton's equation for \eq{\tup{\v}} takes the form \eq{\dot{\tup{\v}}=-\pderiv{\mscr{K}}{\tup{r}} + \tup{a}^\nc =  -\pderiv{V^{0}}{r}\htup{r}  - \pderiv{V^1}{\tup{r}} + \tup{a}^\nc  }, which is equivalent to \eq{\dot{\tup{\v}} = -\pderiv{V^{0}}{r}\htup{r}  + \tup{F} }.  }
    Yet, since all perturbations are completely arbitrary for now, it is easier to lump them all into one object, \eq{\tup{F}}, as in Eq.\eqref{Fcart_0}. 
    \item When the form of \eq{V^{0}(r)} is relevant (for many developments, it is not), then we consider 
    a Kepler-type potential:\footnote{We note that the transformation used in this work is equally effective at linearizing potentials of the more general \textit{Manev}-type, \eq{ V^{0} = -\kconst_1/{r} - \tfrac{1}{2} \kconst_2/{r^2}}, for scalars \eq{\kconst_1,\kconst_2\in\mbb{R}}. This was considered by the authors in \cite{peterson2025prjCoord,peterson2025prjGeomech}. In this work, we limit consideration to Kepler-type dynamics (with perturbations). }
    \begin{small}
    \begin{flalign} \label{Vkep_sum}
     \begin{array}{cc}
     \scrsize{Kepler-type}  \\
     \scrsize{potential} 
    \end{array} 
    &&
        V^{0} \,=\, -\kconst_1 /r 
        \qquad,\qquad 
        \kconst_1 \in \mbb{R}
        &&
    \end{flalign}
    \end{small} 
\end{itemize}
\end{small}

\subsection{A Canonical Transformation for Projective Coordinates} \label{sec:prj_Xform}

The transformation summarized in the following is a modified version of the BF transformation \cite{ferrandiz1987general} that was detailed by the authors in \cite{peterson2025prjCoord,peterson2025phdThesis,peterson2025prjGeomech}. Our transformation differs slightly from the BF transformation at the configuration level, and significantly at the momentum level.

\paragraph{A Canonically-Extended Projective Transformation.}  
We transform from the above cartesian coordinates \eq{(\tup{r},\tup{\v})\in\mbb{R}^6} to
new, redundant, coordinates \eq{(\bartup{q},\bartup{p}) \in \mbb{R}^8} where \eq{\bartup{q}} are configuration coordinates with conjugate momentum coordinates \eq{\bartup{p}}. We partition these coordinates as:\footnote{Later, we will modify the coordinate ordering to \eq{(\tup{q},\tup{p},u,p_\ss{u})} but, for now, this does not matter.}
\begin{small}
\begin{align}
    (\bartup{q},\bartup{p})=(\tup{q},u,\tup{p},p_\ss{u}) \in\mbb{R}^8
    \qquad,\qquad 
    \bartup{q}=(\tup{q},u) \in\mbb{R}^4
    \qquad,\qquad 
    \bartup{p}=(\tup{p},p_\ss{u}) \in\mbb{R}^4
\end{align}
\end{small} 
The transformation begins by specifying a point transformation \eq{\gam:\bartup{q}\mapsto\tup{r}} subject to a constraint: 
\begin{small}
\begin{flalign} \label{PT}
    && \gam:\mbb{R}^4\to\mbb{R}^3 \;,
     \qquad \tup{r} \,=\, \gam(\bartup{q}) \,=\, \tfrac{1}{u} \htup{q}
     \;,
     &&
     \scrsize{subject to} \;\; \varphi(\tup{q}) = \mag{\tup{q}} -1 =0
      \;,
     &&
      \rnk \rmd \gam = 3 
      \quad
\end{flalign}
\end{small}
where \eq{\htup{q}:=\tup{q}/q} with \eq{q:=\mag{\tup{q}}\neq \mag{\bartup{q}}}. 
We \textit{specify} the above point transformation and constraint, which then induce a conjugate momentum coordinate transformation as detailed by the authors in \cite{peterson2025prjCoord,peterson2025phdThesis}. In short:~associating to \eq{\varphi} a Langrage multiplier \eq{\lambda} (which serves like an ``extra'' momentum coordinate), then we obtain a momentum coordinate transformation \eq{\bartup{p}=(\tup{p},p_\ss{u})\leftrightarrow (\tup{\v},\lambda)} as follows \cite{peterson2025prjCoord}:
\begin{small}
\begin{align} \label{Pxform_finaly1}
     \bartup{p} =
    \trn{B}
    \begin{pmatrix}
     \tup{\v} \\ \lambda
    \end{pmatrix} 
\quad \leftrightarrow \quad 
     \begin{pmatrix}
     \tup{\v} \\ \lambda
    \end{pmatrix}  = \invtrn{B} \bartup{p}
\qquad,\qquad
     B =
     \begin{pmatrix}
         \pderiv{\tup{r}}{\bartup{q}}
         \\[5pt] \pderiv{\varphi}{\bartup{q}}
    \end{pmatrix}
    \in \mathbb{R}^{4 \times 4}
\end{align}
\end{small}
The point transformation in Eq.\eqref{PT} then has a ``canonical extension'',  \eq{\Gam:(\bartup{q},\bartup{p})\mapsto (\tup{r},\tup{\v})}, with \eq{\rnk \rmd \Gam = 6}, 
given 
by:\footnote{The point transformation \eq{\gam:\mbb{R}^4\ni(\tup{q},u)\mapsto \tup{r}\in\mbb{R}^3}  is
a submersion; it is  surjective with \eq{\text{rnk} \rmd \gam \equiv \text{rnk} \pderiv{\tup{r}}{(\tup{q},u)}=3}. It is then lifted to a ``canonical submersion'', \eq{\Gam:\mbb{R}^8\ni(\tup{q},u,\tup{p},p_\ss{u})\mapsto (\tup{r},\tup{\v})\in\mbb{R}^6}, which is surjective with \eq{\text{rnk} \rmd \Gam \equiv \rnk \pderiv{(\tup{r},\tup{\v})}{(\tup{q},u,\tup{p},p_\ss{u})} =6}. Note these are actually \textit{local} submersions as the domain and codomains of these maps are not all of the indicated \eq{\mbb{R}^\en}; they are subsets excluding the cases \eq{u=0} and \eq{q=\mag{\tup{q}}=0}.  }
\begin{small}
\begin{flalign} \label{PT_0}
&&
\begin{array}{cc}
   \Gam: \mbb{R}^8\to \mbb{R}^6 
   \\[3pt]
   (\bartup{q},\bartup{p})\mapsto (\tup{r},\tup{\v})
\end{array}
\; \left\{ \;\;
\boxed{\begin{array}{rlllll}
      \tup{r} &\!\!\!=\, \tfrac{1}{u} \htup{q} 
\\[5pt]
     \tup{\v} &\!\!\!=\,  uq   (\imat_3 - \htup{q}\otms\htup{q})\cdot\tup{p} \,-\, u^2 p_\ss{u} \htup{q}
    \;=\, -u (\tup{q}\wdg\tup{p})\cdot\htup{q} - u^2 p_\ss{u} \htup{q}
\end{array} } \right.
&&,&&
\begin{array}{lllll}
     \varphi = q -1 = 0  \\[6pt]
      \lambda = \htup{q}\cdot\tup{p}
\end{array}
\qquad
\end{flalign}
\end{small}
where the constraint \eq{\varphi} is built into the derivation of the momenta transformation, which includes the above Lagrange multiplier \eq{\lambda}
associated with \eq{\varphi}.\footnote{In previous work, the authors instead posed the constraint function as \eq{\varphi = \tfrac{1}{2}(q^2-1)=0}. This changes little other than the fact that the associated \eq{\lambda} is then instead given by \eq{\lambda = \tfrac{1}{q^2}\tup{q}\cdot\tup{p}} (as opposed to the present, equivalent, formulation with \eq{\lambda = \tfrac{1}{q}\tup{q}\cdot\tup{p}=\htup{q}\cdot\tup{p}}). }
A method of inverting the above is given below in Eq.\eqref{qp_rv_0}. 
First, we note that since the transformation in Eq.\eqref{PT_0} is not time-dependent, then it takes the cartesian coordinate Hamiltonian, \eq{\mscr{K}}, to a projective coordinate Hamiltonian, \eq{\mscr{H}}, given simply by direct substitution, \eq{\mscr{H}=\mscr{K}\circ \Gam}. This leads to:
    \begin{small}
    \begin{align} \label{Ham0}
    \mscr{K} = \tfrac{1}{2} \v^2 + V^{0}(r)
    \qquad\Rightarrow \qquad
      \boxed{\mscr{H} \,=\,   \tfrac{1}{2} u^2 \big( \slang^2 +  u^2 p_\ss{u}^2 \big)  \,+\, V^{0}(u) }
      \qquad,\qquad 
      \slang^2 = q^2 p^2 - (\tup{q}\cdot\tup{p})^2
    \end{align}
    \end{small}
    where \eq{\slang^2} is the angular momentum magnitude (noted below), and where our (ab)use of notation  \eq{V^{0}(u)} simply indicates the original \eq{V^{0}(r)}  expressed in the new coordinates using \eq{\tup{r}=\tfrac{1}{u}\htup{q}} and, thus, \eq{r=\mag{\tup{r}}=1/u}.


\begin{small}
\begin{notesq}
    \textit{Angular momentum.} It is worth noting that the 
    angular momentum, \eq{\tup{\slang} = \hdge{\tup{\v}}\cdot \tup{r} = \tup{r}\tms \tup{\v}\in\mbb{R}^3}, takes the same form when expressed in terms of \eq{(\tup{r},\tup{\v})} or \eq{(\bartup{q},\bartup{p})}. It is independent of \eq{(u,p_\ss{u})}, depending only on \eq{(\tup{q},
    \tup{p})}:\footnote{Note \eq{\slang_{ij}  = \lc_{ijk}\slang_k \leftrightarrow \slang_i = \tfrac{1}{2}\lc_{ijk}\slang_{jk}} are Hodge duals of one another. We write simply \eq{\slang_{ij}} rather than \eq{\hdge{\slang}_{ij}}. }
    \begin{small}
    \begin{align}  \label{lang_qp_SUM}
        \hdge{\tup{\slang}} 
    \,=\, \tup{r}\wdg \tup{\v} \,=\, \tup{q}\wdg\tup{p}
    &&,&&
    \tup{\slang} 
    \,=\, \hdge{\tup{\v}}\cdot \tup{r}  \,=\, \hdge{\tup{p}}\cdot \tup{q}
    &&,&&
    \slang^2  \,=\, r^2 \v^2 - (\tup{r}\cdot\tup{\v})^2 \,=\, q^2 p^2 - (\tup{q}\cdot\tup{p})^2
    \end{align}
    \end{small}
     Some useful angular momentum relations are given Appendix \ref{sec:ang_momentum}.   
\end{notesq}
\end{small}

\begin{notesq} \textit{Two ``extra'' integrals of motion.}
    It was shown in \cite{peterson2025prjCoord,peterson2025phdThesis,peterson2022nonminimal,peterson2025prjGeomech} that the new Hamiltonian \eq{\mscr{H}} for the redundant phase space coordinates, \eq{(\bartup{q},\bartup{p})\in\mbb{R}^8}, permits two additional integrals 
    of motion.\footnote{``Integral of motion'' is perhaps misleading terminology in this context; the functions in Eq.\eqref{qlam_0} do not represent a conserved physical quantity for a particle moving in Euclidean 3-space (the actual system we are modeling). They are, rather,  ``kinematic constants'' which arise from artificially increasing the dimension of phase space. However, in a metathetical sense, they are indeed true integrals of motion for the new Hamiltonian system described by \eq{\mscr{H}}.} 
    These are:
    \begin{small}
    \begin{flalign} \label{qlam_0}
         \begin{array}{cc}
         \scrsize{integrals}  \\
         \scrsize{of motion} 
    \end{array} 
    \left. \quad
    \begin{array}{ll}  
        q = \mag{\tup{q}} 
         \\[5pt]
        \lambda =  \htup{q}\cdot\tup{p} 
    \end{array} \right.
    &&
    \begin{array}{ll}
        \dot{q} = \pbrak{q}{\mscr{H}} + \pderiv{q}{\bartup{p}} \cdot \bartup{f}  = 0 
       \\[5pt]
        \dot{\lambda}  = \pbrak{\lambda}{\mscr{H}} + \pderiv{\lambda}{\bartup{p}} \cdot \bartup{f}  = 0
    \end{array} 
    \qquad \Rightarrow \qquad 
     \begin{array}{ll}
         q = q_\zr 
         \\[6pt]
           \lambda = \lambda_\zr  
    \end{array}
    &&
    \end{flalign}
    \end{small}
    where \eq{\bartup{f}=(\tup{f},f_\ss{u})} are arbitrary generalized perturbing forces (discussed soon). 
    That is, the constraint function, \eq{\varphi = q-1}, and Lagrange multiplier, \eq{\lambda=\htup{q}\cdot\tup{p}}, turn out to be 
    integrals of motion of the transformed Hamiltonian system \textit{even in the presence of arbitrary, perhaps nonconservative, forces}.\footnote{Note that \eq{q} being an integral of motion is equivalent to \eq{\varphi = q -1} being an integral of motion. This means that \eq{q=q_\zr = \mrm{const.}} is automatically satisfied. Enforcing the specific value \eq{q=1} amounts to limiting consideration to initial conditions for which \eq{q_\zr=1}.  }
\end{notesq}

\paragraph{Inverse Transformation \& Interpreting the New Coordinates.}
Eq.\eqref{PT_0} is a (local) submersion, \eq{\Gam:\mbb{R}^8 \to \mbb{R}^6}, with \eq{\rnk \rmd \Gam = 6} and no  global unique inverse. However, we may use the integrals of motion in Eq.\eqref{qlam_0} to construct an inverse that is parameterized by some chosen, constant, values of \eq{q=q_\zr>0} and \eq{\lambda=\lambda_\zr}:  
\begin{small}
\begin{align}  \label{qp_rv_0_gen}
\begin{array}{cc}
     \scrsize{restrict to}  \\
     \fnsize{$q=q_\zr$}\\
     \fnsize{$\lambda=\lambda_\zr$} 
\end{array} 
\quad \Rightarrow \qquad
 \inv{\Gam}:(\tup{r},\tup{\v}) \mapsto (\bartup{q},\bartup{p})
\quad \left\{\quad
\begin{array}{llll}
 \tup{q} \,=\,  q_\zr\htup{r} 
      &,\quad 
      \tup{p} \,=\, \tfrac{r}{q_\zr}( \imat_3 -  \htup{r}\otms\htup{r})\cdot \tup{\v} + \lambda_\zr \htup{r}
  \\[5pt]
    u \,=\, 1/r 
   &, \quad 
   p_\ss{u} \,=\, -r^2 \htup{r}\cdot\tup{\v}
\end{array} \right.
\end{align}
\end{small}
The above is \textit{not} a true inverse of Eq.\eqref{PT_0}; it is only unique up to some chosen values \eq{q_\zr>0} and \eq{\lambda_\zr} (set by initial conditions).
We are free to limit consideration to any such values that we please as they place no restrictions on \eq{(\tup{r},\tup{\v})}. 
We only consider the specific values
\eq{q = 1} and \eq{\lambda=0}, in which case the above 
     simplifies to:\footnote{\eq{q=1} ensures that \eq{\tup{q}=\htup{q}=\htup{r}} is the radial unit vector. The value \eq{\lambda=0} is chosen because we see no advantage in considering any other value and because  \eq{\lambda=0} leads to \eq{\tup{p}} having a ``nice'' interpretation.}
\begin{small}
\begin{align}  \label{qp_rv_0}
\begin{array}{cc}
     \scrsize{restrict to}  \\
      \fnsize{$q=1$}\\
     \fnsize{$\lambda=0$} 
\end{array} 
\quad \Rightarrow \qquad
 \inv{\Gam}:(\tup{r},\tup{\v}) \mapsto (\bartup{q},\bartup{p})
\quad \left\{\quad
\boxed{ \begin{array}{llll}
     \tup{q} \,=\,  \htup{r} 
      &,\quad \tup{p} \,=\, -\hdge{\tup{\slang}}\cdot\htup{r} \,=\,  r( \imat_3 -  \htup{r}\otms\htup{r})\cdot \tup{\v} 
  \\[5pt]
  u \,=\, 1/r 
   &, \quad 
   p_\ss{u} \,=\, -r^2 \htup{r}\cdot\tup{\v}
\end{array} }  \right.
 \begin{array}{llll}
          =\, r^2 \dot{\htup{r}}
     \\[6pt]
           =\, -r^2 \dot{r}
    \end{array}
\end{align}
\end{small}
   One does not need to do anything to enforce the ``constraints'' \eq{q=1} and \eq{\lambda=0} other than to transform initial conditions \eq{(\tup{r}_\zr,\tup{\v}_\zr) \mapsto (\bartup{q}_\zr,\bartup{p}_\zr)} using Eq.\eqref{qp_rv_0}. This places no restrictions on \eq{(\tup{r}_\zr,\tup{\v}_\zr)} and it guarantees that  \eq{q_\zr=1} and \eq{\lambda_\zr=0}. Since \eq{q} and \eq{\lambda} are integrals of motion, it also guarantees that  \eq{q=1} and \eq{\lambda=0} hold for all time along any solution curve of the projective coordinate Hamiltonian system. 

\begin{remrm}[\textit{Simplified Relations}]
We use  ``\eq{\simeq}'' to indicate relations which have been simplified using \eq{q=1} and \eq{\lambda=\htup{q}\cdot\tup{p}=0}.
Note the latter is equivalent to \eq{\tup{q}\cdot\tup{p}=0} such that \eq{\tup{p}\simeq \tup{\slang}\tms\htup{q}\simeq \tup{\slang}\tms\tup{q} }. That is, \eq{\tup{q}}, \eq{\tup{p}}, and \eq{\tup{\slang}=\tup{q}\tms\tup{p}} are mutually orthogonal and satisfy: 
\begin{small}
\begin{flalign} \label{qpl_rels}
    \qquad
    \begin{array}{lll}
          \tup{\slang}=\tup{q}\tms \tup{p}
        \\[4pt]
        \slang^2 = q^2 p^2-(\tup{q}\cdot\tup{p})^2 
        \\[4pt]
          \tup{q}\cdot\tup{\slang} = \tup{p}\cdot\tup{\slang}  = 0
    \end{array}
    &&
    \begin{array}{cc}
        \fnsize{using:} \\
         q=1  \\
         \lambda=\htup{q}\cdot\tup{p}=0 
    \end{array}
     \Rightarrow \;\; 
    \left\{ \quad
    \begin{array}{ll}
        \tup{q} \simeq \htup{q} \simeq \htup{p}\tms \htup{\slang} 
      \\[4pt]
        \tup{p} \simeq \tup{\slang}\tms \htup{q} 
    \\[4pt]
          \slang^2 \simeq p^2  
    \end{array} \right.
    \;\;,\quad 
     \begin{array}{lll}
            \hdge{\tup{\slang}}\cdot\tup{q} \simeq -q^2\tup{p} \simeq -\tup{p}
            \\[4pt]
            \hdge{\tup{\slang}}\cdot\tup{p} \simeq p^2\tup{q} \simeq \slang^2 \tup{q} 
            \\[4pt]
              \imat_3 \simeq \htup{q} \otms\htup{q} + \htup{p} \otms\htup{p} + \htup{\slang} \otms\htup{\slang}
        \end{array}
    \qquad
\end{flalign}
\end{small}
In particular, \eq{\tup{q}} and \eq{\tup{p}} directly define the inertial cartesian components of the local vertical local horizontal (LVLH) orthonormal basis associated with the particle's instantaneous position: 
\begin{small}
\begin{align} \label{qpl_rels2}
 \begin{array}{lll}
         \begin{array}{cc}
             \scrsize{orthonormal}  \\
             \scrsize{LVLH basis}
            \end{array}
            \!=\, \{\htup{t}_r,\htup{t}_\tau, \htup{t}_\slang \}
            =  \{\htup{q},-\hdge{\htup{\slang}}\cdot\htup{q},\htup{\slang}\} \simeq \{\htup{q},\htup{p},\htup{\slang}\}
        \qquad\qquad 
        \scrsize{with:} \;\; q^2 \simeq 1\;,\;\; \slang^2 \simeq p^2
\end{array}
\end{align}
\end{small}
\end{remrm}

\begin{remrm} \label{rem:qp=0} 
We reiterate that the relations  \eq{q=1} and \eq{\lambda=\htup{q}\cdot\tup{p}=0} — which further imply those in Eq.\eqref{qpl_rels}-Eq.\eqref{qpl_rels2} —
hold along
any\footnote{Specifically, any solution curve of the transformed Hamiltonian system, including arbitrary perturbations, for \eq{\mscr{H}} in Eq.\eqref{Ham0}.}
solution curve
\eq{(\bartup{q}_t,\bartup{p}_t)} with initial conditions \eq{(\bartup{q}_\zr,\bartup{p}_\zr)} satisfying \eq{q_\zr=1} and \eq{\lambda_\zr=\htup{q}_\zr\cdot\tup{p}_\zr=0}. Equivalently,  they hold along any solution curve with initial conditions \eq{(\bartup{q}_\zr,\bartup{p}_\zr)} satisfying Eq.\eqref{qp_rv_0} for some given \eq{(\tup{r}_\zr,\tup{\v}_\zr)} — this places no restrictions on \eq{(\tup{r},\tup{\v})}. Unless stated otherwise, Eq.\eqref{qp_rv_0} is always chosen as our default inverse transformation \eq{(\tup{r},\tup{\v})\mapsto (\bartup{q},\bartup{p})}. 
\end{remrm}

\paragraph{Transformed Dynamics \& Generalized Forces.}
 When the projective coordinate transformation detailed above is combined with an appropriate evolution parameter transformation, the result fully linearizes Kepler-type dynamics (or Manev-type). First, we quickly note the time-parameterized dynamics and generalized forces for the projective coordinates. 
 The \eq{t}-parameterized Hamiltonian dynamics in projective coordinates — which are still nonlinear — are obtained in he usual way using \eq{\mscr{H}} from Eq.\eqref{Ham0}:
\begin{small}
\begin{align} \label{qpdot_0}
    \mscr{H} =  \tfrac{1}{2} u^2 \big( \slang^2 +  u^2 p_\ss{u}^2 \big)  + V^{0}(u) 
&&,&&
\begin{array}{llll}
     \dot{\tup{q}} = \pderiv{\mscr{H}}{\tup{p}} \,=\,  -u^2 \hdge{\tup{\slang}}\cdot \tup{q}
     &,
\\[6pt]
    \dot{u}  = \pderiv{\mscr{H}}{p_u}  \,=\,  u^4 p_\ss{u} 
    &,
 \end{array}
\quad
\begin{array}{llll}
    \dot{\tup{p}} = -\pderiv{\mscr{H}}{\tup{q}} + \tup{f}
    &\!\!\! =\, - u^2 \hdge{\tup{\slang}}\cdot \tup{p} + \tup{f}
\\[6pt]
     \dot{p}_\ss{u}  = -\pderiv{\mscr{H}}{u} + f_\ss{u}
      &\!\!\!=\,    - u(\slang^2 +  2 u^2 p_\ss{u}^2 )   -  \pderiv{V^{0}}{u}  + f_\ss{u}
\end{array} 
\end{align}
\end{small}
where \eq{V^0} is the transformed central-force potential (using \eq{r=1/u}), and 
where \eq{\bartup{f}=(\tup{f},f_\ss{u})\in\mbb{R}^4} are generalized perturbing forces.
They are obtained from the inertial cartesian components of the total perturbing force, \eq{\tup{F}=-\pderiv{V^1}{\tup{r}}+\tup{a}^\nc \in\mbb{R}^3}, through the usual relations with the Jacobian of \eq{\tup{r}(\bartup{q})}:   
\begin{small}\begin{align} \label{Ftotal_qp}
\tup{F}:=-\pderiv{V^1}{\tup{r}}+\tup{a}^\nc
&&,&&
   \boxed{ 
\begin{array}{lllll}
         \tup{f} :=\, -\pderiv{V^{1}}{\tup{q}} + \tup{\alpha} 
    &\!\! =\,
    \trn{\pderiv{\tup{r}}{\tup{q}}} \cdot  \tup{F}
       \,=\,  \tfrac{1}{u q} ( \imat_3 -  \htup{q}\otms\htup{q} ) \cdot \tup{F}
\\[4pt]
          f_\ss{u} :=\, -\pderiv{V^{1}}{u} + \alpha_\ss{u}
           &\!\! =\,
    \pderiv{\tup{r}}{u} \cdot  \tup{F}
    \,=\,  -\tfrac{1}{u^2}\htup{q}\cdot\tup{F}
\end{array}
 }
&&,&&
\begin{array}{llll}
     \tup{q}\cdot\tup{f} = 0 
\\[3pt]
     \tup{p} \cdot \tup{f} 
     = \tfrac{1}{q} \htup{q}\cdot\hdge{\tup{\slang}} \cdot \tup{f}
     = \tfrac{1}{u q^2}\htup{q}\cdot\hdge{\tup{\slang}} \cdot \tup{F}
     \,\simeq  \tfrac{1}{u}\tup{p}\cdot\tup{F}
 \\[3pt]
     \tup{\slang} \cdot \tup{f} = \tfrac{1}{uq}\tup{\slang}\cdot\tup{F}
\end{array}
\end{align}\end{small}
where \eq{ -\pderiv{V^1}{\bartup{q}} = - \trn{\pderiv{\tup{r}}{\bartup{q}}} \cdot \pderiv{V^1}{\tup{r}} } and \eq{\bartup{\alpha}= \trn{\pderiv{\tup{r}}{\bartup{q}}} \cdot  \tup{a}^\nc} account for arbitrary conservative and nonconservative perturbations, respectively. 
These terms also, individually, satisfy the above relations (e.g., \eq{\tup{q}\cdot \pd_{\tup{q}} V^1 =0} and \eq{\tup{q}\cdot \tup{\alpha} =0}). 
As mentioned at Eq.\eqref{Fcart_0}, any conservative perturbations would usually be accounted for simply by adding \eq{V^1} itself to the Hamiltonian. That is, by adding \eq{V^1(\bartup{q},t)} to \eq{\mscr{H}}, where  \eq{V^1(\bartup{q},t)} is taken to mean the original  \eq{V^1(\tup{r},t)} expressed in the new coordinates using \eq{\tup{r}=\tfrac{1}{u}\htup{q}}. Yet, for the present purposes, it is more convenient to simply absorb any and all arbitrary perturbations into the cartesian components \eq{\tup{F}}, which transforms to the above generalized forces \eq{\bartup{f}=(\tup{f},f_\ss{u})}.


\begin{remrm} \label{rem:Frad}
    If \eq{\tup{F}} is purely radial, that is, if \eq{\tup{F} = F_r \htup{r} = F_r \htup{q}}, then \eq{\tup{f}=0}. As such, any central-forces of any kind — conservative or nonconservative — are absent from the ODEs for \eq{(\tup{q},\tup{p})} (rotational motion), appearing only in those for \eq{(u,p_\ss{u})} (radial motion). 
\end{remrm}

\paragraph{Recovering Cartesian Coordinate Solutions.}
Any dynamics presented in this work using the projective coordinates are for the transformation \eq{(\bartup{q},\bartup{p})\mapsto(\tup{r},\tup{\v})} given in Eq.\eqref{PT_0}.
That same transformation can, of course, be used to recover \eq{(\tup{r},\tup{\v})} solutions from any \eq{(\bartup{q},\bartup{p})} solutions.
However, when converting numerical solutions, the transformation in Eq.\eqref{PT_0} may be simplified considerably using the integrals of motion \eq{q=1} and \eq{\lambda=\htup{q}\cdot\tup{p}=0}, leading to:
\begin{small}
\begin{align} \label{rv_qu_simp}
\begin{array}{lllllll}
      \tup{r} \,=\, \tfrac{1}{u}\htup{q} 
\\[4pt]
      \tup{\v} \,=\, -u(\tup{q}\wdg\tup{p})\cdot\htup{q} - u^2 p_\ss{u} \htup{q} 
\end{array}
\qquad\quad
\xRightarrow[\htup{q}\cdot\tup{p}\,=\,0]{q\,=\,1}
\qquad \quad
 \boxed{ \begin{array}{llll}
    \tup{r}_t  \,\simeq\, \tfrac{1}{u_t}\tup{q}_t
\\[4pt]
    \tup{\v}_t 
    \,\simeq\, 
    u_t\tup{p}_t - w_t \tup{q}_t 
\end{array} }
\qquad w:=u^2 p_\ss{u}
\end{align}
\end{small}
where \eq{w:= u^2 p_\ss{u}} is introduced for later convenience and where the above still holds with \eq{t} exchanged for any other evolution parameter of choice. 
So long as \eq{(\bartup{q}_t , \bartup{p}_t)} is a solution curve starting with \eq{q_\zr=1} and \eq{\lambda_\zr=\htup{q}_\zr\cdot\tup{p}_\zr=0}, then \eq{(\tup{r}_t,\tup{\v}_t)} solutions recovered using the above simplified relations will be \textit{numerically} equivalent to those recovered using the full projective transformation in Eq.\eqref{PT_0}. 

     \textit{Caution:} we stress that the simplified relations in Eq.\eqref{rv_qu_simp} above are \textit{not} the actual map for the projective transformation \eq{(\tup{q},u,\tup{p},p_\ss{u})\mapsto (\tup{r},\tup{\v})} that should be used to transform the Hamiltonian and forces, obtain equations of motion, etc. 
 The actual transformation is given in Eq.\eqref{PT_0}. 
 Yet, when limiting consideration to solution curves satisfying \eq{q=q_\zr=1} and \eq{\htup{q}\cdot\tup{p}=\htup{q}_\zr\cdot\tup{p}_\zr=0} (which we are always free to do), then, along such solution curves, the above is \textit{numerically} equivalent to Eq.\eqref{PT_0}. 

\subsection{Central-Force Dynamics with a Transformation of the Evolution Parameter} \label{sec:ext_cen} 

We now use the projective coordinates described above in conjunction with a transformation the evolution parameter from the time, \eq{t}, to two new evolution parameters, \eq{s} and \eq{\tau}, defined 
by:\footnote{The evolution parameter \eq{s} is the same as used by Burdet \cite{burdet1969mouvement,Burdet+1969+71+84}, \eq{\tau} (true anomaly) is equivalent to the one used by Vitins \cite{vitins1978keplerian},  and Ferrándiz considered both \eq{s} and \eq{\tau} \cite{ferrandiz1987general}.} 
\begin{small}
\begin{align} \label{dtds_0}
      \mrm{d} t \,=\, r^2 \mrm{d} s \,=\, u^{-2} \mrm{d} s
\qquad,\qquad
     \mrm{d} t \,=\, \tfrac{r^2}{\slang} \mrm{d} \tau 
     \,=\, \tfrac{1}{\slang u^2} \mrm{d} \tau 
\qquad,\qquad
    \mrm{d} \tau \,=\, \slang \mrm{d} s
\end{align} 
\end{small}
where \eq{\slang^2 =q^2 p^2 -(\tup{q}\cdot\tup{p})^2 } and 
where \eq{\tau} is equivalent to the true anomaly (up to an additive constant). Using the notation \eq{ \dot{\square} := \diff{\square}{t} }, \eq{ \pdt{\square} := \diff{\square}{s} }, and \eq{ \rng{\square} := \diff{\square}{\tau}}, we note that first-order ODEs transform simply by scaling:
\begin{small}
\begin{align}
     \pdt{\square} := \diff{\square}{s} \,=\, \pdt{t}\dot{\square}
\qquad,\qquad  
     \rng{\square} := \diff{\square}{\tau}  \,=\, \rng{t}\dot{\square} \,=\, \tfrac{1}{\slang} \pdt{\square}
&&
\fnsize{with: } \quad   \boxed{ \pdt{t} = r^2 =  {1}/{u^2} \;\;, \quad  \rng{t} = {r^2}/{\slang} = {1}/{(\slang u^2)} }
\end{align}
\end{small}
That is, using the Hamiltonian dynamics in Eq.\eqref{qpdot_0}, the \eq{s}-parameterized dynamics for \eq{(\bartup{q},\bartup{p})=(\tup{q},u,\tup{p},p_\ss{u})} are given by:
\begin{small}
\begin{flalign} \label{dqp_s_0}
\qquad
\pdt{t} = \tfrac{1}{u^2} 
\;\;,\quad
\left.\begin{array}{lll}
      \pdt{\bartup{q}} = \pdt{t}\pderiv{\mscr{H}}{\bartup{p}}
\\[5pt]
     \pdt{\bartup{p}} = \pdt{t}(-\pderiv{\mscr{H}}{\bartup{q}} + \bartup{f} )
\end{array} \right\}
\quad \Rightarrow  &&
\begin{array}{lllll}
       \pdt{\tup{q}} 
       \,=\,   -\hdge{\tup{\slang}}\cdot\tup{q} 
\\[5pt]  
    \pdt{\tup{p}} 
     \,=\,  -\hdge{\tup{\slang}}\cdot\tup{p}  + \pdt{t}\tup{f} 
\end{array}
\quad,\qquad 
\begin{array}{lllll}
      \pdt{u} 
       \,=\,   u^2 p_\ss{u}
\\[5pt]
      \pdt{p}_\ss{u} 
     \,=\, -\tfrac{1}{u}\big( \slang^2 + 2 u^2 p_\ss{u}^2 \big)   - \pdt{t}\pderiv{V^{0}}{u} + \pdt{t} f_\ss{u}
\end{array}
&&
\end{flalign}
\end{small}
with \eq{\bartup{f}=(\tup{f},f_\ss{u})} as in Eq.\eqref{Ftotal_qp}. 
The \eq{\tau}-parameterized dynamics are obtained similarly, with \eq{\rng{t}} replacing \eq{\pdt{t}}:
\begin{small}
\begin{flalign} \label{dqp_ta_0}
\qquad
\rng{t} = \tfrac{1}{\slang u^2} 
\;\;,\quad 
\left.\begin{array}{lll}
      \rng{\bartup{q}} = \rng{t}\pderiv{\mscr{H}}{\bartup{p}}
\\[5pt]
     \rng{\bartup{p}} = \rng{t}(-\pderiv{\mscr{H}}{\bartup{q}} + \bartup{f} )
\end{array}\right\}
\quad \Rightarrow &&
\begin{array}{lllll}
      \rng{\tup{q}}    \,=\, 
      -\hdge{\htup{\slang}}\cdot\tup{q} 
\\[5pt]  
    \rng{\tup{p}} 
      \,=\,  -\hdge{\htup{\slang}}\cdot\tup{p}   + \rng{t}\tup{f} 
\end{array}
\quad,\qquad 
\begin{array}{lllll}
      \rng{u} \,=\,  \tfrac{1}{\slang} u^2 p_\ss{u}
\\[5pt]
     \rng{p}_\ss{u}  
      \,=\, -\tfrac{1}{\slang u}\big( \slang^2 + 2 u^2 p_\ss{u}^2 \big)  - \rng{t}\pderiv{V^{0}}{u}  + \rng{t} f_\ss{u} 
\end{array}
&&
\end{flalign}
\end{small}
We note that, unlike the the \eq{t}-parameterized ODEs in Eq.\eqref{qpdot_0}, the above \eq{s}- or \eq{\tau}-parameterized 
ODEs in Eq.\eqref{dqp_s_0} or Eq.\eqref{dqp_ta_0} are \textit{not} true Hamiltonian systems (with nonconservative perturbations). They are \textit{conformally}-Hamiltonian systems (with nonconservative perturbations) for the conformal scaling factors \eq{\pdt{t}} and \eq{\rng{t}}, respectively. Yet, by using the extended phase space, one may recover a Hamiltonian structure (with nonconservative perturbations) for the \eq{s}- or \eq{\tau}-parameterized dynamics. While this not overly important for the purposes of this work, the details are given in Appendix \ref{apx:dqp_ds_ext}.

\begin{small}
\begin{itemize}
    \item \textbf{\textit{A quasi-momenta coordinate.}}  
    Though not necessary,
    it is often convenient to exchange the conjugate momentum coordinate \eq{p_\ss{u}} for a quasi-momentum coordinate \eq{w:=u^2 p_\ss{u}}. That is,
    \begin{small}
    \begin{align}
       w  :=\,  u^2 p_\ss{u}  \,=\, \pdt{u} \,=\,  -\dot{r}
         \quad \leftrightarrow \quad
        p_\ss{u} \,=\, w/u^2  \,=\, r^2 w   \,=\,  -\pdt{r}
    \end{align}
    \end{small}
    For instance, if we replace the conjugate pair \eq{(u,p_\ss{u})} with the the non-conjugate pair \eq{(u,w)}, then the above ODEs in Eq.\eqref{dqp_s_0} and Eq.\eqref{dqp_ta_0} are equivalent to the following, more linear, ODEs:
    \begin{small}
    \begin{flalign} \label{uw_eom_gen}
    &&
    \boxed{\begin{array}{llllll}
         \pdt{\tup{q}}  \,=\, 
      -\hdge{\tup{\slang}}\cdot\tup{q} 
      &,\quad 
      \pdt{\tup{p}}  \,=\, 
      -\hdge{\tup{\slang}}\cdot\tup{p} + \pdt{t}\tup{f}
    \\[4pt]
       \pdt{u} \,=\, w
        &,\quad 
         \pdt{w} = -\slang^2 u - \pderiv{V^0}{u} + f_\ss{u}
    \end{array}
    \qquad\quad  \scrsize{or,}  \qquad\quad 
     \begin{array}{llllll}
         \rng{\tup{q}}  \,=\, 
      -\hdge{\htup{\slang}}\cdot\tup{q} 
      &,\quad 
      \rng{\tup{p}}  \,=\, 
      -\hdge{\htup{\slang}}\cdot\tup{p} + \rng{t}\tup{f}
    \\[4pt]
       \rng{u} \,=\,  w/\slang 
        &,\quad 
         \rng{w} = -\slang u - \tfrac{1}{\slang} \pderiv{V^0}{u} +   \tfrac{1}{\slang} f_\ss{u}
    \end{array} }
    &&
    \end{flalign}
    \end{small}
    where the \eq{(\tup{q},\tup{p})} part is unchanged.
    Interestingly, the coordinate \eq{w=u^2 p_\ss{u}} and the conformal scale factor \eq{\pdt{t}=1/u^2} cancel out in such a way that the above \eq{s}-parameterized ODEs for \eq{(u,w)} obey Hamilton's canonical equations in their usual form (without the scale factor). That is, it can be verified that the above ODEs follow from:
    \begin{small}
    \begin{align}
         \mscr{H} = \tfrac{1}{2} u^2 ( \slang^2 +  u^2 p_\ss{u}^2 )  + V^{0} \,=\,  \tfrac{1}{2} ( u^2  \slang^2 +  w^2 )  + V^{0}
         &&,&&
         \begin{array}{llllll}
                \pdt{\tup{q}} = \pdt{t}\pderiv{\mscr{H}}{\tup{p}} 
                 &, \quad
                  \pdt{\tup{p}} = -\pdt{t}\pderiv{\mscr{H}}{\tup{q}} +\pdt{t}\tup{f}
           \\[5pt]
                \pdt{u} = \pderiv{\mscr{H}}{w}
                &, \quad \pdt{w} = -\pderiv{\mscr{H}}{u} + f_\ss{u}
         \end{array}
    \end{align}
    \end{small}
    \item \textbf{\textit{Simplifications.}}
    None of the dynamics so far have been simplified with the integrals of motion \eq{q=1} or \eq{\lambda=\htup{q}\cdot\tup{p}=0}, which imply the relations in Eq.\eqref{qpl_rels} (e.g., \eq{\slang \simeq p}, \eq{-\hdge{\tup{\slang}} \cdot \tup{q} \simeq \tup{p}}, and \eq{\hdge{\tup{\slang}} \cdot \tup{p} \simeq \slang^2 \tup{q}}). 
    As long as we limit consideration of initial conditions to those satisfying \eq{q_\zr=1} and \eq{\lambda_\zr = 0}, then
    we are free to simplify any of the above ODEs using the relations in Eq.\eqref{qpl_rels}. For example, the ODEs for \eq{(\tup{q},\tup{p})} simplify such that Eq.\eqref{uw_eom_gen} is equivalent to:
    \begin{small}
    \begin{align} \label{qp_eom_s_simp_sum}
     \fnsz{\begin{array}{cc}
         q=1  \\[2pt]
        \lambda = \htup{q}\cdot\tup{p}=0 
    \end{array}}
    \Rightarrow \qquad
    \left\{\quad    \begin{array}{lllll}
         \pdt{\tup{q}} \simeq \tup{p} 
         &,\quad 
         \pdt{\tup{p}} 
         \simeq -\slang^2\tup{q} + \pdt{t}\tup{f}
         \\[4pt]
         \pdt{u} = w 
          &,\quad 
          \pdt{w} = -\slang^2 u - \pderiv{V^0}{u} + f_\ss{u}
    \end{array} \right.
      &&\fnsize{or,}&&
     \begin{array}{lllll}
         \rng{\tup{q}} \simeq \tfrac{1}{\slang}\tup{p} \simeq \htup{p}
         &,\quad 
         \rng{\tup{p}} \simeq -\slang \tup{q} + \rng{t}\tup{f}
        \\[4pt]
           \rng{u} \,=\,  w/\slang 
            &,\quad 
             \rng{w} = -\slang u - \tfrac{1}{\slang} \pderiv{V^0}{u} +   \tfrac{1}{\slang} f_\ss{u}
    \end{array}
    \end{align}
    \end{small}
    with \eq{\slang^2 \simeq p^2}.
    We note that simplifying the dynamics such as above is \textit{not} required in order to obtain linear dynamics in the unperturbed case that \eq{\tup{F}=0=(\tup{f},f_\ss{u})}.  When the perturbation terms are discarded, the \textit{un}simplified ODEs in Eq.\eqref{uw_eom_gen} are already fully linear for certain forms of the central-force potential \eq{V^0} (namely, for Kepler or Manev potentials). 
\end{itemize}
\end{small}

\noindent Observe that the angular momentum plays a central role in the projective coordinate dynamics (in any of the above iterations). It is therefore useful to note that the ODEs governing the evolution of the various angular momentum functions from Eq.\eqref{lang_qp_SUM} are given in terms of the projective coordinates 
    as follows:\footnote{Note the relation \eq{\dot{\slang} \simeq \dot{p}\simeq  \tfrac{1}{u} \htup{p} \cdot \tup{F}} in Eq.\eqref{ldot_cen} makes use of \eq{q=1} and \eq{\tup{q}\cdot\tup{p}=0} (thus, \eq{\slang \simeq p}).}
\begin{small}
\begin{align} \label{ldot_cen}
\begin{array}{llll}
     \hdge{\dot{\tup{\slang}}} = \tup{q}\wdg\tup{f}
     \,=\, \tfrac{1}{u} \htup{q} \wdg \tup{F}
     &\!(=\tup{r}\wdg \tup{F})
 \\[4pt]
     \dot{\tup{\slang}} \,=\, \tup{q}\tms\tup{f}
    \,=\, \tfrac{1}{u} \htup{q}\tms\tup{F}
    &\!(= \tup{r}\tms\tup{F})
\end{array}
&&,&&
\begin{array}{lllllll}
     \dot{\slang} &\!\!\!=\,  \tfrac{q^2}{\slang}\tup{p} \cdot\tup{f}  \,=\, \tfrac{1}{u} \htup{q}\cdot\hdge{\htup{\slang}} \cdot\tup{F}
 \\[4pt]
    &\!\!\!\simeq\, \dot{p} \,=\, \htup{p} \cdot\tup{f} \simeq  \tfrac{1}{u} \htup{p} \cdot \tup{F}
\end{array}
\end{align}  
\end{small}
which confirms that angular momentum is conserved for any pure central-force dynamics (in which case \eq{\tup{f}=0}, as per remark \ref{rem:Frad}). 
Reparameterizing the above ODEs by \eq{s} or \eq{\tau} 
simply amounts to scaling by \eq{\pdt{t}=1/u^2} or \eq{\rng{t}=1/(\slang u^2)}.


\paragraph{Second-Order Dynamics.}
Even without making use of the integrals of motion \eq{q=1} or \eq{\htup{q}\cdot\tup{p}=0}, the first-order \eq{s}- or \eq{\tau}-parameterized dynamics in Eq.\eqref{dqp_s_0}, Eq.\eqref{dqp_ta_0}, or Eq.\eqref{uw_eom_gen} are equivalent to the following second-order dynamics for \eq{(\tup{q},u)} \cite{peterson2025prjCoord,peterson2025prjGeomech}:
\begin{small}
\begin{flalign} \label{ddqu_fgen} 
&&
\begin{array}{rllll}
      \pddt{\tup{q}} \,+\, \slang^2\tup{q} &\!\!\! =\, \tfrac{q^2}{u^2}\tup{f}
\\[6pt]
     \pddt{u} + \slang^2 u + \pderiv{V^0}{u}  &\!\!\! =\, f_\ss{u}
\end{array} 
&& \fnsize{or,} &&
 \begin{array}{rlllll}
       \rrng{\tup{q}} \,+\, \tup{q} 
       &\!\!\! =\, \tfrac{q^2}{u^2 \slang^2} ( \imat_3 - \tfrac{1}{\slang}\rng{\tup{q}}\otms\tup{p} ) \cdot \tup{f}
\\[6pt]
     \rrng{u} +  u  +  \tfrac{1}{\slang^2}\pderiv{V^0}{u} 
      &\!\!\! =\,
      \tfrac{1}{\slang^2} ( f_\ss{u} - \tfrac{q^2}{\slang u^2} \rng{u}\tup{p} \cdot \tup{f} )
\end{array} 
&&
\end{flalign}
\end{small}
Or, using using Eq.\eqref{Ftotal_qp} to express the right-hand-side in terms of the inertial cartesian components, \eq{\tup{F}}: 
\begin{small}
\begin{flalign} \label{ddqu_Fcart} 
&&
\begin{array}{rllll}
      \pddt{\tup{q}} \,+\, \slang^2\tup{q} &\!\!\!
     =\,\tfrac{q}{u^3} ( \imat_3 - \htup{q}\otms\htup{q} )\cdot \tup{F}
\\[6pt]
     \pddt{u} + \slang^2 u + \pderiv{V^0}{u}  &\!\!\! 
     =\, -\tfrac{1}{u^2}\htup{q} \cdot \tup{F}
\end{array} 
&& \fnsize{or,} &&
 \begin{array}{rlllll}
       \rrng{\tup{q}} \,+\, \tup{q} 
       &\!\!\! =\, 
       \tfrac{1}{q u^3 \slang^2} ( q^2\imat_3 - \tup{q}\otms\tup{q} -  \rng{\tup{q}} \otms \rng{\tup{q}} ) \cdot \tup{F}
\\[6pt]
     \rrng{u} +  u  +  \tfrac{1}{\slang^2}\pderiv{V^0}{u} 
      &\!\!\! =\, 
    -\tfrac{1}{q u^3 \slang^2} ( u \tup{q} + \rng{u} \rng{\tup{q}} ) \cdot\tup{F}
\end{array} 
&&
\end{flalign}
\end{small}
\begin{small}
\begin{itemize}
    \item \textit{Simplifications.} If we use \eq{q=1} and \eq{\lambda=\htup{q}\cdot\tup{p}=0} (implying the relations in Eq.\eqref{qpl_rels}), then the second-order dynamics in Eq.\eqref{ddqu_fgen} and Eq.\eqref{ddqu_Fcart} simplify to: 
    \begin{small}
    \begin{flalign} 
    \;\;
    \begin{array}{rllll}
              \pddt{\tup{q}} + \slang^2\tup{q} &\!\!\! \simeq\, \tfrac{1}{u^2}\tup{f}
             &\!\! \simeq\, \tfrac{1}{u^3} ( \imat_3 - \tup{q}\otms\tup{q} )\cdot \tup{F}
        \\[6pt]
             \pddt{u} + \slang^2 u + \pderiv{V^0}{u}  &\!\!\! =\, f_\ss{u}
              &\!\! \simeq\, -\tfrac{1}{u^2}\tup{q} \cdot \tup{F}
        \end{array}
    && \fnsize{or,} &&
    \begin{array}{rlllll}
           \rrng{\tup{q}} + \tup{q} 
           &\!\!\! \simeq\, \tfrac{1}{u^2 \slang^2} ( \imat_3 - \rng{\tup{q}}\otms\rng{\tup{q}} ) \cdot \tup{f}
            &\!\!\simeq \,
           \tfrac{1}{u^3\slang^2}(\htup{\slang} \otms \htup{\slang}) \cdot \tup{F}
    \\[6pt]
         \rrng{u} +  u  +  \tfrac{1}{\slang^2}\pderiv{V^0}{u} 
          &\!\!\! \simeq\,  \tfrac{1}{u^2\slang^2} (u^2 f_\ss{u} -   \rng{u} \rng{\tup{q}}\cdot \tup{f} )
         &\!\!\simeq \, -\tfrac{1}{u^3 \slang^2} ( u \tup{q} + \rng{u} \rng{\tup{q}} ) \cdot\tup{F}
    \end{array} 
    \;\;
    \end{flalign}
    \end{small}
    with \eq{  \slang\simeq p \simeq \mag{\pdt{\tup{q}}} } and 
    \eq{ \dot{\slang} \simeq\dot{p} = \htup{p} \cdot\tup{f} \simeq  \tfrac{1}{u} \htup{p} \cdot \tup{F} }. 
\end{itemize}
\end{small}
We note the above second-order dynamics only to clearly illustrate the relation to a perturbed linear harmonic oscillator, and for comparison with previous works. This work is primary concerned with first-order dynamics.

\subsection{Rotational Motion Solutions for Arbitrary Central-Forces} \label{sec:qp_sol_rotate}

The coordinates \eq{(\tup{q},\tup{p})} fully describe the rotational motion of the particle's LVLH basis and closed-form solutions are immediately available for any central-force dynamics; recall from remark \ref{rem:Frad} that central-forces to do not participate in the dynamics for \eq{(\tup{q},\tup{p})}. 
In contrast, the radial motion dynamics — described by the coordinates \eq{(u,p_\ss{u})} or, alternatively, \eq{(u,w:=u^2 p_\ss{u})} — depend on the the particular form of the central force potential, \eq{V^0}, and the radial perturbation term \eq{f_\ss{u}=-\tfrac{1}{u^2}F_r}. Closed-form solutions are only available in certain cases,  including the unperturbed Kepler problem (addressed in section \ref{sec:kep_sol}). Below, we consider the \eq{(\tup{q},\tup{p})} solutions for  arbitrary central-force dynamics.

Let us consider, for the moment, the case that the arbitrary perturbing force, \eq{\tup{F}}, is now purely radial. 
Recall from remark \ref{rem:Frad} that
this leads to \eq{\tup{f}=0} such that angular momentum is preserved:
\begin{small}
\begin{flalign} \label{Frad_f0}
\begin{array}{cc}
   \scrsize{central-force}\\
   \scrsize{dynamics} 
 \end{array} \!:
 &&
    \fnsize{if }  \; \tup{F} = F_r \htup{r}
    \qquad \Rightarrow \qquad 
    \tup{f} = 0 \quad,\quad f_\ss{u} = -\tfrac{1}{u^2} F_r
    \quad,\quad 
    \eq{\tup{\slang}=\tup{\slang}_\zr}
    \quad,\quad 
    \tau = \slang s
&&
\end{flalign}
\end{small}
where \eq{\tau = \slang s} (which assumes \eq{\tau_\zr =s_\zr=0}) follows from preservation of \eq{\slang} and the relation \eq{\mrm{d} \tau = \slang \mrm{d} s}. 
Thus, for any arbitrary central forces,  the first-order ODEs for \eq{(\tup{q},\tup{p})} in Eq.\eqref{dqp_s_0}, Eq.\eqref{dqp_ta_0}, or Eq.\eqref{uw_eom_gen} are linear equations with constant coefficients:
\begin{small}
\begin{flalign} \label{qp_Frad_SUM}
&\begin{array}{cc}
   \scrsize{central-force}\\
   \scrsize{dynamics} 
 \end{array} \!:
 &&
 \begin{array}{lllll}
       \pdt{\tup{q}} = 
      -\hdge{\tup{\slang}}\cdot\tup{q} 
\\[4pt]
      \pdt{\tup{p}} =  -\hdge{\tup{\slang}}\cdot\tup{p} 
\end{array}
\qquad 
\scrsize{or,} \qquad 
\begin{array}{lllll}
       \rng{\tup{q}}    = 
      -\hdge{\htup{\slang}}\cdot\tup{q} 
 \\[4pt]
     \rng{\tup{p}} =  -\hdge{\htup{\slang}}\cdot\tup{p}
\end{array} 
 &&,&&
 \begin{array}{llll}
        \hdge{\tup{\slang}}  = \tup{q}\wdg \tup{p} = \hdge{\tup{\slang}}_\zr \in \somat{3} 
 \end{array}
 &&
\end{flalign}
\end{small}
The solutions to either of  the above coincide and are given by the matrix exponential \eq{\mrm{e}^{-\hdge{\tup{\slang}}s} = \mrm{e}^{-\hdge{\htup{\slang}}\tau}  \in \Somat{3}}, which gives a special orthogonal rotation by \eq{\tau} about \eq{\htup{\slang}=\htup{q}\tms\htup{p}} (the orbit normal direction). Using the
    Rodrigues rotation formula\footnote{For any \eq{\tup{\rho}\in\mbb{R}^3} with \eq{\hdge{\tup{\rho}}\in\somat{3}}, and some parameter \eq{\varep}, then the Rodrigues formula gives \eq{\mrm{e}^{\hdge{\tup{\rho}}\varep}\in \Somat{3}} as follows (with \eq{\rho=\mag{\tup{\rho}}}):
    \begin{align} \nonumber
     \mrm{e}^{\hdge{\tup{\rho}}\varep}  =   \imat_3 + \tfrac{1}{\rho}( \sin{\rho \varep} )\hdge{\tup{\rho}} + \tfrac{1}{\rho^2}(1-\cos{\rho \varep}) \hdge{\tup{\rho}} \cdot \hdge{\tup{\rho}}  
     \,=\,
     \imat_3 + ( \sin{\rho \varep} )\hdge{\htup{\rho}} + (1-\cos{\rho \varep}) \hdge{\htup{\rho}} \cdot \hdge{\htup{\rho}} 
     &&,&&
     \inv{(\mrm{e}^{\hdge{\tup{\rho}}\varep})} = \eq{\mrm{e}^{-\hdge{\tup{\rho}}\varep}  =  \trn{(\mrm{e}^{\hdge{\tup{\rho}}\varep})} }
    \end{align} 
    Note that if \eq{\sigma:=\rho\varep} then \eq{ \mrm{e}^{\hdge{\htup{\rho}}\sigma}= \mrm{e}^{\hdge{\tup{\rho}}\varep}}.  
    },
along with some angular momentum relations in Appendix \ref{sec:ang_momentum}, 
 this leads to solutions:
\begin{small}
\begin{flalign} \label{dqp_Fcen_sol}
 &&
\begin{array}{rllllll}
    \tup{q}_s = \tup{q}_\tau 
      &\!\!\! =\,
      R_\tau(\htup{\slang}) \cdot\tup{q}_\zr \;=\;
      \tup{q}_\zr \csn{\tau} -  \hdge{\htup{\slang}} \cdot \tup{q}_\zr \snn{\tau} 
\\[4pt]
    \tup{p}_s = \tup{p}_\tau 
      &\!\!\! =\, 
      R_\tau(\htup{\slang}) \cdot\tup{p}_\zr \;=\; 
      \tup{p}_\zr \csn{\tau} -  \hdge{\htup{\slang}} \cdot \tup{p}_\zr \snn{\tau} 
\end{array} 
&&,&&
\begin{array}{llll}
        R_\tau(\htup{\slang}) :=
       \mrm{e}^{-\hdge{\htup{\slang}}\tau} \in \Somat{3} 
 \end{array}
\qquad
\end{flalign}
\end{small}
The above verify that \eq{\tup{\slang}=\tup{q}\tms\tup{p}},  \eq{\mag{\tup{q}}}, \eq{\tup{q}\cdot\tup{p}}, and \eq{\mag{\tup{p}}} are preserved (i.e., integrals of motion) for any central-force dynamics:
\begin{small}
\begin{align} \label{qp_sol_constants}
   \tup{\slang}_\tau = \tup{q}_\tau\tms\tup{p}_\tau = \tup{q}_\zr\tms\tup{p}_\zr =\tup{\slang}_\zr
\qquad,\qquad 
    \mag{\tup{q}_\tau} = \mag{\tup{q}_\zr}
\qquad,\qquad 
     \tup{q}_\tau \cdot \tup{p}_\tau = \tup{q}_\zr\cdot \tup{p}_\zr 
 \qquad,\qquad 
     \mag{\tup{p}_\tau} = \mag{\tup{p}_\zr}
\end{align}
\end{small}
As per Eq.\eqref{qlam_0}, preservation of  \eq{\mag{\tup{q}}} and \eq{ \htup{q} \cdot \tup{p}} is guaranteed for any and all dynamics in our projective coordinates (not just for central-forces). Preservation of \eq{\tup{\slang}} and \eq{\mag{\tup{p}}} is unique to central-force dynamics.

\begin{small}
\begin{itemize}
    \item  \textit{Simplifications.} If we use \eq{q=1} and \eq{\lambda=\htup{q}\cdot\tup{p}=0} (implying \eq{\slang\simeq p} along with the other relations in Eq.\eqref{qpl_rels}), then the central-force dynamics and solutions for \eq{(\tup{q},\tup{p})} simplify to:
    \begin{small}
    \begin{flalign} \label{qp_sol_f_m1}
     \qquad 
    \fnsz{\begin{array}{cc}
             q=1  \\
             \htup{q}\cdot\tup{p}=0 
        \end{array}}
        \;\Rightarrow \quad
    \left\{\;\;
    \begin{array}{lllll}
         \pdt{\tup{q}} \simeq \tup{p} 
         \\[4pt]
         \pdt{\tup{p}} 
         \simeq -\slang^2\tup{q} 
    \end{array} \right.
        \;\;\fnsize{ or,}\quad 
     \begin{array}{lllll}
         \rng{\tup{q}} \simeq \tfrac{1}{\slang}\tup{p} 
         \\[4pt]
         \rng{\tup{p}} \simeq -\slang \tup{q} 
    \end{array}
    &&\Rightarrow &&
    \begin{array}{llllll}
           \tup{q}_{\tau}  
           \, \simeq\,  \tup{q}_{\zr}\csn{\tau}  +  \tfrac{1}{\slang}\tup{p}_{\zr}\snn{\tau}
           &\!\!\simeq\,  \tup{q}_{\zr}\csn{\tau}  +  \htup{p}_{\zr}\snn{\tau}
    \\[5pt] 
           \tup{p}_{\tau} 
           \,\simeq\, \tup{p}_{\zr}\csn{\tau}   -\slang\tup{q}_{\zr}\snn{\tau} 
            &\!\!\simeq\, \tup{p}_{\zr}\csn{\tau}   -p_\zr\tup{q}_{\zr}\snn{\tau} 
    \end{array}
     &&
    \end{flalign}
    \end{small}
    where \eq{\tau = \slang s} with \eq{\slang = \slang_\zr \simeq p_\zr =p}. 
    The above solutions also follow immediately from simplifying the solutions in Eq.\eqref{dqp_Fcen_sol}. 
    It is again verified from the above that \eq{\tup{\slang}_\tau=\tup{\slang}_\zr}. 
    Yet, unlike the unsimplified solutions in Eq.\eqref{dqp_Fcen_sol}, the above simplified solutions do not \textit{directly} verify preservation of \eq{\mag{\tup{q}}}, \eq{\tup{q}\cdot\tup{p}}, or \eq{\mag{\tup{p}}}; they are preserved \textit{iff} we limit initial conditions  to \eq{q_\zr=1} and \eq{\lambda_\zr=\htup{q}_\zr\cdot \tup{p}_\zr = 0} (which is what the above already assumes). 
    That is, using the above simplified solutions for \eq{(\tup{q}_\tau,\tup{p}_\tau)}:
    \begin{small}
    \begin{flalign} \label{qp_sol_constants_simp}
        \qquad
         \tup{\slang}_\tau=\tup{\slang}_\zr
         &&,&&
          \fnsz{\begin{array}{cc}
             q_0=1  \\
             \htup{q}_0\cdot\tup{p}_0=0 
        \end{array}}
        \;\Rightarrow \qquad
          \mag{\tup{q}_\tau} \simeq \mag{\tup{q}_\zr}\simeq 1
          \quad,\quad 
          \tup{q}_\tau \cdot \tup{p}_\tau \simeq \tup{q}_\zr\cdot \tup{p}_\zr \simeq 0
          \quad,\quad 
           \mag{\tup{p}_\tau} \simeq \mag{\tup{p}_\zr} \simeq \slang_\zr
           \quad
    \end{flalign}
    \end{small}
\end{itemize}
\end{small}

\paragraph{As a Linear Hamiltonian System.}
 The above developments may also be combined into a single linear ODE for \eq{(\tup{q},\tup{p})\in\mbb{R}^6}:
\begin{small}
\begin{flalign} \label{dz_ds_E3}
   &&
 \begin{array}{llll}
   \diff{}{s} \fnpmat{
        \tup{q} \\
        \tup{p} }
    = L \cdot \fnpmat{
        \tup{q}\\
        \tup{p} }
    \qquad \fnsize{or,} \qquad 
      \diff{}{\tau} \fnpmat{
        \tup{q} \\
        \tup{p} }
    = \hat{L} \cdot \fnpmat{
        \tup{q}\\
        \tup{p} }    
\end{array}
&& \Rightarrow &&
     \fnpmat{
        \tup{q}_s \\
        \tup{p}_s } 
        = \mrm{e}^\fnsz{L s} \cdot  \fnpmat{
        \tup{q}_\zr \\
        \tup{p}_\zr } 
        \;\;=\;\;
        \mrm{e}^\fnsz{\hat{L} \tau} \cdot  \fnpmat{
        \tup{q}_\zr \\
        \tup{p}_\zr } 
         =
        \fnpmat{
        \tup{q}_\tau \\
        \tup{p}_\tau }
        &&
\end{flalign}
\end{small}
where the matrices are given in terms of \eq{\hdge{\tup{\slang}} =\hdge{\tup{\slang}}_\zr } and \eq{\tau =\slang s} by:
\begin{small}
\begin{align} \label{Lmat_spso}
       L\, ,\;  \hat{L}:=\tfrac{1}{\slang} L = \fnpmat{
        -\hdge{\htup{\slang}}  & 0 \\  0 & -\hdge{\htup{\slang}}  }  \in \spmat{6} \cap \somat{6} 
   \qquad,\qquad 
    \mrm{e}^\fnsz{L s} = \mrm{e}^\fnsz{\hat{L} \tau} = \fnpmat{
         R_\tau(\htup{\slang}) & 0 \\
          0 &   R_\tau(\htup{\slang})  }
    \in \Spmat{6} \cap \Somat{6} 
\end{align}
\end{small}
Interestingly, one may also express the above as a different, but equivalent, matrix ODE.
 Using  \eq{-\hdge{\tup{\slang}}\cdot\tup{q} = (q^2 \imat_3 - \tup{q}\otms\tup{q})\cdot\tup{p}} and \eq{-\hdge{\tup{\slang}}\cdot\tup{p} = -(p^2 \imat_3 - \tup{p}\otms\tup{p})\cdot\tup{q}}, the above dynamics for \eq{(\tup{q},\tup{p})} are equivalently expressed as:
\begin{small}
\begin{align} \label{dz_ds_E3_alt}
   \diff{}{s}  \fnpmat{ \tup{q} \\ \tup{p}}
     =  A \cdot   \fnpmat{ \tup{q} \\ \tup{p}}
 \qquad \fnsize{or,} \qquad 
    \diff{}{\tau} 
        \fnpmat{ \tup{q} \\ \tup{p}}
     =  \tfrac{1}{\slang} A \cdot  \fnpmat{ \tup{q} \\ \tup{p}}
    &&\Rightarrow&&
    \fnpmat{ \tup{q}_s \\ \tup{p}_s }
     = \mrm{e}^{A s} \cdot 
    \fnpmat{ \tup{q}_\zr \\ \tup{p}_\zr }
    \;\;=\;\;
    \mrm{e}^{ \frac{1}{\slang} A \tau} \cdot 
     \fnpmat{ \tup{q}_\zr \\ \tup{p}_\zr } 
     =
    \fnpmat{ \tup{q}_\tau\\ \tup{p}_\tau }
\end{align}
\end{small}
where \eq{A=A_\zr \in \spmat{6}} is again a matrix of integrals of motion (for any central-force dynamics), given as follows:
\begin{small}
\begin{align} \label{dz_ds_E3_alt2}
     A := \fnpmat{ -(\tup{q}\cdot\tup{p})\imat_3 & q^2 \imat_3 \\  -p^2 \imat_3 & (\tup{q}\cdot\tup{p})\imat_3 }
     \in \spmat{6}
     \qquad,\qquad 
    \mrm{e}^{A s}  =  \mrm{e}^{ \frac{1}{\slang}A \tau} = 
  \fnpmat{
    \big(\csn{\tau} - \tfrac{1}{\slang}(\tup{q} \cdot \tup{p}) \snn{\tau} \big) \imat_3 
    & \big(\tfrac{1}{\slang}q^2 \snn{\tau} \big) \imat_3 \\
    -\big( \tfrac{1}{\slang}p^2 \snn{\tau} \big) \imat_3 & 
      \big( \csn{\tau} +  \tfrac{1}{\slang}(\tup{q} \cdot \tup{p}) \snn{\tau} \big) \imat_3
    }
    \in \Spmat{6}
\end{align}
\end{small}
One can verify that any of the above matrix ODEs and solutions agree with those in  Eq.\eqref{dqp_Fcen_sol}.
\begin{small}
\begin{itemize}
    \item \sloppy \textit{Eigenvalues.}
    We note that \eq{L\in\spmat{6}} in Eq.\eqref{dz_ds_E3} has eigenvalues \eq{(0,0,\mrm{i}\slang,\mrm{i}\slang,-\mrm{i}\slang,-\mrm{i}\slang)}, 
    whereas \eq{A\in\spmat{6}} in Eq.\eqref{dz_ds_E3_alt} has eigenvalues \eq{(\mrm{i}\slang,\mrm{i}\slang,\mrm{i}\slang,-\mrm{i}\slang,-\mrm{i}\slang,-\mrm{i}\slang)}.
    \item \textit{Simplifications.}  If we use \eq{q=1} and \eq{\lambda=\htup{q}\cdot\tup{p}=0} to simplify the central-force dynamics for \eq{(\tup{q},\tup{p})}, then the above dynamics in the form of Eq.\eqref{dz_ds_E3_alt}-Eq.\eqref{dz_ds_E3_alt2} simplify to:
    \begin{small}
    \begin{align} \label{dqp_Fcen_simp}
           A \,\simeq\, \fnpmat{ 0 &  \imat_3 \\[2pt]
           -\slang^2 \imat_3 & 0 }
           =:  \Lambda_6 
         \in \spmat{6}
         \qquad,\qquad 
         \mrm{e}^{ A s} = \mrm{e}^{ \frac{1}{\slang}A \tau} \simeq \mrm{e}^{\Lambda_6 s} = \mrm{e}^{ \frac{1}{\slang}\Lambda_6 \tau} =  
      \fnpmat{
       \csn{\tau}\,\imat_3 
        & \tfrac{1}{\slang}\snn{\tau}\,\imat_3 \\[2pt]
        - \slang \snn{\tau}\,\imat_3 & 
           \csn{\tau}\,\imat_3
        }
        \in \Spmat{6}
    \end{align}
    \end{small}
    The above leads to the same simplified solutions in Eq.\eqref{qp_sol_f_m1}. 
\end{itemize}
\end{small}
Everything in Eq.\eqref{Frad_f0}-Eq.\eqref{dqp_Fcen_simp} above is valid for any pure central-force dynamics, conservative or nonconservative.

\subsection{Linear Kepler Dynamics \& Closed-Form Solutions} \label{sec:kep_sol}

Kepler-type dynamics correspond to the case that the central-force potential is given by:
\begin{small}
\begin{align}
     V^0 \,=\, - \kconst_1 /r \,=\, -\kconst_1 u
     \qquad,\qquad 
      \pderiv{V^0}{u} \,=\, -\kconst_1 
\end{align}
\end{small}
The projective coordinate Hamiltonian, \eq{\mscr{H}}, is then given by:
\begin{small}
\begin{flalign}
    &&
     \mscr{K} = \tfrac{1}{2} \v^2 - \kconst_1 /r
    \qquad\Rightarrow\qquad
     \boxed{ \mscr{H} =   \tfrac{1}{2} u^2 \big( \slang^2 +  u^2 p_\ss{u}^2 \big) -\kconst_1 u }
     \;\;=\;
     \tfrac{1}{2} \big( u^2 \slang^2 +  w^2 \big) -\kconst_1 u
     &&
\end{flalign}
\end{small}
with \eq{\slang^2 = q^2 p^2 - (\tup{q}\cdot\tup{p})^2}.
The ODEs for perturbed Kepler-type dynamics in the projective coordinates \eq{(\tup{q},u,\tup{p},p_\ss{u})} — or the alternative set \eq{(\tup{q},u,\tup{p},w:=u^2 p_\ss{u})} —  are then easily obtained by the substitution \eq{\pd_{u} V^0 = -\kconst_1 } wherever it appears in the preceding developments of section \ref{sec:ext_cen}. 
For instance, the \eq{s}- or \eq{\tau}-parameterized dynamics for  \eq{(\tup{q},u,\tup{p},w)} follow from Eq.\eqref{uw_eom_gen} as:
\begin{small}
\begin{flalign} \label{uw_eom_kepPert}
&&
\begin{array}{llllll}
     \pdt{\tup{q}}  \,=\, 
  -\hdge{\tup{\slang}}\cdot\tup{q} 
  &,\quad 
  \pdt{\tup{p}}  \,=\, 
  -\hdge{\tup{\slang}}\cdot\tup{p} + \pdt{t} \tup{f}
\\[4pt]
   \pdt{u} \,=\, w
    &,\quad 
     \pdt{w} = -\slang^2 u + \kconst_1  + f_\ss{u}
\end{array}
\qquad\quad \scrsize{or,}  \qquad\quad
 \begin{array}{llllll}
     \rng{\tup{q}}  \,=\, 
  -\hdge{\htup{\slang}}\cdot\tup{q}  
  &,\quad 
  \rng{\tup{p}}  \,=\, 
  -\hdge{\htup{\slang}}\cdot\tup{p}  + \rng{t} \tup{f}
\\[4pt]
   \rng{u} \,=\,  w/\slang 
    &,\quad 
     \rng{w} = -\slang u + \tfrac{\kconst_1}{\slang} + \tfrac{1}{\slang}f_\ss{u}
\end{array} 
&&
\tup{\slang}\neq\tup{\slang}_\zr
\quad
\end{flalign}
\end{small}
where \eq{w:=u^2 p_\ss{u}} is a quasi-momenta coordinate used in place of \eq{p_\ss{u}} 
(see footnote for canonical equations for \eq{(u,p_\ss{u})}\footnote{The perturbed Kepler dynamics for the conjugate pair \eq{(u,p_\ss{u})} are given as follows:
\begin{align}
\begin{array}{llll}
         \pdt{u} = u^2 p_\ss{u} \\[3pt]
         \pdt{p}_\ss{u} = -\tfrac{1}{u}\big( \slang^2 + 2 u^2 p_\ss{u}^2 \big) + \kconst_1 /u^2 + \pdt{t} f_\ss{u}
    \end{array}
    \qquad\quad \fnsize{or,}  \qquad\quad
     \begin{array}{llll}
         \rng{u} = u^2 p_\ss{u}/\slang \\[3pt]
         \rng{p}_\ss{u} = -\tfrac{1}{u \slang}\big( \slang^2 + 2 u^2 p_\ss{u}^2 \big) + \kconst_1 /(\slang u^2) + \rng{t} f_\ss{u}
\end{array}
\end{align} 
In the unperturbed case (\eq{f_\ss{u}=0}), the above is a nonlinear \textit{first}-order system. But, it is indeed equivalent to the linear \textit{second}-order ODEs for \eq{u} in Eq.\eqref{ddqu_kep_sum}.  }).
As noted previously, the particular form of \eq{V^0} has no impact whatsoever on the ``\eq{(\tup{q},\tup{p})}-part'' of the dynamics (i.e., the rotational motion).


\paragraph{Kepler Solutions.}
In the following, we consider only the unperturbed Kepler problem — that is, \eq{\tup{F}=0=(\tup{f},f_\ss{u})}. 
Then, the \eq{s}- or \eq{\tau}-parameterized unperturbed Kepler dynamics are as follows 
(using \eq{w:=u^2 p_\ss{u}}):
\begin{small}
\begin{flalign} \label{uw_eom_kep}
\begin{array}{cc}
     \scrsize{Kepler-type }  \\
      \scrsize{dynamics} 
\end{array} 
&&
\boxed{\begin{array}{llllll}
     \pdt{\tup{q}}  \,=\, 
  -\hdge{\tup{\slang}}\cdot\tup{q} 
  &,\quad 
  \pdt{\tup{p}}  \,=\, 
  -\hdge{\tup{\slang}}\cdot\tup{p} 
\\[4pt]
   \pdt{u} \,=\, w
    &,\quad 
     \pdt{w} = -\slang^2 u + \kconst_1  
\end{array}
\qquad\quad \scrsize{or,}  \qquad\quad
 \begin{array}{llllll}
     \rng{\tup{q}}  \,=\, 
  -\hdge{\htup{\slang}}\cdot\tup{q} 
  &,\quad 
  \rng{\tup{p}}  \,=\, 
  -\hdge{\htup{\slang}}\cdot\tup{p} 
\\[4pt]
   \rng{u} \,=\,  w/\slang 
    &,\quad 
     \rng{w} = -\slang u + {\kconst_1}/{\slang} 
\end{array} }
&&
\tup{\slang}=\tup{\slang}_\zr
\quad
\end{flalign}
\end{small}
As is the case for any cental-force dynamics, 
the angular momentum is conserved, \eq{\tup{\slang}=\tup{\slang}_\zr}, such that the above systems in Eq.\eqref{uw_eom_kep} may be regarded as linear systems.\footnote{More specifically, they are families of linear ODEs, parameterized by values of the (constant) angular momentum, \eq{\tup{\slang}=\tup{\slang}_\zr}.}
Moreover, it follows from  \eq{\tup{\slang}=\tup{\slang}_\zr} that the parameters \eq{s} and \eq{\tau} defined in Eq.\eqref{dtds_0} are related as follows (assuming \eq{\tau_\zr=s_\zr=0}):
\begin{small}
\begin{align}
    \mrm{d} \tau = \slang \mrm{d} s
    &&
\begin{array}{cc}
   \scrsize{central-force}\\
   \scrsize{dynamics} 
\end{array}
 \;\Rightarrow \;\;\quad
 \slang=\slang_\zr
 \quad\;\;\Rightarrow\;\;\quad
 \tau = \slang s
\end{align}
\end{small}
such that it makes little difference whether one wishes to express solutions in terms of \eq{s} or \eq{\tau} as reparameterization between them is trivial. For brevity, we often express solutions in terms of \eq{\tau}. 
For instance, either system in Eq.\eqref{uw_eom_kep} may be solved for solutions in terms of initial conditions and \eq{s} or \eq{\tau=\slang s} as follows (assuming \eq{\tau_\zr=s_\zr=0} for convenience):
\begin{small}
\begin{flalign} \label{qusol_s_2bp}
&&
\boxed{\begin{array}{lllllll}
     \tup{q}_\tau 
      &\!\!\!\! =\,  \tup{q}_\zr \csn{\tau} \,-\,  \hdge{\htup{\slang}} \cdot \tup{q}_\zr \snn{\tau} 
\\[4pt]
      \tup{p}_\tau 
      &\!\!\!\! =\,   \tup{p}_\zr \csn{\tau} \,-\,  \hdge{\htup{\slang}} \cdot \tup{p}_\zr \snn{\tau} 
\\[4pt]
      u_\tau 
      &\!\!\!\! =\,  (u_{\zr}-\tfrac{\kconst_1}{\slang^2})\csn{\tau} \,+\, \tfrac{1}{\slang}w_{\zr}\snn{\tau} \,+\, \tfrac{\kconst_1}{\slang^2}
\\[4pt]
      w_\tau  
      &\!\!\!\! =\,  w_{\zr}\csn{\tau} \,-\, \slang(u_{\zr}-\tfrac{\kconst_1}{\slang^2})\snn{\tau} 
\end{array} }
&&
\begin{array}{llll}
     \tau = \slang s
 \\[3pt]
     \tup{\slang}_\tau =\tup{\slang}_0
\\[3pt]
    \mag{\tup{q}_\tau} = \mag{\tup{q}_\zr}
\\[3pt]
     \tup{q}_\tau \cdot \tup{p}_\tau = \tup{q}_\zr\cdot \tup{p}_\zr 
\\[3pt]
     \mag{\tup{p}_\tau} = \mag{\tup{p}_\zr}
\end{array}
 &&
\end{flalign}
\end{small}
were \eq{\tup{\slang}} and \eq{\slang} really mean \eq{\tup{\slang}_\zr} and \eq{\slang_\zr}. 
The above solutions for \eq{(\tup{q}_\tau,\tup{p}_\tau)} were discussed in more detail in section \ref{sec:qp_sol_rotate}. 
As noted in Eq.\eqref{qp_sol_constants}, these solutions directly verify preservation of \eq{\tup{\slang}=\tup{q}\tms\tup{p}}, \eq{\mag{\tup{q}}}, \eq{\tup{q}\cdot\tup{p}}, and \eq{\mag{\tup{p}}}. 

\begin{small}
\begin{itemize}
    \item \textit{Solution for $p_\ss{u}$.} The Kepler solutions for the conjugate pair  \eq{(u,p_\ss{u})} is easily recovered from Eq.\eqref{qusol_s_2bp} above using \eq{p_\ss{u} = w/u^2}.\footnote{Specifically, \eq{p_\ss{u} = w/u^2\leftrightarrow w=u^2 p_\ss{u}} leads to:  
    \begin{align} \label{upu_KEPsols_sum}
    \begin{array}{lllll}
      u_{\tau} \,=\,  (u_{\zr}-\tfrac{\kconst_1}{\slang^2})\csn{\tau}
        \,+\, \tfrac{1}{\slang}u_{\zr}^2 p_{u_0}\snn{\tau} \,+\, \tfrac{\kconst_1}{\slang^2}
    \qquad,\qquad 
          p_{u_\tau} \,=\,  \dfrac{ -\slang(u_{\zr}-\frac{\kconst_1}{\slang^2})\snn{\tau} \,+\, u_{\zr}^2 p_{u_0}\csn{\tau} }
               {\big[ (u_{\zr}-\frac{\kconst_1}{\slang^2})\csn{\tau}
        \,+\, \frac{1}{\slang}u_{\zr}^2 p_{u_0}\snn{\tau} \,+\, \frac{\kconst_1}{\slang^2} \big]^2}
    \end{array}  
    \end{align}
    }
    \item   \textit{Time solution.} The Kepler solution for \eq{t(\tau)}
    is obtained from that of \eq{u(\tau)} as follows: 
    \begin{small}
    \begin{align}
        \diff{t}{\tau}\,=\, \tfrac{1}{\slang u^2}
        \;\;,\;\; \slang=\slang_\zr 
        && \Rightarrow && 
        \begin{array}{llll}
        t-t_{\zr} \,=\, \int_{\tau_{0}}^{\tau}
         \tfrac{1}{\slang} \Big( (u_{\zr}-\tfrac{\kconst_1 }{\slang^2})\csn{\tau}
        + \tfrac{1}{\slang}w_\zr \snn{\tau} + \tfrac{\kconst_1 }{\slang^2} \Big)^{-2}   \mrm{d} \tau
        \end{array}
    \end{align}
     \end{small}
    \item \textit{Second-order dynamics.} Though we focus on the first-order dynamics in Eq.\eqref{uw_eom_kep}, we note that even \textit{without} simplifying anything with \eq{q=1} or  \eq{\lambda=\htup{q}\cdot\tup{p}=0}, these are equivalent to the following second-order ODEs for a driven linear harmonic oscillator \cite{peterson2025prjCoord,peterson2025prjGeomech}:
    \begin{small}
    \begin{flalign} \label{ddqu_kep_sum}
    &&
    \begin{array}{llll}
           \pddt{\tup{q}} + \slang^2 \tup{q} = 0  
    \\[4pt]
         \pddt{u} + \slang^2 u = \kconst_1  
    \end{array}
     \qquad \scrsize{or,} \qquad 
     \begin{array}{llll}
           \rrng{\tup{q}} + \tup{q}   =  0
    \\[4pt]
         \rrng{u} +  u = \kconst_1 /\slang^2  
    \end{array}
    &&
    \end{flalign}
    \end{small}
    \item  \textit{Simplifications.} 
    Nothing so far has been simplified using the  integrals of motion \eq{q=1} or \eq{\lambda=\htup{q}\cdot\tup{p}=0} (which also imply \eq{\slang \simeq p}). As discussed, we are free to make these simplifications such that, for example, the above already-linear dynamics for \eq{(\tup{q},\tup{p})} in Eq.\eqref{uw_eom_kep} are, after simplifying using Eq.\eqref{qpl_rels}, equivalent to the following still-linear system:  
    \begin{small}
    \begin{flalign} \label{qp_KEPsol_f_m1}
    \;\;
    \fnsz{\begin{array}{cc}
             q=1  \\
             \htup{q}\cdot\tup{p}=0 
        \end{array}}
        \;\Rightarrow \quad
    \left\{\qquad
    \begin{array}{lllll}
         \pdt{\tup{q}} \simeq \tup{p} 
         \\[4pt]
         \pdt{\tup{p}} \simeq -p^2\tup{q} 
         \simeq -\slang^2\tup{q} 
    \end{array} \right.
        \qquad \fnsize{or,}\qquad
     \begin{array}{lllll}
         \rng{\tup{q}} \simeq \tfrac{1}{\slang}\tup{p} 
         \simeq \htup{p}
         \\[4pt]
         \rng{\tup{p}} \simeq -\slang \tup{q} 
         \simeq -p \tup{q}
    \end{array}
    &&
    \begin{array}{lll}
          \slang = \slang_\zr \simeq  p_\zr = p
    \end{array} 
       \qquad
    \end{flalign}
    \end{small}
    The ODEs for \eq{(u,w)} in Eq.\eqref{uw_eom_kep}, could also be rewritten with \eq{\slang \simeq p}. Thus, the Kepler solutions in Eq.\eqref{qusol_s_2bp}, when simplified using the relations in Eq.\eqref{qpl_rels}, are equivalent to:
    \begin{small}
    \begin{flalign} \label{qusol_s_2bp_simp}
    \;\;
    \fnsz{\begin{array}{cc}
             q=1  \\
             \htup{q}\cdot\tup{p}=0 
        \end{array}}
        \Rightarrow \quad
    \left\{\;\;
   \boxed{\begin{array}{lllllll}
         \tup{q}_\tau 
            \simeq\, 
          \tup{q}_\zr \csn{\tau} + \tfrac{1}{\slang} \tup{p}_\zr \snn{\tau} 
            &\!\!\! \simeq\, 
            \tup{q}_\zr \csn{\tau} +  \htup{p}_\zr \snn{\tau} 
    \\[4pt]
          \tup{p}_\tau 
          \simeq\, 
          \tup{p}_{\zr}\csn{\tau} -\slang\tup{q}_{\zr}\snn{\tau} 
          &\!\!\! \simeq\; 
          \tup{p}_{\zr}\csn{\tau} -p_\zr\tup{q}_{\zr}\snn{\tau} 
    \\[4pt]
          u_\tau 
          =\,  (u_{\zr}-\tfrac{\kconst_1}{\slang^2})\csn{\tau} + \tfrac{1}{\slang}w_{\zr}\snn{\tau} + \tfrac{\kconst_1}{\slang^2}
           &\!\!\! \simeq 
            (u_{\zr}-\tfrac{\kconst_1}{p_0^2})\csn{\tau} + \tfrac{1}{p_0}w_{\zr}\snn{\tau} + \tfrac{\kconst_1}{p_0^2}
    \\[4pt]
          w_\tau  
           =\,  w_{\zr}\csn{\tau} - \slang(u_{\zr}-\tfrac{\kconst_1}{\slang^2})\snn{\tau} 
           &\!\!\! \simeq\,  w_{\zr}\csn{\tau} - p_\zr(u_{\zr}-\tfrac{\kconst_1}{p_0^2})\snn{\tau}  
    \end{array} }  \right.
    \quad,&&
    \begin{array}{llll}
         \tau = \slang s \simeq p s
     \\[3pt]
         \tup{\slang} = \tup{\slang}_\zr
    \end{array}
    \end{flalign}
    \end{small}
    where the above simplified solutions for \eq{(\tup{q}_\tau,\tup{p}_\tau)} again directly verify that \eq{\tup{\slang}_\tau = \tup{\slang}_\zr} is preserved. However, as noted in Eq.\eqref{qp_sol_constants_simp}, they do not directly verify preservation of \eq{\mag{\tup{q}}}, \eq{\tup{q}\cdot\tup{p}}, or \eq{\mag{\tup{p}}};  these are preserved \textit{iff} we limit initial conditions  to \eq{q_\zr=1} and \eq{\lambda_\zr=\htup{q}_\zr\cdot \tup{p}_\zr = 0} (which is what the above already assumes), which then leads to \eq{ \mag{\tup{q}_\tau} \simeq \mag{\tup{q}_\zr}\simeq 1}, \eq{\tup{q}_\tau \cdot \tup{p}_\tau \simeq \tup{q}_\zr\cdot \tup{p}_\zr \simeq 0}, and  \eq{ \mag{\tup{p}_\tau} \simeq \mag{\tup{p}_\zr} \simeq \slang_\zr}.
\end{itemize}
\end{small}

\paragraph{Cartesian Coordinate Kepler Solutions.}
The Kepler solutions for cartesian position and velocity coordinates, \eq{(\tup{r},\tup{\v})}, may now be recovered from the above solutions for \eq{(\tup{q},u,\tup{p},w)} using either the full transformation as given in Eq.\eqref{PT_0} or using the simplified relations from Eq.\eqref{rv_qu_simp}. That is, 
\begin{small}
\begin{align} \label{rv_qu_simp_yetagain}
\begin{array}{rlllll}
     \begin{array}{cc}
     \scrsize{unsimplified}
    \end{array} \!\!: 
    &\quad  \tup{r}_\tau \,=\, \tfrac{1}{u_\tau}\htup{q}_\tau 
&,\qquad 
      \tup{\v}_\tau \,=\, -u_\tau (\tup{q}_\tau\wdg\tup{p}_\tau)\cdot \htup{q}_\tau - w_\tau\htup{q}_\tau 
   \;=\, -u_\tau \,\hdge{\tup{\slang}}_\tau \cdot \htup{q}_\tau - w_\tau\htup{q}_\tau 
\\
\begin{array}{cc}
     \; \\
     \scrsize{simplified with}  \\
     \fnsz{q=1, \; \htup{q}\cdot\tup{p}=0}  
\end{array} \!\!: 
&\quad
    \tup{r}_\tau  \,\simeq\, \tfrac{1}{u_\tau}\tup{q}_\tau
&,\qquad 
    \tup{\v}_\tau 
    \,\simeq\, 
    u_\tau \tup{p}_\tau - w_\tau \tup{q}_t 
\end{array}
\end{align}
\end{small}
We first consider the unsimplified relations. For the case at hand (Kepler dynamics), these are equivalent to \eq{\tup{r}_\tau =\tfrac{1}{q_0} \tfrac{1}{u_\tau}\tup{q}_\tau} and \eq{\tup{\v}_\tau =  -\tfrac{1}{q_0} ( u_\tau \hdge{\tup{\slang}}_\zr \cdot \tup{q}_\tau + w_\tau \tup{q}_\tau )} — because \eq{q} is always an integral of motion, and \eq{\hdge{\tup{\slang}}=\tup{q}\wdg\tup{p}} is an integral of motion for any central-force dynamics. 
Substituting into these relations the unsimplified Kepler solutions for \eq{(\tup{q}_\tau,u_\tau,\tup{p}_\tau,w_\tau)} from Eq.\eqref{qusol_s_2bp} leads to:
\begin{small}
\begin{flalign} \label{rv_kepsol_nosimp}
\;\;
\begin{array}{llllll}
     \tup{r}_\tau = \dfrac{\htup{q}_\tau}{u_\tau}
  \,=\, 
   \dfrac{ \htup{q}_\zr \csn{\tau} -  \hdge{\htup{\slang}} \cdot \htup{q}_\zr \snn{\tau} }{ (u_\zr-\frac{\kconst_1}{\slang^2})\csn{\tau} + \frac{1}{\slang}w_\zr\snn{\tau} + \frac{\kconst_1}{\slang^2} }
   \\[4pt]
   \;\;
\end{array} 
\quad,\quad
\begin{array}{llllll}
     \tup{\v}_\tau  = - u_\tau \, \hdge{\tup{\slang}}_\tau \cdot \htup{q}_\tau - w_\tau \htup{q}_\tau 
  &\!\!=\,   \tup{\v}_\zr  - \tfrac{\kconst_1}{\slang^2} \hdge{\tup{\slang}} \cdot \big[\htup{q}_\zr( \csn{\tau}-1) - \hdge{\htup{\slang}}\cdot\htup{q}_\zr \snn{\tau} \big]  
\\[4pt]
      &\!\!\! =\, \tup{\v}_\zr - \tfrac{\kconst_1}{\slang^2} \big[ \hdge{\tup{\slang}}\cdot \htup{q}_\zr(\csn{\tau}-1) +  \slang\htup{q}_\zr\snn{\tau}  \big]
\\[4pt]
    &\!\!=\,  \tup{\v}_\zr  - \tfrac{\kconst_1}{\slang^2} \hdge{\tup{\slang}} \cdot (\htup{q}_\tau - \htup{q}_\zr) 
\end{array} 
\;\;
\end{flalign}
\end{small}
with \eq{ \tup{\v}_\zr = - u_\zr \, \hdge{\tup{\slang}} \cdot \htup{q}_\zr - w_\zr \htup{q}_\zr}
the velocity at \eq{\tau_\zr=0}.\footnote{If \eq{\tau=0} coincides with periapsis, then \eq{\tau} is the true anomaly and \eq{\mag{\tup{\v}_\tau} \leq \mag{\tup{\v}_\zr}}.} 
The above may be written in terms of \eq{(\tup{r}_\zr,\tup{\v}_\zr)} using:
\begin{small}
\begin{align} \label{qp_rv_init_rels}
\begin{array}{llll}
     \htup{q}_\zr = \htup{r}_0
 \\[3pt]
      u_\zr = 1/r_0
\\[3pt]
    w_\zr = -\htup{r}_\zr \cdot \tup{\v}_\zr = - \dot{r}_0
\end{array}
 &&,&&
\begin{array}{llll}
        \hdge{\tup{\slang}} =  \hdge{\tup{\slang}}_\zr = \tup{q}_\zr\wdg \tup{p}_\zr =  \tup{r}_\zr\wdg \tup{\v}_\zr
\\[3pt]
     \tup{\slang}=\tup{\slang}_0 = \hdge{\tup{p}}_\zr\cdot\tup{q}_\zr = \hdge{\tup{\v}}_\zr\cdot\tup{r}_\zr
 \\[3pt]
      \slang^2 = \slang_\zr^2 = q_\zr^2 p_\zr^2 - (\tup{q}_\zr \cdot\tup{p}_\zr)^2 = r_\zr^2 \v_\zr^2 - (\tup{r}_\zr \cdot\tup{\v}_\zr)^2
\end{array}
 &&,&&
\begin{array}{rllll}
      \tup{\v}_\zr &\!\!\!\!=\, - u_\zr  \hdge{\tup{\slang}} \cdot \htup{q}_\zr - w_\zr \htup{q}_\zr
\\[3pt]
     &\!\!\!\!\simeq\,  u_\zr\tup{p}_\zr-w_\zr\tup{q}_\zr
\end{array}
\end{align}
\end{small}
where none of the above require the simplifications \eq{q=1} or \eq{\htup{q}\cdot\tup{p}=0}. 
Eq.\eqref{rv_kepsol_nosimp} then leads to
\begin{small}
\begin{align} \label{rv_kepsol_rv0}
      \tup{r}_\tau  
      \,=\,  \dfrac{ \htup{r}_\zr\csn{\tau}  \,-\,  \hdge{\htup{\slang}} \cdot\htup{r}_\zr \snn{\tau}}{ (1/r_\zr-\frac{\kconst_1}{\slang^2})\csn{\tau} \,-\, \frac{1}{\slang}\htup{r}_\zr \cdot \tup{\v}_\zr\snn{\tau} \,+\, \frac{\kconst_1}{\slang^2}   }  
      \,=\, \dfrac{\htup{r}_\tau}{1/r_\tau}
\qquad,\qquad 
\begin{array}{llllll}
     \tup{\v}_\tau  &\!\!\! =\, \tup{\v}_\zr  - \tfrac{\kconst_1}{\slang^2} \hdge{\tup{\slang}} \cdot \big[\htup{r}_\zr( \csn{\tau}-1) - \hdge{\htup{\slang}}\cdot\htup{r}_\zr \snn{\tau} \big] 
\\[4pt]
      &\!\!\! =\, \tup{\v}_\zr - \tfrac{\kconst_1}{\slang^2} \big[ \hdge{\tup{\slang}}\cdot \htup{r}_\zr(\csn{\tau}-1) +  \slang\htup{r}_\zr\snn{\tau}  \big]
\\[4pt]
   &\!\!\! =\,  \tup{\v}_\zr - \tfrac{\kconst_1}{\slang^2} \hdge{\tup{\slang}}\cdot(\htup{r}_\tau - \htup{r}_\zr )
\end{array}
\end{align}
\end{small}
with \eq{\tup{\slang}=\tup{\slang}_\zr} and \eq{\slang=\slang_\zr} given in terms of \eq{(\tup{r}_\zr,\tup{\v}_\zr)} as in Eq.\eqref{qp_rv_init_rels}. 

\begin{small}
\begin{itemize}
    \item \textit{Simplifications.} Consider the simplified relations in Eq.\eqref{rv_qu_simp_yetagain}. Substitution of the simplified Kepler solutions for \eq{(\tup{q}_\tau,u_\tau,\tup{p}_\tau,w_\tau)} from Eq.\eqref{qusol_s_2bp_simp} then leads to
    \begin{small}
    \begin{flalign} \label{rv_kepsol} 
    \qquad
    \begin{array}{llllll}
          \tup{r}_\tau \simeq \dfrac{\tup{q}_\tau }{u_\tau} 
          &\!\!\! \simeq\,  \dfrac{\tup{q}_\zr\csn{\tau}  \,+\,  \frac{1}{\slang} \tup{p}_\zr\snn{\tau}}{ (u_\zr-\frac{\kconst_1}{\slang^2})\csn{\tau} + \frac{1}{\slang}w_\zr\snn{\tau} + \frac{\kconst_1}{\slang^2}   }  
    \end{array} 
    &&,&&
    \begin{array}{llllll}
         \tup{\v}_\tau  \simeq u_\tau\tup{p}_\tau - w_\tau\tup{q}_\tau 
        &\!\!\! \simeq\,  \tup{\v}_\zr + \tfrac{\kconst_1}{\slang^2} \big[ \tup{p}_\zr(\csn{\tau}-1) -  \slang\tup{q}_\zr\snn{\tau}  \big]
        \\[4pt]
        &\!\! \simeq\,  \tup{\v}_\zr + \tfrac{\kconst_1}{\slang^2} (\tup{p}_\tau - \tup{p}_\zr )
    \end{array}
    \;\;
    \end{flalign}
    \end{small}
    where  \eq{\tup{\v}_\zr \simeq  u_\zr\tup{p}_\zr-w_\zr\tup{q}_\zr}. The above agrees with Eq.\eqref{rv_kepsol_nosimp} and Eq.\eqref{rv_kepsol_rv0} using the following relations (cf.~Eq.\eqref{qp_rv_0}):
    \begin{small}
    \begin{align}
          \tup{q}\simeq \htup{q} = \htup{r}
          \qquad,\qquad 
          \tup{p} \simeq -\hdge{\tup{\slang}}\cdot \tup{q} \simeq -\hdge{\tup{\slang}}\cdot \htup{q} =  -\hdge{\tup{\slang}}\cdot \htup{r} 
    \end{align}
    \end{small}
\end{itemize}
\end{small}

\paragraph{Relations with the Perifocal Basis.} 
The close relation between the projective coordinates, \eq{(\tup{q},u,\tup{p},p_\ss{u})}, and the LVLH basis was discussed in Eq.\eqref{qpl_rels}-Eq.\eqref{qpl_rels2}. 
Let us now consider another set of three  mutually orthogonal vectors often encountered in Kepler dynamics; the Laplace-Runge-Lenz (LRL) vector/eccentricity vector, \eq{\tup{e}}, and the Hamilton vector, \eq{\tup{h}}, along with the usual specific angular momentum 
vector, \eq{{\tup{\slang}}}.\footnote{The Laplace-Runge-Lenz (LRL) vector and the Hamilton vector are often defined as \eq{\tup{E}=\tup{\v}\tms {\tup{\slang}}-\kconst_1 \htup{r}} and \eq{\tup{H}=\tfrac{1}{\slang^2}{\tup{\slang}}\tms \tup{E}}, respectively.   The eccentricity vector \eq{\tup{e}} and Hamilton vector \eq{\tup{h}} seen in Eq.\eqref{LRL_e} are simply a scaling by \eq{\kconst_1}, that is, \eq{\tup{e}=\tfrac{1}{\kconst_1}\tup{E}} and \eq{\tup{h}=\tfrac{1}{\kconst_1}\tup{H}}. We should also not that, in the present context, all of these ``vectors'' are really just the inertial cartesian components (\eq{\mbb{R}^3}-valued functions). }
Their cartesian components are given in terms of the inertial cartesian coordinates, \eq{(\tup{r},\tup{\v})}, and in terms of the projective coordinates (using Eq.\eqref{PT_0} or Eq.\eqref{rv_qu_simp}) as 
follows:\footnote{It has already been shown that \eq{\tup{\slang} = \tup{r}\tms \tup{\v} = \tup{q}\tms \tup{p}}. The expressions for \eq{ \kconst_1 \tup{e}} and \eq{\kconst_1  \tup{h}} in terms of \eq{(\bartup{q},\bartup{p})} are seen as follows using the relations in Eq.\eqref{PT_0} or Eq.\eqref{rv_qu_simp}:\\
        \eq{\qquad\qquad\qquad \kconst_1 \tup{e} \,=\
        \hdge{\tup{\slang}}\cdot\tup{\v} - \kconst_1 \htup{r} 
        \,=\, 
         \hdge{\tup{\slang}}\cdot (- u \hdge{\tup{\slang}}\cdot \htup{q} - w \htup{q}) 
         - \kconst_1 \htup{q} 
         \,=\, 
         \slang^2 u \htup{q} - w \hdge{\tup{\slang}}\cdot \htup{q}   - \kconst_1 \htup{q}
         \,=\, 
         \slang^2(u -\tfrac{\kconst_1 }{\slang^2} )\htup{q} -   w \hdge{\tup{\slang}}\cdot\htup{q}
          \,\simeq \,
          \slang^2(u -\tfrac{\kconst_1 }{\slang^2} )\tup{q} +   w \tup{p}
        } \\
        \eq{\qquad\qquad\qquad 
         \kconst_1  \tup{h} 
        \,=\, \tup{\v} + \tfrac{\kconst_1 }{\slang^2} \hdge{\tup{\slang}}\cdot \htup{r} 
        \,=\, 
        - u \hdge{\tup{\slang}}\cdot \htup{q} - w \htup{q} + + \tfrac{\kconst_1 }{\slang^2} \hdge{\tup{\slang}}\cdot \htup{q}
        \,=\,
        -(u - \tfrac{\kconst_1 }{\slang^2})\hdge{\tup{\slang}}\cdot \htup{q} - w \htup{q}
        \,\simeq\,
        (u - \tfrac{\kconst_1 }{\slang^2})\tup{p} - w \tup{q}
        } . 
    }
\begin{small}
\begin{flalign} \label{LRL_e}
&&
\begin{array}{rllllll}
      \kconst_1 \tup{e}
      &\!\!\! =\, \hdge{\tup{\slang}}\cdot\tup{\v}  - \kconst_1 \htup{r} 
      &=\, 
       \slang^2(u -\tfrac{\kconst_1 }{\slang^2} )\htup{q} \,-\,   w \hdge{\tup{\slang}}\cdot\htup{q}
      & \simeq\,  \slang^2(u -\tfrac{\kconst_1 }{\slang^2} )\tup{q} \,+\,   w \tup{p} 
\\[4pt]
    \kconst_1\tup{h} =\, -\tfrac{\kconst_1 }{\slang^2} \hdge{\tup{\slang}}\cdot \tup{e}
    &\!\!\! =\, \tup{\v} + \tfrac{\kconst_1}{\slang^2} \hdge{\tup{\slang}}\cdot \htup{r}
    & =\, 
     -(u -\tfrac{\kconst_1 }{\slang^2} )\hdge{\tup{\slang}}\cdot\htup{q} \,-\,   w \htup{q} 
     & \simeq\, (u - \tfrac{\kconst_1 }{\slang^2})\tup{p} \,-\,  w \tup{q} 
\\[4pt]
    \tup{\slang} &\!\!\! =\, \hdge{\tup{\v}}\cdot\tup{r} 
     & =\,  \hdge{\tup{p}}\cdot\tup{q}
\end{array}
&&
w:=u^2 p_\ss{u}
\qquad
\end{flalign}
\end{small}
where ``\eq{\simeq}'' denotes expressions simplified using \eq{q=1} and \eq{\htup{q}\cdot\tup{p}=0}.
It is well-known that the above \eq{\mbb{R}^3}-valued functions are all conserved for pure Kepler dynamics.
Their normalization defines an orthonormal basis, \eq{\{\htup{e},\htup{h},\htup{\slang}\}=:\{\htup{o}_e,\htup{o}_h,\htup{o}_\slang\}}, often called the \textit{perifocal basis}. For Keplerian motion, this basis is 
constant/stationary.\footnote{with  \eq{\htup{e}} and \eq{\htup{h}} defining the orbital plane;  \eq{\htup{e}} directed towards periapsis,  \eq{\htup{h}} directed along the velocity at periapsis (direction tangent to the orbit), and \eq{\hat{\tup{\slang}}} directed normal to the orbit plane, completing the right-handed triad. The magnitude \eq{e=\mag{\tup{e}}} is the usual dimensionless eccentricity of the orbit. }
We may verify this by substituting into the above the simplified Kepler solutions from Eq.\eqref{qusol_s_2bp_simp}, leading to:
\begin{small}\begin{align} \label{LRL_e_qp0}
\begin{array}{rllllll}
      \kconst_1 \tup{e}_\tau
      & \simeq\, 
      \slang^2_\tau (u_\tau -\tfrac{\kconst_1 }{\slang^2_\tau} )\tup{q}_\tau \,+\,   w_\tau \tup{p}_\tau 
      &\simeq\, \slang^2_\zr (u_\zr - \tfrac{\kconst_1}{\slang^2_0}) \tup{q}_\zr \,+\, w_\zr \tup{p}_\zr
\\[4pt]
    \kconst_1\tup{h}_\tau 
     & \simeq\, (u_\tau - \tfrac{\kconst_1 }{\slang^2_\tau})\tup{p}_\tau \,-\,  w_\tau \tup{q}_\tau 
     & \simeq\, (u_\zr - \tfrac{\kconst_1 }{\slang^2_0})\tup{p}_\zr \,-\,  w_\zr \tup{q}_\zr 
\\[4pt]
    \tup{\slang}_\tau 
     & =\,  \hdge{\tup{p}}_\tau\cdot\tup{q}_\tau 
     & =\, \hdge{\tup{p}}_\zr \cdot\tup{q}_\zr 
\end{array}
\end{align}\end{small}
verifying that \eq{(\tup{e},\tup{h},\tup{\slang})=(\tup{e}_\zr,\tup{h}_\zr,\tup{\slang}_\zr)} are indeed conserved for pure Kepler motion. 
Furthermore, if \eq{\tau_\zr=0} coincides with periapsis such that \eq{\tau} is the actual true anomaly, then \eq{\dot{r}_\zr=0} and \eq{w_\zr = \pdt{u}_\zr = -\dot{r}_\zr = 0 } vanishes and the first two of the above simplify to the following (with \eq{\slang=\slang_\zr}):
\begin{small}
\begin{flalign} \label{LRL_e_qp0_periapsis}
\quad
\begin{array}{cccc}
     \scrsize{if  $\tau_\zr=0$} \\
     \scrsize{is periapsis}
\end{array}  
\!\!\!:
&&
w_\zr = -\dot{r}_\zr = 0  
\qquad \Rightarrow \qquad 
\begin{array}{rllllll}
      \kconst_1 \tup{e}
       &\!\!\! \simeq\, \slang^2 (u_\zr - \tfrac{\kconst_1}{\slang^2}) \tup{q}_\zr 
\\[4pt]
    \kconst_1\tup{h} 
      &\!\!\! \simeq\, (u_\zr - \tfrac{\kconst_1 }{\slang^2})\tup{p}_\zr 
\end{array}
&&
\end{flalign}
\end{small}
Note that \eq{\kconst_1/\slang^2}, which appears frequently, is the inverse of the well-known \textit{semilatus rectum}, \eq{P_\ss{\mrm{slr}} = \slang^2/\kconst_1}. Still assuming \eq{\tau_\zr=0} coincides with periapsis (i.e., \eq{r_\zr = r_\ss{\mrm{min}}}), then we may use the conic section formula, \eq{r_\tau = P_\ss{\mrm{slr}}/(1+e\csn{\tau})}, to obtain:
\begin{small}
\begin{flalign}
\quad
\begin{array}{cccc}
     \scrsize{if  $\tau_\zr=0$} \\
     \scrsize{is periapsis}
\end{array}  
\!\!\!:
&&
\begin{array}{llll}
    u_\tau -  \tfrac{\kconst_1}{\slang^2}  =  \tfrac{1}{r} - \tfrac{1}{P_{\mrm{slr}}} 
     =  \tfrac{\kconst_1}{\slang^2} e \csn{\tau}
&,\qquad 
  u_\zr -  \tfrac{\kconst_1}{\slang^2}  =  \tfrac{1}{r_{\mrm{p}}} - \tfrac{1}{P_{\mrm{slr}}} 
   =  \tfrac{\kconst_1}{\slang^2} e 
   \,=\, \tfrac{e}{(1+e) r_{\mrm{p}} }
   \,=\, \tfrac{e}{(1+e)  } u_\ss{\mrm{p}}
\end{array}
&&
\end{flalign}
\end{small}
where \eq{e=\mag{\tup{e}}} is the classic eccentricity and where \eq{(\cdot)_\zr = (\cdot)_\ss{\mrm{p}}} corresponds to periapsis (i.e., \eq{r_\zr = r_\ss{\mrm{p}}=r_\ss{\mrm{min}} = P_\ss{\mrm{slr}}/(1+e)} and \eq{u_\zr = u_\ss{\mrm{p}}=u_\ss{\mrm{max}} =(1+e)/P_\ss{\mrm{slr}}} correspond to the closest approach). 
We may also consider the above within specific cases (e.g., elliptic, parabolic, etc.), for which we note:
\begin{small}
\begin{flalign}
\quad
\begin{array}{cccc}
     \scrsize{if  $\tau_\zr=0$} \\
     \scrsize{is periapsis}
\end{array}  
\!\!\!:
&&
\begin{array}{rlll}
     \scrsize{circular } (e =0): &\quad
  u_\zr -  \tfrac{\kconst_1}{\slang^2} =  u_\tau -  \tfrac{\kconst_1}{\slang^2} \,=\, 0
\\[4pt]
     \scrsize{elliptic } (e < 1): &\quad
   u_\zr -  \tfrac{\kconst_1}{\slang^2}  =  \tfrac{1}{2} \tfrac{r_{\mrm{a}} - r_{\mrm{p}} }{ r_{\mrm{a}}r_{\mrm{p}}}
    =  \tfrac{1}{2}( \tfrac{1}{r_{\mrm{p}}} - \tfrac{1}{r_{\mrm{a}}})
    \;= 
   \tfrac{1}{2}( u_\ss{\mrm{p}} - u_\ss{\mrm{a}})
\\[4pt]
  \scrsize{parabolic } (e =1): &\quad
  u_\zr -  \tfrac{\kconst_1}{\slang^2} = \tfrac{1}{2 r_{\mrm{p}}} \,=\, \tfrac{1}{2} u_\zr \,=\, \tfrac{1}{2} u_\ss{\mrm{p}}
\\[4pt]
  \scrsize{hyperbolic } (e>1): &\quad
  u_\zr -  \tfrac{\kconst_1}{\slang^2} =
  \tfrac{\kconst_1}{\slang^2} e 
   \,=\, \tfrac{e}{(1+e)  } u_\ss{\mrm{p}}
   \;=\;
   \begin{matrix}
        \scrsize{same as above} \\[-1pt]
         \scrsize{general relations}
   \end{matrix}
\end{array}
&&
\end{flalign}
\end{small}
The equalities indicated above for the elliptic  case (\eq{e<1}) have also used the well-known elliptic relations \eq{P_\ss{\mrm{slr}} = \slang^2/\kconst_1 =2 \tfrac{ r_{\mrm{a}}r_{\mrm{p}}}{r_{\mrm{a}} + r_{\mrm{p}} }  } and \eq{e = \tfrac{ r_{\mrm{a}} - r_{\mrm{p}}}{r_{\mrm{a}} + r_{\mrm{p}} } }, where \eq{r_\ss{\mrm{a}}=r_\ss{\mrm{max}}} and \eq{u_\ss{\mrm{a}} = 1/r_\ss{\mrm{a}}=u_\ss{\mrm{min}}} corresponds to apoapsis (farthest approach of an elliptic/closed orbit).
Although apoapsis only exists in truth for elliptic/closed orbits, note that the above relations for the elliptic case (\eq{e<1}) also hold in the parabolic case (\eq{e=1}) using the limits \eq{r_\ss{\mrm{a}}\mapsto \infty} and, thus, \eq{u_\ss{\mrm{a}}=1/r_\ss{\mrm{a}}\mapsto 0}.

\section{NEW ORBIT ELEMENTS FOR ARBITRARILY-PERTURBED KEPLER DYNAMICS} \label{sec:VOP_2}

We now pick back up with the closed-form \eq{s}- or \eq{\tau}-parameterized Kepler solutions in projective coordinates given in section \ref{sec:kep_sol} (also in appendix \ref{sec:prj_STM_new}).
More specifically, solutions in our ``\eq{w}-modified'' projective coordinates wherein the conjugate pair \eq{(u,p_\ss{u})} is replaced by the quasi-conjugate pair \eq{(u,w:=u^2 p_\ss{u})}. 
We will use the closed-form solutions for \eq{(\tup{q},u,\tup{p},w)} to define a set of orbit elements, \eq{(\tup{Q},U,\tup{P},W)}, along with their governing equations of motion,  which correspond to constant initial conditions in the unperturbed Kepler case and which are osculating elements in the general perturbed case. 

Some of these developments will use the matrix formulation of Kepler dynamics in projective coordinates detailed in appendix \ref{sec:prj_STM_new}.

\subsection{Variation of Parameters for ``Initial Conditions as Elements''}


Before considering the problem at hand (orbital dynamics in projective coordinates), let us first outline the general method of variation of parameters (VOP) for ``initial conditions as elements'' for any finite-dimensional autonomous ODE on \eq{\mbb{R}^\en}.

\paragraph{General Method.}


Though our notation here may overlap with specific notation used elsewhere in this work, the following general theory is in no way specific to orbital dynamics or projective coordinates. It pertains to any finite-dimensional first order system of autonomous ODEs. 
Here, we will use \eq{\rng{\square}:=\diff{}{\tau}} where \eq{\tau} is any arbitrary evolution parameter. We also use \eq{\tup{x}\in\mbb{R}^\en} to denote any arbitrary set of state space coordinates.
The general method is as follows:
\begin{small}
\begin{enumerate}[left=1ex]
    \item Staring with some unperturbed dynamical system, let \eq{\rng{\tup{x}} = \tup{X}(\tup{x})} be an autonomous first-order ODE on \eq{\mbb{R}^\en}. Assume that this has known closed-form solution, \eq{\tup{x}_\tau = \varphi_\tau (\tup{x}_\zr)}, for arbitrary initial conditions \eq{\tup{x}_\zr\in\mbb{R}^\en}. Here, we may view the vector field simply as a smooth map, \eq{\tup{X}:\mbb{R}^\en \to \tsp\pmb{.}\mbb{R}^\en \cong \mbb{R}^\en}, with \eq{\tau}-parameterized flow \eq{\varphi_\tau:\mbb{R}^\en \to \mbb{R}^\en} and inverse \eq{ \inv{\varphi_\tau} = \varphi_{-\tau}} which, by definition, satisfies:
    \begin{small}
    \begin{align}
         \pderiv{}{\tau} \varphi_\tau = \tup{X}\circ {\varphi_\tau}
    \end{align}
    \end{small}
    \item  We now consider the perturbed ODE described by \eq{\rng{\tup{x}} = \tup{X}(\tup{x}) + \tup{Y}(\tup{x})}, where \eq{\tup{Y}} models the perturbations. 
    \item Using the same \textit{un}perturbed flow, \eq{\varphi_\tau}, of \eq{\tup{X}}, we define new coordinates, \eq{\tup{\xi}}, by a \eq{\tau}-dependent transformation:
    \begin{small}
    \begin{align} \label{vopGen_xform}
        \tup{\xi}:=  \inv{\varphi_\tau}(\tup{x}) 
        \quad \leftrightarrow \quad 
        \tup{x}=\varphi_\tau (\tup{\xi})
    \end{align}
    \end{small}
    \item  We wish to find the governing equations for \eq{\tup{\xi}} such that \eq{\tup{x}_\tau=\varphi_\tau (\tup{\xi}_\tau)} is a solution to the perturbed ODE \eq{\rng{\tup{x}} = \tup{X}(\tup{x}) + \tup{Y}(\tup{x})}.  In the unperturbed case \eq{\tup{Y}=0}, then we should get constant solutions \eq{\tup{\xi}_\tau = \tup{x}_\zr}. That is, \eq{\tup{\xi}} defined by Eq.\eqref{vopGen_xform} are ``slow'' coordinates (often called \textit{elements} in the context of celestial mechanics); they evolve slowly in the presence of perturbations, and remain constant for the original, unperturbed, dynamics. In a bit more detail: 
    \begin{small}
    \begin{itemize}[nosep]
        \item We wish to obtain some (generally, nonautonomous) ODE \eq{\rng{\tup{\xi}} = \tup{E}_\tau(\tup{\xi})} whose solutions \eq{\tup{\xi}_\tau=\rho_{\tau,\tau_0}(\tup{\xi}_\zr)} are such that \eq{\tup{x}_\tau=\varphi_\tau (\tup{\xi}_\tau)} is a solution to the perturbed ODE \eq{\rng{\tup{x}} = \tup{X}(\tup{x}) + \tup{Y}(\tup{x})}. 
         Assuming \eq{\tau_\zr=0}, then  define \eq{\rho_\tau:=\rho_{\tau,0}} and note that \eq{\tup{\xi}_\zr = \inv{\varphi_\zr}(\tup{x}_\zr) = \tup{x}_\zr} such that 
        \eq{\tup{x}_\tau=\varphi_\tau (\tup{\xi}_\tau) = \varphi_\tau \circ \rho_\tau (\tup{\xi}_\zr) = \varphi_\tau \circ \rho_\tau (\tup{x}_\zr) }.
        I.e., the flow of the perturbed dynamics, \eq{\tup{X}+\tup{Y}}, is given by \eq{\varphi_\tau \circ \rho_\tau}, where \eq{\varphi_\tau} is the original unperturbed flow of \eq{\tup{X}} (and also the map for the coordinate transformation \eq{\tup{x}\leftrightarrow \tup{\xi}}), and where \eq{\rho_{\tau}=\rho_{\tau,0}} is the flow of the yet-to-be-determined vector field \eq{\tup{E}_\tau} for \eq{\rng{\tup{\xi}} = \tup{E}_\tau(\tup{\xi})}.
     \end{itemize}
    \end{small}
    \item Applying the chain rule to \eq{\tup{x}=\varphi_\tau (\tup{\xi})}, and making use of the relation \eq{\pderiv{}{\tau} \varphi_\tau = \tup{X}\circ {\varphi_\tau}}, we have:
    \begin{small}
    \begin{align}
         \rng{\tup{x}} = \diff{}{\tau} \varphi_\tau(\tup{\xi}) \,=\,  \pderiv{\varphi_\tau}{\tup{\xi}} \cdot\rng{\tup{\xi}} \,+\, \pderiv{\varphi_\tau}{\tau}(\tup{\xi}) 
         \,=\, \pderiv{\varphi_\tau}{\tup{\xi}} \cdot\rng{\tup{\xi}} \,+\, \tup{X}(\varphi_\tau(\tup{\xi}))
         \,=\, \pderiv{\varphi_\tau}{\tup{\xi}} \cdot\rng{\tup{\xi}} \,+\, \tup{X}(\tup{x})
    \end{align}
    \end{small}
    \item Equating the above with  \eq{\rng{\tup{x}} = \tup{X}(\tup{x}) + \tup{Y}(\tup{x})} and solving for \eq{\rng{\tup{\xi}}} leads to:
    \begin{small}
    \begin{align} \label{vopGen_ode1}
        \begin{array}{lllll}
             \rng{\tup{x}} = \tup{X}(\tup{x}) + \tup{Y}(\tup{x})
         \;\;=\;\; 
              \rng{\tup{x}} =  \pderiv{\varphi_\tau}{\tup{\xi}} \cdot\rng{\tup{\xi}} \,+\, \tup{X}(\tup{x})
        \end{array}
          \qquad \Rightarrow \qquad 
            \rng{\tup{\xi}} \,=\, \inv{( \pderiv{\varphi_\tau}{\tup{\xi}} )} \cdot \tup{Y}(\tup{x}) \;=\; \inv{( \pderiv{\varphi_\tau}{\tup{\xi}} )} \cdot \tup{Y}( \varphi_\tau(\tup{\xi}))
    \end{align}
    \end{small}
    For our purposes, the partials of \eq{\varphi_\tau} are simply the Jacobian of the coordinate transformation in Eq.\eqref{vopGen_xform}. That is, 
    adopting the common informal notation \eq{\pderiv{\varphi_\tau}{\tup{\xi}} \equiv \pderiv{\tup{x}}{\tup{\xi}}} and  \eq{\inv{(\pderiv{\varphi_\tau}{\tup{\xi}})} \equiv \inv{(\pderiv{\tup{x}}{\tup{\xi}})} \equiv \pderiv{\tup{\xi}}{\tup{x}} \equiv \pderiv{(\inv{\varphi_\tau})}{\tup{x}}} — note these Jacobians are \eq{\tau}-dependent — then we may write:
    \begin{small}
    \begin{align} \label{vopGEN_ode_inf}
         \boxed{ \rng{\tup{\xi}} \,=\, \inv{(\pderiv{\tup{x}}{\tup{\xi}})} \cdot\tup{Y} \,=\, \pderiv{\tup{\xi}}{\tup{x}} \cdot \tup{Y} }
         \qquad\qquad 
         \fnsize{i.e.,} \quad
            \rng{\tup{\xi}} \,=\, \tup{E}_{\tau}
        \quad,\quad
         \tup{E}_\tau = \inv{(\pderiv{\tup{x}
         }{\tup{\xi}})} \cdot\tup{Y} = \pderiv{\tup{\xi}}{\tup{x}} \cdot \tup{Y}
    \end{align}
    \end{small}
    Note we have omitted some compositions; the above should really be  \eq{\rng{\tup{\xi}} = \inv{(\pderiv{\tup{x}}{\tup{\xi}})} \cdot \tup{Y}( \varphi_\tau(\tup{\xi})) = (\pderiv{\tup{\xi}}{\tup{x}} \cdot \tup{Y})\circ \varphi_\tau(\tup{\xi})}, where the composition ensures that the right-hand-side of the above is expressed in terms of \eq{\tup{\xi}}. 
        A more formal statement of Eq.\eqref{vopGEN_ode_inf} would 
        be:\footnote{Note that Eq.\eqref{vopGen_ode1} is equivalent to:   
        \begin{align}
            \rng{\tup{\xi}} \,=\, \inv{( \pderiv{\varphi_\tau}{\tup{\xi}} )} \cdot \tup{Y}(\tup{x}) \;=\; \inv{( \pderiv{\varphi_\tau}{\tup{\xi}} )} \cdot \tup{Y}( \varphi_\tau(\tup{\xi}))
            \;\;\equiv\, \inv{(\dif \varphi_\tau(\tup{\xi}))} \cdot \tup{Y}(\varphi_\tau(\tup{\xi}))
            \;=\; 
            (\dif \inv{\varphi_\tau} \cdot  \tup{Y})_{\varphi_\tau (\tup{\xi})} 
            \;=\; (\varphi_\tau^* \tup{Y})_{\tup{\xi}}
        \end{align} }
        \begin{small}
        \begin{align}
         \rng{\tup{\xi}} = \tup{E}_\tau(\tup{\xi})
            \qquad,\qquad 
            \tup{E}_\tau := \varphi_\tau^* \tup{Y} \,=\,  (\dif \inv{\varphi_\tau} \cdot  \tup{Y})\circ {\varphi_\tau} 
        \end{align}
        \end{small}
    Yet, given the informal notational conventions used throughout this work, the less precise version in Eq.\eqref{vopGEN_ode_inf} will be sufficient for our purposes.  
\end{enumerate}
\end{small}



\paragraph{Applied to the Problem at Hand.}
Now return to the problem at hand:~Kepler dynamics expressed in our ``\eq{w}-modified'' projective coordinates \eq{\tup{x}= (\tup{q},u,\tup{p},w)\in\mbb{R}^8} as detailed in section \ref{sec:kep_sol} (or, equivalently, in appendix \ref{sec:prj_STM_new}). Recall that \eq{w:=u^2 p_\ss{u}} is a quasi-momentum coordinate that we use in place of the conjugate momentum coordinate \eq{p_\ss{u}} 
(this makes the \textit{first}-order Kepler dynamics fully linear\footnote{We saw in section \ref{sec:prj_sum} that the \textit{first}-order \eq{s}- or \eq{\tau}-parameterized Kepler dynamics for \eq{(\tup{q},\tup{p},u,w)} are fully linear, while those of the true canonical coordinate set \eq{(\tup{q},\tup{p},u,p_\ss{u})} are not fully linear due to nonlinearity in the \eq{(u,p_\ss{u})} dynamics.}).

\begin{notesq}
   For the following, it will be convenient to re-order our phase space coordinates \eq{(\tup{q},u,\tup{p},w)} to the modified order \eq{(\tup{q},\tup{p},u,w)}. That is,  we split them into a ``\eq{(\tup{q},\tup{p})}-part'' (rotational motion) and ``\eq{(u,w)}-part'' (radial motion):
\begin{small}
\begin{align}
\begin{array}{cc}
      \scrsize{standard}  \\
     \scrsize{ordering}
\end{array}\!\!\!:
\left.\quad 
\begin{array}{llll}
      \tup{x} = (\tup{q},u,\tup{p},w) 
      \simeq (\bartup{q},\pdt{\bartup{q}})
\end{array}\right.
&&,&&
\begin{array}{cc}
      \scrsize{modified}  \\
     \scrsize{ordering}
\end{array}\!\!\!:
\left.\quad 
\begin{array}{llll}
       \tup{x} = (\tup{q},\tup{p},u,w) 
      \simeq (\tup{q},\pdt{\tup{q}},u,\pdt{u})
\end{array}\right.
\end{align}
Unless indicated otherwise, any matrices appearing in the following are in relation to the modified ordering  \eq{ \tup{x} = (\tup{q},\tup{p},u,w) }. 
\end{small}
\end{notesq}

\vspace{1ex}
\noindent 
So far, we have been simultaneously considering the \eq{s}- and \eq{\tau}-parameterized perturbed (or not) Kepler dynamics.  
We will now stick with \eq{\tau} as our chosen evolution parameter, where \eq{\mrm{d} t = (r^2/\slang)\mrm{d}\tau = 1/(\slang u^2)\mrm{d}\tau}, making \eq{\tau} the true anomaly up to an additive constant. 
Everything that follows could be also carried out using \eq{s} as the evolution parameter rather than \eq{\tau} (where \eq{\mrm{d}t = r^2 \mrm{d}s = u^\ss{-2}\mrm{d} s}). However, the \eq{\tau}-parameterized Kepler dynamics have unit natural frequency \eq{1}, whereas the  \eq{s}-parameterized dynamics have natural frequency \eq{\slang} (angular momentum magnitude) which is generally \textit{not} constant in the perturbed case and the simple relation \eq{\tau=\slang s} does \textit{not} hold. We therefore choose to use \eq{\tau} rather than \eq{s} for simplicity. In particular, to simplify the variation of parameters (VOP) 
procedure.\footnote{Taking partials w.r.t \eq{(\tup{q},\tup{p})} of combinations of \eq{\snn{\slang s}} and \eq{\csn{\slang s}} is more burdensome than taking partials of combinations of \eq{\snn{\tau}} and \eq{\csn{\tau}}.}
The resulting ODEs can always be reparameterized by an appropriate conformal scaling.

The \eq{\tau}-parameterized perturbed Kepler dynamics are given as a first-order matrix ODE as follows:
\begin{small}
\begin{flalign}
\qquad
   \tup{x} =  (\tup{q},\tup{p},u,w)
&&,&&
    \rng{\tup{x}} = \tup{X} + \tup{Y} \,=\, \tfrac{1}{\slang}M(\tup{\slang})\cdot\tup{x} + \tfrac{1}{\slang} \tup{k}  \,+\, \tup{Y}
&&,&&
     \tup{Y} = (\tup{0},\rng{t} \tup{f}, 0 , \tfrac{1}{\slang} f_\ss{u})
&&,&&
\rng{t}=\tfrac{1}{\slang u^2}
\end{flalign}
\end{small}
where the term \eq{\tup{X} = \tfrac{1}{\slang}M(\tup{\slang})\cdot\tup{x} + \tfrac{1}{\slang} \tup{k}} corresponds to the unperturbed Kepler dynamics — with \eq{M\in\spmat{6;2}\subset \mbb{R}^{8 \times 8}} and \eq{\tup{k}\in\mbb{R}^8} defined in section \ref{sec:prj_STM_unsimp} — and where the perturbations, \eq{\tup{Y}\in\mbb{R}^8}, are given in terms of the generalized forces \eq{(\tup{f},f_\ss{u})\in\mbb{R}^4}  as seen above (with \eq{(\tup{f},f_\ss{u})} defined in Eq.\eqref{Ftotal_qp}). 

We denote the elements corresponding to initial conditions of \eq{\tup{x}} by 
\eq{\tup{\xi}=(\tup{Q},\tup{P},U,W)}. That is, the elements \eq{\tup{\xi}} are defined from the coordinates \eq{\tup{x}} using the \eq{\tau}-parameterized Kepler solutions from section \ref{sec:kep_sol} (equivalently, section \ref{sec:prj_STM_new}). We may, or may not, choose to simplify this transformation using the integrals of motion \eq{q=1} and \eq{\lambda=\htup{q}\cdot\tup{p}=0}.
We will consider first the unsimplified case, then the simplified case. 
In \textit{either} case, we note that \eq{ \pderiv{\tup{\xi}}{\tup{x}}} is partitioned as:
\begin{small}
\begin{flalign} \label{elem_jacob_genKep}
\qquad
\begin{array}{llll}
      \tup{x} =  (\tup{q},\tup{p},u,w)
\\[3pt]
     \tup{\xi}=(\tup{Q},\tup{P},U,W) 
\end{array}
&&
  \pderiv{\tup{\xi}}{\tup{x}}
     \,=
 \fnpmat{ 
     \pderiv{(\tup{Q},\tup{P})}{(\tup{q},\tup{p})}
     &    \pderiv{(\tup{Q},\tup{P})}{(u,w)} 
     \\[6pt]
     \pderiv{(U,W)}{(\tup{q},\tup{p})}
     &
      \pderiv{(U,W)}{(u,w)}
     }
\;=\;
 \fnpmat{ 
     \fnpmat{ 
        \pderiv{\tup{Q}}{\tup{q}}
     & \pderiv{\tup{Q}}{\tup{p}}
     \\[4pt]
      \pderiv{\tup{P}}{\tup{q}}
     & \pderiv{\tup{P}}{\tup{p}}    
     }_{6\times 6}
     & 
     \fnpmat{
     \tup{0} &  \tup{0}  \\[4pt] \tup{0} &  \tup{0}
     }_{6\times 2}
 \\[14pt]
    \fnpmat{ 
       \trn{\pderiv{U}{\tup{q}}}
     &  \trn{\pderiv{U}{\tup{p}}}
     \\[4pt]
       \trn{\pderiv{W}{\tup{q}}}
     &  \trn{\pderiv{W}{\tup{p}}}
    }_{2\times 6}
 &
    \fnpmat{ 
     \pderiv{U}{u}
     &\pderiv{U}{w} 
     \\[4pt]
     \pderiv{W}{u}
     &\pderiv{W}{w}
    }_{2\times 2}
 }  
&&
\end{flalign}
\end{small}
such that the general VOP formula in Eq.\eqref{vopGEN_ode_inf} leads to:
\begin{small}
\begin{align}
    \rng{\tup{\xi}} \,=\, \pderiv{\tup{\xi}}{\tup{x}} \cdot \tup{Y}
    \qquad \Rightarrow \qquad
   \begin{array}{llllll}
        \rng{\tup{Q}} \,=\, \pderiv{\tup{Q}}{\tup{p}} \cdot \rng{t}\tup{f}
        &,\quad 
         \rng{U} = \pderiv{U}{\tup{p}} \cdot \rng{t}\tup{f} \,+\, \pderiv{U}{w}\tfrac{1}{\slang} f_\ss{u}
    \\[4pt]
         \rng{\tup{P}} \,=\, \pderiv{\tup{P}}{\tup{p}} \cdot \rng{t}\tup{f}
          &,\quad 
           \rng{W} = \pderiv{W}{\tup{p}} \cdot \rng{t}\tup{f} \,+\, \pderiv{W}{w}\tfrac{1}{\slang} f_\ss{u}
   \end{array}
\end{align}
\end{small}

\subsection{Elements from Kepler Solutions} \label{sec:VOP_prj_nonsimp}


\noindent 
We define elements 
\eq{\tup{\xi}=(\tup{Q},\tup{P},U,W)} from the coordinates \eq{\tup{x}=(\tup{q},\tup{p},u,w)}
using the unsimplified Kepler flow, \eq{\phi_\tau:\mbb{R}^8\to\mbb{R}^8} with \eq{\inv{\phi_{\tau}} =\phi_{-\tau} }, from Eq.\eqref{qusol_MAT_2bp_inv_nonsimp} in Appx.~\ref{sec:prj_STM_new}: 
\begin{small}
\begin{flalign} \label{100th_time_elems}  
&&
\begin{array}{ccccc}
     \boxed{ \tup{x} = \phi_\tau (\tup{\xi}) = \Sig_\tau(\tup{L}) \cdot \tup{\xi} + \tup{\sig}_\tau(L) 
    \qquad \leftrightarrow \qquad 
    \tup{\xi} := \inv{\phi_{\tau}} (\tup{x}) = \Sig_{-\tau}(\tup{\slang}) \cdot \tup{x} + \tup{\sig}_{-\tau}(\slang) }
\\[10pt]
\begin{array}{lllllll}
    \tup{q}
       \,=\, 
      \tup{Q} \csn{\tau} - \hdge{\htup{L}} \cdot\tup{Q} \snn{\tau} 
\\[4pt]
       \tup{p}
       \,=\, 
      \tup{P} \csn{\tau} - \hdge{\htup{L}} \cdot\tup{P} \snn{\tau} 
\\[4pt]
      u
          \,=\,  (U-\tfrac{\kconst_1}{L^2})\csn{\tau} \,+\, \tfrac{1}{L} W \snn{\tau} \,+\, \tfrac{\kconst_1}{L^2}
\\[4pt]
       w 
         \,=\,  -L(U-\tfrac{\kconst_1}{L^2})\snn{\tau} \,+\, W \csn{\tau}
\end{array}
 \qquad \leftrightarrow \qquad 
\begin{array}{llll}
     \tup{Q} \,:=\, \tup{q}\csn{\tau} + \hdge{\htup{\slang}}\cdot\tup{q} \snn{\tau} 
\\[4pt]
      \tup{P} \,:=\, \tup{p}\csn{\tau} + \hdge{\htup{\slang}}\cdot\tup{p} \snn{\tau}
\\[4pt]
     U 
      \,:=\, (u-\tfrac{\kconst_1}{\slang^2})\csn{\tau} -  \tfrac{1}{\slang}w\snn{\tau} + \tfrac{\kconst_1}{\slang^2}
\\[4pt]
      W 
      \,:=\, \slang(u-\tfrac{\kconst_1}{\slang^2})\snn{\tau} + w\csn{\tau} 
\end{array}
\end{array}
 && \tup{L} \equiv \tup{\slang}
 \quad
\end{flalign}
\end{small}
where \eq{\tup{\slang}=\tup{q}\tms\tup{p}} and where we have temporarily introduced \eq{\tup{L}=\tup{Q}\tms\tup{P}} only to clarify "what is a function of what". 
Yet, all angular momentum functions are the same in terms of either \eq{\tup{x}} or \eq{\tup{\xi}} for the following reason. Recall that \eq{\tup{\slang}=\tup{q}\tms\tup{p}} — as well as \eq{\mag{\tup{q}}}, \eq{\tup{q}\cdot\tup{p}}, and \eq{\mag{\tup{p}}} — are all integrals of motion of the Kepler solution flow, \eq{\phi_\tau}. Clearly, the above transformation then also preserves the form of these functions. Indeed, it is quick to verify from the above that:
\begin{small}
\begin{align} \label{obvious}
\begin{array}{rclllll}
      \hdge{\tup{\slang}} = \tup{q}\wdg \tup{p} 
      &=& \tup{Q}\wdg \tup{P} =  \hdge{\tup{L}}
\\[4pt]
    \tup{\slang} = \tup{q}\tms \tup{p} 
    &=& \tup{Q}\tms \tup{P} = \tup{L}
\\[4pt]
    \slang^2 = q^2 p^2 - (\tup{q}\cdot\tup{p})^2
    &=&
     Q^2 P^2 - (\tup{Q}\cdot\tup{P})^2 = L^2
\\[4pt]
     \imat_3 = \htup{q}\otms \htup{q} + (\hdge{\htup{\slang}}\cdot\htup{q}) \otms (\hdge{\htup{\slang}}\cdot\htup{q}) + \htup{\slang}\otms \htup{\slang} 
     &=&
    \htup{Q}\otms \htup{Q} + (\hdge{\htup{L}}\cdot\htup{Q}) \otms (\hdge{\htup{L}}\cdot\htup{Q}) + \htup{L}\otms \htup{L}
\\[4pt]
     \imat_3 = \htup{p}\otms \htup{p} + (\hdge{\htup{\slang}}\cdot\htup{p}) \otms (\hdge{\htup{\slang}}\cdot\htup{p}) + \htup{\slang}\otms \htup{\slang} 
     &=&
    \htup{P}\otms \htup{P} + (\hdge{\htup{L}}\cdot\htup{P}) \otms (\hdge{\htup{L}}\cdot\htup{P}) + \htup{L}\otms \htup{L}
\end{array}
&&,&&
\begin{array}{lllll}
      q^2 = Q^2 
\\[4pt]
   \tup{q}\cdot\tup{p} = \tup{Q} \cdot \tup{P} 
\\[4pt]
    p^2 = P^2 
\\[4pt]
    \vphantom{\hdge{\htup{L}}}
\\[4pt]
    \vphantom{\hdge{\htup{L}}}
\end{array}
\end{align}
\end{small}
    We will now dispense with \eq{\tup{L}}, using only \eq{\tup{\slang}} with the understanding that it can always be interpreted as a function of either \eq{(\tup{q},\tup{p})} or \eq{(\tup{Q},\tup{P})} as indicated above. 

\begin{notesq}
   \textit{Integrals of motion.} 
   Recall that \eq{\tup{\slang}} and \eq{p=P} are integrals of motion only in the case of pure central-force dynamics, but not in the general perturbed case. However, 
    recall from Eq.\eqref{qlam_0} that, for any and all arbitrary forces, \eq{q} and \eq{\lambda=\htup{q}\cdot\tup{p}} are \textit{always} integrals of motion. It follows from Eq.\eqref{obvious} that \eq{Q} and \eq{\htup{Q}\cdot\tup{P}} are likewise integrals of motion for all arbitrary perturbations. 
    As previously discussed, we are always free to limit consideration to the values \eq{q=Q=1} and \eq{\lambda=\htup{q}\cdot\tup{p}=\htup{Q}\cdot\tup{P}=0}, and simplify governing equations accordingly. We will make these simplifications later on.  
\end{notesq}

\paragraph{Element ODEs.}
The equations of motion for \eq{\tup{\xi}=(\tup{Q},\tup{P},U,W}) resulting from the VOP method are obtained using the general formula \eq{ \rng{\tup{\xi}} = \pderiv{\tup{\xi}}{\tup{x}} \cdot \tup{Y}} from Eq.\eqref{vopGEN_ode_inf}, with \eq{\tup{Y}} given for the problem at hand:
\begin{small}
\begin{flalign} \label{vopKep_ode_nosimpMAT}
&&
    \rng{\tup{\xi}} \,=\, \pderiv{\tup{\xi}}{\tup{x}} \cdot \tup{Y}
    \qquad,\qquad 
    \tup{Y}=(\tup{0},\rng{t} \tup{f}, 0 , \tfrac{1}{\slang} f_\ss{u}) \,=\, \rng{t} (\tup{0},\tup{f}, 0 , u^2 f_\ss{u})
&&
\rng{t}=\tfrac{1}{\slang u^2}
\qquad
\end{flalign}
\end{small}
where \eq{(\tup{f},f_\ss{u})} are the generalized forces defined in Eq.\eqref{Ftotal_qp} 
and where \eq{\pderiv{\tup{\xi}}{\tup{x}} } follows immediately from what we called \eq{\pderiv{\tup{x}_0}{\tup{x}_\tau}} in the Kepler STMs of Eq.\eqref{STM_prj_nosimp}:  
\begin{small}
\begin{align} \label{Jacob_elem_prj_nosimp}
   \pderiv{\tup{\xi}}{\tup{x}}
&\,=\,
\fnpmat{ 
    \fnpmat{ 
    \imat_3 \csn{\tau} + \tfrac{1}{\slang} \big( \hdge{\tup{\slang}} - \hdge{\tup{q}}\cdot\hdge{\tup{p}} -   (\hdge{\htup{\slang}}\cdot\tup{q})\otms (\hdge{\htup{\slang}}\cdot\tup{p}) \big) \snn{\tau} 
    &  -\tfrac{q^2}{\slang}   \htup{\slang}\otms\htup{\slang} \snn{\tau} 
    \\[3pt]
    \tfrac{p^2}{\slang}   \htup{\slang}\otms\htup{\slang} \snn{\tau} 
    &  \imat_3 \csn{\tau} +   \tfrac{1}{\slang} \big( \hdge{\tup{\slang}} +  \hdge{\tup{p}}\cdot\hdge{\tup{q}} +   (\hdge{\htup{\slang}}\cdot\tup{p})\otms (\hdge{\htup{\slang}}\cdot\tup{q}) \big) \snn{\tau}  
    } 
    &
    \fnpmat{ \tup{0} & \tup{0} \\[3pt] \tup{0} & \tup{0} }
\\[18pt]
    \fnpmat{ 
         \tfrac{1}{\slang}\big( \tfrac{2\kconst_1}{\slang^2}(1-\csn{\tau}) - \tfrac{1}{\slang}w \snn{\tau} \big) \tup{p}\cdot\hdge{\htup{\slang}}
         &  -\tfrac{1}{\slang}\big( \tfrac{2\kconst_1}{\slang^2} (1-\csn{\tau}) - \tfrac{1}{\slang}w \snn{\tau} \big) \tup{q}\cdot\hdge{\htup{\slang}}
     \\[3pt]
          - (u + \tfrac{\kconst_1}{\slang^2})\snn{\tau}\,   \tup{p}\cdot\hdge{\htup{\slang}}
         &   (u + \tfrac{\kconst_1}{\slang^2})\snn{\tau}\,  \tup{q}\cdot\hdge{\htup{\slang}}
     }
    &
    \fnpmat{
    \csn{\tau} & -\tfrac{1}{\slang}\snn{\tau}
    \\[3pt]
    \slang \snn{\tau} & \csn{\tau}
    }  
} 
\\[8pt] \label{Jacob_elem_prj_simp}
    &\,\simeq\,
        \fnpmat{ 
        \fnpmat{ 
        \imat_3 \csn{\tau} + \hdge{\htup{\slang}} \snn{\tau} 
        &  -\tfrac{1}{\slang}\htup{\slang}\otms\htup{\slang} \snn{\tau} 
        \\[3pt]
       \slang \htup{\slang}\otms\htup{\slang}  \snn{\tau}
        &  \imat_3 \csn{\tau} +   \hdge{\htup{\slang}} \snn{\tau} 
        } 
        &
        \fnpmat{ \tup{0} & \tup{0} \\[3pt] \tup{0} & \tup{0} }
    \\[18pt]
         \fnpmat{ 
             -\big( \tfrac{2\kconst_1}{\slang^2}(1-\csn{\tau}) - \tfrac{1}{\slang}w \snn{\tau} \big) \trn{\tup{q}}
             &  -\tfrac{1}{\slang}\big( \tfrac{2\kconst_1}{\slang^2}(1-\csn{\tau}) - \tfrac{1}{\slang}w \snn{\tau} \big) \trn{\htup{p}}
         \\[3pt]
               \slang (u + \tfrac{\kconst_1}{\slang^2})\snn{\tau}\,  \trn{\tup{q}}
             &   (u + \tfrac{\kconst_1}{\slang^2})\snn{\tau}\, \trn{\htup{p}}
         }
        &
        \fnpmat{
        \csn{\tau} & -\tfrac{1}{\slang}\snn{\tau}
        \\[3pt]
        \slang \snn{\tau} & \csn{\tau}
        }  
    }  \qquad,\qquad \slang \simeq p
\end{align}
\end{small}
where, as usual,  ``\eq{\simeq}'' indicates simplified expressions using the integrals of motion \eq{q=1} and \eq{\lambda=\htup{q}\cdot\tup{p}=0}. We first consider the unsimplified case. 
For the lower left sub-matrix, we note the following relations (verified from Eq.\eqref{100th_time_elems}):\footnote{The relations in Eq.\eqref{some_UW_rels} show that the ODEs for \eq{(U,W)} presented in this work are equivalent to those given previously by the authors in \cite{peterson2025phdThesis,peterson2023regularized}.}
\begin{small}
\begin{align}\label{some_UW_rels}
\begin{array}{lllll}
       \tfrac{2\kconst_1}{\slang^2} (1-\csn{\tau}) - \tfrac{1}{\slang}w \snn{\tau} 
     \;=\;  U + \tfrac{\kconst_1}{\slang^2}  - (u + \tfrac{\kconst_1}{\slang^2})\csn{\tau}
\qquad,\qquad 
       (u+\tfrac{\kconst_1}{\slang^2})\snn{\tau} \,=\, \tfrac{1}{\slang}W - \tfrac{1}{\slang}w \csn{\tau} +  \tfrac{2\kconst_1}{\slang^2}\snn{\tau} 
\end{array}
\end{align}
\end{small}
Then, \textit{without} simplifying, 
the element ODEs follow from Eq.\eqref{vopKep_ode_nosimpMAT} with \eq{ \pderiv{\tup{\xi}}{\tup{x}}} 
from Eq.\eqref{Jacob_elem_prj_nosimp}:\footnote{The relations in Eq.\eqref{Ftotal_qp_elemSec}-Eq.\eqref{Ftotal_qp_elemSec_2} show that the ODEs in Eq.\eqref{QPelems_ODE_VOP2} are also equivalent to: 
\begin{align} \nonumber
 \begin{array}{lllll}
      \rng{\tup{Q}} =
         -\rng{t}\tfrac{q^2}{\slang}  \snn{\tau} \Big(  \imat_3 - (\hdge{\htup{\slang}}\cdot\htup{q})\otms (\hdge{\htup{\slang}}\cdot\htup{q}) \Big) \cdot \tup{f} 
         \;=
          -\rng{t}\tfrac{q^2}{\slang}  \snn{\tau} ( \htup{\slang}\otms \htup{\slang}) \cdot \tup{f} 
\\[8pt]
     \rng{\tup{P}} =
       \rng{t}\Big(\imat_3 \csn{\tau} +   \tfrac{1}{\slang} \snn{\tau} \big[ \hdge{\tup{\slang}} +  \hdge{\tup{p}}\cdot\hdge{\tup{q}} -   \tfrac{q^2}{\slang} (\hdge{\htup{\slang}}\cdot\tup{p})\otms \tup{p} \big]  \Big)  \cdot \tup{f}
\end{array}
&&,&&
\begin{array}{lllll}
     \rng{U} =
     -\rng{t}\tfrac{1}{\slang} \Big( u^2 f_\ss{u} \snn{\tau} +  \tfrac{q^2}{\slang}\big[U + \tfrac{\kconst_1}{\slang^2}  - (u + \tfrac{\kconst_1}{\slang^2})\csn{\tau}  \big] \tup{p} \cdot \tup{f}  \Big)
\\[8pt]
     \rng{W} = 
     \rng{t} \Big( u^2 f_\ss{u} \csn{\tau}  +  \tfrac{q^2}{\slang} (u + \tfrac{\kconst_1}{\slang^2})\snn{\tau}\, \tup{p} \cdot \tup{f}  \Big)
\end{array} 
\end{align}
}
\begin{small}
\begin{flalign} \label{QPelems_ODE_VOP2}
\;
\begin{array}{llllll}
     \rng{\tup{Q}} \,=\, 
          -\rng{t}\tfrac{Q^2}{\slang}  \snn{\tau} (\htup{\slang}\otms \htup{\slang}) \cdot \tup{f} 
          &,
\\[8pt]
    \rng{\tup{P}} \,=\,  \rng{t}\Big(\imat_3 \csn{\tau} +   \tfrac{1}{\slang} \snn{\tau} \big[ \hdge{\tup{\slang}} +  \hdge{\tup{p}}\cdot\hdge{\tup{q}} +   (\hdge{\htup{\slang}}\cdot\tup{p})\otms (\hdge{\htup{\slang}}\cdot\tup{q}) \big]  \Big)  \cdot \tup{f}
    &,
\end{array}
&&
\begin{array}{lllll}
    \rng{U} \,=\, -\rng{t}\tfrac{1}{\slang} \Big( u^2 f_\ss{u} \snn{\tau} +  \big[U + \tfrac{\kconst_1}{\slang^2}  - (u + \tfrac{\kconst_1}{\slang^2})\csn{\tau}  \big] (\tup{q}\cdot\hdge{\htup{\slang}}) \cdot \tup{f}  \Big)
\\[8pt]
    \rng{W} \,=\, \rng{t} \Big( u^2 f_\ss{u} \csn{\tau} +  (u + \tfrac{\kconst_1}{\slang^2})\snn{\tau}\, (\tup{q}\cdot\hdge{\htup{\slang}} ) \cdot \tup{f} \Big)
\end{array} 
\end{flalign}
\end{small}
where the relations in Eq.\eqref{obvious} hold. 
We may also express the right-hand-side of the ODEs in Eq.\eqref{QPelems_ODE_VOP2} in terms of the three inertial cartesian components, \eq{\tup{F}}, of the total perturbing forces. This leads to the following (using Eq.\eqref{Ftotal_qp_elemSec}-Eq.\eqref{Ftotal_qp_elemSec_2} below):
\begin{small}
\begin{flalign} \label{QPelems_ODE_VOP2_Ftot}
&&
\boxed{ \begin{array}{rlllll}
     \rng{\tup{Q}} &\!\!\!=\, 
    -\rng{t}\tfrac{Q}{\slang u }  \snn{\tau}\,( \htup{\slang} \otms \htup{\slang}) \cdot \tup{F} 
   \quad =\,
   -\rng{t}\tfrac{Q}{\slang u }  F_{\slang} \snn{\tau}\, \htup{\slang}
\\[10pt]
    \rng{\tup{P}} 
    &\!\!\!=\,-\rng{t}\tfrac{1}{u q} \Big(  (\imat_3 -  \htup{q} \otms \htup{q})\csn{\tau} 
     -
       \tfrac{1}{\slang} \snn{\tau} \big[ 
    \htup{q}\otms (\hdge{\tup{\slang}}\cdot\htup{q})
    -  \hdge{\tup{p}}\cdot\hdge{\tup{q}} -   (\hdge{\htup{\slang}}\cdot\tup{p})\otms (\hdge{\htup{\slang}}\cdot\tup{q}) \big] \Big)  \cdot \tup{F}
\\[10pt]
    \rng{U} &\!\!\!=\, 
     \rng{t}\tfrac{1}{\slang u}\Big(
            u \htup{q} \snn{\tau} -  \big[ U + \tfrac{\kconst_1}{\slang^2}  - (u + \tfrac{\kconst_1}{\slang^2})\csn{\tau}  \big] (\htup{q}\cdot\hdge{\htup{\slang}}) 
     \Big) \cdot \tup{F}
     \\[6pt]
    &\!\!\!=\,
     \rng{t}\tfrac{1}{\slang u}\Big(
            F_r u \snn{\tau} -  F_{\tau} \big[ U + \tfrac{\kconst_1}{\slang^2}  - (u + \tfrac{\kconst_1}{\slang^2})\csn{\tau}  \big]  
     \Big)
\\[10pt]
    \rng{W} &\!\!\!=\, 
    -\rng{t} \tfrac{1}{u}\Big(
        u \htup{q} \csn{\tau}
        - (u + \tfrac{\kconst_1}{\slang^2})\snn{\tau} \,(\htup{q}\cdot\hdge{\htup{\slang}})
    \Big) \cdot\tup{F}
    \\[6pt]
    &\!\!\!=\,
     -\rng{t} \tfrac{1}{u}\Big(
        F_r u  \csn{\tau}
        - F_{\tau} (u + \tfrac{\kconst_1}{\slang^2})\snn{\tau} 
    \Big) 
\end{array} }
\qquad 
\begin{array}{lll}
     \rng{t} = 1/(\slang u^2)
\end{array}
\quad
\end{flalign}
\end{small}
where the second equalities for \eq{\rng{\tup{Q}}}, \eq{\rng{U}}, and \eq{\rng{W}} follow from the fact that
\eq{\{\htup{q},-\hdge{\htup{\slang}}\!\cdot\htup{q},\,\htup{\slang}\} = \{\htup{t}_r,\htup{t}_\tau,\htup{t}_{\slang}\}} 
is the LVLH basis (rather, its inertial cartesian components)
and where we have denoted by \eq{(F_r,F_\tau,F_{\slang})} the LVLH components of the total perturbing force.\footnote{The force components \eq{(F_r,F_\tau,F_{\slang})\neq \tup{F}} in Eq.\eqref{QPelems_ODE_VOP2_Ftot} are the components of the total perturbing force in the LVLH basis \eq{\{\htup{t}_r,\htup{t}_\tau,\htup{t}_{\slang}\} = \{\htup{q},-\hdge{\htup{\slang}}\!\cdot\htup{q}, \,\htup{\slang}\} }. They are clearly \textit{not} the inertial cartesian components which we denote by \eq{\tup{F}=(F_{r_1},F_{r_2},F_{r_3})}.}
The same can be done for the above \eq{\rng{\tup{P}}} equation, though the result is not particularly helpful at this point (see   Eq.\eqref{QPelems_ODE_VOP2_Ftot_simp} below).

\begin{small}
\begin{itemize}
    \item \textit{Generalized force relations.}
    Recall the generalized forces \eq{(\tup{f},f_\ss{u})} are given in terms of the inertial cartesian components, \eq{\tup{F}}, as in Eq.\eqref{Ftotal_qp}:
    \begin{small}
    \begin{align} \label{Ftotal_qp_elemSec}
    \begin{array}{lllll}
             \tup{f} :=\, 
        \trn{\pderiv{\tup{r}}{\tup{q}}} \cdot  \tup{F}
           \,=\,  \tfrac{1}{u q} ( \imat_3 -  \htup{q}\otms\htup{q} ) \cdot \tup{F}
           \;=\,  -\tfrac{1}{u q} \hdge{\htup{q}}\cdot\hdge{\htup{q}}\cdot \tup{F}
    \\[4pt]
              f_\ss{u} :=\, 
        \pderiv{\tup{r}}{u} \cdot  \tup{F}
        \,=\,  -\tfrac{1}{u^2}\htup{q}\cdot\tup{F}
    \end{array}
    &&,&&
    \begin{array}{llll}
         \tup{q}\cdot\tup{f} = 0 
    \\[3pt]
         \tup{p} \cdot \tup{f} 
         = \tfrac{1}{q} \htup{q}\cdot\hdge{\tup{\slang}} \cdot \tup{f}
         = \tfrac{1}{u q^2}\htup{q}\cdot\hdge{\tup{\slang}} \cdot \tup{F}
         \,\simeq  \tfrac{1}{u}\tup{p}\cdot\tup{F}
     \\[3pt]
         \tup{\slang} \cdot \tup{f} = \tfrac{1}{uq}\tup{\slang}\cdot\tup{F}
    \end{array}
    \end{align}
    \end{small}
    from which we then note:\footnote{Some relations have used the
        identities \eq{\hdge{\tup{n}}\cdot\hdge{\tup{n}}\cdot\hdge{\tup{n}} = -n^2 \hdge{\tup{n}} } and \eq{\hdge{\tup{n}}\cdot\hdge{\tup{n}}\cdot\hdge{\tup{n}}\cdot\hdge{\tup{n}} = -n^2 \hdge{\tup{n}}\cdot\hdge{\tup{n}} } for any \eq{\tup{n}\in\mbb{R}^3}.  }
    \begin{small}
    \begin{flalign} \label{Ftotal_qp_elemSec_2}
    \;\;
    \begin{array}{llll}
        \hdge{\tup{\slang}} \cdot \tup{f} = (\tup{q}\otms\tup{p}) \cdot \tup{f} 
        \quad =  \tfrac{1}{u q}(\htup{q}\cdot\hdge{\tup{\slang}} \cdot \tup{F}) \htup{q}
        \,= - \tfrac{1}{u q} \big( \htup{q}\otms(\hdge{\tup{\slang}} \cdot \htup{q}) \big) \cdot \tup{F}
    \\[3pt]
        (\hdge{\htup{\slang}}\cdot\tup{q})\cdot\tup{f} = 
        -\tup{q} \cdot \hdge{\htup{\slang}}\cdot \tup{f}
        = -\tfrac{q^2}{\slang} \tup{p}\cdot\tup{f}
        \quad = -\tfrac{1}{u}\htup{q}\cdot\hdge{\htup{\slang}} \cdot \tup{F}
        \,=
        \tfrac{1}{u}(\hdge{\htup{\slang}} \cdot \htup{q}) \cdot \tup{F}
    \\[3pt]
       \big(  \imat_3 - (\hdge{\htup{\slang}}\cdot\htup{q})\otms (\hdge{\htup{\slang}}\cdot\htup{q}) \big) \cdot \tup{f} 
         \;=\; ( \htup{\slang}\otms \htup{\slang}) \cdot \tup{f} 
    \end{array}
    &&,&&
    \begin{array}{llll}
        \hdge{\tup{q}} \cdot \tup{f}  
        &=  \tfrac{1}{u}  \hdge{\htup{q}} \cdot \tup{F}
    \\[3pt]
        \hdge{\tup{q}}\cdot  \hdge{\tup{q}} \cdot \tup{f} = -q^2 \tup{f} 
        &=  \tfrac{1}{u} \hdge{\tup{q}}\cdot  \hdge{\htup{q}} \cdot \tup{F}
    \\[3pt]
        \hdge{\tup{p}}\cdot  \hdge{\tup{q}} \cdot \tup{f}  
        & = \tfrac{1}{u} \hdge{\tup{p}}\cdot  \hdge{\htup{q}} \cdot \tup{F}
    \end{array}
    \;
    \end{flalign}
    \end{small}
\end{itemize}
\end{small}

\paragraph{Simplified ODEs for the Elements.}
As mentioned, we may simplify the dynamics using the integrals of motion \eq{q=Q=1} and \eq{\lambda=\htup{q}\cdot\tup{p}=\htup{Q}\cdot\tup{P}=0} (implying the relations in Eq.\eqref{qpl_rels}). This means that,
in addition to the relations in Eq.\eqref{obvious}, we now further have that \eq{(\tup{q},\tup{p})} and \eq{(\tup{Q},\tup{P})} satisfy:
\begin{small}
\begin{flalign}
\qquad
     \begin{array}{cc}
        \fnsize{using:} \\
         q=1  \\[1pt]
         \lambda=\htup{q}\cdot\tup{p}=0 
    \end{array}
     \Rightarrow \;\;
    \left\{ \qquad\qquad
    \begin{array}{ll}
        Q^2 = q^2 \simeq 1
    \\[4pt]
        \tup{Q}\cdot\tup{P} = \tup{q}\cdot\tup{p} \simeq 0
    \\[4pt]
        P^2 = p^2 \simeq \slang^2
    \end{array} \right.
    \qquad,\qquad 
     \begin{array}{rcllll}
            \hdge{\tup{\slang}}\cdot\tup{q} \simeq -\tup{p}
            &,& 
             \hdge{\tup{\slang}}\cdot\tup{Q} \simeq -\tup{P}
            \\[4pt]
            \hdge{\tup{\slang}}\cdot\tup{p} 
            \simeq \slang^2 \tup{q} 
             &,& 
            \hdge{\tup{\slang}}\cdot\tup{P} 
            \simeq \slang^2 \tup{Q} 
            \\[4pt]
              \imat_3 \simeq\, \htup{q} \otms\htup{q} + \htup{p} \otms\htup{p} + \htup{\slang} \otms\htup{\slang}
              &\simeq& \htup{Q} \otms\htup{Q} + \htup{P} \otms\htup{P} + \htup{\slang} \otms\htup{\slang}
    \end{array}
    \quad 
\end{flalign}
\end{small}
Additionally,  the matrix \eq{\pderiv{\tup{\xi}}{\tup{x}}} simplifies to that seen in Eq.\eqref{Jacob_elem_prj_simp}
such that the element ODEs in Eq.\eqref{QPelems_ODE_VOP2} simplify to:
\begin{small}
\begin{align} \label{QPelems_ODE_VOP2_simp}
 \begin{array}{rlllll}
     \rng{\tup{Q}}  &\!\!\!\simeq\, -\rng{t}\tfrac{1}{\slang}  \snn{\tau}\, (\htup{\slang}\otms\htup{\slang}) \cdot \tup{f} 
\\[6pt]
    \rng{\tup{P}}  &\!\!\!\simeq\,  \rng{t} (\imat_3 \csn{\tau} +   \hdge{\htup{\slang}} \snn{\tau} )  \cdot \tup{f}
       \\[4pt]
       &\!\!\! \simeq\,   \rng{t} ( \htup{P}\otms\htup{p} \,+\, \htup{\slang}\otms\htup{\slang}\csn{\tau}  )  \cdot \tup{f}
 \end{array}
\qquad,\qquad  
\begin{array}{rlllll}
    \rng{U}  &\!\!\!\simeq\, -\rng{t}\tfrac{1}{\slang} \Big( u^2 f_\ss{u} \snn{\tau} \,+\, \big[U + \tfrac{\kconst_1}{\slang^2}  - (u + \tfrac{\kconst_1}{\slang^2})\csn{\tau}  \big] \htup{p} \cdot \tup{f}  \Big)
\\[6pt]
    \rng{W} &\!\!\!\simeq\, 
     \rng{t} \Big( u^2 f_\ss{u} \csn{\tau}  \,+\,   (u + \tfrac{\kconst_1}{\slang^2})\snn{\tau}\, \htup{p} \cdot \tup{f}  \Big)
     \\[4pt]
    \vphantom{ \tfrac{1}{\slang}\tup{q}\otms\tup{p} }
\end{array} 
\end{align}
\end{small}
    where, for the \eq{\rng{\tup{P}}} equation, we have used
    \eq{\hdge{\htup{\slang}}\cdot\tup{f}=\tfrac{1}{\slang}(\tup{q}\otms\tup{p})\cdot\tup{f}}
   from Eq.\eqref{Ftotal_qp_elemSec_2}, along with the following:\footnote{More explicitly: \\
        \eq{
         \rng{\tup{P}}  \simeq  \rng{t} (\imat_3 \csn{\tau} +   \hdge{\htup{\slang}} \snn{\tau} )  \cdot \tup{f}
         =   \rng{t} (\imat_3 \csn{\tau} +   \tfrac{1}{\slang}\tup{q}\otms\tup{p} \snn{\tau} )  \cdot \tup{f}
         \simeq   \rng{t} \Big( \tfrac{1}{\slang^2}\tup{P}\otms\tup{p} +  (\imat_3 - \tfrac{1}{\slang^2}\tup{p}\otms\tup{p})\csn{\tau}  \Big)  \cdot \tup{f}
         \simeq   \rng{t} \Big( \htup{P}\otms\htup{p} +  (\imat_3 -\htup{p}\otms\htup{p})\csn{\tau}  \Big)  \cdot \tup{f}
           \simeq   \rng{t} ( \htup{P}\otms\htup{p} + \htup{\slang}\otms\htup{\slang}\csn{\tau}  )  \cdot \tup{f}
        }.
    }
\begin{small}
\begin{align}
     \tup{P} \,:=\, \tup{p}\csn{\tau} + \hdge{\htup{\slang}}\cdot\tup{p} \snn{\tau}
     \,\simeq\, \tup{p}\csn{\tau} + \slang \tup{q} \snn{\tau}
     \qquad \Rightarrow \qquad 
     \tup{q}\snn{\tau} \simeq \tfrac{1}{\slang}(\tup{P}-\tup{p}\csn{\tau}) \simeq  \htup{P}-\htup{p}\csn{\tau}
\end{align}
\end{small}
Or, the simplified ODEs expressed in terms of \eq{\tup{F}} using Eq.\eqref{Ftotal_qp_elemSec}-Eq.\eqref{Ftotal_qp_elemSec_2}:
\begin{small}
\begin{align} \label{QPelems_ODE_VOP2_Ftot_simp}
\boxed{ \begin{array}{rlllll}
     \rng{\tup{Q}} &\!\!\!\simeq\, -\rng{t}\tfrac{1}{\slang u}  \snn{\tau}\, (\htup{\slang}\otms\htup{\slang}) \cdot \tup{F} 
     &\quad \simeq\,
      -\rng{t}\tfrac{1}{P u} F_{\slang}  \snn{\tau}\, \htup{\slang}
\\[6pt]
    \rng{\tup{P}} 
     &\!\!\!\simeq\,   \rng{t} \tfrac{1}{u} (  \htup{P}\otms\htup{p} \,+\, \htup{\slang}\otms\htup{\slang}\csn{\tau} 
     )  \cdot \tup{F}
      &\quad \simeq\,
      \rng{t} \tfrac{1}{u} ( F_{\tau} \htup{P} \,+\, F_{\slang}  \csn{\tau} \, \htup{\slang} 
     ) 
\\[6pt]
    \rng{U} &\!\!\!\simeq\,
    \rng{t}\tfrac{1}{\slang u} \Big( 
      u \htup{q} \snn{\tau} - 
    \big[U + \tfrac{\kconst_1}{\slang^2}  - (u + \tfrac{\kconst_1}{\slang^2})\csn{\tau}  \big] \htup{p} \Big)\cdot \tup{F} 
     &\quad \simeq\,
     \rng{t}\tfrac{1}{P u} \Big( 
      F_r u  \snn{\tau} - 
    F_{\tau} \big[U + \tfrac{\kconst_1}{P^2}  - (u + \tfrac{\kconst_1}{P^2})\csn{\tau}  \big] \Big) 
\\[6pt]
    \rng{W} &\!\!\!\simeq\, 
   -\rng{t} \tfrac{1}{u} \Big( 
            u \htup{q} \csn{\tau} - (u + \tfrac{\kconst_1}{\slang^2})\snn{\tau}\, \htup{p}
    \Big) \cdot\tup{F}
     &\quad \simeq\,
     -\rng{t} \tfrac{1}{u} \Big( 
            F_r u \csn{\tau} - F_{\tau} (u + \tfrac{\kconst_1}{P^2})\snn{\tau} 
    \Big) 
\end{array} }
&& 
\begin{array}{lll}
     \slang \simeq P  \\[3pt]
     \rng{t} \simeq 1/(P u^2)
\end{array}
\end{align}
\end{small}
where the second equalities follow from the fact that \eq{\{\tup{q},\htup{p},\htup{\slang}\} \simeq \{\htup{t}_r,\htup{t}_\tau,\htup{t}_{\slang}\}} 
is the LVLH basis (rather, its inertial cartesian components)
and where we have denoted by \eq{(F_r,F_\tau,F_{\slang})} the LVLH components of the total perturbing force.

\begin{notesq}
    Note from Eq.\eqref{QPelems_ODE_VOP2_Ftot_simp}  that if the perturbing force is also some central-force of any kind (i.e., if  \eq{\tup{F}=F_r\htup{r}=F_r\tup{q}}), then  \eq{\rng{\tup{Q}}=\rng{\tup{P}}=0} and these elements remain constant (as does \eq{\tup{\slang}=\hdge{\tup{P}}\cdot\tup{Q}}). Similarly, if \eq{\tup{F}=F_{\slang} \htup{\slang}} is  purely in the orbit normal direction, then \eq{\rng{U} =\rng{W} =0}. 
\end{notesq}


Finally, we note that one could always choose to use the time, \eq{t}, as the evolution parameter. The \eq{t}-parameterized element ODEs follow from Eq.\eqref{QPelems_ODE_VOP2_Ftot_simp} as:
\begin{small}
\begin{align} \label{QPelems_ODE_t_simp}
\begin{array}{rlllll}
     \dot{\tup{Q}} &\!\!\!\simeq\, 
      -\tfrac{1}{P u} F_{\slang}  \snn{\tau}\, \htup{\slang}
      &,
\\[6pt]
    \dot{\tup{P}} &\!\!\!\simeq\, 
       \tfrac{1}{u} ( F_{\tau} \htup{P} \,+\, F_{\slang}  \csn{\tau} \, \htup{\slang} 
     ) 
     &,
 \end{array}
\qquad 
 \begin{array}{rlllll}
    \dot{U} &\!\!\!\simeq\,
     \tfrac{1}{P u} \Big( 
      F_r u  \snn{\tau} - 
    F_{\tau} \big[U + \tfrac{\kconst_1}{P^2}  - (u + \tfrac{\kconst_1}{P^2})\csn{\tau}  \big] \Big) 
\\[6pt]
    \dot{W} &\!\!\!\simeq\, 
     - \tfrac{1}{u} \Big( 
            F_r u \csn{\tau} - F_{\tau} (u + \tfrac{\kconst_1}{P^2})\snn{\tau} 
    \Big) 
\end{array} 
\quad,\qquad 
\dot{\tau} \,=\, \slang u^2 \simeq P u^2
\end{align}
\end{small}
where \eq{\tau} itself is included as a ninth coordinate in the same way that \eq{t} was previously included as a ninth coordinate (though we have not always been indicating this explicitly).




\paragraph{Elements Defined from Simplified Kepler Solutions.}

The above simplified ODEs for the elements in Eq.\eqref{QPelems_ODE_VOP2_simp} and Eq.\eqref{QPelems_ODE_VOP2_Ftot_simp} are still for \eq{\tup{\xi}:=\inv{\phi_\tau}(\tup{x})} defined from \eq{\tup{x}} using the \eq{\tup{x}} coordinate \textit{unsimplified} Kepler solution flow, \eq{\phi_\tau}, as in Eq.\eqref{100th_time_elems}. Only the final ODEs for \eq{\tup{\xi}} are simplified, but not their definition. We now simplify their definition to begin with. The resulting ODEs are the same as those already given in Eq.\eqref{QPelems_ODE_VOP2_simp}-Eq.\eqref{QPelems_ODE_VOP2_Ftot_simp}.

Using the \textit{simplified} Kepler flow, \eq{\theta_\tau:\mbb{R}^8\to\mbb{R}^8}, from Eq.\eqref{qusol_MAT_2bp_inv} — that is, simplified using the integrals of motion \eq{q=1} and \eq{\lambda=\htup{q}\cdot\tup{p}=0}, along with the subsequent relations in Eq.\eqref{qpl_rels} — then the elements 
\eq{\tup{\xi}=(\tup{Q},\tup{P},U,W)} are defined from \eq{\tup{x}=(\tup{q},\tup{p},u,w)} as follows:
\begin{small}
\begin{flalign} \label{prjElems_def_simp} 
&&
\begin{array}{ccccc}
     \tup{x} = \theta_\tau(\tup{\xi}) = E_\tau(\slang) \cdot \tup{\xi} + \tup{\sig}_\tau(\slang) 
    \qquad \leftrightarrow \qquad 
    \tup{\xi} := \inv{\theta_\tau}(\tup{x}) = E_{-\tau}(\slang) \cdot \tup{x} + \tup{\sig}_{-\tau}(\slang)  
\\[10pt]
\begin{array}{lllllll}
    \tup{q} 
       \,=\, 
      \tup{Q} \csn{\tau} + \tfrac{1}{\slang} \tup{P} \snn{\tau} 
\\[4pt]
       \tup{p} 
       \,=\, 
      -\slang\tup{Q} \snn{\tau}  +  \tup{P} \csn{\tau}
\\[4pt]
      u 
         \,=\,  (U-\tfrac{\kconst_1}{\slang^2})\csn{\tau} \,+\, \tfrac{1}{\slang} W \snn{\tau} \,+\, \tfrac{\kconst_1}{\slang^2}
\\[4pt]
       w  
         \,=\,  -\slang(U-\tfrac{\kconst_1}{\slang^2})\snn{\tau} \,+\, W\csn{\tau}
\end{array}
 \qquad \leftrightarrow \qquad 
\begin{array}{lll}
     \tup{Q} :=\, \tup{q}\csn{\tau} - \tfrac{1}{\slang}\tup{p}\snn{\tau} 
\\[4pt]
      \tup{P} :=\, \slang\tup{q}\snn{\tau} +  \tup{p}\csn{\tau}    
\\[4pt]
     U :=\, (u-\tfrac{\kconst_1}{\slang^2})\csn{\tau} -  \tfrac{1}{\slang}w\snn{\tau} + \tfrac{\kconst_1}{\slang^2}
\\[4pt]
      W :=\, \slang(u-\tfrac{\kconst_1}{\slang^2})\snn{\tau} + w\csn{\tau} 
\end{array}
\end{array}
&&
\end{flalign}
\end{small}
where \eq{\slang \simeq p = P}. That is, the above definition using the simplified Kepler flow could be written more explicitly as:
\begin{small}
\begin{align} \label{prjElems_def_simp_2}
 \begin{array}{lllllll}
    \tup{q}  \,=\, 
      \tup{Q} \csn{\tau} + \tfrac{1}{P} \tup{P} \snn{\tau} 
\\[3pt]
       \tup{p}    \,=\, 
      -P\tup{Q} \snn{\tau}  +  \tup{P} \csn{\tau}
\\[3pt]
      u \,=\,  (U-\tfrac{\kconst_1}{P^2})\csn{\tau} \,+\, \tfrac{1}{P} W \snn{\tau} \,+\, \tfrac{\kconst_1}{P^2}
\\[3pt]
       w      \,=\,  -P(U-\tfrac{\kconst_1}{P^2})\snn{\tau} \,+\, W\csn{\tau}
\end{array}
 \qquad \leftrightarrow \qquad 
\begin{array}{lll}
     \tup{Q} :=\, \tup{q}\csn{\tau} - \tfrac{1}{p}\tup{p}\snn{\tau} 
\\[3pt]
      \tup{P} :=\, p\tup{q}\snn{\tau} +  \tup{p}\csn{\tau}    
\\[3pt]
     U :=\, (u-\tfrac{\kconst_1}{p^2})\csn{\tau} -  \tfrac{1}{p}w\snn{\tau} + \tfrac{\kconst_1}{p^2}
\\[3pt]
      W :=\, p(u-\tfrac{\kconst_1}{p^2})\snn{\tau} + w\csn{\tau} 
\end{array} 
\end{align}
\end{small}
For the above, the matrix \eq{\pderiv{\tup{\xi}}{\tup{x}}} is given as follows (note this is not the same \eq{\pderiv{\tup{\xi}}{\tup{x}}} as given previously in Eq.\eqref{Jacob_elem_prj_nosimp} or Eq.\eqref{Jacob_elem_prj_simp}): 
\begin{small}
\begin{gather} \label{Jacob_elem_prj_SUPERsimp}
\pderiv{\tup{\xi}}{\tup{x}} \,=\,
\fnpmat{ 
    \fnpmat{ 
    \csn{\tau} \, \imat_3
    &  -\tfrac{1}{p}(\imat_3-\htup{p}\otms\htup{p})\snn{\tau} 
    \\[3pt]
    p \snn{\tau} \,\imat_3
    & \csn{\tau} \, \imat_3 + \tup{q}\otms \htup{p} \snn{\tau}
    } 
    &
    \fnpmat{ \tup{0} & \tup{0} \\[3pt] \tup{0} & \tup{0} }
\\[14pt]
   \fnpmat{ 
         0 &
         -\tfrac{1}{p}\big[\tfrac{2\kconst_1}{p^2}(1- \csn{\tau}) - \tfrac{1}{p}w \snn{\tau} \big] \trn{\htup{p}}
         \\[3pt]
          0
         &  (u + \tfrac{\kconst_1}{p^2})\snn{\tau} \,\trn{\htup{p}}
     }
    &
    \fnpmat{
    \csn{\tau} & -\tfrac{1}{p}\snn{\tau}
    \\[3pt]
    p \snn{\tau} & \csn{\tau}
    }  
}
\end{gather}
\end{small}
where part of the upper and lower sub-matrices can be rewritten using:
\begin{small}
\begin{align}
    \csn{\tau} \, \imat_3 + \tup{q}\otms \htup{p} \snn{\tau}  = \tfrac{1}{p}\tup{P} \otms \htup{p} + (\imat_3 - \htup{p}\otms\htup{p})  \csn{\tau} 
\qquad,\qquad
      \tfrac{2\kconst_1}{\slang^2} (1-\csn{\tau}) - \tfrac{1}{\slang}w \snn{\tau} 
     \;=\;  U + \tfrac{\kconst_1}{\slang^2}  - (u + \tfrac{\kconst_1}{\slang^2})\csn{\tau}
\end{align}
\end{small}
The governing equations for the elements defined in Eq.\eqref{prjElems_def_simp} or Eq.\eqref{prjElems_def_simp_2} are again found using the VOP equation \eq{\rng{\tup{\xi}} =\pderiv{\tup{\xi}}{\tup{x}}\cdot\tup{Y}}. With \eq{\pderiv{\tup{\xi}}{\tup{x}}} given as above in Eq.\eqref{Jacob_elem_prj_SUPERsimp}, this leads to precisely the same simplified ODEs as in Eq.\eqref{QPelems_ODE_VOP2_simp}-Eq.\eqref{QPelems_ODE_VOP2_Ftot_simp} (after appropriate application of various relations given in section \ref{sec:VOP_prj_nonsimp}).

\begin{notesq}
    Although we have used the same notation, \eq{(\tup{Q},\tup{P},U,W)}, the elements defined in Eq.\eqref{prjElems_def_simp} or Eq.\eqref{prjElems_def_simp_2} are not necessarily the same as the elements defined from the unsimplified solution as in Eq.\eqref{100th_time_elems}. However, when one limits consideration of the integrals of motion \eq{q} and \eq{\lambda=\htup{q}\cdot\tup{p}} to the values \eq{q=Q=1} and \eq{\lambda=\htup{q}\cdot\tup{p}=\htup{Q}\cdot\tup{P}=0} (which is the most convenient choice), then these definitions of the elements coincide, as do their governing equations of motion which are given by Eq.\eqref{QPelems_ODE_VOP2_Ftot_simp}.  
\end{notesq}


\paragraph{Elements $\mapsto$ Cartesian Coordinates.}
    The transformation for inertial cartesian coordinates \eq{(\tup{r},\tup{\v})} in terms of the elements follows immediately from the (unsimplified) definition of the elements in Eq.\eqref{100th_time_elems}, along with Eq.\eqref{rv_kepsol_nosimp}: 
    \begin{small}
    \begin{flalign} \label{rv_prjElems_nosimp} 
    \;\;
    \begin{array}{llllll}
         \tup{r} = \tfrac{1}{u}\htup{q}
      \,=\, 
       \dfrac{ \htup{Q} \csn{\tau} -  \hdge{\htup{\slang}} \cdot \htup{Q} \snn{\tau} }{ (U-\frac{\kconst_1}{\slang^2})\csn{\tau} + \frac{1}{\slang} W\snn{\tau} + \frac{\kconst_1}{\slang^2} }
       \\[4pt]
        \;\;
    \end{array} 
    \qquad,\qquad
    \begin{array}{llllll}
         \tup{\v}  = - u \hdge{\tup{\slang}} \cdot \htup{q} - w \htup{q} 
      &\!\!=\,   \tup{V}   - \tfrac{\kconst_1}{\slang^2} \hdge{\tup{\slang}} \cdot (\htup{q} - \htup{Q}) 
    \\[4pt]
        &\!\!=\,  \tup{V}   - \tfrac{\kconst_1}{\slang^2} \hdge{\tup{\slang}} \cdot \big[\htup{Q}( \csn{\tau}-1) - \hdge{\htup{\slang}}\cdot\htup{Q} \snn{\tau} \big]  
     \\[4pt]
          &\!\!\! =\, \tup{V}  - \tfrac{\kconst_1}{\slang^2} \big[ \hdge{\tup{\slang}}\cdot \htup{Q}(\csn{\tau}-1) +  \slang\htup{Q}\snn{\tau}  \big]
    \end{array} 
    \;\;
    \end{flalign}
    \end{small}
    with \eq{\tup{V}:= -U \hdge{\tup{\slang}} \cdot \htup{Q} - W \htup{Q}}
    and where \eq{\tup{\slang}=\tup{Q}\wdg\tup{P}} and \eq{\slang^2 = Q^2 P^2 - (\tup{Q}\cdot\tup{P})^2}. 
    The above does not make use of the integrals of motion. 
    If we simplify the above using  the integrals of motion \eq{q=Q=1} and \eq{\lambda=\htup{q}\cdot\tup{p}=\htup{Q}\cdot\tup{P}=0}, then:
    \begin{small}
    \begin{flalign} \label{rv_prjElems_simp}
    \;\;
    \begin{array}{llllll}
         \tup{r}  
          \,\simeq\, \tfrac{1}{u}\tup{q}
          \,\simeq\,  \dfrac{ \tup{Q} \csn{\tau}  \,+\,  \tfrac{1}{\slang}\tup{P}\snn{\tau}}{ (U-\frac{\kconst_1}{\slang^2})\csn{\tau} \,+\, \frac{1}{\slang} W \snn{\tau} \,+\, \frac{\kconst_1}{\slang^2}   }  
    \end{array} 
    \qquad,\qquad
    \begin{array}{llllll}
         \tup{\v}  \,\simeq\,  u \tup{p} - w \tup{q}
         \,\simeq\, 
         \tup{V} + \tfrac{\kconst_1}{\slang^2} \big[ \tup{P}(\csn{\tau}-1) -  \slang\tup{Q}\snn{\tau}  \big] 
    \end{array} 
    \;\;
    \end{flalign}
    \end{small}
    with \eq{\tup{V} \simeq U\tup{P} - W\tup{Q}}
   and \eq{\slang^2 \simeq P^2}.


\subsection{Example:~\txi{J2} Gravitational Perturbation} \label{sec:j2}

\begin{notation}
    Here, the symbol \eq{J_2} is the ``\eq{J_2} constant" (second zonal harmonic) for the leading-order gravitational correction for oblate bodies — a numeric constant depending on shape and mass distribution of the body. It should not be confused with the canonical symplectic \eq{2\tms 2} matrix (also denoted \eq{J_2} but not used here). 
\end{notation}

\noindent  As an example of the perturbed Kepler problem, consider the two-body problem for masses \eq{m_a} and \eq{m_b}, where \eq{m_b} is a uniform sphere but \eq{m_a} is some oblate body with equatorial radius \eq{R_a} whose \eq{J_2} term is a significant gravitational perturbation. The inertial cartesian coordinate Hamiltonian, with \eq{V^1} modeling the \eq{J_2} term, is given by
\begin{small}\begin{flalign} \label{Vj2_r}
&&
    \mscr{K} = \tfrac{1}{2} \v^2 - \tfrac{\kconst_1}{r} + V^{1}
\qquad,\qquad
     V^{1}(\tup{r})\,=\, \tfrac{1}{3}\tfrac{\til{J}_2}{r^3}(3 \hat{r}_3^2 - 1)
&&
    ,\quad \til{J}_2 := \tfrac{3}{2} J_2 \kconst_1 R_a^2
\quad 
\end{flalign}\end{small}
where \eq{\hat{r}_i=r_i/r} and we have absorbed various numeric constants into \eq{\til{J}_2} defined above. In the following, we assume no other perturbations are present. That is, \eq{\tup{a}^\nc=0}  such that \eq{\tup{F}=-\pderiv{V^1}{\tup{r}}}, and therefore \eq{(\tup{\alpha},\alpha_\ss{u})=0} such that \eq{(\tup{f},f_\ss{u})=(-\pderiv{V^1}{\tup{q}},-\pderiv{V^1}{u})}.

There are two approaches we may take when calculating the partials  \eq{\pderiv{V^{1}}{\tup{q}}} and \eq{\pderiv{V^{1}}{u}} appearing in the equations of motion for the projective coordinates. One method is to first obtain the cartesian components \eq{\pderiv{V^1}{\tup{r}}},
which is found from the above as follows:
\begin{small}\begin{align} \label{j2_a}
  \tup{F} \,=\,   -\pderiv{V^{1}}{\tup{r}} 
   \,=\,
    \tfrac{\til{J}_2}{r^4}
     \begin{pmatrix}
        (5\hat{r}_3^2 -1)\hat{r}_1 \\[3pt]
         (5\hat{r}_3^2-1)\hat{r}_2 \\[3pt]
          (5\hat{r}_3^2-3)\hat{r}_3 
    \end{pmatrix}
    \,=\,  
    \tfrac{\til{J}_2}{r^4} \big( (5 \hat{r}_3^2 -1)\htup{r} - 2 \hat{r}_3 \ibase_3 \big)
    \,=\, 
    \til{J}_2  u^4 \big( (5 \hat{q}_3^2 -1)\htup{q} - 2 \hat{q}_3 \ibase_3 \big)
    \qquad \;\; \hat{q}_i\simeq q_i
\end{align}\end{small}
where \eq{\ibase_3 = (0,0,1)} and
where the form on the right has been expressed in terms of \eq{(\tup{q},u)} using  \eq{\hat{r}_i= \hat{q}_i} and \eq{r=1/u} (it could also be simplified using \eq{q=1}). We may then find \eq{(\tup{f},f_\ss{u})=-(\pderiv{V^{1}}{\tup{q}}, \pderiv{V^{1}}{u}) }  
from the above by using Eq.\eqref{Ftotal_qp}:
\begin{small}\begin{align}  \label{sum_dV_j2}
    \tup{f} \,=\,  -\pderiv{V^{1}}{\tup{q}} \,=\, -\trn{\pderiv{\tup{r}}{\tup{q}}}\cdot \pderiv{V^{1}}{\tup{r}}  
     \,=\,  \tfrac{1}{u q} (\imat_3 - \htup{q}\otms\htup{q}) \cdot \tup{F}
\quad, && 
   f_\ss{u} \,=\,  -\pderiv{V^{1}}{u} \,=\, -\pderiv{\tup{r}}{u}\cdot \pderiv{V^{1}}{\tup{r}}  
   \,=\, -\tfrac{1}{u^2} \htup{q} \cdot \tup{F}
\end{align}\end{small}
Alternatively, we may simply use the point transformation, \eq{\tup{r} =\tfrac{1}{u}\htup{q}}, to express \eq{V^{1}} in terms of \eq{(\tup{q},u)} by using  \eq{\hat{r}_i= \hat{q}_i} and \eq{r=1/u}. Substituting this into \eq{V^{1}(\tup{r})} given by Eq.\eqref{Vj2_r} leads to the projective coordinate Hamiltonian with \eq{V^{1}(\bartup{q})}:
\begin{small}\begin{align} \label{dVj2}
V^1(\bartup{q}) = \tfrac{1}{3}\til{J}_2 u^3(3\hat{q}_3^2 - 1)
\qquad \Rightarrow \qquad
& \mscr{H} \,=\,
     \tfrac{1}{2}u^2\big( \slang^2  + u^2 p_u^2 \big)  - \kconst_1 u 
     +  \tfrac{1}{3}\til{J}_2 u^3(3\hat{q}_3^2 - 1) 
\end{align}\end{small}
Note that we do \textit{not}
 simplify the potential \eq{V^{1}} using \eq{q=1} as this would change the functional dependence on \eq{\tup{q}} and thus lead to an incorrect expression for \eq{\pderiv{V^{1}}{\tup{q}}}. From the above, we find that the correct \eq{J_2} generalized forces are:
\begin{small}\begin{align} \label{dvdq_j2}
   \tup{f} \,=\, -\pderiv{V^{1}}{\tup{q}} \,=\,
    \tfrac{2}{q}\til{J}_2 u^3 ( \hat{q}_3^2 \htup{q} - \hat{q}_3 \ibase_3 ) 
    \,\simeq\,  2\til{J}_2  u^3 ( q_3^2\tup{q} - q_3 \ibase_3 )
&&, &&
   f_\ss{u} \,=\,  -\pderiv{V^{1}}{u} \,=\, -\til{J}_2  u^2(3\hat{q}_3^2 - 1)
    \,\simeq\, -\til{J}_2  u^2(3q_3^2 - 1) 
\end{align}\end{small}
where only after differentiating do we   simplify the above using \eq{q=1}. It can be verified that substituting Eq.\eqref{j2_a} into Eq.\eqref{sum_dV_j2} leads to the same result as given above.

\begin{notesq}
    Assuming no other perturbations,  
    the equations of motion for the \eq{J_2}-perturbed Kepler problem in projective coordinates, or the associated orbit elements, are obtained simply be substituting Eq.\eqref{dvdq_j2} for \eq{(\tup{f},f_\ss{u})} — or Eq.\eqref{j2_a} for \eq{\tup{F}} — wherever these terms appear in the perturbed Kepler dynamics given previously. 
\end{notesq}


\paragraph{Numerical Verification with \txi{J2} Perturbation.}
The following plots serve as numerical verification of the dynamics given in
Eq.\eqref{QPelems_ODE_VOP2_Ftot_simp},
including the \eq{J_2} perturbation detailed above.
The following are simulations run for the \eq{J_2}-perturbed Kepler problem in the ``projective elements'' defined in 
Eq.\eqref{100th_time_elems},
with initial conditions given in terms of classic orbit elements \eq{(a,e,i,\omega,\Omega,f= \tau)}, with \eq{\tau} the true anomaly, by \eq{a_\zr = 8.59767038 \cdot 10^{3} (\mrm{km})}, \eq{e_\zr=0.2}, \eq{i_\zr=20 (\mrm{deg.})}, \eq{\omega_\zr=70 (\mrm{deg.})},  \eq{\Omega_\zr = 135 (\mrm{deg.})}, and true anomaly \eq{f_\zr = \tau_\zr = 0}. The numerical integration is carried out using ode45 in \sc{matlab} with units scaled such that the Earth's radius is \eq{R_\mrm{e}=1}, Earth's gravitational parameter is \eq{\kconst_1 = 1}, and Earths second zonal harmonic is \eq{J_2=1.082638\cdot 10^{-3}}.

\begin{figure}[!h]
	\centering
	\includegraphics[scale=0.5]{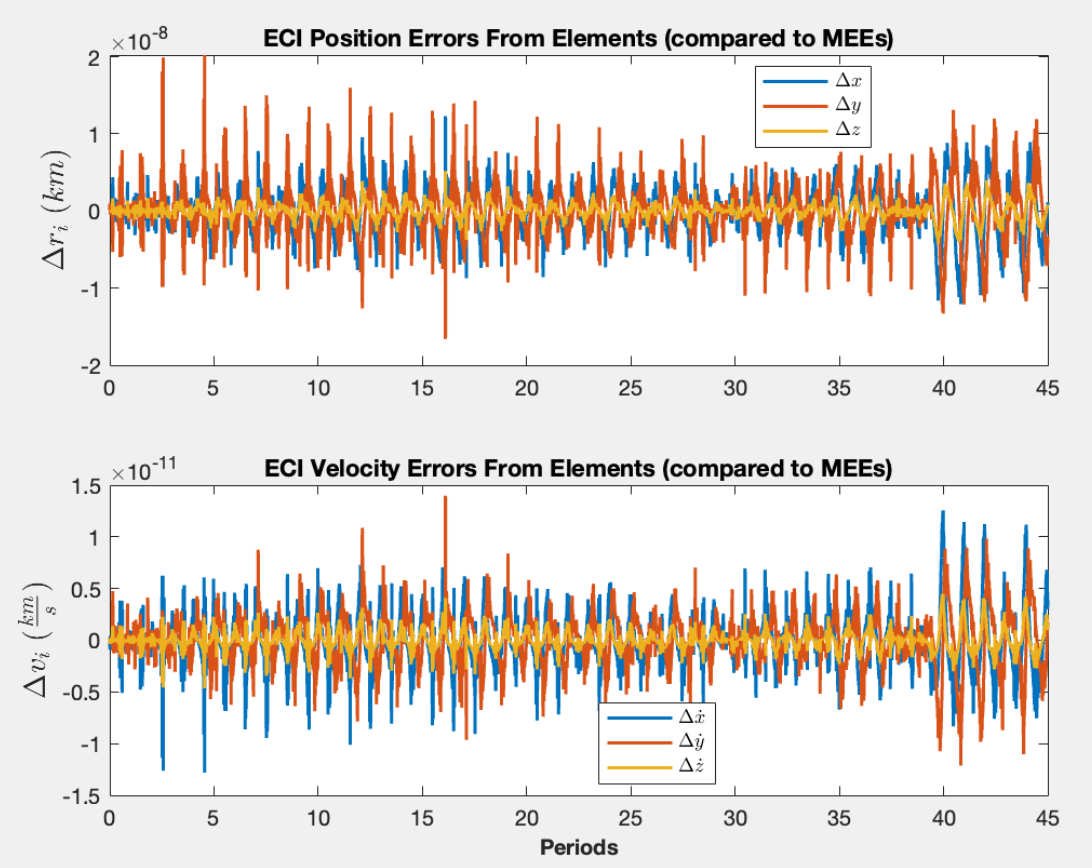}
	\caption{Verification of the ODEs for the new elements, showing cartesian ECI position and velocity errors as compared to propagating the modified equinoctial elements (MEEs). ($J_2$ gravitational term included in dynamics.)}
\end{figure}

\begin{figure}[!h]
	\centering
	\includegraphics[scale=0.4]{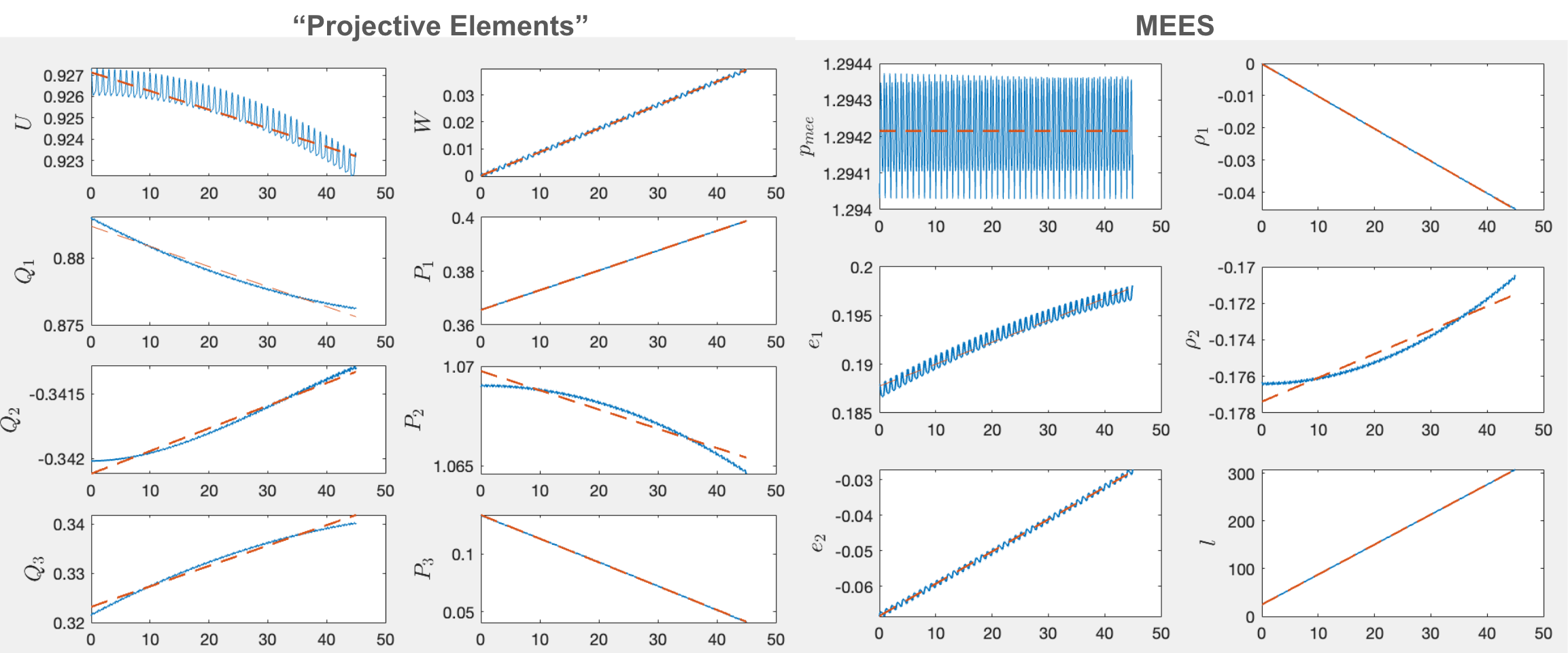}
	\caption{Mean rates, estimated by a simple least squares fit, of new elements compared to mean rates of the modified equinoctial elements (MEEs). Y-axes in dimensionless units. ($J_2$ gravitational term included in dynamics.)}
\end{figure}




\newpage
\phantomsection
\addcontentsline{toc}{section}{REFERENCES}
\begin{small}

\end{small}

\begin{appendices}
\renewcommand{\appendixname}{Appx.}
\titleformat{\section}{\centering\Large\bfseries\sffamily}{\textmd{\large{Appendix \thesection}}}{0em}{\vskip .1\baselineskip\centering}  


\section{SOME ANGULAR MOMENTUM RELATIONS} \label{sec:ang_momentum}

We collect some useful relations involving projective coordinates and the angular momentum, which plays a central role throughout this  work. 
Recall that the angular momentum coordinate vector, \eq{\tup{\slang}\in\mbb{R}^3}, the  antisymmetric matrix, \eq{\hdge{\tup{\slang}}\in\somat{3}},  and  the magnitude, \eq{\slang}, are given in terms of inertial cartesian coordinates \eq{(\tup{r},\tup{\v})\in\mbb{R}^6} by the usual relations:
\begin{small}\begin{align} \label{l_rv_general}
      \tup{\slang} \,=\, \tup{r}\tms \tup{\v} \,=\, \hdge{\tup{\v}}\cdot\tup{r}
 &&,&&
       \hdge{\tup{\slang}} \,=\,  \tup{r} \wdg \tup{\v}  \,=\, \tup{r}\otms \tup{\v}-\tup{\v}\otms \tup{r}
 &&,&&
        \slang^2  \,=\, \mag{\tup{\slang}} 
      \,=\, r^2 \v^2  - (\tup{r}\cdot\tup{\v})^2
\end{align}\end{small}
For the following, it may be helpful to recall the relations for the Hodge dual on \eq{\mbb{R}^3} given in Eq.\eqref{hodge_cord}.

\paragraph{General Relations.}
The functions in Eq.\eqref{l_rv_general} are invariant under a family a projective transformations (including that used in this work detailed in section \ref{sec:prj_Xform}, as well as the BF transformation \cite{ferrandiz1987general}).  
With inertial cartesian coordinates, \eq{(\tup{r},\tup{\v})\in\mbb{R}^6}, 
consider  a  family of ``canonically-extended'' projective transformations, \eq{\mbb{R}^8\ni(\bartup{q},\bartup{p})\mapsto (\tup{r},\tup{\v})\in \mbb{R}^6}, for projective coordinates \eq{(\bartup{q},\bartup{p})=(\tup{q},u,\tup{p},p_\ss{u})} given as follows for any real numbers \eq{n\neq0,m\in\mbb{R}} \cite{peterson2025prjCoord,peterson2022nonminimal}:\footnote{The transformation used in this work corresponds to the case \eq{n=m=-1}.}
\begin{small}
\begin{align} \label{prj_family}
\; \left. \;\;
\begin{array}{lllll}
      \tup{r} \,=\, u^n q^m \tup{q} 
\\[4pt]
     \tup{\v} \,=\,  \tfrac{1}{u^n q^m} \big(  (\imat_3 - \htup{q}\otms\htup{q})\cdot\tup{p} \,+\, \tfrac{u}{n q} p_\ss{u}\htup{q}\big)
\end{array} \right.
&&,&&
  \left(\begin{array}{lllll}
         \varphi = q -1  = 0
     \\[3pt]
         \lambda = \htup{q}\cdot\tup{p} - \tfrac{m+1}{n} \tfrac{1}{q} u p_\ss{u}
    \end{array}\right)
\end{align}
\end{small}
Only the  transformation  \eq{(\bartup{q},\bartup{p})\mapsto (\tup{r},\tup{\v})} for \eq{\tup{r}(\bartup{q},\bartup{p})} and \eq{\tup{\v}(\bartup{q},\bartup{p})} seen above is needed for the angular momentum relations given below in Eq.\eqref{l_qp_general}-Eq.\eqref{angmoment_rels_crd2}.
Yet, for completeness, we note that the above is subject to a constraint \eq{\varphi} with associated Lagrange multiplier \eq{\lambda}, and that one may restrict consideration to \eq{\varphi=0} (i.e., \eq{q=1}) and \eq{\lambda=0} in order to obtain an inverse transformation for \eq{\mbb{R}^6\ni(\tup{r},\tup{\v}) \mapsto (\bartup{q},\bartup{p})\in\mbb{R}^8 } as follows  (see \cite{peterson2025prjCoord} for details):
\begin{small}
\begin{align} \label{prj_inv_family}
\fnsz{\begin{array}{cc}
     \text{with:}  \\
     q = 1 \\[1pt]
      \lambda = 0
\end{array}}
\;\;\Rightarrow 
\qquad
\; \left\{ \;\;
\begin{array}{lllll}
       \tup{q} \,=\,  \htup{r} 
       &,\qquad  u \,=\, r^\ss{1/n}  
 \\[3pt]
     \tup{p}  \,=\,  r( \imat_3 + m \htup{r}\otms\htup{r})\cdot \tup{\v}
     \,=\, -\hdge{\tup{\slang}}\cdot\htup{r} + (m+1)(\htup{r}\cdot\tup{\v})\tup{r}
     &,\qquad 
       p_\ss{u} \,=\, 
      n r^\ss{-1/n } \tup{r}\cdot\tup{\v} 
\end{array} \right.
\end{align}
\end{small}

\noindent 
Now, we have introduced the general family of transformations in Eq.\eqref{prj_family} because, it turns out, the angular momentum functions in Eq.\eqref{l_rv_general} are invariant under any such transformation. 
It can be verified by direct substitution of \eq{\tup{r}(\bartup{q},\bartup{p})} and \eq{\tup{\v}(\bartup{q},\bartup{p})} from Eq.\eqref{prj_family} that the functions in Eq.\eqref{l_rv_general} are unchanged in the new coordinates:
\begin{small}\begin{align} \label{l_qp_general}
\begin{array}{rllllll}
      \tup{\slang} &\!\!\!=\, \tup{r}\tms \tup{\v} \,=\, \hdge{\tup{\v}}\cdot\tup{r}
  \\[3pt]
     &\!\!\!=\, \tup{q}\tms \tup{p}  \,=\, \hdge{\tup{p}}\cdot\tup{q}
\end{array}
 &&,&&
\begin{array}{rllllll}
       \hdge{\tup{\slang}} &\!\!\!=\,  \tup{r} \wdg \tup{\v}  \,=\, \tup{r}\otms \tup{\v}-\tup{\v}\otms \tup{r}
  \\[3pt]
     &\!\!\!=\,  \tup{q} \wdg \tup{p}  \,=\, \tup{q}\otms \tup{p}-\tup{p}\otms \tup{q}
\end{array}
 &&,&&
\begin{array}{rllllll}
        \slang^2  = \mag{\tup{\slang}}^2
      &\!\!\!=\, r^2 \v^2  - (\tup{r}\cdot\tup{\v})^2
  \\[3pt]
     &\!\!\!=\, q^2 p^2  - (\tup{q}\cdot\tup{p})^2
\end{array}
\end{align}\end{small}
We further note the following useful relations (where \eq{\slang_{ij}=q_i p_j-p_i q_j=-\slang_{ji}}):
\begin{small}
\begin{align} \label{angmoment_rels_crd} 
\begin{array}{lllllll}
    \slang^2 \,=\,  \slang_i \slang_i = \tfrac{1}{2} \slang_{ij} \slang_{ij} 
    = \tup{q}\cdot \hdge{\tup{\slang}} \cdot\tup{p}  = q^2 p^2 - (\tup{q}\cdot\tup{p})^2
\\[4pt]
     0 \,=\,  \slang_{ij}q_i q_j = \slang_{ij}p_i p_j  = \tup{q}\cdot \hdge{\tup{\slang}} \cdot\tup{q} = \tup{p}\cdot \hdge{\tup{\slang}} \cdot\tup{p}
\end{array} 
&&,&&
\begin{array}{lll}
     \hdge{\tup{\slang}} \cdot\hdge{\tup{\slang}} \cdot\tup{q} = -\slang^2 \tup{q}
\\[4pt]
     \hdge{\tup{\slang}} \cdot\hdge{\tup{\slang}} \cdot\tup{p} = -\slang^2 \tup{p}
\end{array}
&&,&&
\begin{array}{lll}
     \hdge{\tup{\slang}} \cdot\tup{q} = -(q^2 \imat_3 - \tup{q}\otms\tup{q}) \cdot\tup{p} 
     =  \hdge{\tup{q}} \cdot \hdge{\tup{q}} \cdot\tup{p}
\\[4pt]
     \hdge{\tup{\slang}} \cdot\tup{p} = (p^2 \imat_3 - \tup{p}\otms\tup{p}) \cdot\tup{q} 
     = -\hdge{\tup{p}}\cdot\hdge{\tup{p}}\cdot\tup{q}
\end{array}
\end{align}
\end{small} 
as well as the following relations for various partial derivatives (where \eq{\pd^2_{\tup{b}\tup{a}} := \pd_{\tup{b}}\pd_{\tup{a}}}):
\begin{small}
\begin{align} \label{angmoment_rels_crd2}
 &\begin{array}{llllll}
      \pd_{\tup{q}}
     \tfrac{1}{2}\slang^2 \,=\,  \hdge{\tup{\slang}}\cdot\tup{p} 
\\[4pt]
      \pd_{\tup{p}}
     \tfrac{1}{2}\slang^2 \,=\,   -\hdge{\tup{\slang}}\cdot\tup{q}
\end{array}
&&,&&
\begin{array}{rllllll}
    \pd^2_{\tup{q}\tup{q}} \tfrac{1}{2}\slang^2  =&
  \pd_{\tup{q}} (\hdge{\tup{\slang}}\cdot\tup{p}) \,=\,  - \hdge{\tup{p}}\cdot\hdge{\tup{p}}
\\[4pt] 
    -\pd^2_{\tup{p}\tup{p}} \tfrac{1}{2}\slang^2 =&
    \pd_{\tup{p}} (\hdge{\tup{\slang}}\cdot\tup{q}) \,=\, \hdge{\tup{q}}\cdot\hdge{\tup{q}}  
\\[4pt]
     -\pd^2_{\tup{q}\tup{p}} \tfrac{1}{2}\slang^2 =&
   \pd_{\tup{q}}
      (\hdge{\tup{\slang}}\cdot\tup{q}) \,=\, \hdge{\tup{\slang}} - \hdge{\tup{q}}\cdot\hdge{\tup{p}} 
  \\[4pt]
    \pd^2_{\tup{p}\tup{q}} \tfrac{1}{2}\slang^2 =&
     \pd_{\tup{p}} (\hdge{\tup{\slang}}\cdot\tup{p}) \,=\, \hdge{\tup{\slang}} +  \hdge{\tup{p}}\cdot\hdge{\tup{q}}
\end{array}
\\[6pt] \nonumber 
&\begin{array}{llllll}
  \pd_{\tup{q}}
     \slang \,=\,   \hdge{\htup{\slang}}\cdot\tup{p}
\\[4pt]
      \pd_{\tup{p}} \slang \,=\,   -\hdge{\htup{\slang}}\cdot\tup{q}
\end{array}
&&,&& 
\begin{array}{rllllll}
      \pd^2_{\tup{q}\tup{q}} \slang =&
     \pd_{\tup{q}} (\hdge{\htup{\slang}}\cdot\tup{p}) 
     \,=\,
      \tfrac{p^2}{\slang} \big( \imat_3 - \htup{p}\otms\htup{p} - (\hdge{\htup{\slang}}\cdot\htup{p})\otms (\hdge{\htup{\slang}}\cdot\htup{p}) \big)
     &\!\! = \tfrac{p^2}{\slang} \htup{\slang} \otms \htup{\slang}
\\[4pt] 
    -\pd^2_{\tup{p}\tup{p}} \slang =&
   \pd_{\tup{p}} (\hdge{\htup{\slang}}\cdot\tup{q})
   \,=\, 
   - \tfrac{q^2}{\slang} \big( \imat_3 - \htup{q}\otms\htup{q} - (\hdge{\htup{\slang}}\cdot\htup{q})\otms (\hdge{\htup{\slang}}\cdot\htup{q}) \big)
    &\!\! =- \tfrac{q^2}{\slang} \htup{\slang} \otms \htup{\slang}
\\[4pt]
     -\pd^2_{\tup{q}\tup{p}} \slang =&
   \pd_{\tup{q}}
      (\hdge{\htup{\slang}}\cdot\tup{q}) \,=\, \tfrac{1}{\slang} \big( \hdge{\tup{\slang}} - \hdge{\tup{q}}\cdot\hdge{\tup{p}} -   (\hdge{\htup{\slang}}\cdot\tup{q})\otms (\hdge{\htup{\slang}}\cdot\tup{p}) \big)
\\[4pt]
    \pd^2_{\tup{p}\tup{q}} \slang =&
     \pd_{\tup{p}} (\hdge{\htup{\slang}}\cdot\tup{p}) \,=\, \tfrac{1}{\slang} \big( \hdge{\tup{\slang}} +  \hdge{\tup{p}}\cdot\hdge{\tup{q}} +   (\hdge{\htup{\slang}}\cdot\tup{p})\otms (\hdge{\htup{\slang}}\cdot\tup{q}) \big)
 \end{array}
\end{align}
\end{small}
where we have 
used\footnote{In more detail:
\begin{align} \nonumber
\begin{array}{rllll}
     \pd^2_{\tup{q}\tup{q}} \slang =&\!\!\!\!
     \pd_{\tup{q}} (\hdge{\htup{\slang}}\cdot\tup{p}) \,=\,
     -\tfrac{1}{\slang} \big(  \hdge{\tup{p}}\cdot\hdge{\tup{p}} +  (\hdge{\htup{\slang}}\cdot\tup{p})\otms (\hdge{\htup{\slang}}\cdot\tup{p}) \big)
     &=\, 
     \tfrac{p^2}{\slang} \big( \imat_3 - \htup{p}\otms\htup{p} - (\hdge{\htup{\slang}}\cdot\htup{p})\otms (\hdge{\htup{\slang}}\cdot\htup{p}) \big)
     \,=\,  \tfrac{p^2}{\slang} \htup{\slang} \otms \htup{\slang}
\\[4pt]
     -\pd^2_{\tup{p}\tup{p}} \slang =&\!\!\!\!
   \pd_{\tup{p}} (\hdge{\htup{\slang}}\cdot\tup{q}) \,=\,
   \tfrac{1}{\slang} \big(  \hdge{\tup{q}}\cdot\hdge{\tup{q}} + (\hdge{\htup{\slang}}\cdot\tup{q})\otms (\hdge{\htup{\slang}}\cdot\tup{q}) \big)
   &=\, 
   - \tfrac{q^2}{\slang} \big( \imat_3 - \htup{q}\otms\htup{q} - (\hdge{\htup{\slang}}\cdot\htup{q})\otms (\hdge{\htup{\slang}}\cdot\htup{q}) \big)
   \,=\,  - \tfrac{q^2}{\slang} \htup{\slang} \otms \htup{\slang}
\end{array}
\end{align}
}
the below relation, following from the fact that \eq{\{\htup{q}, -\hdge{\htup{\slang}}\cdot\htup{q}, \htup{\slang}\} } and \eq{\{\htup{p}, -\hdge{\htup{\slang}}\cdot\htup{p}, \htup{\slang}\} } are orthonormal bases: 
\begin{small}
\begin{align} \label{angmoment_rels_iden}
\imat_3 \,=\,  
     \htup{q}\otms\htup{q} + (\hdge{\htup{\slang}}\cdot\htup{q})\otms (\hdge{\htup{\slang}}\cdot\htup{q}) +  \htup{\slang} \otms \htup{\slang}
    \,=\,  \htup{p}\otms\htup{p} + (\hdge{\htup{\slang}}\cdot\htup{p})\otms (\hdge{\htup{\slang}}\cdot\htup{p}) +  \htup{\slang} \otms \htup{\slang}
\end{align}
\end{small}

\noindent For the above angular momentum relations in Eq.\eqref{l_qp_general}-Eq.\eqref{angmoment_rels_iden}, we note:
\begin{small}
\begin{enumerate}
    \item Everything holds exactly the same with \eq{(\tup{q},\tup{p})} replaced by inertial cartesian coordinates \eq{(\tup{r},\tup{\v})}. 
    \item More generally, if for any \eq{\tup{x},\tup{y}\in\mbb{R}^3} we define \eq{\hdge{\tup{z}}:=\tup{x}\wdg\tup{y}} (i.e, \eq{\tup{z}:=\tup{x}\tms\tup{y}=\hdge{\tup{y}}\cdot\tup{x}}), then everything holds with \eq{(\tup{q},\tup{p},\tup{\slang})} replaced by \eq{(\tup{x},\tup{y},\tup{z})}.  
    \item Nothing depends on \eq{(u,p_\ss{u})} nor the choices of \eq{n\neq 0,m\in\mbb{R}} in the family of transformations in Eq.\eqref{prj_family}.
    \item Nothing involves the functions \eq{\varphi} and \eq{\lambda} in Eq.\eqref{prj_family}. That is, nothing is simplified using \eq{q=1} or \eq{\lambda=0}. 
\end{enumerate}
\end{small}

\paragraph{Simplified Relations (in the case \eq{m=-1}).}
 As mentioned, the above angular momentum relations do not involve the functions \eq{\varphi} or \eq{\lambda} in Eq.\eqref{prj_family}. 
 Yet, one \textit{could}, if they wished, rewrite some of the above using \eq{\varphi=0} (i.e., \eq{q=1}) and \eq{\lambda=0}. The reasons for this is the same as mentioned in section \ref{sec:prj_Xform} 
 (reviewed in the footnote\footnote{It was shown in \cite{peterson2025prjCoord,peterson2025prjGeomech} that \eq{q=\mag{\tup{q}}} and \eq{\lambda} are integrals of motion of the Hamiltonian system in any projective coordinates defined by Eq.\eqref{prj_family} for any \eq{n,m\in\mbb{R}} (note \eq{q} being an integral of motion is equivalent to \eq{\varphi} being an integral of motion). This holds for any arbitrary forces, conservative or nonconservative. Furthermore, we may choose to limit consideration to the values \eq{q=1} and \eq{\lambda =0} as this places no restrictions on the cartesian coordinates. In particular, if Eq.\eqref{prj_inv_family} is used to transform initial conditions \eq{(\tup{r}_\zr,\tup{\v}_\zr)\mapsto (\bartup{q}_\zr,\bartup{p}_\zr)}, then it holds that \eq{q_\zr=1} and \eq{\lambda_\zr =0} and, therefore, it holds for all time along any solution curve in projective coordinates that \eq{q=1} and \eq{\lambda =0}. }). 
For the above angular momentum relations, this does not lead to anything particularly interesting until specific values of \eq{n,m\in\mbb{R}} are chosen. In particular, only for the choice \eq{m=-1} does simplifying the above with \eq{q=1} and \eq{\lambda=0} lead to notable changes, detailed below.

This work uses a projective transformation detailed in section \ref{sec:prj_Xform} which corresponds to \eq{m=n=-1} in Eq.\eqref{prj_family}.\footnote{That is, we use a transformation given as follows: 
\begin{align} \label{prj__family}
\begin{array}{lllll}
      \tup{r} \,=\, \tfrac{1}{u} \htup{q} 
      &,\quad 
      ( \varphi = q -1  = 0)
\\[4pt]
     \tup{\v} \,=\,  u q (\imat_3 - \htup{q}\otms\htup{q})\cdot\tup{p} - u^2 p_\ss{u}\htup{q} 
     &,\quad (\lambda = \htup{q}\cdot\tup{p})
\end{array} 
&& \longleftrightarrow &&
\begin{array}{llll}
     \tup{q} \,=\,  \htup{r} 
     &,\quad 
      u \,=\, 1/r  
 \\[4pt]
     \tup{p}  
     \,=\, -(\tup{r}\wdg\tup{\v})\cdot\htup{r}
     \,=\, -\hdge{\tup{\slang}}\cdot\htup{r}
     &,\quad
      p_\ss{u} \,=\,  - r^2 \htup{r}\cdot\tup{\v}
\end{array}
\end{align}
}
Although the value of \eq{n} is inconsequential for the present discussion, the value of \eq{m} is not: with \eq{m=-1}, it is noteworthy that the multiplier \eq{\lambda} is then given by \eq{\lambda=\htup{q}\cdot\tup{p}} such that limiting consideration to \eq{\lambda=0} means that \eq{(\tup{q},\tup{p},\tup{\slang})} are all mutually orthogonal (this holds only for \eq{m=-1}). Together with \eq{q=1}, this means \eq{\tup{q}} and \eq{\tup{p}} satisfy the below relations (from Eq.\eqref{qpl_rels}): 
\begin{small}
\begin{flalign} \label{qpl_rels_apx}
    \qquad
    \begin{array}{lll}
          \tup{\slang}=\tup{q}\tms \tup{p}
        \\[4pt]
        \slang^2 = q^2 p^2-(\tup{q}\cdot\tup{p})^2 
        \\[4pt]
          \tup{q}\cdot\tup{\slang} = \tup{p}\cdot\tup{\slang}  = 0
    \end{array}
    &&
    \begin{array}{cc}
        \fnsize{using:} \\
         q=1  \\[1pt]
         \lambda=\htup{q}\cdot\tup{p}=0 
    \end{array}
     \Rightarrow \;\; 
    \left\{ \quad
    \begin{array}{ll}
        \tup{q} \simeq \htup{q} \simeq \htup{p}\tms \htup{\slang} 
      \\[4pt]
        \tup{p} \simeq \tup{\slang}\tms \htup{q} 
    \\[4pt]
          \slang^2 \simeq p^2  
    \end{array} \right.
    \;\;,\quad 
     \begin{array}{lll}
            \hdge{\tup{\slang}}\cdot\tup{q} \simeq -q^2\tup{p} \simeq -\tup{p}
            \\[4pt]
            \hdge{\tup{\slang}}\cdot\tup{p} \simeq p^2\tup{q} \simeq \slang^2 \tup{q} 
            \\[4pt]
              \imat_3 \simeq \htup{q} \otms\htup{q} + \htup{p} \otms\htup{p} + \htup{\slang} \otms\htup{\slang}
        \end{array}
    \quad
\end{flalign}
\end{small}
In particular, \eq{ \{\htup{q},\htup{p},\htup{\slang}\} \simeq \{\htup{t}_r,\htup{t}_\tau, \htup{t}_\slang \}} are precisely the inertial cartesian components of the LVLH basis. 
Using the above relations, the derivatives in Eq.\eqref{angmoment_rels_crd2} simplify to:
\begin{small}
\begin{align} \label{angmoment_rels_simplified}
 &\begin{array}{llllll}
      \pd_{\tup{q}}
     \ttfrac{1}{2}\slang^2 \,=\, \hdge{\tup{\slang}}\cdot\tup{p} 
     &\simeq\, \slang^2 \tup{q}
\\[4pt]
      \pd_{\tup{p}}
     \ttfrac{1}{2}\slang^2 \,=\, -\hdge{\tup{\slang}}\cdot\tup{q}
     &\simeq\,  \tup{p}
\end{array}
&&,&&
\begin{array}{llllll}
      \pd_{\tup{q}} (\hdge{\tup{\slang}}\cdot\tup{p}) \,=\,  - \hdge{\tup{p}}\cdot\hdge{\tup{p}} 
\\[4pt] 
   \pd_{\tup{p}} (\hdge{\tup{\slang}}\cdot\tup{q}) \,=\, \hdge{\tup{q}}\cdot\hdge{\tup{q}} 
\\[4pt]
   \pd_{\tup{q}}
      (\hdge{\tup{\slang}}\cdot\tup{q}) \,=\, \hdge{\tup{\slang}} - \hdge{\tup{q}}\cdot\hdge{\tup{p}} 
      &\simeq\, \hdge{\tup{\slang}} - \tup{p}\otms\tup{q}
\\[4pt]
     \pd_{\tup{p}} (\hdge{\tup{\slang}}\cdot\tup{p}) \,=\, \hdge{\tup{\slang}} +  \hdge{\tup{p}}\cdot\hdge{\tup{q}}
     &\simeq\, \hdge{\tup{\slang}} + \tup{q}\otms\tup{p}
\end{array}
\\[6pt] \nonumber 
&\begin{array}{llllll}
 \pd_{\tup{q}}
     \slang \,=\, \hdge{\htup{\slang}}\cdot\tup{p}
     &\simeq\, \slang \tup{q}
\\[4pt]
      \pd_{\tup{p}} \slang \,=\, -\hdge{\htup{\slang}}\cdot\tup{q} 
      &\simeq\, \htup{p}
\end{array}
&&,&& 
\begin{array}{llllll}
      \pd_{\tup{q}} (\hdge{\htup{\slang}}\cdot\tup{p}) 
      \,=\, \tfrac{p^2}{\slang}\htup{\slang}\otms\htup{\slang}
       &\simeq\,
       \slang  \htup{\slang}\otms\htup{\slang}
        &\simeq\, 
      \slang ( \imat_3  - \htup{q}\otms \htup{q}  - \htup{p}\otms\htup{p} )
\\[4pt] 
         \pd_{\tup{p}} (\hdge{\htup{\slang}}\cdot\tup{q}) 
         \,=\, - \tfrac{q^2}{\slang}\htup{\slang}\otms\htup{\slang}
    &\simeq\,  - \tfrac{1}{\slang} \htup{\slang} \otms \htup{\slang}
     &\simeq\,   
       - \tfrac{1}{\slang} ( \imat_3 - \htup{q}\otms\htup{q} - \htup{p}\otms \htup{p} )
\\[4pt]
  \pd_{\tup{q}}
      (\hdge{\htup{\slang}}\cdot\tup{q}) 
      \,=\, \fnsize{Eq.\eqref{angmoment_rels_crd2}}
       &\simeq\,  \hdge{\htup{\slang}}
\\[4pt]
     \pd_{\tup{p}} (\hdge{\htup{\slang}}\cdot\tup{p}) 
     \,=\, \fnsize{Eq.\eqref{angmoment_rels_crd2}}
     &\simeq\, \hdge{\htup{\slang}}
 \end{array}
\end{align}
\end{small}
(the last two relations are shown in the footnote\footnote{Using the relations in Eq.\eqref{qpl_rels_apx}: 
\begin{align} \nonumber 
\begin{array}{llll}
     \pd_{\tup{q}}
      (\hdge{\htup{\slang}}\cdot\tup{q}) \,=\, \tfrac{1}{\slang} \big( \hdge{\tup{\slang}} - \hdge{\tup{q}}\cdot\hdge{\tup{p}} -   (\hdge{\htup{\slang}}\cdot\tup{q})\otms (\hdge{\htup{\slang}}\cdot\tup{p}) \big)
      &\simeq\, 
      \hdge{\htup{\slang}} - \hdge{\tup{q}}\cdot\hdge{\htup{p}} + \htup{p}\otms \tup{q}
       &=\,  \hdge{\htup{\slang}} + (\tup{q}\cdot\htup{p})\imat_3 
       &\simeq\,  \hdge{\htup{\slang}}
\\[4pt]
     \pd_{\tup{p}} (\hdge{\htup{\slang}}\cdot\tup{p}) \,=\, \tfrac{1}{\slang} \big( \hdge{\tup{\slang}} +  \hdge{\tup{p}}\cdot\hdge{\tup{q}} +   (\hdge{\htup{\slang}}\cdot\tup{p})\otms (\hdge{\htup{\slang}}\cdot\tup{q}) \big)
     &\simeq\, 
     \hdge{\htup{\slang}} +  \hdge{\htup{p}}\cdot\hdge{\tup{q}}
     - \tup{q}\otms\htup{p}
     &=\,  \hdge{\htup{\slang}}  - (\tup{q}\cdot\htup{p})\imat_3
     &\simeq\, \hdge{\htup{\slang}} 
\end{array}
\end{align} }).
The above relations denoted with ``\eq{\simeq}'' follow from using \eq{\lambda=\htup{q}\cdot\tup{p}=0} and \eq{q=1} (and thus \eq{\slang^2\simeq p^2}). \textit{These relations are specific to projective coordinates defined using \eq{m=-1} in Eq.\eqref{prj_family}.} 
We further note that these relations denoted with ``\eq{\simeq}'' are valid specifically along solution curves of the projective coordinate Hamiltonian system that start with initial conditions satisfying \eq{q_\zr=1} and \eq{\lambda_\zr =\htup{q}_\zr \cdot\tup{p}_\zr =0} (we are always free to limit consideration to such initial conditions).

\section{KEPLER STATE TRANSITION MATRICES IN PROJECTIVE COORDINATES} \label{sec:prj_STM_new}


Kepler dynamics and solutions in projective coordinates were given in section \ref{sec:kep_sol}. We now re-frame those results with more emphasis on their matrix structure as a linear ODE and present the corresponding Kepler state transition matrix (STM). 
First, we recall some basics of linear systems on real coordinate space, \eq{\mbb{R}^\en}.

\paragraph*{Some Review.}
Consider some \eq{\en}-dim linear autonomous and \textit{in}homogeneous ODE, \eq{\diff{}{\varep}\tup{x} = M\cdot\tup{x} + \tup{k}}, where \eq{M\in\mbb{R}^{\en\times\en}} and \eq{\tup{k}\in\mbb{R}^\en} are constant and were \eq{\varep} is some evolution parameter (independent variable). The solution to any such system may be expressed in terms of initial conditions \eq{\tup{x}_{\zr}} at \eq{\varep=0} as:
\begin{small}
\begin{flalign} \label{lin_system_gen}
\qquad
   \diff{}{\varep}\tup{x} = \tup{X}(\tup{x}) \,=\, M\cdot\tup{x} + \tup{k}
   \quad\;\; \Rightarrow \quad\;\; 
    \tup{x}_{\varep} = \phi_\varep(\tup{x}_{\zr}) \,=\,\Sigma_\varep  \cdot \tup{x}_{\zr} + \tup{\sig}_\varep
    &&,&&
    \begin{array}{llll}
        \Sigma_\varep := \mrm{e}^{M \varep} \in\Glmat{\en}
         \;\;,\quad 
         \inv{\Sigma_\varep} =  \Sigma_{-\varep} 
     \\[4pt] 
     \tup{\sig}_{\varep} := \int_{\zr}^{\varep}  \inv{\Sigma_\varep} \cdot \tup{k}\, \mrm{d} \varep 
    \end{array}
\end{flalign}
\end{small}
where \eq{\tup{X}} is the linear inhomogeneous vector field\footnote{In this work, a vector field \eq{\tup{X}} on \eq{\mbb{R}^\en} may be regarded simply as a smooth map \eq{\tup{X}:\mbb{R}^\en\to\mbb{R}^\en}. }
for \eq{M} and \eq{\tup{k}},
with \eq{\varep}-parameterized solution flow \eq{\phi_\varep:\mbb{R}^\en\to\mbb{R}^\en} given in terms of \eq{\Sigma_\varep} and \eq{\tup{\sig}_\varep} defined as above. 
The matrix \eq{\Sigma_\varep =\mrm{e}^{M \varep} } is sometimes referred to as the STM. Yet, this term is only truly accurate in the linear \textit{homogeneous} case, \eq{ \diff{}{\varep}\tup{x} = M\cdot\tup{x}} (when \eq{\tup{k}=0=\tup{\sig}_\varep}), which has solution \eq{\tup{x}_{\varep} = \Sigma_\varep   \cdot \tup{x}_{\zr}}. The term STM is also used more generally to refer to
a matrix \eq{\Phi_\varep:=\pderiv{\phi_\varep}{\tup{x}}\big|_{\tup{x}_0}} which, by abuse of notation, one might write as \eq{\pderiv{\tup{x}_{\varep}}{\tup{x}_0}}. 
This matrix itself satisfies the following ODE:
\begin{small}
\begin{flalign} \label{STM_ODE}
  \Phi_\varep:=\pderiv{\phi_\varep}{\tup{x}}\big|_{\tup{x}_0} \equiv   \pderiv{\tup{x}_{\varep}}{\tup{x}_0} \in\Glmat{\en}
\qquad,\qquad
  \diff{}{\varep} \Phi_\varep \,=\, \pderiv{\tup{X}}{\tup{x}} \cdot \Phi_\varep
\qquad,\qquad
  \Phi_\zr \,=\, \imat_\en
\end{flalign}
\end{small}
We use the term STM to refer to \eq{\Phi_\varep =\pderiv{\tup{x}_{\varep}}{\tup{x}_0}}, not \eq{\Sigma_\varep =\mrm{e}^{M \varep}}. 
In the case that \eq{M} and \eq{\tup{k}} in Eq.\eqref{lin_system_gen} are truly numeric constants, then one has \eq{\Phi_\varep = \Sigma_\varep } and \eq{\pderiv{\tup{X}}{\tup{x}} = M}. However, in the following, \eq{M} will instead be a matrix of integrals of motion. While this still allows for \eq{M} to be treated as constant and to use Eq.\eqref{lin_system_gen}, it means that  \eq{\Phi_\varep \neq \Sigma_\varep } and \eq{\pderiv{\tup{X}}{\tup{x}} \neq M}. 

\subsection{Unsimplified Matrix Equations and STM} \label{sec:prj_STM_unsimp}


Returning to the problem at hand, 
note the \eq{s}- or \eq{\tau}-parameterized unperturbed
Kepler dynamics from Eq.\eqref{uw_eom_kep} can indeed be posed in the above general form of  Eq.\eqref{lin_system_gen} — though the matrix \eq{M} in Eq.\eqref{lin_system_gen} will be a matrix of integrals of motion, not literal real numbers. 
We already examined this for the ``\eq{(\tup{q},\tup{p})}-part'' of the dynamics in Eq.\eqref{dz_ds_E3}-Eq.\eqref{dz_ds_E3_alt2}. 
On that note, it will now be convenient to re-order our eight  phase space coordinates by splitting them into a ``\eq{(\tup{q},\tup{p})}-part'' (rotational motion) and ``\eq{(u,p_\ss{u})}-part'' (radial motion), where the latter may also be replaced by the pair \eq{(u,w:=u^2 p_\ss{u})}.

\begin{notesq}
    \textit{Coordinate ordering.}
    Rather than the standard configuration-momentum split ordering, we will adopt a modified ordering:
\begin{small}
\begin{align}
\begin{array}{cc}
      \fnsize{standard}  \\
     \fnsize{ordering}
\end{array}
\left\{\quad 
\begin{array}{llll}
      \tup{z} =  (\tup{q},u,\tup{p},p_\ss{u}) =(\bartup{q},\bartup{p}) 
\\[4pt]
      \tup{x} = (\tup{q},u,\tup{p},w) 
      \simeq (\bartup{q},\pdt{\bartup{q}})
\end{array}\right.
&&,&&
\begin{array}{cc}
      \fnsize{modified}  \\
     \fnsize{ordering}
\end{array}
\left\{\quad 
\begin{array}{llll}
      \tup{z} =  (\tup{q},\tup{p},u,p_\ss{u}) 
\\[4pt]
       \tup{x} = (\tup{q},\tup{p},u,w) 
      \simeq (\tup{q},\pdt{\tup{q}},u,\pdt{u})
\end{array}\right.
\end{align}
\end{small}
where \eq{\tup{z}\in\mbb{R}^8} denotes the set of ``canonical projective coordinates'' and \eq{\tup{x}\in\mbb{R}^8} denotes said coordinates with \eq{p_\ss{u}} replaced by the quasi-momentum coordinate \eq{w=u^2 p_\ss{u}}. 
The following developments will mostly use \eq{ \tup{x} = (\tup{q},\tup{p},u,w)}.
\end{notesq}

\noindent Recall that \textit{first}-order Kepler dynamics for \eq{\tup{z}} are not fully linear due to nonlinearity in the \eq{(u,p_\ss{u})} dynamics, but that those of \eq{\tup{x}} are indeed fully linear. They are fully linear whether or not one simplifies the ODEs using the integrals of motion \eq{q=1} and \eq{\lambda=\htup{q}\cdot\tup{p}=0}. We consider first the unsimplified case, then the simplified case.



The ODEs in Eq.\eqref{uw_eom_kep} and solutions in Eq.\eqref{qusol_s_2bp} describe unperturbed (and unsimplified) Kepler dynamics using the coordinates \eq{\tup{x}=(\tup{q},\tup{p},u,w)}. 
Recall these ODEs from Eq.\eqref{uw_eom_kep} were given by:
\begin{small}
\begin{flalign} \label{uw_eom_kep_yetagain}
\begin{array}{cc}
     \fnsize{Kepler-type }  \\
      \fnsize{dynamics} 
\end{array} 
&&
\begin{array}{llllll}
     \pdt{\tup{q}}  \,=\, 
  -\hdge{\tup{\slang}}\cdot\tup{q} 
  &,\quad 
  \pdt{\tup{p}}  \,=\, 
  -\hdge{\tup{\slang}}\cdot\tup{p} 
\\[4pt]
   \pdt{u} \,=\, w
    &,\quad 
     \pdt{w} = -\slang^2 u + \kconst_1  
\end{array}
\qquad\quad\fnsize{or,}  \qquad\quad
 \begin{array}{llllll}
     \rng{\tup{q}}  \,=\, 
  -\hdge{\htup{\slang}}\cdot\tup{q} 
  &,\quad 
  \rng{\tup{p}}  \,=\, 
  -\hdge{\htup{\slang}}\cdot\tup{p} 
\\[4pt]
   \rng{u} \,=\,  w/\slang 
    &,\quad 
     \rng{w} = -\slang u + {\kconst_1}/{\slang} 
\end{array} 
&&
\end{flalign}
\end{small}
with \eq{\tup{\slang}=\tup{\slang}_\zr} an integral of motion. 
The above, and their solutions, are equivalent to the following: 
\begin{small}
\begin{align} \label{dqp_mat}
   \diff{}{s} \fnpmat{
        \tup{q} \\[2pt]
        \tup{p} }
    = L \cdot \fnpmat{
        \tup{q}\\[2pt]
        \tup{p} }
    \quad &\fnsize{or,} \quad 
      \diff{}{\tau} \fnpmat{
        \tup{q} \\[2pt]
        \tup{p} }
    = \tfrac{1}{\slang} L \cdot \fnpmat{
        \tup{q}\\[2pt]
        \tup{p} }    
&& \Rightarrow &&
     \fnpmat{
        \tup{q}_s \\[2pt]
        \tup{p}_s } 
        = \mrm{e}^{L s} \cdot  \fnpmat{
        \tup{q}_\zr \\[2pt]
        \tup{p}_\zr } 
        \;\;=\;\;
        \mrm{e}^{\frac{1}{\slang} L \tau} \cdot  \fnpmat{
        \tup{q}_\zr \\[2pt]
        \tup{p}_\zr } 
         =
        \fnpmat{
        \tup{q}_\tau \\[2pt]
        \tup{p}_\tau }
\\[6pt] \nonumber
      \diff{}{s} \fnpmat{u \\ w}
    = \Lambda_2 \cdot \fnpmat{u \\ w} +  \fnpmat{0 \\ \kconst_1}
    \quad &\fnsize{or,} \quad
     \diff{}{\tau} \fnpmat{u \\ w}
    = \tfrac{1}{\slang} \Lambda_2 \cdot \fnpmat{u \\ w} +  \tfrac{1}{\slang}\fnpmat{0 \\ \kconst_1}
   && \Rightarrow &&
    \fnpmat{u_s \\[2pt] w_s} = \mrm{e}^{\Lambda_2 s} \cdot \fnpmat{u_\zr \\[2pt] w_\zr } 
    + \fnpmat{ \sig^u_s \\[2pt] \sig^w_s}
  \;\;=\;\;
   \mrm{e}^{\frac{1}{\slang}\Lambda_2 \tau} \cdot \fnpmat{u_\zr \\[2pt] w_\zr } 
    + \fnpmat{ \sig^u_\tau \\[2pt] \sig^w_\tau }
    =   \fnpmat{u_\tau \\[2pt] w_\tau} 
\end{align}
\end{small}
where \eq{\tau=\slang s} (assuming \eq{\tau_\zr=s_\zr=0}) and where \eq{L} and \eq{\Lambda_2} are matrices of integrals of motion that depend only on \eq{\tup{\slang}=\tup{\slang}_\zr}. They are given as follows, along with the other terms appearing above:
\begin{small}
\begin{align} \label{Lmat_spso_again}
\begin{array}{llllll}
      L := \fnpmat{
        -\hdge{\tup{\slang}}  & 0 \\  0 & -\hdge{\tup{\slang}}  }  \in \spmat{6} \cap \somat{6} 
&,\qquad 
     \mrm{e}^{L s} = \mrm{e}^{\frac{1}{\slang}L \tau} = \fnpmat{
          R_{\tau}(\htup{\slang}) & 0 \\
          0 &    R_{\tau}(\htup{\slang})  }
    \in \Spmat{6} \cap \Somat{6}  
    &,\qquad
     R_{\tau}(\htup{\slang}):= \mrm{e}^{-\hdge{\htup{\slang}} \tau} 
     \in \Somat{3}
\\[12pt]
         \Lambda_2 := \fnpmat{ 0 & 1 \\  -\slang^2  & 0 } \in\spmat{2}
 &,\qquad 
        \mrm{e}^{\Lambda_2 s} =  \mrm{e}^{ \frac{1}{\slang} \Lambda_2 \tau} = 
        \fnpmat{
            \csn{\tau}  & \tfrac{1}{\slang}\snn{\tau}  \\ 
            -\slang \snn{\tau}  &  \csn{\tau} } 
            \in \Spmat{2}
&,\qquad 
      \fnpmat{ \sig^u_\tau \\[2pt] \sig^w_\tau }
     = \fnpmat{ 
        \tfrac{\kconst_1}{\slang^2}(1-\csn{\tau}) \\
          \tfrac{\kconst_1}{\slang}\snn{\tau}  }
\end{array}
\end{align}
\end{small}
where the inhomogeneous term \eq{(\sig^u ,\sig^w)} is defined as usual 
(see footnote\footnote{As per the general formulas in Eq.\eqref{lin_system_gen}, \eq{(\sig^u,\sig^w)} is defined by:
     \eq{\quad  \fnpmat{ \sig^u_s \\[2pt] \sig^w_s } := \int_0^s \big[ \mrm{e}^{-\Lambda_2 s} \cdot \fnpmat{0 \\ \kconst_1} \big] \mrm{d} s
     \,=\,  \fnpmat{ \sig^u_\tau \\[2pt] \sig^w_\tau } = \int_0^\tau \big[ \mrm{e}^{-\Lambda_2 \tau /\slang} \cdot \fnpmat{0 \\ \kconst_1/\slang} \big] \mrm{d} \tau } .
}),
and where we have specified the dimension, 2, on \eq{\Lambda_2} because this matrix reappears several times with different dimensions
 (see footnote\footnote{For any finite dimension \eq{\en}, we define \eq{\Lambda_{2\en}\in\spmat{2\en}} and \eq{\mrm{e}^{\Lambda_{2\en}\tau/\slang} \in \Spmat{2\en} } as follows: 
\begin{align} \label{ThisMat_Ngen}
   \Lambda_{2\en} := \fnpmat{0 & \imat_\en \\ -\slang^2 \imat_\en & 0 }
     \in \spmat{2\en}
    \qquad,\qquad 
    \mrm{e}^{\Lambda_{2\en} s} = \mrm{e}^{\Lambda_{2\en}\tau/\slang}  
    = \fnpmat{ \csn{\tau} \,\imat_\en & \tfrac{1}{\slang}\snn{\tau}\,\imat_\en \\ 
            -\slang \snn{\tau}\,\imat_\en &  \csn{\tau}\,\imat_\en  }
            \in \Spmat{2\en}
\end{align}
where \eq{\Lambda_{2\en}} has eigenvalues \eq{\pm\mrm{i}\slang} (\eq{\en} of each sign)
}). 
The above \eq{L}, \eq{\mrm{e}^{Ls}=\mrm{e}^{\frac{1}{\slang} L \tau}}, and \eq{R_{\tau}(\htup{\slang})}  were discussed in more detail around Eq.\eqref{Lmat_spso}. 
The dynamics in Eq.\eqref{dqp_mat} can then be combined as a single linear inhomogeneous  ODE:
\begin{small}
\begin{flalign} \label{qusol_MAT_2bp_nonsimp}
\quad
\tup{x}=(\tup{q},\tup{p},u,w)
&&
   \boxed{
\begin{array}{rlll}
        &\pdt{\tup{x}} \,=\, M\cdot\tup{x} + \tup{k} 
\\[5pt] 
   \fnsize{or,} &\rng{\tup{x}} \,=\, \tfrac{1}{\slang}M \cdot \tup{x} \,+\, \tfrac{1}{\slang}\tup{k} 
\end{array} }
    \quad  \Rightarrow \qquad 
     \tup{x}_\tau \,=\, \Sig_\tau \cdot \tup{x}_\zr \,+\, \tup{\sig}_\tau 
&&
\tau = \slang s
\quad
\end{flalign} 
\end{small}
with \eq{\tup{k}\in\mbb{R}^8} a numeric constant, with \eq{M = M_\zr\in \mbb{R}^{8 \times 8}} a matrix of integrals of motion given below, with \eq{\Sig_\tau := \mrm{e}^{\frac{1}{\slang}M \tau} = \mrm{e}^{M s}},  and with \eq{\tup{\sig}_\tau := \int_{\zr}^{\tau} \inv{\Sig_\tau} \cdot \tup{k}\, \mrm{d} \tau  \in \mbb{R}^8}  defined as in  Eq.\eqref{lin_system_gen}. These are given for the ordering \eq{\tup{x}=(\tup{q},\tup{p},u,w)} 
as:\footnote{Using the submatrices defined in Eq.\eqref{Lmat_spso_again}, \eq{M} and \eq{\Sig_\tau} are given explicitly by: 
\begin{align} \nonumber
\begin{array}{lllll}
        M 
=
\fnpmat{
        \fnpmat{
        -\hdge{\tup{\slang}}  & 0_{3\times 3} \\  0_{3\times 3} & -\hdge{\tup{\slang}}  }
      & 0_{6\times 2}
  \\
        0_{2\times 6}  
        & \fnpmat{ 0 & 1 \\  -\slang^2  & 0 }
    }     
        \in \spmat{6;2}
   \qquad,\qquad 
       \Sig_\tau := \mrm{e}^{\frac{1}{\slang}M \tau}
 =
 \fnpmat{ 
         \fnpmat{
          R_{\tau}(\htup{\slang}) & 0_{3\times 3} \\
          0_{3\times 3} &    R_{\tau}(\htup{\slang})  }
             & 0_{6\times 2}
     \\
        0_{2\times 6}
        & \fnpmat{
            \csn{\tau}  & \tfrac{1}{\slang}\snn{\tau}  \\ 
            -\slang \snn{\tau}  &  \csn{\tau} }
     }
         \in \Spmat{6;2}
\end{array}
\end{align}
}
\begin{small}
\begin{align} \label{KepMats_nonsimp}
\begin{array}{lllll}
        M :=  \fnpmat{ L & 0 \\ 0 & \Lambda_2 }   
        \in \spmat{6;2}
    &,\qquad 
      \Sig_\tau := \mrm{e}^{\frac{1}{\slang}M \tau}
     =  \fnpmat{
           \mrm{e}^{\frac{1}{\slang}L \tau}   & 0 \\
           0 & \mrm{e}^{\frac{1}{\slang}\Lambda_2 \tau} 
           }
            \in \Spmat{6;2}
    &,\qquad
    \tup{k} := \fnpmat{  \tup{0} \\  \tup{0} \\ 0  \\ \kconst_1}
    &,\qquad
     \tup{\sig}_\tau = \fnpmat{ \tup{0} \\
        \tup{0} \\
         \tfrac{\kconst_1}{\slang^2}(1-\csn{\tau}) \\
          \tfrac{\kconst_1}{\slang}\snn{\tau}  }
\end{array}
\end{align}
\end{small}
where \eq{\Spmat{6;2}} and \eq{\spmat{6;2}} denote, respectively, symplectic and Hamiltonian matrices with regards to \eq{\fnpmat{ \jmat_6 &0 \\ 0 & \jmat_2} } 
    (see footnote\footnote{We define \eq{\Spmat{6;2}} and \eq{\spmat{6;2}} as follows,  where \eq{\jmat_{2\en}\in \Spmat{2\en}} denotes the standard symplectic matrix on \eq{\mbb{R}^{2\en}}: 
    \begin{align} \label{Sp_alt_def}
       \Spmat{6;2} :=\; \big\{\, S \in \mbb{R}^{8\times 8}    \;\big|\;  \trn{S}\jmat_\ss{6;\!2} S = \jmat_\ss{6;\!2} \, \big\}
    \qquad,\qquad 
       \spmat{6;2}  :=\; \big\{\, H \in \mbb{R}^{8\times 8}    \;\big|\;  \jmat_\ss{6;\!2} H + \trn{H} \jmat_\ss{6;\!2} = 0 \,\big\}
        \qquad,\qquad 
        \jmat_\ss{6;\!2} := \jmat_6\oplus \jmat_2 = \fnpmat{
        \jmat_6 &0 \\ 0 & \jmat_2}
    \end{align} }).
Note that \eq{M} has eigenvalues \eq{(0,0,\mrm{i}\slang,\mrm{i}\slang,-\mrm{i}\slang,-\mrm{i}\slang)} (from \eq{L}) and \eq{(\mrm{i}\slang,-\mrm{i}\slang)} (from \eq{\Lambda_2}). 
It can be verified that the solution in Eq.\eqref{qusol_MAT_2bp_nonsimp}, when written out explicitly, agrees with that given previously in Eq.\eqref{qusol_s_2bp}: 
\begin{small}
\begin{flalign}  \label{qusol_MAT_2bp_inv_nonsimp}
\qquad
\begin{array}{ccccc}
    \boxed{ \tup{x}_\tau = \phi_\tau (\tup{x}_\zr) = \Sig_\tau \cdot \tup{x}_\zr + \tup{\sig}_\tau 
    \qquad \leftrightarrow \qquad 
    \tup{x}_\zr = \inv{\phi_\tau} (\tup{x}_\tau) = \inv{\Sig_\tau} \cdot \tup{x}_\tau + \tup{\varsig}_\tau }
\\[10pt]
\begin{array}{lllllll}
    \tup{q}_\tau
       \,=\, 
      \tup{q}_\zr \csn{\tau} - \hdge{\htup{\slang}}_\zr\cdot\tup{q}_\zr \snn{\tau} 
\\[4pt]
       \tup{p}_\tau
       \,=\, 
      \tup{p}_\zr \csn{\tau} - \hdge{\htup{\slang}}_\zr \cdot\tup{p}_\zr \snn{\tau} 
\\[4pt]
      u_\tau
      \,=\,  u_{\zr} \csn{\tau} + \tfrac{1}{\slang_0}w_{\zr}\snn{\tau} + \tfrac{\kconst_1}{\slang_0^2}(1-\csn{\tau})
\\[4pt]
       w_\tau 
      \,=\, w_{\zr}\csn{\tau} -\slang_\zr u_{\zr} \snn{\tau} + \tfrac{\kconst_1}{\slang_0}\snn{\tau}
\end{array}
 \qquad \leftrightarrow \qquad 
\begin{array}{llll}
     \tup{q}_\zr \,=\, \tup{q}_\tau\csn{\tau} + \hdge{\htup{\slang}}_\tau\cdot\tup{q}_\tau \snn{\tau} 
\\[4pt]
      \tup{p}_\zr \,=\, \tup{p}_\tau\csn{\tau} + \hdge{\htup{\slang}}_\tau\cdot\tup{p}_\tau \snn{\tau}
\\[4pt]
     u_\zr \,=\,  u_\tau \csn{\tau} - \tfrac{1}{\slang_\tau}w_\tau\snn{\tau} + \tfrac{\kconst_1}{\slang_\tau^2}(1-\csn{\tau})
\\[4pt]
      w_\zr \,=\,    w_\tau\csn{\tau} + \slang_\tau u_\tau \snn{\tau}  - \tfrac{\kconst_1}{\slang_\tau}\snn{\tau}
\end{array}
\end{array}
&& \tup{\slang}_\tau =\tup{\slang}_\zr
\end{flalign}
\end{small}
where we have denoted by \eq{\phi_\tau:\mbb{R}^8\to\mbb{R}^8} the (unsimplified) Kepler flow in coordinates \eq{\tup{x}}, and where we note:
\begin{small}
\begin{align} \label{qusol_MAT_invTerm}
    \inv{\phi_\tau} =  \phi_{-\tau}
\qquad,\qquad 
    \inv{\Sig_\tau} =  \Sig_{-\tau}
\qquad,\qquad 
     \tup{\varsig}_\tau := -\inv{\Sig_\tau} \cdot \tup{\sig}_\tau  
           = \tup{\sig}_{-\tau}
\end{align}
\end{small}

\begin{small}
\begin{notesq}
   \textit{Angular momentum dependence.} Though not explicit in our notation, many terms in the above Kepler dynamics depend on the angular momentum, \eq{\tup{\slang}}. In particular:
    \begin{small}
    \begin{align} \label{KepODE_ang_depend1}
         M = M(\tup{\slang}) 
         \qquad,\qquad 
         \Sig_\tau = \Sig_\tau(\tup{\slang}) 
         \qquad,\qquad 
         \tup{\sig}_\tau = \tup{\sig}_\tau(\tup{\slang})
    \end{align}
    \end{small}
    with \eq{\tup{\slang} = \tup{\slang}(\tup{x}) = \tup{q}\tms\tup{p}} a function of the coordinates. 
    Yet, since \eq{\tup{\slang}=\tup{\slang}_\zr} is conserved for any central-force dynamics, we have treated it as a constant when solving the Kepler ODEs. Still, it should be noted that Eq.\eqref{qusol_MAT_2bp_nonsimp} is really a family of linear ODEs parameterized by angular momentum values (set by initial conditions).
    For instance, for a given \eq{\tup{\slang}_\zr}, it would be more accurate to write the ODE and the solution in  Eq.\eqref{qusol_MAT_2bp_nonsimp} as:
    \begin{small}
    \begin{align} \label{KepODE_ang_depend2}
        \rng{\tup{x}} \,=\, \tfrac{1}{\slang_0} M(\tup{\slang}_\zr) \cdot \tup{x} \,+\, \tfrac{1}{\slang_0}\tup{k} 
    \qquad  \Rightarrow \qquad 
         \tup{x}_\tau \,=\, \phi_\tau (\tup{x}_\zr) \,=\, \Sig_\tau(\tup{\slang}_\zr) \cdot \tup{x}_\zr \,+\, \tup{\sig}_\tau(\tup{\slang}_\zr)
         &&,&&
         \pderiv{\tup{x}_\tau}{\tup{x}_0} \neq \Sig_\tau
    \end{align}
    \end{small}
    This dependence on \eq{\tup{\slang}} of the solution flow, \eq{\phi_\tau}, is  reflected in the way we have written the explicit solutions in Eq.\eqref{qusol_MAT_2bp_inv_nonsimp}.
\end{notesq}
\end{small}

\paragraph*{Kepler STM in Projective Coordinates.}
As indicated above, the matrices \eq{\Sig_\tau} and \eq{\pderiv{\tup{x}_\tau}{\tup{x}_0}} — both of which are sometimes referred to as the STM — do \textit{not} coincide. When perturbations are considered and \eq{\tup{\slang}} is no longer conserved, then the object of interest is the more general version of the STM given by \eq{\Phi_\tau := \pderiv{\tup{x}_\tau}{\tup{x}_0}} (one might write this more precisely as \eq{ \pderiv{\phi_\tau}{\tup{x}}\big|_{\tup{x}_0}}), which we note is partitioned as:
\begin{small}
\begin{flalign} 
\qquad
\begin{array}{llll}
      \tup{x} =  (\tup{q},\tup{p},u,w)
\end{array}
&&
 \Phi_\tau := \pderiv{\phi_\tau}{\tup{x}}\big|_{\tup{x}_0} \equiv  \pderiv{\tup{x}_\tau}{\tup{x}_0}
     \,=
 \pmat{ 
     \pderiv{(\tup{q}_\tau,\tup{p}_\tau)}{(\tup{q}_0,\tup{p}_0)}
     &    \pderiv{(\tup{q}_\tau,\tup{p}_\tau)}{(u_0,w_0)} 
     \\[6pt]
     \pderiv{(u_\tau,w_\tau)}{(\tup{q}_0,\tup{p}_0)}
     &
      \pderiv{(u_\tau,w_\tau)}{(u_0,w_0)}
     }
\;=\;
 \fnpmat{ 
     \fnpmat{ 
        \pderiv{\tup{q}_\tau}{\tup{q}_0}
     & \pderiv{\tup{q}_\tau}{\tup{p}_0}
     \\[4pt]
      \pderiv{\tup{p}_\tau}{\tup{q}_0}
     & \pderiv{\tup{p}_\tau}{\tup{p}_0}    
     }_{6\times 6}
     & 
     \fnpmat{
     \tup{0} &  \tup{0}  \\[4pt] \tup{0} &  \tup{0}
     }_{6\times 2}
 \\[14pt]
    \fnpmat{ 
       \trn{\pderiv{u_\tau}{\tup{q}_0}}
     &  \trn{\pderiv{u_\tau}{\tup{p}_0}}
     \\[4pt]
       \trn{\pderiv{w_\tau}{\tup{q}_0}}
     &  \trn{\pderiv{w_\tau}{\tup{p}_0}}
    }_{2\times 6}
 &
    \fnpmat{ 
     \pderiv{u_\tau}{u_0}
     &\pderiv{u_\tau}{w_0} 
     \\[4pt]
     \pderiv{w_\tau}{u_0}
     &\pderiv{w_\tau}{w_0}
    }_{2\times 2}
 }  
&&   
\end{flalign}
\end{small}
with \eq{\phi_\tau} the Kepler flow in Eq.\eqref{qusol_MAT_2bp_inv_nonsimp} and
where \eq{(u_\tau,w_\tau)} depend on \eq{(\tup{q}_\zr,\tup{p}_\zr)} only through \eq{\slang=\slang_\zr}. 
Using the solutions in Eq.\eqref{qusol_MAT_2bp_inv_nonsimp} — along with angular momentum relations from Eq.\eqref{l_qp_general}-Eq.\eqref{angmoment_rels_crd2} of Appx.~\ref{sec:ang_momentum} — the above leads to:
\begin{small}
\begin{align} \label{STM_prj_nosimp}
  \Phi_\tau = \pderiv{\tup{x}_\tau}{\tup{x}_0}
   &\,=\,\left.\fnpmat{ 
    \fnpmat{ 
    \imat_3 \csn{\tau} - \tfrac{1}{\slang} \big( \hdge{\tup{\slang}} - \hdge{\tup{q}}\cdot\hdge{\tup{p}} -   (\hdge{\htup{\slang}}\cdot\tup{q})\otms (\hdge{\htup{\slang}}\cdot\tup{p}) \big) \snn{\tau} 
    &  \tfrac{q^2}{\slang}  \htup{\slang}\otms\htup{\slang}  \snn{\tau} 
    \\[3pt]
    -\tfrac{p^2}{\slang}  \htup{\slang}\otms\htup{\slang}  \snn{\tau} 
    &  \imat_3 \csn{\tau} -   \tfrac{1}{\slang} \big( \hdge{\tup{\slang}} +  \hdge{\tup{p}}\cdot\hdge{\tup{q}} +   (\hdge{\htup{\slang}}\cdot\tup{p})\otms (\hdge{\htup{\slang}}\cdot\tup{q}) \big) \snn{\tau}  
    }
    &
    \fnpmat{ \tup{0} & \tup{0} \\[3pt] \tup{0} & \tup{0} }
\\[18pt]
    \fnpmat{ 
         \tfrac{1}{\slang}\big( \tfrac{2\kconst_1}{\slang^2}(1-\csn{\tau}) + \tfrac{1}{\slang}w \snn{\tau} \big)\tup{p}\cdot\hdge{\htup{\slang}}
         &\;\;  -\tfrac{1}{\slang}\big( \tfrac{2\kconst_1}{\slang^2} (1-\csn{\tau}) + \tfrac{1}{\slang}w \snn{\tau} \big) \tup{q}\cdot\hdge{\htup{\slang}}
     \\[3pt]
          (u + \tfrac{\kconst_1}{\slang^2})\snn{\tau}\,   \tup{p}\cdot\hdge{\htup{\slang}}
         &   -(u + \tfrac{\kconst_1}{\slang^2})\snn{\tau}\, \tup{q}\cdot\hdge{\htup{\slang}}
     }
    &
    \fnpmat{
    \csn{\tau} & \tfrac{1}{\slang}\snn{\tau}
    \\[3pt]
    -\slang \snn{\tau} & \csn{\tau}
    }
} \right|_{\tup{x}=\tup{x}_0}
\\[8pt] \nonumber
   \inv{\Phi_\tau} =  \pderiv{\tup{x}_0}{\tup{x}_\tau} 
&\,=\,
\left.\fnpmat{ 
    \fnpmat{ 
    \imat_3 \csn{\tau} + \tfrac{1}{\slang} \big( \hdge{\tup{\slang}} - \hdge{\tup{q}}\cdot\hdge{\tup{p}} -   (\hdge{\htup{\slang}}\cdot\tup{q})\otms (\hdge{\htup{\slang}}\cdot\tup{p}) \big) \snn{\tau} 
    &  -\tfrac{q^2}{\slang}  \htup{\slang}\otms\htup{\slang}  \snn{\tau}  
    \\[3pt]
     \tfrac{p^2}{\slang}  \htup{\slang}\otms\htup{\slang}  \snn{\tau}  
    &  \imat_3 \csn{\tau} +   \tfrac{1}{\slang} \big( \hdge{\tup{\slang}} +  \hdge{\tup{p}}\cdot\hdge{\tup{q}} +   (\hdge{\htup{\slang}}\cdot\tup{p})\otms (\hdge{\htup{\slang}}\cdot\tup{q}) \big) \snn{\tau}  
    } 
    &
    \fnpmat{ \tup{0} & \tup{0} \\[3pt] \tup{0} & \tup{0} }
\\[18pt]
    \fnpmat{ 
         \tfrac{1}{\slang}\big( \tfrac{2\kconst_1}{\slang^2}(1-\csn{\tau}) - \tfrac{1}{\slang}w \snn{\tau} \big) \tup{p}\cdot\hdge{\htup{\slang}}
         &\;\;  -\tfrac{1}{\slang}\big( \tfrac{2\kconst_1}{\slang^2} (1-\csn{\tau}) - \tfrac{1}{\slang}w \snn{\tau} \big) \tup{q}\cdot\hdge{\htup{\slang}}
     \\[3pt]
          - (u + \tfrac{\kconst_1}{\slang^2})\snn{\tau}\,   \tup{p}\cdot\hdge{\htup{\slang}}
         &   (u + \tfrac{\kconst_1}{\slang^2})\snn{\tau}\,  \tup{q}\cdot\hdge{\htup{\slang}}
     }
    &
    \fnpmat{
    \csn{\tau} & -\tfrac{1}{\slang}\snn{\tau}
    \\[3pt]
    \slang \snn{\tau} & \csn{\tau}
    }  
} \right|_{\tup{x}=\tup{x}_\tau}
=\, \Phi_{-\tau}\big|_{\tup{x}=\tup{x}_\tau}
\end{align}
\end{small}
The above follow from the unsimplified Kepler flow, \eq{\phi_\tau}, in Eq.\eqref{qusol_MAT_2bp_inv_nonsimp}.
If we now use the integrals of motion \eq{q=1} and  and \eq{\lambda=\htup{q}\cdot\tup{p}=0} — implying the relations in Eq.\eqref{qpl_rels_apx}-Eq.\eqref{angmoment_rels_simplified} of Appx.~\ref{sec:ang_momentum} —  to simplify the above, we obtain:
\begin{small}
\begin{align} \label{STM_prj_somesimp}
    \Phi_\tau 
    &\,\simeq\,
    \left.\fnpmat{ 
    \fnpmat{ 
    \imat_3 \csn{\tau} -  \hdge{\htup{\slang}} \snn{\tau} 
    & \tfrac{1}{\slang} \hdge{\htup{\slang}}\otms \hdge{\htup{\slang}} \snn{\tau} 
    \\[3pt]
    -\slang \hdge{\htup{\slang}}\otms \hdge{\htup{\slang}} \snn{\tau} 
    &     \imat_3 \csn{\tau} -  \hdge{\htup{\slang}} \snn{\tau} 
    } 
    &
    \fnpmat{ \tup{0} & \tup{0} \\[3pt] \tup{0} & \tup{0} }
\\[18pt]
   \fnpmat{ 
         -\big( \tfrac{2\kconst_1}{\slang^2}(1-\csn{\tau}) + \tfrac{1}{\slang}w \snn{\tau} \big) \trn{\tup{q}}
         &\;\;  -\tfrac{1}{\slang}\big( \tfrac{2\kconst_1}{\slang^2}(1-\csn{\tau}) + \tfrac{1}{\slang}w \snn{\tau} \big) \trn{\htup{p}}
     \\[3pt]
           -\slang (u + \tfrac{\kconst_1}{\slang^2})\snn{\tau}\, \trn{\tup{q}}
         &  - (u + \tfrac{\kconst_1}{\slang^2})\snn{\tau}\, \trn{\htup{p}}
     }
    &
    \fnpmat{
    \csn{\tau} & \tfrac{1}{\slang}\snn{\tau}
    \\[3pt]
    -\slang \snn{\tau} & \csn{\tau}
    }  
      }\right|_{\tup{x}=\tup{x}_0}
      \quad,\quad \slang \simeq p
\\[8pt] \nonumber 
   \inv{\Phi_\tau} 
   &\,\simeq\,
    \left.\fnpmat{ 
    \fnpmat{ 
    \imat_3 \csn{\tau} + \hdge{\htup{\slang}} \snn{\tau} 
    &  -\tfrac{1}{\slang}\htup{\slang}\otms\htup{\slang} \snn{\tau} 
    \\[3pt]
   \slang \htup{\slang}\otms\htup{\slang}  \snn{\tau}
    &  \imat_3 \csn{\tau} +   \hdge{\htup{\slang}} \snn{\tau} 
    } 
    &
    \fnpmat{ \tup{0} & \tup{0} \\[3pt] \tup{0} & \tup{0} }
\\[18pt]
     \fnpmat{ 
         -\big( \tfrac{2\kconst_1}{\slang^2}(1-\csn{\tau}) - \tfrac{1}{\slang}w \snn{\tau} \big) \trn{\tup{q}}
         &\;\;  -\tfrac{1}{\slang}\big( \tfrac{2\kconst_1}{\slang^2}(1-\csn{\tau}) - \tfrac{1}{\slang}w \snn{\tau} \big) \trn{\htup{p}}
     \\[3pt]
           \slang (u + \tfrac{\kconst_1}{\slang^2})\snn{\tau}\,  \trn{\tup{q}}
         &   (u + \tfrac{\kconst_1}{\slang^2})\snn{\tau}\, \trn{\htup{p}}
     }
    &
    \fnpmat{
    \csn{\tau} & -\tfrac{1}{\slang}\snn{\tau}
    \\[3pt]
    \slang \snn{\tau} & \csn{\tau}
    }  
} \right|_{\tup{x}=\tup{x}_\tau}
\end{align}
\end{small}
In all the above, the upper left sub-matrices have used the relation: 
\begin{small}
\begin{align}
    \htup{\slang}\otms \htup{\slang} 
    \;=\; \imat_3 - \htup{p}\otms\htup{p} - (\hdge{\htup{\slang}}\cdot\htup{p})\otms (\hdge{\htup{\slang}}\cdot\htup{p})
    \;=\; \imat_3 - \htup{q}\otms\htup{q} - (\hdge{\htup{\slang}}\cdot\htup{q})\otms (\hdge{\htup{\slang}}\cdot\htup{q})
    \;\simeq\; 
    \imat_3 - \htup{q}\otms\htup{q} - \htup{p}\otms\htup{p}
\end{align}
\end{small}

\begin{notesq}
    \textit{Canonical coordinate STM.} Let us denote by \eq{\Psi_\tau} the STM for the canonical coordinates \eq{\tup{z}=(\tup{q},\tup{p},u,p_\ss{u})}. This may be recovered from the STM \eq{\Phi_\tau} given above for \eq{\tup{x}=(\tup{q},\tup{p},u,w)} using:
    \begin{small}
    \begin{align}
        \Psi_\tau \,=\,
        \pderiv{\tup{z}_\tau}{\tup{z}_0} 
         \,=\, 
         \inv{(\pderiv{\tup{x}_\tau}{\tup{z}_\tau})} \cdot   \pderiv{\tup{x}_\tau}{\tup{x}_0}\cdot   \pderiv{\tup{x}_0}{\tup{z}_0}
         \,=\, 
         \pderiv{\tup{z}_\tau}{\tup{x}_\tau} \cdot   \Phi_{\tau} \cdot   \pderiv{\tup{x}_0}{\tup{z}_0}
    \end{align}
    \end{small}
    where the only transformation is \eq{w=u^2 p_\ss{u} \leftrightarrow p_\ss{u} = w/u^2} such that \eq{\pderiv{\tup{z}}{\tup{x}}} and  \eq{\pderiv{\tup{x}}{\tup{z}}} are given simply by
    \begin{small}
    \begin{align}
          \pderiv{\tup{x}}{\tup{z}} \,=\, 
          \fnpmat{ 
            \imat_6 & 0_{6\times 2} 
            \\
            0_{2 \times 6} 
            & 
            \fnpmat{1 &  0 \\
                   2 u p_\ss{u} & u^2
            } }
     \qquad,\qquad 
             \pderiv{\tup{z}}{\tup{x}}  = \inv{( \pderiv{\tup{x}}{\tup{z}})} 
          \,=\, 
          \fnpmat{ 
            \imat_6 & 0_{6\times 2} 
            \\
            0_{2 \times 6} 
            & 
            \fnpmat{1 &  0 \\
                   -2 p_\ss{u}/u & 1/u^2
            } }
    \end{align}
    \end{small}
\end{notesq}

\subsection{Simplified Matrix Equations and STM}

The above matrix ODEs and solutions are without simplifying anything using the integrals of motion \eq{q=1} or \eq{\lambda=\htup{q}\cdot\tup{p}=0}. 
The simplified STM in Eq.\eqref{STM_prj_somesimp} is still obtained from the unsimplified Kepler solutions; only the resulting \eq{\Phi_\tau =\pderiv{\tup{x}_\tau}{\tup{x}_0} } is simplified after differentiating.

We will now use \eq{q=1} and \eq{\htup{q}\cdot\tup{p}=0} (and thus \eq{\slang^2 \simeq p^2}) to simply the dynamics and solutions to begin with. 
The Kepler dynamics in Eq.\eqref{uw_eom_kep} then simplify to:
\begin{small}
\begin{flalign} \label{uw_eom_kep_simp}
\begin{array}{cc}
     \fnsize{Kepler-type }  \\
      \fnsize{dynamics} 
\end{array} 
&&
\begin{array}{llllll}
     \pdt{\tup{q}}  
     \,\simeq\, \tup{p}
  &,\quad 
  \pdt{\tup{p}}  
  \,\simeq\, - p^2 \tup{q}
  \,\simeq\, - \slang^2 \tup{q}
\\[4pt]
   \pdt{u} \,=\, w
    &,\quad 
     \pdt{w} = -\slang^2 u + \kconst_1  
\end{array}
\qquad\quad \fnsize{or,}  \qquad\quad
 \begin{array}{llllll}
     \rng{\tup{q}} 
     \,\simeq\, \tfrac{1}{\slang}\tup{p} \,\simeq\, \htup{p}
  &,\quad 
  \rng{\tup{p}}  
  \,\simeq\, -\slang \tup{q}
\\[4pt]
   \rng{u} \,=\,  w/\slang 
    &,\quad 
     \rng{w} = -\slang u + {\kconst_1}/{\slang} 
\end{array} 
&&
\end{flalign}
\end{small}
where \eq{\slang=\slang_\zr\simeq p_\zr = p} is preserved. The above is equivalent to the following matrix ODE:
\begin{small}
\begin{flalign} \label{qusol_MAT_2bp}
\quad
\tup{x}=(\tup{q},\tup{p},u,w)
&&
   \boxed{
\begin{array}{rlll}
        &\pdt{\tup{x}} \,\simeq\, \til{\Lambda} \cdot \tup{x} \,+\, \tup{k} 
\\[5pt] 
   \fnsize{or,} & \rng{\tup{x}} \,\simeq\, \tfrac{1}{\slang}\til{\Lambda} \cdot \tup{x} \,+\, \tfrac{1}{\slang}\tup{k} 
\end{array} }
    \quad  \Rightarrow \qquad 
    \tup{x}_\tau \simeq E_\tau \cdot \tup{x}_\zr \,+\, \tup{\sig}_\tau  
&&
\tau = \slang s
\quad
\end{flalign} 
\end{small}
with \eq{\tup{k}} and \eq{\tup{\sig}_\tau} the same as in Eq.\eqref{KepMats_nonsimp}, and 
where \eq{\til{\Lambda}=\til{\Lambda}_\zr} is a matrix of integrals of motion given as follows, along with \eq{E_\tau}:
\begin{small}
\begin{align}
        \til{\Lambda} := \fnpmat{ 
        \Lambda_6 & 0 \\
        0 & \Lambda_2 }
          \in \spmat{6;2}
\qquad,\qquad 
       E_\tau := \mrm{e}^{\frac{1}{\slang}\til{\Lambda} \tau} =  \mrm{e}^{\til{\Lambda} s}
        = \fnpmat{ \mrm{e}^{\frac{1}{\slang}\Lambda_6 \tau}  & 0 \\ 0 & \mrm{e}^{\frac{1}{\slang}\Lambda_2 \tau}  }
          \in \Spmat{6;2}
\qquad,\qquad
\begin{array}{llll}
     \tup{k} \,=\, \fnsize{Eq.\eqref{KepMats_nonsimp}}
     \\[3pt]
     \tup{\sig}_\tau \,=\, \fnsize{Eq.\eqref{KepMats_nonsimp}}
     \\[3pt]
      \tup{\varsig}_\tau := -\inv{E_\tau} \cdot \tup{\sig}_\tau = \tup{\sig}_{-\tau}
\end{array}
\end{align}
\end{small}
with \eq{\Lambda_{2\en}\in\spmat{2\en}} and \eq{\mrm{e}^{\Lambda_{2\en}\tau/\slang} \in \Spmat{2\en} } defined as in Eq.\eqref{ThisMat_Ngen} for any finite dimension \eq{\en}. 
When written out explicitly,  the solution in Eq.\eqref{qusol_MAT_2bp} agrees with that given previously in Eq.\eqref{qusol_s_2bp_simp}: 
\begin{small}
\begin{flalign}  \label{qusol_MAT_2bp_inv}
\quad
\begin{array}{ccccc}
     \boxed{ \tup{x}_\tau \simeq \theta_\tau(\tup{x}_\zr) = E_\tau \cdot \tup{x}_\zr + \tup{\sig}_\tau 
    \qquad \leftrightarrow \qquad 
    \tup{x}_\zr \simeq \inv{\theta_\tau}(\tup{x}_\tau) = \inv{E_\tau} \cdot \tup{x}_\tau + \tup{\varsig}_\tau }  
\\[10pt]
\begin{array}{lllllll}
    \tup{q}_\tau
       \,\simeq\, 
      \tup{q}_\zr \csn{\tau} + \tfrac{1}{\slang_0}\tup{p}_\zr \snn{\tau} 
\\[4pt]
       \tup{p}_\tau
       \,\simeq\, 
      \tup{p}_\zr \csn{\tau} - \slang_\zr \tup{q}_\zr \snn{\tau} 
\\[4pt]
      u_\tau
      \,=\,  u_{\zr} \csn{\tau} + \tfrac{1}{\slang_0}w_{\zr}\snn{\tau} + \tfrac{\kconst_1}{\slang_0^2}(1-\csn{\tau})
\\[4pt]
       w_\tau 
      \,=\,  w_{\zr}\csn{\tau}  -\slang_\zr u_{\zr} \snn{\tau} + \tfrac{\kconst_1}{\slang_0}\snn{\tau}
\end{array}
 \qquad \leftrightarrow \qquad 
\begin{array}{lll}
     \tup{q}_\zr \,\simeq\,  
     \tup{q}_\tau \csn{\tau} - \tfrac{1}{\slang_\tau}\tup{p}_\tau \snn{\tau} 
\\[4pt]
      \tup{p}_\zr \,\simeq\,  \tup{p}_\tau \csn{\tau} + \slang_\tau \tup{q}_\tau \snn{\tau} 
\\[4pt]
     u_\zr \,=\,  u_\tau \csn{\tau} - \tfrac{1}{\slang_\tau}w_\tau\snn{\tau} + \tfrac{\kconst_1}{\slang_\tau^2}(1-\csn{\tau})
\\[4pt]
      w_\zr \,=\,  w_\tau\csn{\tau} + \slang_\tau u_\tau \snn{\tau}  - \tfrac{\kconst_1}{\slang_\tau}\snn{\tau}
\end{array}
\end{array}
&&
\begin{array}{lll}
     \slang_\tau = \slang_\zr  \\[2pt]
     \slang \simeq p
\end{array}
\end{flalign}
\end{small}
where we have defined \eq{\theta_\tau} as the Kepler flow in coordinates \eq{\tup{x}} that has been simplified with \eq{q=1} and \eq{\htup{q}\cdot\tup{p}=0}. 
As before, 
note \eq{\inv{\theta_\tau} =  \theta_{-\tau}}, \eq{\inv{E_\tau} =  E_{-\tau}}, and \eq{\tup{\varsig}_\tau = \tup{\sig}_{-\tau}}.
\begin{small}
\begin{itemize}
      \item \textit{Angular momentum dependence.} The comments on angular momentum dependence around Eq.\eqref{KepODE_ang_depend1}-Eq.\eqref{KepODE_ang_depend2} still apply. Yet, for the above simplified relations, this dependence is only on the \textit{magnitude}, \eq{\slang}, which simplifies to \eq{\slang\simeq p}. 
      That is, \eq{\til{\Lambda} = \til{\Lambda}(\slang) }, \eq{ E_\tau = E_\tau(\slang)}, and \eq{\tup{\sig}_\tau = \tup{\sig}_\tau(\slang)}, with \eq{\slang\simeq p}. 
      To clarify, we could write the solutions in Eq.\eqref{qusol_MAT_2bp_inv} as:
      \begin{small}
    \begin{flalign}  \label{qusol_2bp_just_p}
    \begin{array}{ccccc}
    \begin{array}{lllllll}
        \tup{q}_\tau
           \simeq 
          \tup{q}_\zr \csn{\tau} + \tfrac{1}{p_0}\tup{p}_\zr \snn{\tau} 
    \\[4pt]
           \tup{p}_\tau
           \simeq 
          \tup{p}_\zr \csn{\tau} - p_\zr \tup{q}_\zr \snn{\tau} 
    \\[4pt]
          u_\tau
          \simeq  u_{\zr} \csn{\tau} + \tfrac{1}{p_0}w_{\zr}\snn{\tau} + \tfrac{\kconst_1}{p_0^2}(1-\csn{\tau})
    \\[4pt]
           w_\tau 
          \simeq  w_{\zr}\csn{\tau}  -p_\zr u_{\zr} \snn{\tau} + \tfrac{\kconst_1}{p_0}\snn{\tau}
    \end{array}
     \quad \leftrightarrow \quad 
    \begin{array}{lll}
         \tup{q}_\zr \simeq  
         \tup{q}_\tau \csn{\tau} - \tfrac{1}{p_\tau}\tup{p}_\tau \snn{\tau} 
    \\[4pt]
          \tup{p}_\zr \simeq  \tup{p}_\tau \csn{\tau} + p_\tau \tup{q}_\tau \snn{\tau} 
    \\[4pt]
         u_\zr \simeq  u_\tau \csn{\tau} - \tfrac{1}{p_\tau}w_\tau\snn{\tau} + \tfrac{\kconst_1}{p_\tau^2}(1-\csn{\tau})
    \\[4pt]
          w_\zr \simeq  w_\tau\csn{\tau} + p_\tau u_\tau \snn{\tau}  - \tfrac{\kconst_1}{p_\tau}\snn{\tau}
    \end{array}
    \end{array}
    &&,&&
    \begin{array}{lll}
        p_\tau = p_\zr  \\[3pt]
         \tup{p}_\tau \neq \tup{p}_\zr 
    \end{array}
    \end{flalign}
    \end{small}
    \item \textit{Back to ``standard'' ordering.} We have been using the modified coordinate ordering \eq{\tup{x} = (\tup{q},\tup{p},u,w)} which is best for illustrating the structure of the \textit{un}simplified dynamics. For the \textit{simplified} dynamics given above, one could just as easily return to the standard coordinate ordering, \eq{\tup{x} = (\tup{q},u,\tup{p},w)}, for which the above simplified matrix ODE takes the form
    \begin{small}
    \begin{flalign} \label{qusol_MAT_2bp_qupw}
    \qquad
    \tup{x}=(\tup{q},u,\tup{p},w)
    &&
    \begin{array}{rlll}
            &\pdt{\tup{x}} \,\simeq\, \Lambda_8 \cdot \tup{x} \,+\, \tup{k} 
    \\[5pt] 
       \fnsize{or,} & \rng{\tup{x}} \,\simeq\, \tfrac{1}{\slang}\Lambda_8 \cdot \tup{x} \,+\, \tfrac{1}{\slang}\tup{k} 
    \end{array}
        \qquad  \Rightarrow \qquad 
        \tup{x}_\tau \simeq E_\tau \cdot \tup{x}_\zr \,+\, \tup{\sig}_\tau  
    &&
    \end{flalign} 
    \end{small}
    where everything is the ``same'' of what was given in Eq.\eqref{qusol_MAT_2bp}-Eq.\eqref{qusol_MAT_2bp_inv}, up to a re-ordering:
    \begin{small}
    \begin{align}
             \Lambda_8 = \fnsize{ Eq.\eqref{ThisMat_Ngen} }
             \in \spmat{8}
    &&,&& 
           E_\tau = \mrm{e}^{\frac{1}{\slang}\Lambda_8 \tau}
           = \fnsize{ Eq.\eqref{ThisMat_Ngen} }
                \in \Spmat{8}
    &&,&&
        \tup{\sig}_\tau = \fnpmat{ \tup{0} \\
            \tfrac{\kconst_1}{\slang^2}(1-\csn{\tau}) \\
            \tup{0} \\
              \tfrac{\kconst_1}{\slang}\snn{\tau}  }
    &&,&&
    \tup{\varsig}_\tau = \tup{\sig}_{-\tau}
    \end{align}
    \end{small}
    where \eq{\Lambda_8} and \eq{\til{\Lambda}} both have eigenvalues \eq{\pm\mrm{i}\slang} (four of each sign). Note we have re-used some notation (\eq{\tup{x}}, \eq{E_\tau}, \eq{\tup{\sig}_\tau}, and \eq{\tup{\varsig}_\tau}) but now with a slightly different meaning (just re-ordering of indices). 
\end{itemize}
\end{small}
\vspace{1ex}

\noindent  Continuing with the modified ordering \eq{\tup{x}=(\tup{q},\tup{p},u,w)}, note that if we use the simplified Kepler flow, \eq{\theta_\tau}, from Eq.\eqref{qusol_MAT_2bp_inv} or Eq.\eqref{qusol_2bp_just_p} — treating \eq{\slang} as \eq{\slang\simeq p} —  to define an STM \eq{\Theta_\tau :=\pderiv{\theta_\tau}{\tup{x}}\big|_{\tup{x}_0} \equiv \pderiv{\tup{x}_\tau}{\tup{x}_0}}, then this would lead to:
\begin{small}
\begin{align}
  \Theta_\tau := \pderiv{\theta_\tau}{\tup{x}}\big|_{\tup{x}_0} \equiv \pderiv{\tup{x}_\tau}{\tup{x}_0}
   &\,\simeq\,
     \left.\fnpmat{ 
    \fnpmat{ 
    \csn{\tau} \, \imat_3
    &  \tfrac{1}{p}(\imat_3-\htup{p}\otms\htup{p})\snn{\tau} 
    \\[3pt]
   -p \snn{\tau} \,\imat_3
    & \csn{\tau} \, \imat_3 - \tup{q}\otms \htup{p} \snn{\tau}
    } 
    &
    \fnpmat{ \tup{0} & \tup{0} \\[3pt] \tup{0} & \tup{0} }
\\[14pt]
   \fnpmat{ 
         0 &
         -\tfrac{1}{p}\big( \tfrac{2\kconst_1}{p^2}(1- \csn{\tau}) + \tfrac{1}{p}w \snn{\tau} \big) \trn{\htup{p}}
         \\[3pt]
          0
         &  -(u + \tfrac{\kconst_1}{p^2})\snn{\tau} \,\trn{\htup{p}}
     }
    &
    \fnpmat{
    \csn{\tau} & \tfrac{1}{p}\snn{\tau}
    \\[3pt]
    -p \snn{\tau} & \csn{\tau}
    }  
} \right|_{\tup{x}=\tup{x}_0}
\\[8pt] \nonumber 
    \inv{\Theta_\tau} =  \pderiv{\tup{x}_0}{\tup{x}_\tau}
     &\,\simeq\, 
      \left.\fnpmat{ 
    \fnpmat{ 
    \csn{\tau} \, \imat_3
    &  -\tfrac{1}{p}(\imat_3-\htup{p}\otms\htup{p})\snn{\tau} 
    \\[3pt]
    p \snn{\tau} \,\imat_3
    & \csn{\tau} \, \imat_3 + \tup{q}\otms \htup{p} \snn{\tau}
    } 
    &
    \fnpmat{ \tup{0} & \tup{0} \\[3pt] \tup{0} & \tup{0} }
\\[14pt]
   \fnpmat{ 
         0 &
         -\tfrac{1}{p}\big( \tfrac{2\kconst_1}{p^2}(1- \csn{\tau}) - \tfrac{1}{p}w \snn{\tau} \big) \trn{\htup{p}}
         \\[3pt]
          0
         &  (u + \tfrac{\kconst_1}{p^2})\snn{\tau} \,\trn{\htup{p}}
     }
    &
    \fnpmat{
    \csn{\tau} & -\tfrac{1}{p}\snn{\tau}
    \\[3pt]
    p \snn{\tau} & \csn{\tau}
    }  
} \right|_{\tup{x}=\tup{x}_\tau}
\end{align}
\end{small}
The above \eq{\Theta_\tau} 
is not the same as the full STM \eq{\Phi_\tau}, from Eq.\eqref{STM_prj_nosimp} — which was obtained from the unsimplified Kepler flow — nor is it the same as the simplified form of \eq{\Phi_\tau} given in Eq.\eqref{STM_prj_somesimp}.

\section{ANOTHER APPROACH TO THE ORBIT ELEMENTS} \label{sec:VOP}


The following does not, for the most part, yield anything new beyond what was given in section \ref{sec:VOP_2}. In particular, the result is the same as Eq.\eqref{QPelems_ODE_VOP2_simp} and Eq.\eqref{QPelems_ODE_VOP2_Ftot_simp} (though the approach is a bit different).  There is quite a bit of redundancy between the following and section \ref{sec:VOP_2}. 
We include it here mostly for comparison with previous work by the authors in \cite{peterson2025phdThesis,peterson2023regularized} (where something similar to the following was presented) and with earlier work by Schumacher \cite{schumacher1987results}.
Some of the following expressions for the element ODEs may differ slightly from section \ref{sec:VOP_2}, though they are indeed equivalent. We also give an alternative element set, and their governing equations, which is not addressed in  section \ref{sec:VOP_2}.

\begin{notation}
    We have been using ``\eq{\simeq}'' to indicate relations that have been simplified using the integrals of motion \eq{q=1} and \eq{\lambda=\htup{q}\cdot\tup{p}=0}. Below, we will make free use of these relations and stop using the symbol ``\eq{\simeq}''. Although this means \eq{\slang=p}, we will often continue to write \eq{\slang} to keep in mind that this is the specific angular momentum magnitude. 
\end{notation}

\noindent 
We pick up  where section \ref{sec:kep_sol} left off, with closed-form Kepler solutions in projective coordinates. 
We will stick with \eq{\tau} as our chosen evolution parameter, where \eq{\mrm{d} t = (r^2/\slang)\mrm{d}\tau = 1/(\slang u^2)\mrm{d}\tau}, making \eq{\tau} the true anomaly up to an additive constant. 
Everything that follows could be also carried out using \eq{s} as the evolution parameter rather than \eq{\tau} (where \eq{\mrm{d}t = r^2 \mrm{d}s = u^\ss{-2}\mrm{d} s}). However, the \eq{\tau}-parameterized Kepler dynamics have unit natural frequency \eq{1}, whereas the  \eq{s}-parameterized dynamics have natural frequency \eq{\slang} which is generally \textit{not} constant in the perturbed case and the simple relation \eq{\tau=\slang s} does \textit{not} hold. We therefore choose to use \eq{\tau} rather than \eq{s} for simplicity (in particular, to simplify the variation of parameters procedure).

We now consider the perturbed Kepler problem, with \eq{\tau}-parameterized ODEs for \eq{\tup{x} =(\tup{q},u,\tup{p},w) \simeq (\bartup{q},\pdt{\bartup{q}}) } given as in Eq.\eqref{uw_eom_kepPert}. Simplifying them using \eq{q=1} and \eq{\lambda=\htup{q}\cdot\tup{p}=0} leads to: 
\begin{small}\begin{align} \label{qpw_eom_f_2bp}
\begin{array}{llll}
     \rng{\tup{q}} \,=\, -  \hdge{\htup{\slang}}\cdot\tup{q} 
     &\!\! =\, \tfrac{1}{\slang}\tup{p}
     &,
 \\[4pt]
     \rng{\tup{p}}  \,=\, -\hdge{\htup{\slang}}\cdot\tup{p}  + \rng{t} \tup{f} 
      &\!\! =\, -\slang \tup{q} + \rng{t}\tup{f}
     &,
\end{array}
\quad
\begin{array}{llll}
     \rng{u}   \,=\,  w/ \slang 
 \\[4pt]
       \rng{w} \,=\, -\slang u + \tfrac{1}{\slang}\kconst_1 + \tfrac{1}{\slang} f_\ss{u}
\end{array} 
\end{align}\end{small}
where \eq{\rng{t}=\tfrac{1}{\slang u^2}}, and where the generalized perturbation terms \eq{\tup{f}=\tup{F}\cdot\pderiv{\tup{r}}{\tup{q}}} and \eq{f_\ss{u}=\tup{F}\cdot\pderiv{\tup{r}}{u}}  were given in Eq.\eqref{Ftotal_qp}.



\subsection{Elements from Simplified Kepler Solutions}

 For the above perturbed Kepler dynamics, we  seek a solution that has the \textit{same algebraic form} as the unperturbed 
 (and simplified\footnote{That is, simplified using the integrals of motion \eq{q=1} and \eq{\lambda=\htup{q}\cdot\tup{p}=0}.}) 
 Kepler solutions in Eq.\eqref{qusol_s_2bp_simp} or, equivalently, Eq.\eqref{qusol_MAT_2bp}.  In order to do this, we no longer treat  \eq{\tup{x}_{\zr}}  as a set of constants/initial conditions but, instead, as a set of slowly evolving/osculating elements, \eq{\tup{x}_{\zr}(\tau)}. 
 More precisely, we introduce a set of eight elements \eq{\tup{\xi}:=(\tup{Q},U,\tup{P},W)}, defined by a 
 \eq{\tau}-dependent coordinate transformation \eq{\tup{x} \leftrightarrow \tup{\xi} } using the same (simplified) Kepler solution flow \eq{\theta_\tau:\mbb{R}^8\to\mbb{R}^8} from Eq.\eqref{qusol_MAT_2bp}-Eq.\eqref{qusol_MAT_2bp_inv}:
\begin{small}
\begin{gather} \nonumber 
           \tup{x} \,=\, \theta_\tau(\tup{\xi}) \,=\, E_\tau(\slang) \cdot \tup{\xi} \,+\, \tup{\sig}_\tau (\slang)
    \qquad \leftrightarrow \qquad 
    \tup{\xi} :=\, \inv{\theta_\tau}(\tup{x}) \,=\,  E_{-\tau}(\slang) \cdot \tup{x}\,+\, \tup{\sig}_{-\tau}(\slang) 
\\ \label{Xsol_kep2}
 \begin{array}{lllllll}
    \tup{q} 
       \,=\, 
      \tup{Q} \csn{\tau} + \tfrac{1}{\slang} \tup{P} \snn{\tau} 
\\[4pt]
       \tup{p} 
       \,=\, 
      -\slang\tup{Q} \snn{\tau}  +  \tup{P} \csn{\tau}
\\[4pt]
      u 
         \,=\,  (U-\tfrac{\kconst_1}{\slang^2})\csn{\tau} \,+\, \tfrac{1}{\slang} W \snn{\tau} \,+\, \tfrac{\kconst_1}{\slang^2}
\\[4pt]
       w  
         \,=\,  -\slang(U-\tfrac{\kconst_1}{\slang^2})\snn{\tau} \,+\, W\csn{\tau}
\end{array}
 \qquad \leftrightarrow \qquad 
\begin{array}{lll}
     \tup{Q} :=\, \tup{q}\csn{\tau} - \tfrac{1}{\slang}\tup{p}\snn{\tau} 
\\[4pt]
      \tup{P} :=\, \slang\tup{q}\snn{\tau} +  \tup{p}\csn{\tau}    
\\[4pt]
     U :=\, (u-\tfrac{\kconst_1}{\slang^2})\csn{\tau} -  \tfrac{1}{\slang}w\snn{\tau} + \tfrac{\kconst_1}{\slang^2}
\\[4pt]
      W :=\, \slang(u-\tfrac{\kconst_1}{\slang^2})\snn{\tau} + w\csn{\tau} 
\end{array} 
\end{gather}
\end{small}
It can be verified from the above, along with Eq.\eqref{qpl_rels}, that \eq{\tup{Q}}, \eq{\tup{P}}, and \eq{\tup{\slang}} satisfy the same relations as \eq{\tup{q}}, \eq{\tup{p}}, and \eq{\tup{\slang}}:
\begin{small}\begin{align} \label{qp0_rels2}
\begin{array}{ll}
      \tup{Q}\cdot\tup{Q} \,=\, \tup{q}\cdot\tup{q} \,=\, 1 
\\[4pt]
    \tup{Q}\cdot\tup{P} \,=\,  \tup{q}\cdot\tup{p} \,=\, 0
\\[4pt]
    \tup{P}\cdot\tup{P} \,=\, \tup{p}\cdot\tup{p} \,=\, \slang^2
\end{array}
&&
\begin{array}{ll}
     \tup{Q}\wdg \tup{P} \,=\, \tup{q}\wdg \tup{p} \,=\,   \hdge{\tup{\slang}}
\\[4pt]
    \tup{Q}\tms \tup{P} \,=\, \tup{q}\tms \tup{p} \,=\,   \tup{\slang}
\\[4pt] 
    \htup{P}\tms \htup{\slang} \,=\, \tup{Q}
\\[4pt] 
    \tup{\slang}\tms \tup{Q} \,=\, \tup{P}
\end{array}
&&
\begin{array}{lll}
    \slang^2 = q^2 p^2 - (\tup{q}\cdot\tup{p})^2 \,=\, Q^2 P^2 - (\tup{Q}\cdot\tup{P})^2  \quad = p^2 = P^2 
        \\[4pt]
      \imat_3 = \htup{q}\otms\htup{q} + \htup{p}\otms\htup{p} + \htup{\slang}\otms\htup{\slang}
    \,=\, \htup{Q}\otms\htup{Q} + \htup{P}\otms\htup{P} + \htup{\slang}\otms\htup{\slang}
\end{array}
\end{align}\end{small}
Note in the transformation Eq.\eqref{Xsol_kep2} that the \eq{\tau}-dependent terms  \eq{E_\tau}, \eq{\tup{\sig}_\tau}, and \eq{\tup{\varsig}_\tau} all also depend on \eq{\slang} (no longer constant) and thus should be regarded as functions of \eq{\tup{x}} in the transformation \eq{ \tup{\xi} = \inv{\theta_\tau}(\tup{x})}, or as functions of  \eq{\tup{\xi}} in the transformation \eq{\tup{x} = \theta_\tau(\tup{\xi})}. More specifically, these terms depend on \eq{\slang=\slang(\tup{q},\tup{p})=\slang(\tup{Q},\tup{P})} as given in Eq.\eqref{qp0_rels2}, which simplifies to \eq{\slang = p = P}. To clarify, we could write Eq.\eqref{Xsol_kep2} more explicitly as:
\begin{small}
\begin{align} \label{elems_kep_def2}
\boxed{ \begin{array}{lllllll}
    \tup{q}  \,=\, 
      \tup{Q} \csn{\tau} + \tfrac{1}{P} \tup{P} \snn{\tau} 
\\[3pt]
       \tup{p}    \,=\, 
      -P\tup{Q} \snn{\tau}  +  \tup{P} \csn{\tau}
\\[3pt]
      u \,=\,  (U-\tfrac{\kconst_1}{P^2})\csn{\tau} \,+\, \tfrac{1}{P} W \snn{\tau} \,+\, \tfrac{\kconst_1}{P^2}
\\[3pt]
       w      \,=\,  -P(U-\tfrac{\kconst_1}{P^2})\snn{\tau} \,+\, W\csn{\tau}
\end{array}
 \qquad \leftrightarrow \qquad 
\begin{array}{lll}
     \tup{Q} :=\, \tup{q}\csn{\tau} - \tfrac{1}{p}\tup{p}\snn{\tau} 
\\[3pt]
      \tup{P} :=\, p\tup{q}\snn{\tau} +  \tup{p}\csn{\tau}    
\\[3pt]
     U :=\, (u-\tfrac{\kconst_1}{p^2})\csn{\tau} -  \tfrac{1}{p}w\snn{\tau} + \tfrac{\kconst_1}{p^2}
\\[3pt]
      W :=\, p(u-\tfrac{\kconst_1}{p^2})\snn{\tau} + w\csn{\tau} 
\end{array} }
\end{align}
\end{small}
By way of Eq.\eqref{rv_kepsol} and Eq.\eqref{Xsol_kep2}, we also have the cartesian coordinates \eq{(\tup{r},\tup{\v})} are given in terms of the elements by:
\begin{small}
\begin{align} \label{rv_QPelems}
\begin{array}{llllll}
      \tup{r}  \,=\, \tfrac{1}{u}\htup{q} 
      &\!\! \simeq\, \tfrac{1}{u}\tup{q}
      &=\,  \dfrac{ \tup{Q} \csn{\tau}  \,+\,  \tfrac{1}{\slang}\tup{P}\snn{\tau}}{ (U-\frac{\kconst_1}{\slang^2})\csn{\tau} \,-\, W_\zr\snn{\tau} \,+\, \frac{\kconst_1}{\slang^2}   }  
\\[14pt] 
     \tup{\v}  
     \,=\, -u(\tup{q}\wdg\tup{p})\cdot\htup{q} - w \htup{q}
     &\!\!\simeq\, u \tup{p} - w \tup{q}
     &=\, 
     U\tup{P} - W\tup{Q} + \tfrac{\kconst_1}{\slang^2} \big[ \tup{P}(\csn{\tau}-1) -  \slang\tup{Q}\snn{\tau}  \big] 
\end{array}
\end{align}
\end{small}
with \eq{\slang^2\simeq p^2 \simeq P^2} as in E.\eqref{qp0_rels2}

\begin{notesq}
    The orbit elements defined above are not action-angle coordinates obtained from the Hamilton-Jacobi equation, nor are they any other type of canonical/symplectic coordinates. 
\end{notesq}

\paragraph{Interpretation of the Elements.}
Eq.\eqref{Xsol_kep2} defines the elements \eq{\tup{\xi}:=(\tup{Q},U,\tup{P},W)} as a new set of ``slow'' coordinates in such a way that, in the perturbed case, the solutions for the ``fast'' coordinates, \eq{\tup{x}_\tau}, are given in terms of the solutions \eq{\tup{\xi}_\tau} by the same relation as they are given in terms of \eq{\tup{x}_\zr} in the unperturbed case: 
\begin{small}\begin{align} \label{vop_idea}
    \begin{array}{cc}
     \scrsize{perturbed solution:}  \\[3pt]
     \tup{x}_\tau \,=\,  \theta_\tau (\tup{\xi}_{\tau} )  
     \end{array}
&&,&&  
     \begin{array}{cc}
     {\scrsize{unperturbed solution:}}  \\[3pt]
     \tup{x}_\tau \,=\, \theta_\tau (\tup{\xi}_{\tau} )   \,=\, \theta_\tau(\tup{x}_{\zr})
     \;\;,\;\;  \tup{\xi}_{\tau} = \tup{x}_{\zr}
\end{array}
\end{align}\end{small}
where \eq{\theta_\tau:\mbb{R}^8\to\mbb{R}^8} is the same map in both cases (the flow for the unperturbed Kepler dynamics). 
Notably, in the unperturbed Kepler case, then  \eq{\tup{\xi}} has trivial solution \eq{\tup{\xi}_{\tau}=\tup{x}_\zr}, coinciding with the constant initial conditions. 
We have \textit{required} the perturbed solution satisfy the algebraic relations seen in Eq.\eqref{Xsol_kep2}. 
Only for  unperturbed Keplerian motion does it hold that  \eq{\tup{\xi}_\tau = \tup{x}_{\zr}} are constants coinciding with initial conditions. 
For general perturbed motion, \eq{\tup{\xi}} is now interpreted as a new set of ``slow'' coordinates that are \textit{defined} in terms of the ``fast'' coordinates, \eq{\tup{x}}, by Eq.\eqref{Xsol_kep2}.

\paragraph{Equations of Motion for the Elements $\tup{\xi}$.} 
Using  variation of parameters, the governing equations of motion for the elements \eq{\tup{\xi}=(\tup{Q},U,\tup{P},W)} are found to be as follows (this will be derived shortly):
\begin{small}
\begin{align} \label{dX0_2}
 \begin{array}{rllll}
      \rng{\tup{Q}} &\!\!\! =\, 
       -\tfrac{\snn{\tau}}{\slang^2u^2} ( \imat_3 - \htup{p}\otms\htup{p} ) \cdot \tup{f}
       & =\, -\tfrac{\snn{\tau}}{\slang^2u^2} (\htup{\slang}\otms\htup{\slang} ) \cdot \tup{f}
\\[8pt]
     \rng{\tup{P}} &\!\!\! =\, 
     \tfrac{1}{\slang u^2}\Big( \htup{P} \otms \htup{p}  \,+\,  \csn{\tau} (\imat_3 - \htup{p} \otms  \htup{p} ) \Big) \cdot \tup{f}
     &=\, \tfrac{1}{\slang u^2}\Big( \htup{P} \otms \htup{p} \,+\, \csn{\tau}\, \htup{\slang}\otms\htup{\slang} \Big) \cdot \tup{f}
\\[8pt] 
     \rng{U}  &\!\!\! =\,
     -\tfrac{1}{\slang^2 u^2}\Big( u^2 f_\ss{u} \snn{\tau} 
     \,+\, \big[  U + \tfrac{\kconst_1}{\slang^2}  - (u + \tfrac{\kconst_1}{\slang^2})\csn{\tau} \big] \htup{p} \cdot \tup{f} \Big)
\\[8pt]
    \rng{W}  &\!\!\! =\, 
    \tfrac{1}{\slang u^2}\Big( 
     u^2 f_\ss{u} \csn{\tau}   \,+\, (u+\tfrac{\kconst_1}{\slang^2})\snn{\tau}\, \htup{p}\cdot\tup{f} \Big)
\end{array} 
\end{align}
\end{small}
where \eq{\htup{P}=\tfrac{1}{P}\tup{P}=\tfrac{1}{\slang}\tup{P}}.
It may be more useful to express the right-hand-side of the above using the cartesian components of the perturbing forces, \eq{\tup{F}=-\pderiv{V^1}{\tup{r}}+\tup{a}^\nc}. To do so, we may substitute  \eq{\tup{f}=\tup{F}\cdot\pderiv{\tup{r}}{\tup{q}}} and \eq{f_\ss{u}=\tup{F}\cdot\pderiv{\tup{r}}{u}}  from Eq.\eqref{Ftotal_qp} and use:
\begin{small}
\begin{align}
\htup{q}\cdot\tup{f}=0
\qquad,\qquad 
\htup{p}\cdot\tup{f} = \tfrac{1}{u p}\htup{q}\cdot \hdge{\tup{\slang}}\cdot\tup{F} = \tfrac{1}{u}\htup{p}\cdot \tup{F} 
\qquad,\qquad 
  \htup{\slang}\cdot\tup{f}  = \tfrac{1}{u}\htup{\slang}\cdot \tup{F}
\qquad,\qquad 
    f_\ss{u} = -\tfrac{1}{u^2} \htup{q} \cdot \tup{F}
\end{align}
\end{small}
where some of the above have used \eq{q=1} and \eq{\tup{q}\cdot\tup{p}=0}. 
Equation Eq.\eqref{dX0_2} then leads
to:\footnote{The equations for \eq{(U,W)} are equivalently written as:
\begin{align} 
\begin{array}{lllllll}
     \rng{U} \,=\,
     \tfrac{1}{\slang^2 u^3} \Big( (u\tup{q} + \tfrac{w}{\slang} \htup{p}) \snn{\tau}  \,-\,  \tfrac{2\kconst_1}{\slang^2}(1-\csn{\tau}) \htup{p} \Big) \cdot \tup{F} 
      & =\,  
     \tfrac{1}{\slang^2 u^3} \Big( u \snn{\tau}\,\tup{q} \,-\,  \big[\tfrac{2\kconst_1}{\slang^2}(1-\csn{\tau}) - \tfrac{w}{\slang}\snn{\tau}  \big] \htup{p} \Big) \cdot \tup{F} 
\\[4pt]
     \rng{W} \,=\,
      -\tfrac{1}{\slang u^3} \Big(  (u\tup{q} + \tfrac{w}{\slang} \htup{p})  \csn{\tau} \,-\, (\tfrac{W }{\slang}+ \tfrac{2\kconst_1}{\slang^2}\snn{\tau}) \htup{p} \big) \cdot \tup{F}
      &=\, 
     -\tfrac{1}{\slang u^3} \Big( u \csn{\tau}\,\tup{q} \,-\,  \big[  \tfrac{W}{\slang} - \tfrac{w}{\slang}\csn{\tau}  + \tfrac{2\kconst_1}{\slang^2}\snn{\tau} \big] \htup{p} \Big)\cdot \tup{F}
\end{array}
\end{align}
}
\begin{small}\begin{align} \label{vop_X2}
\boxed{ \begin{array}{rlllll}
      \rng{\tup{Q}} 
     &\!\!\! =\, -\tfrac{\snn{\tau}}{\slang^2 u^3}(\htup{\slang} \otms \htup{\slang})  \cdot \tup{F}
     &  =\, -\tfrac{\snn{\tau}}{\slang^2 u^3} F_{\slang} \htup{\slang} 
\\[8pt]
      \rng{\tup{P}}  
      &\!\!\! =\,  \tfrac{1}{\slang u^3} \big(  \htup{P} \otms \htup{p} \,+\, \csn{\tau}\,\htup{\slang} \otms \htup{\slang} \big)  \cdot \tup{F}
      &=\,
      \tfrac{1}{\slang u^3} \big( F_{\tau} \htup{P} \,+\, F_{\slang} \csn{\tau}\,  \htup{\slang} \big) 
\\[8pt] 
       \rng{U}   &\!\!\! =\, 
         \tfrac{1}{\slang^2 u^3} \Big( u \snn{\tau}\,\tup{q} \,-\,  \big[ U + \tfrac{\kconst_1}{\slang^2}  - (u + \tfrac{\kconst_1}{\slang^2})\csn{\tau}  \big] \htup{p} \Big) \cdot \tup{F} 
        &=\, 
         \tfrac{1}{\slang^2 u^3} \Big(  F_r u \snn{\tau} \,-\,  F_{\tau} \big[ U + \tfrac{\kconst_1}{\slang^2}  - (u + \tfrac{\kconst_1}{\slang^2})\csn{\tau}  \big]  \Big) 
\\[8pt]
    \rng{W}  
    &\!\!\! =\,
     -\tfrac{1}{\slang u^3} \Big( u \csn{\tau}\,\tup{q} \,-\,   (u+\tfrac{\kconst_1}{\slang^2})\snn{\tau}\, \htup{p} \Big)\cdot \tup{F}
     &=\,
      -\tfrac{1}{\slang u^3} \Big( F_r u \csn{\tau} \,-\,  F_{\tau}  (u+\tfrac{\kconst_1}{\slang^2})\snn{\tau}  \Big)
\end{array} }
\end{align}\end{small}
where the second equalities follow from the fact that \eq{\{\tup{q},\htup{p},\htup{\slang}\}=\{\htup{t}_r,\htup{t}_\tau,\htup{t}_{\slang}\}} 
is the LVLH basis (rather, its inertial cartesian components)
and where we have denoted by \eq{(F_r,F_\tau,F_{\slang})} the LVLH components of the total perturbing force.\footnote{The force components \eq{(F_r,F_\tau,F_{\slang})} in Eq.\eqref{vop_X2} are the components of the total perturbing force in the LVLH basis \eq{\{\tup{q},\htup{p},\htup{\slang}\}=\{\htup{t}_r,\htup{t}_\tau,\htup{t}_{\slang}\}}. They are clearly \textit{not} the inertial cartesian components which we denote by \eq{\tup{F}:= -\pderiv{V^1}{\tup{r}}+\tup{a}^\nc =(F_{r_1},F_{r_2},F_{r_3})}.}

\begin{notesq}
    Note from Eq.\eqref{dX0_2} or Eq.\eqref{vop_X2} that if the perturbing force is also some central-force of any kind (i.e., if  \eq{\tup{F}=F_r\htup{r}=F_r\tup{q}}), then  \eq{\rng{\tup{Q}}=\rng{\tup{P}}=0} and these elements remain constant. Similarly, if \eq{\tup{F}=F_{\slang} \htup{\slang}} is  only in the orbit normal direction, then \eq{\rng{U} =\rng{W} =0}. 
\end{notesq}

\noindent Although we have expressed the right-hand-side of the element's governing equations using a mix of the ``fast'' coordinates \eq{\tup{x}=(\tup{q},u,\tup{p},w)} and the ``slow'' elements \eq{\tup{\xi}=(\tup{Q},U,\tup{P},W)}, these equations can be expressed entirely in terms of the elements themselves using the relations in Eq.\eqref{Xsol_kep2} and Eq.\eqref{qp0_rels2}.
For instance:
\begin{small}
\begin{gather} \label{qpl_eye}
\begin{array}{rclllllll}
      \htup{\slang} \;=&\!\!
      \htup{q}\tms \htup{p} &=&  \htup{Q}\tms \htup{P} 
\\[3pt]
     \htup{\slang}\otms\htup{\slang} \;=&\!\!
     \imat_3 - \htup{q}\otms\htup{q} - \htup{p}\otms\htup{p}
     &=&
    \imat_3 - \htup{Q}\otms\htup{Q} - \htup{P}\otms\htup{P}
\\[3pt]
     \slang^2 \;=&\!\!
     q^2 p^2 - (\tup{q}\cdot\tup{p})^2 = p^2
     &=&  Q^2 P^2 - (\tup{Q}\cdot\tup{P})^2  = P^2
\end{array}
\\[6pt] \nonumber 
 u\tup{q} + \tfrac{w}{\slang} \htup{p} \,=\, U \tup{Q} + \tfrac{W }{\slang} \htup{P} + \tfrac{\kconst_1}{\slang^2}(\tup{q}-\tup{Q})
     \;=\; (U-\tfrac{\kconst_1}{\slang^2})\tup{Q} + \tfrac{W }{\slang} \htup{P} + \tfrac{\kconst_1}{\slang^2}\tup{q}
     \quad =\; 
     \tfrac{\kconst_1}{\slang^2}(\tup{e} + \tup{q})
\end{gather}
\end{small}
where, in the last line,  \eq{\tup{e}} is the eccentricity (see Eq.\eqref{LRL_e_qp0}).

\begin{footnotesize}
\begin{itemize}[topsep=2ex]
    \item[]  \textit{Derivation of Eq.\eqref{dX0_2}.}
To obtain the governing equations of motion for the elements \eq{\tup{\xi}}, we use the fact that the Kepler solution form — given by Eq.\eqref{Xsol_kep2}  — but with \eq{\tup{\xi}} no longer constant, must  satisfy the \textit{perturbed} equations of motion for \eq{\tup{x}} given by Eq.\eqref{qpw_eom_f_2bp}.
Differentiating Eq.\eqref{Xsol_kep2} with respect to \eq{\tau}  yields:
\begin{align} \nonumber
    \begin{array}{rl}
         \rng{\tup{q}} & =    \big[ -\tup{Q}\snn{\tau}   +   \tfrac{1}{\slang}\tup{P}\csn{\tau} \big]  +  \rng{\tup{Q}}\csn{\tau}  +  \tfrac{1}{\slang}(\rng{\tup{P}} - \tup{P} \tfrac{\rng{\slang}}{\slang}) \snn{\tau} 
\quad =    \tfrac{1}{\slang}\tup{p}  + \rng{\tup{Q}}\csn{\tau}  +  \tfrac{1}{\slang}(\rng{\tup{P}} - \tup{P}\tfrac{\rng{\slang}}{\slang}) \snn{\tau} 
\\[8pt]
  \rng{\tup{p}}  & =   -\big[\slang\tup{Q}\csn{\tau}
             +   \tup{P}\snn{\tau}\big]
             -  (\slang\rng{\tup{Q}}+\rng{\slang}\tup{Q})\snn{\tau}
             +   \rng{\tup{P}}\csn{\tau}  
\quad\;\;\; =  -\slang\tup{q}  -  (\slang\rng{\tup{Q}}+\rng{\slang}\tup{Q})\snn{\tau}
             +   \rng{\tup{P}}\csn{\tau}  
\\[8pt]
      \rng{u} & =  \big[-(U -\tfrac{\kconst_1}{\slang^2})\snn{\tau}
    + \tfrac{1}{\slang}W \csn{\tau} \big] + (\rng{U}  +\tfrac{2\kconst_1}{\slang^3}\rng{\slang})\csn{\tau} + \tfrac{1}{\slang}(\rng{W} -\tfrac{\rng{\slang}}{\slang}W )\snn{\tau} - \tfrac{2\kconst_1}{\slang^3}\rng{\slang}
 \\[3pt]
    & =   \tfrac{1}{\slang}w  +  (\rng{U}  + \tfrac{2\kconst_1}{\slang^3}\rng{\slang})\csn{\tau}  +  \tfrac{1}{\slang}(\rng{W} -\tfrac{\rng{\slang}}{\slang}W )\snn{\tau}  -  \tfrac{2\kconst_1}{\slang^3}\rng{\slang}
\\[8pt]
  \rng{w} & =  -\big[\slang(U -\tfrac{\kconst_1}{\slang^2})\csn{\tau} + W \snn{\tau}\big]  -   \slang(\rng{U}  + \tfrac{2\kconst_1}{\slang^3}\rng{\slang})\snn{\tau}  -  \rng{\slang}(U -\tfrac{\kconst_1}{\slang^2})\snn{\tau}     +  \rng{W} \csn{\tau}
 \\[3pt] 
 & =  -\slang( u - \tfrac{\kconst_1}{\slang^2})  -   \slang(\rng{U}  + \tfrac{2\kconst_1}{\slang^3}\rng{\slang})\snn{\tau}  -  \rng{\slang}(U -\tfrac{\kconst_1}{\slang^2})\snn{\tau}     +  \rng{W} \csn{\tau}
\end{array} 
\end{align}
where the second equality for each of the above follows from the requirement that the algebraic relations in Eq.\eqref{Xsol_kep2} must still hold. 
Equating the above to the perturbed equations of motion given in Eq.\eqref{qpw_eom_f_2bp}, we find 
\begin{align} \label{rels2_vop}
&\begin{array}{lll}
      \rng{\tup{q}}=\tfrac{1}{\slang}\tup{p} 
\\[4pt]
    \rng{\tup{p}}+\slang\tup{q} = \tfrac{1}{\slang u^2} \tup{f}
\\[4pt]
       \rng{u} = \tfrac{1}{\slang}w 
\\[4pt]
    \rng{w} + \slang u -\tfrac{\kconst_1}{\slang} = \tfrac{1}{\slang} f_\ss{u} 
\end{array}
&& \Rightarrow &&
\begin{array}{lll}
      \rng{\tup{Q}}\csn{\tau}  +  \tfrac{1}{\slang}(\rng{\tup{P}} - \tup{P}\tfrac{\rng{\slang}}{\slang}) \snn{\tau}  = \tup{0}
\\[4pt]
    -(\slang \rng{\tup{Q}} + \rng{\slang}\tup{Q})\snn{\tau}
             +   \rng{\tup{P}}\csn{\tau}  \,=\, \tfrac{1}{\slang u^2} \tup{f}
\\[4pt]
      (\rng{U}  + \tfrac{2\kconst_1}{\slang^3}\rng{\slang})\csn{\tau}  +  \tfrac{1}{\slang}(\rng{W} -\tfrac{\rng{\slang}}{\slang}W )\snn{\tau}  -  \tfrac{2\kconst_1}{\slang^3}\rng{\slang}  =  0
\\[4pt]
    -\slang(\rng{U}  + \tfrac{2\kconst_1}{\slang^3}\rng{\slang})\snn{\tau}  -  \rng{\slang}(U -\tfrac{\kconst_1}{\slang^2})\snn{\tau}     +  \rng{W} \csn{\tau}  =  \tfrac{1}{\slang} f_\ss{u}
\end{array}
\end{align}
The first two relations may be solved for \eq{\rng{\tup{Q}}} and \eq{\rng{\tup{P}}}, and the last two relations for \eq{\rng{U} } and \eq{\rng{W} }, leading to:
\begin{align} \label{dX0_2.1}
\begin{array}{llll}
      \rng{\tup{Q}} \,=\, 
       -\tfrac{1}{\slang}(\tfrac{1}{\slang u^2}\tup{f} - \tfrac{\rng{\slang}}{\slang}\tup{p})\snn{\tau}
\\[4pt]
     \rng{\tup{P}} \,=\,  (\tfrac{1}{\slang u^2}\tup{f}-\tfrac{\rng{\slang}}{\slang}\tup{p})\csn{\tau} \,+\,  \tfrac{\rng{\slang}}{\slang}\tup{P}
 \end{array}
\quad,\quad 
\begin{array}{llll}
     \rng{U}  \,=\, -\tfrac{1}{\slang}(\tfrac{1}{\slang}f_\ss{u}  \,-\, \tfrac{\rng{\slang}}{\slang}w)\snn{\tau}
     -   \tfrac{2\kconst_1}{\slang^3}\rng{\slang}(1-\csn{\tau})
\\[4pt]
    \rng{W}  \,=\, (\tfrac{1}{\slang}f_\ss{u}-\tfrac{\rng{\slang}}{\slang}w)\csn{\tau} \,+\, \tfrac{\rng{\slang}}{\slang}W   \,+\, \tfrac{2\kconst_1}{\slang^2}\rng{\slang} \snn{\tau} 
\end{array}
&&,&&
    \rng{\slang} = \rng{p} 
    =  \tfrac{1}{\slang u^2}\htup{p}\cdot\tup{f} = \tfrac{1}{\slang u^3}\htup{p}\cdot\tup{F}
\end{align}
After substitution of \eq{\rng{\slang}} as given above (from Eq.\eqref{ldot_cen}), and rearranging some terms in the ODEs for \eq{(U,W)}, we obtain:
\begin{align} \label{dX0_99}
 \begin{array}{rllll}
      \rng{\tup{Q}} &\!\!\! =\, 
       -\tfrac{\snn{\tau}}{\slang^2u^2}\big( \imat_3 - \htup{p}\otms\htup{p} \big) \cdot \tup{f}
\\[4pt]
     \rng{\tup{P}} &\!\!\! =\, 
     \tfrac{1}{\slang u^2}\Big( \csn{\tau} (\imat_3 - \htup{p} \otms  \htup{p} ) + \htup{P} \otms \htup{p} \Big) \cdot \tup{f}
 \end{array}
\quad,\quad 
\begin{array}{rllll}
     \rng{U}  &\!\!\! =\,
     -\tfrac{1}{\slang^2 u^2}\Big( (u^2\snn{\tau}) f_\ss{u} 
     \,+\, \big[ \tfrac{2\kconst_1}{\slang^2}(1-\csn{\tau}) - \tfrac{w}{\slang} \snn{\tau} \big] \htup{p} \cdot \tup{f} \Big)
\\[4pt]
    \rng{W}  &\!\!\! =\, 
    \tfrac{1}{\slang u^2}\Big( 
     (u^2 \csn{\tau}) f_\ss{u}  \,+\, \big[ \tfrac{W }{\slang} - \tfrac{w}{\slang} \csn{\tau} + \tfrac{2\kconst_1}{\slang^2} \snn{\tau} \big] \htup{p}\cdot\tup{f} \Big)
\end{array} 
\end{align}
The above is equivalent to the ODEs that were claimed in Eq.\eqref{dX0_2}. This equivalence is seen once one makes use of the following relations: 
\begin{align}
     (\imat_3 - \htup{p}\otms\htup{p}) \cdot \tup{f} \,=\, (\htup{\slang}\otms\htup{\slang})\cdot \tup{f}
     \qquad,\qquad 
 \begin{array}{lllll}
       \tfrac{2\kconst_1}{\slang^2} (1-\csn{\tau}) - \tfrac{1}{\slang}w \snn{\tau} 
     \;=\;  U + \tfrac{\kconst_1}{\slang^2}  - (u + \tfrac{\kconst_1}{\slang^2})\csn{\tau}
\\[4pt]
       \tfrac{1}{\slang}W - \tfrac{1}{\slang}w \csn{\tau} +  \tfrac{2\kconst_1}{\slang^2}\snn{\tau} 
    \;=\; (u+\tfrac{\kconst_1}{\slang^2})\snn{\tau} 
 \end{array}
\end{align}
The first of the above is explained in the footnote.\footnote{We have indicated that \eq{(\imat_3 - \htup{p}\otms\htup{p})\cdot\tup{f}= (\htup{\slang}\otms\htup{\slang})\cdot\tup{f} }. This is verified as follows. It was shown previously that \eq{\htup{q}\cdot \tup{f}=0} always holds (see Eq.\eqref{Ftotal_qp}). Also, recall that \eq{\{\htup{q}=\tup{q},\htup{p},\htup{\slang}\}=\{\htup{t}_r,\htup{t}_\tau,\htup{t}_{\slang}\}} are the inertial cartesian components of the LVLH basis such that \eq{\imat_3 = \htup{q}\otms\htup{q} + \htup{p}\otms\htup{p} + \htup{\slang}\otms\htup{\slang} }. Therefore:\\ 
   \eq{\qquad\qquad\qquad\qquad\qquad\qquad (\htup{\slang}\otms\htup{\slang})\cdot \tup{f} \,=\, (\imat_3 - \htup{q}\otms\htup{q} - \htup{p}\otms\htup{p}) \cdot \tup{f}
    \,=\,  (\imat_3 - \htup{p}\otms\htup{p}) \cdot \tup{f} } .
}
\end{itemize}
\end{footnotesize}

\subsection{Alternative Elements}
Our element set \eq{\tup{\xi} =(\tup{Q},U,\tup{P},W)} was defined from the coordinates \eq{\tup{x} =(\tup{q},u,\tup{p},w)}
using the unperturbed \eq{\tau}-parameterized Kepler flow. 
However, one could, if they wish, make any number of adjustments to these elements (so long as they still fully define the state \eq{(\tup{r},\tup{\v})}). Here, we will briefly present an alternative choice for the ``\eq{W} element'' and an alternative choice for the ``\eq{U} element''. But, first, we address what would seem to be an intuitive alternative element set, but which turns out to be flawed.

\paragraph{An \textit{Incomplete} Element Set.}
Given that the preceding developments used  \eq{\tau}, not \eq{s}, as the evolution parameter, it may seem odd that we used the state coordinates \eq{(\tup{q},u,\tup{p},w)=(\bartup{q},\diff{\bartup{q}}{s})} to define associated elements \eq{(\tup{Q},U,\tup{P},W)} and their \eq{\tau}-parameterized governing equations. It would seem more fitting to instead use coordinates \eq{(\bartup{q},\diff{\bartup{q}}{\tau})=(\tup{q},u,\htup{p},\til{w})} — where \eq{\htup{p}:=\tup{p}/p=\tup{p}/\slang=\rng{\tup{q}}} and \eq{\til{w}:=w/\slang = \tfrac{u^2}{\slang}p_u =\rng{u}} — to define associated elements \eq{(\tup{Q},U,\htup{P},\til{W})}, where \eq{\htup{P}:=\tup{P}/P=\tup{P}/\slang} and \eq{\til{W}:=W/\slang}. 
 After going through the variation of parameters procedure to find the governing equations for \eq{(\tup{Q},U,\htup{P},\til{W})}, 
 the result
 is:\footnote{Deriving the governing equations of \eq{(\tup{Q},U,\htup{P},\til{W})} mirrors the  preceding developments for deriving those of \eq{(\tup{Q},U,\tup{P},W)}. 
 We would simply replace \eq{w} and \eq{W} with \eq{w=\slang \til{w}} and \eq{W=\slang \til{W}} and replace \eq{\tup{p}} and \eq{\tup{P}} with \eq{\tup{p}=\slang\htup{p}} and \eq{\tup{P}=\slang\htup{P}}, in Eq.\eqref{Xsol_kep2}. The equations of motion for the ``fast'' coordinates would simply be  \eq{\rng{\htup{p}}=\rrng{\tup{q}}} and \eq{\rng{\til{w}}=\rrng{u}} given by Eq.\eqref{ddqu_fgen} or Eq.\eqref{ddqu_Fcart}.}
\begin{small}
\begin{align} \label{dX_vop11}
\begin{array}{lllll}
     \rng{\htup{P}}
     \,=\, \tfrac{\csn{\tau}}{\slang^2 u^3}(\htup{\slang} \otms \htup{\slang})  \cdot \tup{F}
     &\quad,
\end{array}
\qquad 
\begin{array}{lllll}
    \rng{\til{W}} 
    \,=\,  -\tfrac{1}{\slang^2 u^3} \Big( (u\tup{q} + \til{w} \htup{p})\csn{\tau} \,-\, \tfrac{2\kconst_1}{\slang^2} \snn{\tau}\,\htup{p}   \Big)\cdot \tup{F}
\end{array}
\end{align}
\end{small}
with \eq{(\rng{\tup{Q}},\rng{U})} the same as in Eq.\eqref{vop_X2}. 
The elements \eq{(\tup{Q},U,\htup{P},\til{W})} and their governing equations were also derived by Schumacher in \cite{schumacher1987results}.
Although the above equations for \eq{\rng{\htup{P}}} and \eq{\rng{\til{W}}} are slightly simpler than those for \eq{\rng{\tup{P}}} and \eq{\rng{W}} seen in equation  Eq.\eqref{vop_X2}, there is problem: \textit{the set \eq{(\tup{Q},U,\htup{P},\til{W})} is not a complete element set}. 
The coordinates \eq{(\tup{q},u,\htup{p},\til{w})} do \textit{not} fully define the state \eq{(\tup{r},\tup{\v})} and, thus, neither does the set \eq{(\tup{Q},U,\htup{P},\til{W})}. The issue lies in replacing \eq{\tup{p}} with \eq{\htup{p}} and, likewise, replacing \eq{\tup{P}} with \eq{\htup{P}}. For instance, there is no way to express the angular momentum in terms of \eq{(\tup{q},\htup{p})} or \eq{(\tup{Q},\htup{P})}. If we wished to use \eq{(\tup{q},u,\htup{p},\til{w})} or \eq{(\tup{Q},U,\htup{P},\til{W})}, then we would need to include the angular momentum \eq{\slang} (or some equivalent) as a ninth element with \eq{\rng{\slang}} governed by Eq.\eqref{dX0_2.1}. While there is nothing inherently wrong with adding \eq{\slang} as a ninth element, it is unnecessary given the availability of the complete element set \eq{(\tup{Q},U,\tup{P},W)}.

\paragraph{Alternative $(U,W)$ Elements.}
 One \textit{could} exchange the pair \eq{(U,W)} for some alternative \eq{(\til{U},\til{W})} in order to slightly simply the transformation and the dynamics.
 First, we note that the ODE for \eq{(U,W)} in Eq.\eqref{vop_X2} are equivalently written as:
\begin{small}
\begin{align} \label{UW_ode_again}
\begin{array}{lllllll}
     \rng{U} \,=\,
     \tfrac{1}{\slang^2 u^3} \Big( (u\tup{q} + \tfrac{w}{\slang} \htup{p}) \snn{\tau}  \,-\,  \tfrac{2\kconst_1}{\slang^2}(1-\csn{\tau}) \htup{p} \Big) \cdot \tup{F} 
    &=\,
     \tfrac{1}{\slang^2 u^3} \Big(  F_r u \snn{\tau} \,-\,  F_{\tau} \big[ U + \tfrac{\kconst_1}{\slang^2}  - (u + \tfrac{\kconst_1}{\slang^2})\csn{\tau}  \big]  \Big) 
\\[4pt]
     \rng{W} \,=\,
      -\tfrac{1}{\slang u^3} \Big(  (u\tup{q} + \tfrac{w}{\slang} \htup{p})  \csn{\tau} \,-\, (\tfrac{W }{\slang}+ \tfrac{2\kconst_1}{\slang^2}\snn{\tau}) \htup{p} \big) \cdot \tup{F}
     &=\,
      -\tfrac{1}{\slang u^3} \Big( F_r u \csn{\tau} \,-\,  F_{\tau}  (u+\tfrac{\kconst_1}{\slang^2})\snn{\tau}  \Big)
\end{array}
\end{align}
\end{small}
\begin{small}
\begin{itemize}
    \item \textit{An Alternative ``$W$ Element''.}
    While replacing \eq{\tup{P}} with \eq{\htup{P}} is problematic, there is no problem in replacing \eq{W} with  \eq{\til{W}:=W/\slang}. That is, one could use \eq{(\tup{Q},U,\tup{P},\til{W})} as a valid set of orbit elements. The ``\eq{(u,w)}-part'' of the transformation in Eq.\eqref{Xsol_kep2} would then be replaced by
    \begin{small}
    \begin{align} \label{Walt}
    \begin{array}{lll}
         \til{w} := w/\slang  \\[4pt]
         \til{W} := W/\slang
    \end{array}
    &&,&&
    \begin{array}{lllllll}
          u  \,=\,  (U-\tfrac{\kconst_1}{\slang^2})\csn{\tau} \,+\,  \til{W} \snn{\tau} \,+\, \tfrac{\kconst_1}{\slang^2}
    \\[4pt]
           \til{w}  \,=\,  -(U-\tfrac{\kconst_1}{\slang^2})\snn{\tau} \,+\,\til{W} \csn{\tau}
    \end{array}
     \quad \leftrightarrow \quad 
    \begin{array}{lll}
         U :=\, (u-\tfrac{\kconst_1}{\slang^2})\csn{\tau} - \til{w}\snn{\tau} + \tfrac{\kconst_1}{\slang^2}
    \\[4pt]
          \til{W} :=\, (u-\tfrac{\kconst_1}{\slang^2})\snn{\tau} + \til{w}\csn{\tau} 
    \end{array} 
    \end{align}
    \end{small}
    If using the element set \eq{(\tup{Q},U,\tup{P},\til{W})}, the governing equation for \eq{W} in Eq.\eqref{UW_ode_again} would be replaced by the following for \eq{\til{W}}:
    \begin{small}
    \begin{align} \label{dX_vop11_alt}
    \begin{array}{lllll}
        \rng{\til{W}} 
        \,=\,  -\tfrac{1}{\slang^2 u^3} \Big( (u\tup{q} + \til{w} \htup{p})\csn{\tau} \,-\, \tfrac{2\kconst_1}{\slang^2} \snn{\tau} \,\htup{p}   \Big)\cdot \tup{F}
    \end{array} 
    \end{align}
    \end{small}
    \item \textit{An Alternative ``$U$ Element''.}
    Looking at the transformation in Eq.\eqref{Xsol_kep2}, 
    one might be inspired to replace the element \eq{U} (corresponding to \eq{u_\zr=1/r_\zr}) with the alternative \eq{\til{U}:= U - \kconst_1/\slang^2} such that the ``\eq{(u,w)}-part'' of the transformation in Eq.\eqref{Xsol_kep2} would then be replaced by 
    \begin{small}
    \begin{align} \label{Ualt}
     \til{U} := U - \kconst_1/\slang^2
     &&,&&
    \begin{array}{lllllll}
          u   \,=\,  \til{U}\csn{\tau} \,+\, \tfrac{1}{\slang} W \snn{\tau} \,+\, \tfrac{\kconst_1}{\slang^2}
    \\[4pt]
           w   \,=\,  -\slang \til{U} \snn{\tau} \,+\, W\csn{\tau}
    \end{array}
     \quad \leftrightarrow \quad 
    \begin{array}{lll}
         \til{U} \,=\, (u-\tfrac{\kconst_1}{\slang^2})\csn{\tau} -  \tfrac{1}{\slang}w\snn{\tau} 
    \\[4pt]
          W \,=\, \slang(u-\tfrac{\kconst_1}{\slang^2})\snn{\tau} + w\csn{\tau} 
    \end{array} 
    \end{align}
    \end{small}
    If using the element set \eq{(\tup{Q},
    \til{U},\tup{P},W)}, the governing equation for \eq{U} in Eq.\eqref{UW_ode_again} would be replaced by the following 
   for \eq{\til{U}}:\footnote{The governing equation for \eq{\til{U}= U - \kconst_1/\slang^2} is found by straightforward differentiation and substitution of \eq{\rng{U}} and \eq{\rng{\slang}}. That is,  \eq{\ring{\til{U}} = \rng{U} + 2\tfrac{\kconst_1}{\slang^3}\rng{\slang} = \rng{U} + 2\tfrac{\kconst_1}{\slang^4 u^2}\htup{p}\cdot\tup{f} }. }
    \begin{small}
    \begin{align}
    \begin{array}{lllll}
          \rng{\til{U}} 
          \,=\, 
          \tfrac{1}{\slang^2 u^3} \Big( (u\tup{q} + \tfrac{w}{\slang} \htup{p}) \snn{\tau}  \,+\,  \tfrac{2\kconst_1}{\slang^2}\csn{\tau}\,\htup{p} \Big) \cdot \tup{F} 
    \end{array}
    \end{align}
    \end{small}
\end{itemize}
\end{small}
One could combine the above and simultaneously replace the pair \eq{(U,W)} with \eq{(\til{U},\til{W})} as defined above. The ``\eq{(u,w)}-part'' of the transformation in Eq.\eqref{Xsol_kep2} would then be replaced by
\begin{small}
\begin{align}
    \begin{array}{lllllll}
      u  \,=\,  \til{U}\csn{\tau} \,+\, \til{W}  \snn{\tau} \,+\, \tfrac{\kconst_1}{\slang^2}
\\[4pt]
       \til{w}   \,=\,  - \til{U} \snn{\tau} \,+\,\til{W} \csn{\tau}
\end{array}
 \quad \leftrightarrow \quad 
\begin{array}{lll}
     \til{U} \,=\, (u-\tfrac{\kconst_1}{\slang^2})\csn{\tau} - \til{w}\snn{\tau} 
\\[4pt]
      \til{W} \,=\, (u-\tfrac{\kconst_1}{\slang^2})\snn{\tau} + \til{w}\csn{\tau} 
\end{array} 
\end{align}
\end{small}
with the governing equations for \eq{(\til{U},\til{W})} given by
\begin{small}
\begin{align}  \label{UWalt_eom}
\begin{array}{lllll}
     \rng{\til{U}} 
     \,=\,  \tfrac{1}{\slang^2 u^3} \Big( (u\tup{q} + \til{w}\htup{p}) \snn{\tau}  \,+\,  \tfrac{2\kconst_1}{\slang^2}\csn{\tau}\,\htup{p} \Big) \cdot \tup{F}
       &=\; 
     \tfrac{1}{\slang^2 u^3} \Big( F_r u \snn{\tau} \,-\,  F_\tau  \big(  \til{U}  - (u + \tfrac{\kconst_1}{\slang^2})\csn{\tau}  \big) \Big)
\\[6pt]
    \rng{\til{W}}
    \,=\,  -\tfrac{1}{\slang^2 u^3} \Big( (u\tup{q} + \til{w} \htup{p})\csn{\tau} \,-\, \tfrac{2\kconst_1}{\slang^2} \snn{\tau} \,\htup{p}   \Big)\cdot \tup{F}
      &=\; 
     -\tfrac{1}{\slang^2 u^3} \Big( F_r u \csn{\tau} \,+\,  F_\tau \big(  \til{W}  - (u+\tfrac{\kconst_1}{\slang^2})\snn{\tau}  \big) \Big)
\end{array} 
\end{align}
\end{small}
where, for any of the above dynamics in Eq.\eqref{Walt}-Eq.\eqref{UWalt_eom}, the right-hand-side may be expressed in various ways using:
\begin{small}
\begin{gather}
   u\tup{q} + \tfrac{w}{\slang} \htup{p} \,=\, (U-\tfrac{\kconst_1}{\slang^2})\tup{Q} + \tfrac{W }{\slang} \htup{P} + \tfrac{\kconst_1}{\slang^2}\tup{q}
   \quad=\quad
   u\tup{q} + \til{w} \htup{p} \,=\,  \til{U}\tup{Q} + \til{W}\htup{P} + \tfrac{\kconst_1}{\slang^2}\tup{q}
    \;=\; 
     \tfrac{\kconst_1}{\slang^2}(\tup{e} + \tup{q})
\\ \nonumber
\begin{array}{llll}
      U + \tfrac{\kconst_1}{\slang^2}  - (u + \tfrac{\kconst_1}{\slang^2})\csn{\tau} 
     \,=\,
    \tfrac{2\kconst_1}{\slang^2} (1-\csn{\tau}) - \tfrac{1}{\slang}w \snn{\tau} 
\\[3pt]
     (U - \tfrac{\kconst_1}{\slang^2})  - (u + \tfrac{\kconst_1}{\slang^2})\csn{\tau} 
    \,=\, 
     -\tfrac{1}{\slang}w \snn{\tau} - \tfrac{2\kconst_1}{\slang^2} \csn{\tau} 
 \\[3pt]
  (U - \tfrac{\kconst_1}{\slang^2})  - (u - \tfrac{\kconst_1}{\slang^2})\csn{\tau} 
    \,=\,   -\tfrac{1}{\slang}w \snn{\tau} 
\end{array}
\qquad,\qquad 
\begin{array}{llll}
   (u+\tfrac{\kconst_1}{\slang^2})\snn{\tau} = \tfrac{W}{\slang} - \tfrac{w}{\slang} \csn{\tau} +  \tfrac{2\kconst_1}{\slang^2}\snn{\tau} 
\\[3pt]
    (u-\tfrac{\kconst_1}{\slang^2})\snn{\tau} \,=\, \tfrac{W}{\slang} - \tfrac{w}{\slang} \csn{\tau}
\\[3pt]
     \tfrac{1}{\slang}w \csn{\tau} - \tfrac{2\kconst_1}{\slang^2} \snn{\tau} 
     \,=\, \tfrac{1}{\slang}W -    (u+\tfrac{\kconst_1}{\slang^2})\snn{\tau}
\end{array}
\end{gather}
\end{small}
.

\section{REPARAMETERIZING HAMILTONIAN DYNAMICS: CONFORMAL SCALING vs. EXTENDED PHASE SPACE} \label{apx:dqp_ds_ext}



What follows is an extension of section \ref{sec:ext_cen} and is specific to the reparameterized Hamiltonian dynamics expressed in our projective coordinates. While not critical for the main purposes of this work, the following is included for completeness and to clarify any confusion of how transforming the evolution parameter affects the "Hamiltonian-ness'' of a Hamiltonian system. 

 For the ``canonical projective coordinates'', \eq{(\bartup{q},\bartup{p})=(\tup{q},u,\tup{p},p_\ss{u})}, we mentioned that the \eq{t}-parameterized ODEs in Eq.\eqref{qpdot_0} are a Hamiltonian system (with nonconservative perturbations). However, the \eq{s}- or \eq{\tau}-parameterized 
ODEs in Eq.\eqref{dqp_s_0} or Eq.\eqref{dqp_ta_0} are \textit{not} true Hamiltonian systems (with nonconservative perturbations). Recall these were given by:
\begin{small}
\begin{flalign} \label{dqp_s_apx}
\quad
\pdt{t} = \tfrac{1}{u^2} 
\;\;,\quad
\left.\begin{array}{lll}
      \pdt{\bartup{q}} = \pdt{t}\pderiv{\mscr{H}}{\bartup{p}}
\\[5pt]
     \pdt{\bartup{p}} = \pdt{t}(-\pderiv{\mscr{H}}{\bartup{q}} + \bartup{f} )
\end{array} \right\}
\quad \Rightarrow  &&
\begin{array}{lllll}
       \pdt{\tup{q}} 
       \,=\,   -\hdge{\tup{\slang}}\cdot\tup{q} 
\\[5pt]  
    \pdt{\tup{p}} 
     \,=\,  -\hdge{\tup{\slang}}\cdot\tup{p}  + \pdt{t}\tup{f} 
\end{array}
\quad,\qquad 
\begin{array}{lllll}
      \pdt{u} 
       \,=\,   u^2 p_\ss{u}
\\[5pt]
      \pdt{p}_\ss{u} 
     \,=\, -\tfrac{1}{u}\big( \slang^2 + 2 u^2 p_\ss{u}^2 \big)   - \pdt{t}\pderiv{V^{0}}{u} + \pdt{t} f_\ss{u}
\end{array}
&&
\\[8pt]  \label{dqp_ta_apx}
\quad
\rng{t} = \tfrac{1}{\slang u^2} 
\;\;,\quad 
\left.\begin{array}{lll}
      \rng{\bartup{q}} = \rng{t}\pderiv{\mscr{H}}{\bartup{p}}
\\[5pt]
     \rng{\bartup{p}} = \rng{t}(-\pderiv{\mscr{H}}{\bartup{q}} + \bartup{f} )
\end{array}\right\}
\quad \Rightarrow &&
\begin{array}{lllll}
      \rng{\tup{q}}    \,=\, 
      -\hdge{\htup{\slang}}\cdot\tup{q} 
\\[5pt]  
    \rng{\tup{p}} 
      \,=\,  -\hdge{\htup{\slang}}\cdot\tup{p}   + \rng{t}\tup{f} 
\end{array}
\quad,\qquad 
\begin{array}{lllll}
      \rng{u} \,=\,  \tfrac{1}{\slang} u^2 p_\ss{u}
\\[5pt]
     \rng{p}_\ss{u}  
      \,=\, -\tfrac{1}{\slang u}\big( \slang^2 + 2 u^2 p_\ss{u}^2 \big)  - \rng{t}\pderiv{V^{0}}{u}  + \rng{t} f_\ss{u} 
\end{array}
&&
\end{flalign}
\end{small}
The above are \textit{conformally} Hamiltonian systems — with possibly-nonconservative perturbations \eq{(\tup{f},f_\ss{u})}  as described in Eq.\eqref{Ftotal_qp} — for the conformal scaling factors \eq{\pdt{t}=1/u^2} and \eq{\rng{t}=1/(\slang u^2)}, respectively. Even if we remove the nonconservative perturbations, the above will not have all the usual properties of Hamiltonian systems. E.g., volume preservation (Liouville's theorem) is generally not satisfied and the state transition matrix will generally not be a symplectic matrix.   However, there is a way to recover all the usual Hamiltonian properties.

\paragraph{Recovering Hamiltonian Structure Using Extended Phase Space.}
The above reparameterized dynamics may indeed be realized as a Hamiltonian system  (with nonconservative perturbations) by making use of the \textit{extended phase space}, where \eq{t} is added as configuration coordinate with conjugate momenta coordinate denoted \eq{p_t} (see \cite{peterson2025prjCoord,peterson2025phdThesis,struckmeier2005hamiltonian,lanczos2012variational}). 
The extended Hamiltonian \eq{\wtscr{H}(\bartup{q},\bartup{p},t,p_t)} for \eq{s} as the evolution parameter is then defined as \eq{\wtscr{H} := \pdt{t}(\mscr{H} + p_t )}, where \eq{\pdt{t}=\diff{t}{s}=1/u^2}. 
The \eq{s}-parameterized Hamiltonian dynamics are given in projective coordinates using Hamilton's canonical equations with \eq{\wtscr{H}}:
\begin{small}
\begin{gather}  \nonumber 
   \wtscr{H} :=\, \tfrac{1}{u^2}(\mscr{H} + p_t ) 
    \,=\, \tfrac{1}{2}\big(\slang^2 + u^2 p_\ss{u}^2 \big)  \,+\, \wt{V}^\zr(u)  \,+\,  
 u^\ss{-2} p_t 
 \phantom{\qquad -\tfrac{1}{u}\big( \slang^2 + 2 u^2 p_\ss{u}^2 \big)   - \pdt{t}\pderiv{V^{0}}{u} + \pdt{t} f_\ss{u}}
\\ \label{dqp_s_SUM}
\begin{array}{lllll}
       \pdt{\tup{q}} = \pderiv{\wtscr{H}}{\tup{p}}   \,=\, 
      -\hdge{\tup{\slang}}\cdot\tup{q} 
\\[5pt]
       \pdt{u}  = \pderiv{\wtscr{H}}{p_u}  \,=\,   u^2 p_\ss{u}
\\[5pt]  
     \pdt{t}   = \pderiv{\wtscr{H}}{p_t} \,=\,  1/u^2
\end{array}
\quad,\qquad 
\begin{array}{rlllll}
     \pdt{\tup{p}}  = -\pderiv{\wtscr{H}}{\tup{q}} + \pdt{t}\tup{f} 
      &\!\!\!=\,  -\hdge{\tup{\slang}}\cdot\tup{p}  + \pdt{t}\tup{f} 
\\[5pt]
      \pdt{p}_\ss{u}  = -\pderiv{\wtscr{H}}{u} + \pdt{t} f_\ss{u}
     &\!\!\!=\, -u p_\ss{u}^2    -  \pderiv{\wt{V}^0}{u} +  \tfrac{2}{u^3} p_t  + \pdt{t} f_\ss{u}
       &= - \tfrac{1}{u}(\slang^2 +  2 u^2 p_\ss{u}^2 )   -  \pdt{t} \pderiv{V^{0}}{u} + \tfrac{2}{u} \wtscr{H}  +  \pdt{t} f_\ss{u}
\\[5pt]
      \pdt{p}_t  = -\pderiv{\wtscr{H}}{t} + \pdt{t} f_t  
      &\!\!\!=\, -\pdt{t}(\pd_t V^1 + \bartup{\alpha}\cdot \dot{\bartup{q}})  
\end{array}
\end{gather}
\end{small}
where \eq{\wt{V}^0:= \pdt{t}V^0=  u^\ss{-2} V^0} still depends only on \eq{u}, where \eq{V^1} is a potential accounting for conservative perturbations, and where \eq{\bartup{\alpha}=(\tup{\alpha},\alpha_\ss{u})} account for nonconservative perturbations (all perturbations are combined together in \eq{\bartup{f}=(\tup{f},f_\ss{u})}). 
Note that only the above \eq{p_\ss{u}} equation differs from the \eq{s}-parameterized conformally-Hamiltonian  dynamics in Eq.\eqref{dqp_s_apx} (because \eq{\pdt{t}=u^\ss{-2}} depends only on \eq{u}).


 We may also consider the \eq{\tau}-parameterized Hamiltonian dynamics on extended phase space. These are obtained in a similar manner as Eq.\eqref{dqp_s_SUM}, but using an extended Hamiltonian \eq{\whscr{H}:=\rng{t}(\mscr{H}+p_t)=\tfrac{1}{\slang}\wtscr{H}}, where \eq{\rng{t}=\diff{t}{\tau}=\tfrac{1}{\slang u^2}}. This leads to the following Hamiltonian dynamics:
\begin{small}
\begin{gather} \label{dqp_TA_SUM}
    \whscr{H}\,=\,  \tfrac{1}{\slang}\wtscr{H} 
\qquad,\qquad 
\begin{array}{lllll}
      \rng{\tup{q}} \,=\, \pderiv{\whscr{H}}{\tup{p}}   \,=\, 
      -(1 - \tfrac{\whscr{H}}{\slang})  \hdge{\htup{\slang}}\cdot\tup{q} 
\\[5pt]
      \rng{u} \,=\, \pderiv{\whscr{H}}{p_u}  \,=\,   \tfrac{1}{\slang} u^2 p_\ss{u}
\\[5pt]  
    \rng{t}  \,=\, \pderiv{\whscr{H}}{p_t} \,=\,  \tfrac{1}{\slang u^2}
\end{array}
\;\;,\qquad 
\begin{array}{rlllll}
    \rng{\tup{p}} \,=\, -\pderiv{\whscr{H}}{\tup{q}} +\rng{t}\tup{f} 
      &\!\!\!=\,  -(1- \tfrac{\whscr{H}}{\slang} ) \hdge{\htup{\slang}}\cdot\tup{p}  
      +\rng{t}\tup{f} 
\\[5pt]
     \rng{p}_\ss{u} \,=\, -\pderiv{\whscr{H}}{u} +\rng{t} f_\ss{u}
      &\!\!\!=\, - \tfrac{1}{\slang u}(\slang^2 +  2 u^2 p_\ss{u}^2 )   -  \rng{t} \pderiv{V^{0}}{u} + \tfrac{2}{u} \whscr{H}  +  \rng{t} f_\ss{u}
\\[5pt]
     \rng{p}_t \,=\, -\pderiv{\whscr{H}}{t} + \rng{t} f_t  
      &\!\!\!=\, -\rng{t}(\pd_t V^1 + \bartup{\alpha}\cdot \dot{\bartup{q}})  
\end{array}
\end{gather}
\end{small}
which differ notably from the \eq{\tau}-parameterized conformally-Hamiltonian dynamics given in Eq.\eqref{dqp_ta_apx}.
Yet, somehow, they must be equivalent. Indeed they are, as described below.

It can be shown (cf.~\cite{struckmeier2005hamiltonian,lanczos2012variational}) that, along a solution curve in extended phase space, \eq{p_t} always has a \textit{value} of \eq{p_t=-\mscr{H}} such that the extended Hamiltonians have values \eq{\wtscr{H}=0=\whscr{H}} (when evaluated along a solution curve). 
This relation, when substituted back into the above extended phase space dynamics, lead to ODEs for \eq{(\bartup{q},\bartup{p})} that are equivalent to the conformal scaling by \eq{\pdt{t}} or \eq{\rng{t}} given in Eq.\eqref{dqp_s_apx} or Eq.\eqref{dqp_ta_apx}.
For instance, for the \eq{\tau}-parameterized dynamics in Eq.\eqref{dqp_TA_SUM}, we note: 
\begin{small}
\begin{flalign} \label{qpdot_ext_alt_apx}
\quad
\begin{array}{llll}
      \rng{\bartup{q}}= \pderiv{\whscr{H}}{\bartup{p}} 
      &\!\! =\; \rng{t}\pderiv{\mscr{H}}{\bartup{p}} + (\mscr{H}+p_t) \pderiv{}{\bartup{p}} \rng{t}
\\[5pt]
     \rng{\bartup{p}} = -\pderiv{\whscr{H}}{\bartup{q}} + \rng{t}\bartup{f}
      &\!\! =\;  \rng{t}(-\pderiv{\mscr{H}}{\bartup{q}} + \bartup{f}) - (\mscr{H}+p_t) \pderiv{}{\bartup{q}} \rng{t} 
\end{array}
&&,&&
p_t=-\mscr{H} \quad\Rightarrow \qquad 
\begin{array}{llll}
      \rng{\bartup{q}} \,=\, \rng{t}\pderiv{\mscr{H}}{\bartup{p}} 
      &=\; \rng{t}\dot{\tup{q}} 
\\[5pt]
     \rng{\bartup{p}} \,=\,  \rng{t}(-\pderiv{\mscr{H}}{\bartup{q}} + \bartup{f} )
     &=\; \rng{t}\dot{\bartup{p}} 
\end{array}
\qquad
\end{flalign}
\end{small}
and similarly for the \eq{s}-parameterized dynamics. 
That is, using \eq{p_t=-\mscr{H}} is equivalent to \eq{\wtscr{H}=0=\whscr{H}} in the ODEs of Eq.\eqref{dqp_s_SUM} and  Eq.\eqref{dqp_TA_SUM}. The result is equivalent to the conformally-Hamiltonian dynamics given in Eq.\eqref{dqp_s_apx} or Eq.\eqref{dqp_ta_apx}.

\begin{notesq}
    To clarify, the \eq{10} ODEs in Eq.\eqref{dqp_s_SUM} and  Eq.\eqref{dqp_TA_SUM} are true Hamiltonian systems (with nonconservative perturbations), whereas the \eq{8} ODEs in Eq.\eqref{dqp_s_apx} or Eq.\eqref{dqp_ta_apx} are \textit{not}; they are conformally Hamiltonian systems (with nonconservative perturbations). These ODEs and their solution curves agree using \eq{p_t=-\mscr{H}}. 
\end{notesq}




\end{appendices}




\end{document}